\documentclass[12pt,a4paper,twoside,nofootinbib,british]{book}
\usepackage[UKenglish]{babel}
\usepackage[T1]{fontenc}
\usepackage[math]{anttor}
\usepackage[utf8]{inputenc}
\usepackage{babel}
\usepackage{latexsym,amsmath,amssymb}
\usepackage{graphicx,overpic}
\usepackage{bm}
\usepackage{textcomp}
\usepackage{graphicx}
\usepackage{enumerate}
\usepackage{float} 
\usepackage{bm}
\usepackage{verbatim}
\usepackage{enumerate}
\usepackage{color}
\usepackage{dcolumn}
\usepackage{geometry}
\usepackage{textcomp}
\usepackage{mathrsfs}
\usepackage{cite}
\usepackage{hyperref}
\usepackage[stable]{footmisc}
\usepackage[T1]{fontenc}
\usepackage[footnotesize,bf,it]{caption}

\def\b{\begin{equation}}
\def\e{\end{equation}}

\newcommand{\ket}[1]{\left| {#1} \right\rangle}
\newcommand{\bra}[1]{\left\langle {#1} \right|}
\newcommand{\ematriz}[3]{\left\langle {#1} \left|{#2}\right|{#3}\right\rangle}	
\newcommand{\braket}[2]{\left\langle {#1}\left|{#2}\right.\right\rangle}
\newcommand{\proj}[2]{\left| {#1} \right\rangle\!\left\langle {#2} \right|}

\newcommand{\biket}[2]{\left| {#1} \right\rangle_{\text{I}}\left| {#2} \right\rangle_{\text{II}}}
\newcommand{\bikete}[2]{\left| {#1} \right\rangle_{\text{B}}\left| {#2} \right\rangle_{\bar {\text{B}}}}
\newcommand{\biketn}[2]{\left| {#1}_\omega \right\rangle_{\text{I}}^+\left| {#2}_\omega \right\rangle_{\text{II}}^-}
\newcommand{\pa}{p}
\newcommand{\tr}{\operatorname{Tr}}
\newcommand{\omegar}{{\omega_\text{R}}}
\newcommand{\diff}{\text{d}}

\newcommand{\dimctwo}{l}
\newcommand{\BbbR}{\mathbb{R}}

\newcommand{\qr}{q_\text{R}}
\newcommand{\ro}{r_{\omega}}
\newcommand{\eq}[1]{(\ref{#1})}

\newcommand{\ql}{q_\text{L}}

\newcommand{\eo}{\eta_{\text{out}}}
\newcommand{\uin}[1]{{u_{{#1}}^{\text{in}}}}
\newcommand{\uout}[1]{{u_{{#1}}^{\text{out}}}}
\newcommand{\uh}[1]{{u_{{#1}}^{{\text{hor}}}}}
\newcommand{\ain}[1]{{a_{{#1}}^{\text{in}}}}
\newcommand{\aout}[1]{{a_{{#1}}^{\text{out}}}}
\newcommand{\ah}[1]{{a_{{#1}}^{{\text{hor}}}}}

\def\b{\begin{equation}}
\def\e{\end{equation}}

\def\openone{\leavevmode\hbox{\small$1$\normalsize\kern-.37em$1$}}

\allowdisplaybreaks

\makeatletter
\def\cleardoublepage{\clearpage\if@twoside \ifodd\c@page\else
    \hbox{}
    \thispagestyle{empty}
    \newpage
    \if@twocolumn\hbox{}\newpage\fi\fi\fi}
\makeatother \clearpage{\pagestyle{empty}\cleardoublepage}

\numberwithin{equation}{section}

\newcommand{\qed}{\nobreak \ifvmode \relax \else
      \ifdim\lastskip<1.5em \hskip-\lastskip
      \hskip1.5em plus0em minus0.5em \fi \nobreak
      \vrule height0.75em width0.5em depth0.25em\fi}
      
      \usepackage{fancyhdr}
\begin{document}
\thispagestyle{empty}

\include{0a-portada-bis}

\thispagestyle{empty}
\begin{flushright}
\ \\ \par
\vspace{7cm}
{\it A Mar\'ia Mart\'inez L\'opez},\\
{\it y a Modesto Mart\'in Clara}
\vskip 0.2cm
que no solo me han dado la vida,\\
sino que han sacrificado parte de las suyas\\
para hacer esto posible.
\end{flushright}

\chapter*{Agradecimientos}
\thispagestyle{empty}

\qquad\!\! No hay palabras para agradecer suficientemente a Juan Le\'on,  con el que he compartido los a\~nos que ha llevado producir esta tesis. Juan ha sido un mentor y un amigo, lleno de paciencia y comprensi\'on infinitas. Igualmente quiero agradecer muy cari\~nosamente a Luis Garay, un buen amigo y uno de los mejores profesores que he conocido, cuya dedicaci\'on y genio sirven de inspiraci\'on a sus estudiantes.

\medskip

A Miguel Montero, colega y amigo y sin duda uno de los estudiantes m\'as brillantes y prometedores que he conocido, le quiero agradecer su inestimable ayuda en la redacci\'on de esta memoria de Tesis Doctoral.

\medskip

Agradezco tambi\'en a mis compa\~neros del CSIC por todo su apoyo, tanto a los que me encontr\'e al inicio de este bonito viaje, Carlos, Borja, Jaime y Juanjo, como  a los que tuve el inmenso placer de conocer mas tarde, Emilio, Marco y Jordi (a qui a m\'es agraeixo que m'ajud\`es amb el meu precari catal\`a). Igualmente a tantos otros compa\~neros de profesi\'on con los que he compartido conferencias, ca\~nas y charlas, formales e informales, especialmente a Jar\'in, a Juanma, a Javier Cerrillo y a Doug Plato.

\medskip

Estoy en deuda con Ivette Fuentes, con Gerardo Adesso y con Jorma Louko, tres buenos amigos a los que conoc\'i a mi llegada a Nottingham y que tanto han aportado a mi formaci\'on como cient\'ifico y como persona. Tambi\'en quiero agradecer a Andrzej Dragan, con el que es un autentico placer hablar de F\'isica, por esas conversaciones y discusiones en las que siempre se aprende algo. Mis agradecimientos tambi\'en van al profesor Robert Mann por todo aquello que he aprendido trabajando con \'el.

\medskip

Con mucho cari\~no agradezco a Antony Lee toda su ayuda y los buenos tiempos que con \'el he pasado en Nottingham, as\'i como a Karen Stung, Merlijn van Horssen, Nicolai Friis, David Bruschi y a todos los estudiantes de la `Physics and Mathematics School' de la Universidad de Nottingham. 

\medskip

A mis padres les estar\'e eternamente agradecido por todos los sacrificios que han hecho por mi, su valor y su lucha diaria durante tanto tiempo para que yo tuviera una educaci\'on. Sin su apoyo nada de esto hubiera sido posible. 

\medskip

Todo lo que pueda decir es poco para mis grand\'isimos y querid\'isimos amigos y compa\~neros Javier Burgos, Christian, Diego, Pablo y Javier  Cu\'ellar (``Plocas''), que han sido y son como hermanos para m\'i, y que como tal han compartido conmigo todas las penas y alegr\'ias de mi vida durante la elaboraci\'on de esta tesis. Tambi\' en para mi vieja amiga Tamara, que sin dudarlo me ha prestado su sabidur\'ia y su comprensi\'on cada vez que la he necesitado.
\medskip

Tengo palabras de agradecimiento para toda la gente de la Universidad que durante este tiempo me han acompa\~nado en aventuras y desventuras, en especial para toda la gente de Hypatia, y para mis viejas amigas Tere, Luci, Elena y Cristina. Menci\'on especial merecen mis amigas Andrea, Ana Dolcet, Alba y Adela  que tan pacientemente me han escuchado siempre hablar sobre mi trabajo, as\'i como mis compa\~neros Jacobo y Bea. Tambi\'en Clara, por el apoyo, cari\~no y comprensi\'on que me dedic\'o durante la elaboraci\'on de esta tesis, y por todos los dem\'as agradecimientos que se han quedado en el tintero.

\medskip

Y por \'ultimo, mi m\'as profundo agradecimiento a Paloma y a la fortuna que ha entrelazado nuestros caminos. Te doy las gracias por el impagable apoyo que me has dado durante el tramo final de esta tesis, por tu valiosa ayuda, por el tiempo que hemos compartido y por reconciliarme con mi querido Madrid. Muchas gracias por ense\~narme a tener fe en el futuro, por el cine y por todas y cada una de las maravillosas sonrisas que me has regalado durante este tiempo y las que est\'an por venir. 

\medskip

 Durante el desarrollo de esta tesis he recibido financiaci\'on del programa CSIC JAE-PREDOC2007, del proyecto FIS2008-05705/FIS  del Ministerio de Ciencia de Espa\~na y del consorcio QUITEMAD.

\thispagestyle{empty}

\cleardoublepage

\chapter*{Acknowledgements}
\thispagestyle{empty}

I do not have the words to thank my supervisor Juan Le\'on with whom I have shared the long 3 years it took to produce this thesis. Juan has been a mentor and a friend, always full of patience and comprehension. Likewise, I would like to kindly thank Luis Garay, a good friend and one of the best lecturers I have ever encountered, whose dedication and genius provide inspiration for all his students.

\medskip

A special thanks to Miguel Montero, colleague and friend and undoubtedly one of the most brilliant and promising students I have ever met, for his invaluable help with writing this dissertation.

\medskip

I also thank my colleagues at CSIC for all their support, both those that I met at the beginning of this journey (Carlos, Borja, Jaime and Juanjo) and those that I had the great pleasure to meet later on (Emilio, Marco and Jordi, who helped me with my poor catalan). Likewise, I thank all the colleages with whom I shared conferences, drinks and  talks (formal and informal), especially Jar\'in, Juanma, Javier Cerrillo and Doug Plato.

\medskip

I am indebted to Ivette Fuentes, Gerardo Adesso and Jorma Louko, good friends whom I met when I arrived in Nottingham and that have so much contributed to my formation as a scientist and as a person.  Also to Andrzej Dragan, with whom sharing a conversation about Physics is a real pleasure, for those discussions where I always learnt something new. I would also like to thank Professor Robert Mann, for all that I learnt working with him.

\medskip

I wish to fondly thank Antony Lee, a good friend always ready to lend a hand when I needed it, for his help and hospitality and for all the good times we had in Nottingham, as well as Merlijn van Horssen, Karen Stung, Nicolai Friis, David Bruschi and all the students at the Physics and Mathematics School of the University of Nottingham. 

\medskip

I will always be thankful to my parents, for all the sacrifices they have made for me, their courage and their determination to give me an education. Without them this thesis would have never been possible.
 
 \medskip
 
 There are no words to thank my beloved and great friends and colleagues Javier Burgos, Christian, Diego, Pablo and Javier Cu\'ellar (``Plocas''), who are  brothers to me in all but blood, and who, as such, have shared all my joys and sorrows during my PhD. Also to my old friend Tamara, always supportive, who lent me her wisdom anytime I needed it.
 
\medskip
 
I am grateful to all the people at the University that have been with me all over these years. Especially to all my fellows of Hypatia and my old friends  Tere, Luci, Elena and Cristina. My friends Andrea, Ana Dolcet, Alba and Adela deserve  a special mention for always being there to patiently listen to my speaking about my work, along with my friends and classmates Jacobo and Bea. Also to Clara, for all the support, affection and comprehension that she devoted to me during the best part of my doctorate, and for all the thanks that were left unsaid.

\medskip

And last but definitely not least, my deepest thanks to Paloma and the fortune that has entangled our paths. Thank you very much for the priceless support that I received from you at the end of my PhD, for all the moments we have shared and for a timely reconciliation with my beloved Madrid. Many thanks for showing me how to have faith in the future, for the cinema and for every single beautiful smile that you have given me during this time and those yet to come.

\medskip

During the development of this thesis I was supported by a CSIC JAE - PREDOC2007 Grant, the Spanish MICINN Project FIS2008-05705/FIS  and the QUITEMAD consortium.

\thispagestyle{empty}

\setcounter{page}{1}
\thispagestyle{empty}
\tableofcontents

\newpage


{\renewcommand{\thechapter}{}\renewcommand{\chaptername}{}
\addtocounter{chapter}{0}
\section*{Published articles derived from this thesis}\markboth{\sl PUBLICATIONS}{\sl PUBLICATIONS}}
\addcontentsline{toc}{chapter}{Publications} 

The research included in this thesis has given rise to the following articles:
\vspace{0.5cm}
\begin{itemize}
 \item  \emph{Redistribution of particle and anti-particle entanglement in non-inertial frames}.
    E. Mart\'in-Mart\'inez and I. Fuentes. Phys. Rev. A 83, 052306 (2011)
    \vspace{0.3cm} 
\item   \emph{Using Berry's phase to detect the Unruh effect at lower accelerations}.
    E. Mart\'in-Mart\'inez and I. Fuentes, R. B. Mann. arXiv:1012.2208. Submitted to Phys. Rev. Letters
    \vspace{0.3cm}
\item   \emph{The entangling side of the Unruh-Hawking effect}
    M. Montero and E. Mart\'in-Mart\'inez. arXiv:1011.6540. Submitted to JHEP
\vspace{0.3cm}
\item   \emph{The Unruh effect in quantum information beyond the single-mode approximation}
    D. E. Bruschi, J. Louko, E. Mart\'in-Mart\'inez, A. Dragan and I. Fuentes. Phys. Rev. A, 82, 042332 (2010)
\vspace{0.3cm}
\item   \emph{Quantum entanglement produced in the formation of a black hole}
    E. Mart\'in-Mart\'inez, L. J. Garay and J. Le\'on. Phys. Rev. D, 82, 064028 (2010)
\vspace{0.3cm}
\item   \emph{Unveiling quantum entanglement degradation near a Schwarzschild black hole}
    E. Mart\'in-Mart\'inez, L. J. Garay and J. Le\'on. Phys. Rev. D, 82, 064006 (2010)
\vspace{0.3cm}
\item   \emph{Entanglement of Dirac fields in an expanding spacetime}
    I. Fuentes, R.B. Mann, E. Mart\'in-Mart\'inez and S. Moradi. Phys. Rev. D, 82, 045030 (2010)
\vspace{0.3cm}
\item   \emph{Population bound effects on bosonic correlations in non-inertial frames}
    E. Mart\'in-Mart\'inez and J. Le\'on, Phys. Rev. A, 81, 052305 (2010)
\vspace{0.3cm}
\item   \emph{Quantum correlations through event horizons: Fermionic vs bosonic entanglement}
    E. Mart\'in-Mart\'inez and J. Le\'on, Phys. Rev. A, 81, 032320 (2010)
\vspace{0.3cm}
\item   \emph{Fermionic entanglement that survives a black hole}
    E. Mart\'in-Mart\'inez and J. Le\'on, Phys. Rev. A, 80, 042318 (2009)
\vspace{0.3cm}
\item   \emph{Spin and occupation number entanglement of Dirac fields for non-inertial observers}
    J. Le\'on and E. Mart\'in-Mart\'inez, Phys. Rev. A, 80, 012314 (2009)
\end{itemize}

Also, during the development of this thesis another article  has been published by its author

\begin{itemize}
\item   \emph{Physical qubits from charged particles: IR divergences in quantum information}.
    J. Le\'on and E. Mart\'in-Mart\'inez, Phys. Rev. A, 79, 052309 (2009)
\end{itemize}

\cleardoublepage

{\renewcommand{\thechapter}{}\renewcommand{\chaptername}{}
\addtocounter{chapter}{0}
\chapter*{Introduction}\markboth{\sl Introduction}{\sl Introduction}}
\addcontentsline{toc}{chapter}{Introduction}

\section*{Introduction and objectives}
\addcontentsline{toc}{section}{Introduction and objectives} 

General relativity is the theory that describes gravity which is currently accepted in the frame of modern physics. It fits all the phenomenology previously observed and successfully predicted a plethora of experimental results. Not only has it been experimentally tested a number of times but its application has been instrumental in leading to day-to-day modern technology as, for instance, the Global Positioning System (GPS). The theory basically consists of  a geometric description of gravity: mass and energy move in a curved spacetime and the spacetime is curved by the presence of mass and energy. However, it is far from being complete. General relativity allows ill-defined objects such as singularities, and in the presence of a singularity it loses its predictive power. These problems are strongly related with the classicality of the theory: general relativity is classical and close to a singularity the energies and distances involved reach the Planck scale. A quantum description of gravity is still to appear, being one of the most important (if not the most) challenges of modern theoretical physics. In the absence of a full quantum theory for gravity, quantum field theory in curved spacetimes, which describe the interaction of quantum fields with this classical (but relativistic) gravity, is the most complete theory so far.

Quantum information theory, on the other hand, deals with problems in information theory when the information is stored in and managed with quantum systems. Quantum mechanics allows us to carry out tasks that were considered impossible in the classical world: we can use quantum simulators to find solutions to quantum dynamical problems that would take too long for classical computers; we can store a large amount of information in quantum memories taking advantage of the superposition principle; we can implement completely secure communication using quantum key distribution protocols (quantum cryptography) and much more. Arguably, the most important of these achievements is being able to construct and implement quantum algorithms that transform quantum mechanical systems into quantum computers that can, for instance, factorise prime numbers in a time that grows polynomially with their lengths \cite{Shor} or find elements in a non-indexed list in a time that grows as the square root of the number of elements \cite{Grover}. Again, this is one of the challenges of modern physics: to tame the laws of quantum mechanics and use this new quantum physics knowledge to build new technology and solve problems which are practically unsolvable otherwise.

Despite their apparently separated application areas, general relativity and quantum information are not disjoint research fields. On the contrary, following the pioneering work of Alsing and Milburn \cite{Alsingtelep} a wealth of works considered different situations in which entanglement was studied in a general relativistic setting, for instance, quantum information tasks influenced by black holes  \cite{TeraUedaSpd,PanBlackHoles,ManSchullBlack,Adeschul}, entanglement in an expanding universe \cite{caball,Steeg} and entanglement with non-inertial partners \cite{Alicefalls,AlsingSchul,TeraUeda2,KBr}.

Even though many of
the systems used in the implementation of quantum information
involve relativistic systems such as photons, the vast majority of
investigations on entanglement assume that the Universe is flat and
non-relativistic. Understanding entanglement in general spacetimes is
ultimately necessary because the world is fundamentally
relativistic. Moreover, entanglement  plays a prominent
role in black hole thermodynamics \cite{bombelli,Canent,Terashima,Emparan,levay,Cadoni,Hu2,NavarroSalas} and in the
information loss problem \cite{Maldacena,Preskill,Lloyd2,Ahn1,ManSchullBlack}.

Entanglement behaviour in non-inertial frames was first considered in \cite{Alsingtelep} where the fidelity of teleportation between relative accelerated partners was analysed. After this, occupation number entanglement degradation of scalar \cite{Alicefalls} and Dirac \cite{AlsingSchul} fields due to Unruh effect was shown. 

 In particular, the Unruh effect \cite{DaviesUnr,Unruh0,Takagi,Crispino} --which consists in the emergence of noise when an accelerated observer is describing Minkowski vacuum from his proper frame-- affects the possible entanglement that an accelerated observer Rob would share with an inertial observer Alice. 

To analyse quantum correlations in non-inertial settings it is necessary to combine knowledge from different branches of physics; quantum field theory in curved spacetimes and quantum information theory. This combination of disciplines became known as relativistic quantum information, which is developing at an accelerated pace. It also provides novel tools for the analysis of the Unruh and Hawking effects \cite{DaviesUnr,Unruh,Takagi,Crispino,Hawking} allowing us to study the behaviour of the correlations shared between non-inertial observers.

Recently, there has been increased interest in understanding entanglement and
quantum communication in black hole spacetimes
\cite{Xian,Pan2,Ahntropez} and in using quantum information techniques
to address questions in gravity \cite{Ternada,Ternada2}. Studies on
relativistic entanglement show the emergence of conceptually important
qualitative differences to a non-relativistic treatment. For
instance, entanglement was found to be an observer-dependent
property that is degraded from the perspective of accelerated
observers moving in flat spacetime \cite{Alicefalls,AlsingSchul,Adeschul,Villalba}.
These results show that entanglement in curved spacetime might
not be an invariant concept. Relativisitic quantum information theory uses well-known tools coming from quantum information and quantum optics to study quantum effects provoked by gravity to learn information about the spacetime. We can take advantage of our knowledge about quantum correlations and effects produced by the gravitational interaction to set the basis for experimental proposals ultimately aiming at finding corrections due to quantum gravity effects, too mild to be directly observed.

The differences found between bosonic \cite{Alicefalls} and fermionic \cite{AlsingSchul} entanglement leave an open question about the origin of this distinct behaviour. How can it be possible that bosonic entanglement quickly dies as the acceleration of a non-inertial observer increases while some amount of fermionic entanglement survives even in the limit of infinite acceleration?.

First answers given in the literature by the pioneers who discovered the phenomenon pointed at the difference in the dimension of each system Hilbert space as a possible responsible for these discrepancies, but the question remained open. In this thesis we will demonstrate the strong relationship between statistics and entanglement in non-inertial frames. We will prove that the huge differences between bosonic and fermionic non-inertial entanglement behaviour are related to the counting statistics of the field and have little to do with the Hilbert space dimension for each field mode. This result banishes previous ideas, that were extended in the literature, about the origin of those differences.

Entanglement behaviour in the presence of black holes had not been thoroughly analysed previously. The few studies about entanglement degradation focused on the asymptotically flat region of Schwarzschild spacetime. It would be much more interesting to have results about  entanglement behaviour in the proximities of the event horizon. In this thesis we will present the way to export the results obtained in the frame of uniformly accelerated observers to proper curved space times and black holes scenarios with event horizons. We will also develop a formalism to account for the behaviour of entanglement as a function of the observer's distance to the event horizon of a black hole,  going beyond the analysis in the asymptotically flat region of Schwarzschild spacetime made in previous literature. Here we will provide a rigorous study about what happens when entangled pairs are at small distances from the event horizon.

Almost all the previous work on field entanglement in non-inertial settings made use of what is known as `single mode approximation'. This approximation has allowed pioneering studies of correlations for non inertial observers, but it is based on misleading assumptions about the change of basis between inertial and uniformly accelerated observers and it is partially flawed. In this thesis, we will discuss how this approximation has been misinterpreted since its inception \cite{Alsingtelep,AlsingMcmhMil} and thereafter in all the subsequent works.  We will see the proper physical meaning of such an approximation and will learn to what extent it is valid and how to relax it. We will show that going beyond the single mode approximation will allows us to reach a better understanding of the phenomenon of fermionic entanglement survival in the limit of infinite acceleration \cite{AlsingSchul} and find that the Unruh effect can amplify entanglement and not only destroy it as it was thought before.

There are very few works on field entanglement in general relativistic scenarios for non-stationary spacetimes. Only for bosonic fields and expanding universes some work exists \cite{caball}. As a part of this thesis we will analyse the behaviour of entanglement in non-stationary scenarios. The objective is to prove that the gravitational interaction induces non-classical effects in quantum fields that can be useful in a dual sense: account for quantum effects of the gravitational interaction and provide a basis to obtain information about the nature of gravity in real and analog gravity systems. In simple words, we will analyse how the vacuum state of a field evolves --under the gravitational interaction-- to states that present quantum entanglement. Once again we will see that  huge differences between fermions and bosons appear in a very relevant way in this context. We will prove that fermions are more useful in order to experimentally account for this entanglement and suggest how one can take advantage of these differences to extract information about the underlying background geometry in analog gravity experiments or in cosmology.

Last but not least, using the knowledge gained from other disciplines (in particular tools coming from quantum optics and solid state physics) we will confront the problem of directly measuring the Unruh effect. Experimental detection of the Unruh effect  \cite{ChenTaj,Crispino} required accelerations of order $10^{25}g$ where $g$ is the surface gravity of the Earth. We prove that a detector moving in a flat spacetime acquires a global geometric phase, which is the same for any inertial detector but differs, due to the Unruh effect, for accelerated ones. Taking advantage of this phenomenon we will propose a general experimental setting to detect this effect where the accelerations needed are $10^9$ times smaller than previous proposals, sustained only for a few nanoseconds' time.

\section*{Structure of the thesis}
\addcontentsline{toc}{section}{Structure of the thesis dissertation} 
\begin{list}{\labelitemi}{\leftmargin=1em}
 \item The first section of this thesis (Preliminaries) intends to serve as a brief and notational introduction to the formalism of quantum information theory (chapter \ref{QIi}) and quantum field theory in curved spacetimes (chapter \ref{INTROU}). In these two chapters we present the basic concepts that serve as building blocks for the rest of the original content presented in this thesis. We will also present in this section the problem of the single mode approximation used in previous literature.\end{list}
 
\noindent The research presented here is structured in three blocks that conform the three parts of this thesis:

\begin{list}{\labelitemi}{\leftmargin=1em}
\item  Part \ref{part1}: The relationship between statistics and entanglement in non-inertial frames is studied, disproving the previous idea that the dimensionality of the Hilbert  space controls entanglement behaviour and  obtaining universal laws (only dependent on statistics) for non-inertial entanglement. This part consists of a brief discussion about previous results and the following 7 chapters:
\begin{list}{\labelitemii}{\leftmargin=1em}
\item In chapter \ref{onehalf} we investigate the Unruh effect on entanglement taking into account the spin degree of freedom of the Dirac field. Previous works only explored spinless fermionic fields\footnote{See Grassmann scalar fields in Appendix \ref{appB}}. We go beyond earlier results and we also  analyse spin Bell states, obtaining their entanglement dependence on the acceleration of one of the partners. Then, we consider simple analogs to the occupation number entangled state $\left|00\right\rangle+\left|11\right\rangle$ but with spin quantum numbers for $\left|11\right\rangle$. We show that entanglement degradation in terms of the acceleration happens to be the same for both cases and, furthermore, it coincides with that of the spinless fermionic field despite the different Hilbert space dimension in each case. This is a  first hint against the idea that dimension rules entanglement behaviour. We also introduce a procedure to consistently erase the spin information from our setting, being able to account for correlations present only in the occupation number degree of freedom.  
\item In Chapter \ref{multimode} we introduce an explicitly multimode formalism considering an arbitrary number of accessible modes when analysing bipartite entanglement degradation due to Unruh effect.  A single frequency mode of a fermion field only has a few accessible levels due to Pauli exclusion principle, conversely to bosonic fields which had an infinite number of excitable levels. This was argued to justify fermionic entanglement survival in the infinite acceleration limit. Here we consider entangled states that mix different frequency modes. Hence, the dimension of the Hilbert space in the accelerated observer basis can grow unboundedly, even for a fermion field. We will prove that, despite this analogy with the bosonic case, entanglement loss is limited. We will show that this comes from fermionic statistics through the characteristic structure it imposes on the system's density matrix regardless of its dimension. The surviving entanglement is shown to be independent of the specific maximally entangled state chosen, the kind of fermionic field analysed, and the number of accessible modes considered.
\item In Chapter \ref{etanthrough} we  disclose the behaviour of quantum and classical correlations among all the different spatial-temporal regions of a spacetime with apparent horizons, comparing fermionic with bosonic fields. We show the emergence of conservation laws for entanglement and classical correlations, pointing out the crucial role that statistics plays in the information exchange (and more specifically, the entanglement tradeoff) across the horizon. 
\item In Chapter \ref{boundedpop} we analyse the effect of bounding the occupation number of bosonic field modes on the correlations among inertial and non-inertial observers in a spacetime with apparent horizons. We show that the behaviour of finite-dimensional bosonic fields is qualitatively similar to standard bosonic fields and not to fermionic fields. This completely banishes the notion that dimension rules entanglement behaviour. We show that the main differences between bosonic fields and fermionic fields are still there even if we impose the same dimension for both: for bosonic fields no entanglement is created in the physical subsystems whatever the values of the dimension bound and the acceleration. Moreover, entanglement is very quickly lost as acceleration increases for both finite and infinite dimension. We study in detail the mutual information conservation law found before for bosons and fermions. We will show that for bosons this law stems from classical correlations while for fermions it has a quantum origin. Finally, we will also discuss the entanglement across the causally disconnected regions comparing the fermionic cases with their finite occupation number bosonic analogs. 
\item In Chapter \ref{blackhole1} we analyse the entanglement degradation provoked by the Hawking effect in a bipartite system Alice-Rob when Rob is in the proximities of a Schwarzschild black hole while Alice is free-falling into it. As a result, we will be able to determine the degree of entanglement as a function of the distance of Rob to the event horizon, the mass of the black hole, and the frequency of Rob's entangled modes. By means of this analysis we will show that all the interesting phenomena occur in the vicinity of the event horizon and that, in fact,  Rob has to be very close to the the black hole to see appreciable effects. The universality of the phenomenon is presented: there are not fundamental differences for different masses when working in the natural unit system adapted to each black hole. We also discuss some aspects of the localization of Alice and Rob states. 
\end{list}
\item Part \ref{part2}: We explore the so-called single mode approximation, finding its appropriate physical interpretation and correcting previous statements and uses of such an approximation in the literature. We will see how one can go beyond it, obtaining striking results: on the one hand, we will gain a  deeper understanding about the strange entanglement behaviour of fermionic fields in the infinite acceleration limit and on the other hand we will see how to implement techniques to amplify entanglement by means of the Unruh and Hawking effects. This part consists of the following 3 chapters:
\begin{list}{\labelitemii}{\leftmargin=1em}
\item In Chapter \ref{sma} we address the validity of the single-mode approximation that is commonly invoked in the analysis of entanglement in non-inertial frames and in other relativistic quantum information scenarios. We show that the single-mode approximation is not valid for arbitrary states, finding corrections to previous studies beyond such an approximation in the bosonic and fermionic cases. We also exhibit a class of wave packets for which the single-mode approximation is justified subject to the peaking constraints set by an appropriate Fourier transform. This will give us the proper physical frame of such an approximation.
\item In Chapter \ref{parantpar} we show that going beyond the single mode approximation allows us to analyse the entanglement tradeoff between particle and anti-particle modes of a Dirac field from the perspective of inertial and uniformly accelerated observers. Our results show that a redistribution of entanglement between particle and anti-particle modes plays a key role in the survival of fermionic field entanglement in the infinite acceleration limit.
\item In Chapter \ref{generatio}  going beyond the single mode approximation we show that the Unruh effect can create net quantum entanglement between inertial and accelerated observers, depending on the choice of the inertial state. For the first time, it is shown that the Unruh effect not only destroys entanglement, but may also create it. This opens a new and unexpected resource for finding experimental evidence of the Unruh and Hawking effects.
\end{list}
\item  Part \ref{part3}:   In this last part we study entanglement creation due to the gravitational interaction in two dynamical physically interesting scenarios: the formation of a black hole due to stellar collapse and the expansion of the Universe. We end this thesis presenting a proposal of dectection of the Unruh and Hawking effect by means of the geometric phase acquired by moving detectors. This part consists of the following 3 chapters:
\begin{list}{\labelitemii}{\leftmargin=1em}
\item Chapter \ref{stellarcollapse} shows that a field in the vacuum state, which is  in principle  separable, can evolve to an entangled state  induced by a  gravitational collapse. We will study, quantify, and discuss the origin of this entanglement, showing that it could even reach the maximal entanglement limit for low frequencies or very small black holes, with consequences in micro-black hole formation and the final stages of evaporating black holes. This entanglement provides quantum information resources between the modes that escape to the asymptotic future (thermal Hawking radiation) and those which fall into the event horizon. We will also show that  fermions are more sensitive than bosons to this quantum entanglement generation. This fact could be helpful in finding experimental evidence of the genuine quantum Hawking effect in analog models.
\item In chapter \ref{expandingU}  we study the entanglement generated between Dirac modes in a 2-dimensional conformally flat Robertson-Walker universe showing that inflation-like expansion generates quantum entanglement.  We find radical qualitative differences between the bosonic and fermionic entanglement generated by the expansion. The particular way in which fermionic fields become entangled encodes more information about the underlying spacetime than in the bosonic case, thereby allowing us to reconstruct the history of the expansion. This highlights, once again, the importance of bosonic/fermionic statistics to account for relativistic effects on the entanglement of quantum fields.
\item In chapter \ref{BerryPh} we show that a detector acquires a Berry phase due to its motion in spacetime. The phase is different for the inertial and accelerated detectors as a direct consequence of the Unruh effect. We exploit this fact to design a novel method to measure the Unruh effect.  Surprisingly, the effect is detectable for accelerations $10^9$ times smaller than previous proposals, sustained only for times of nanoseconds.
\end{list}
\end{list}

 The main results of this thesis are summarised in the conclusions section.

In appendix \ref{appB} we present the standard formalism of Klein-Gordon and Dirac equation in curved spacetimes.

\cleardoublepage

\addcontentsline{toc}{part}{Preliminaries}
\part*{Preliminaries}

\chapter{Quantum entanglement and mutual information}\label{QIi}

This chapter is intended as a very brief presentation (almost merely notational) to the concept of quantum entanglement. The motivation for this section is to introduce the concept and to present two state functionals that we will use to measure the `amount' of entanglement and correlations of a bipartite quantum system. Nevertheless, the matter of entanglement is not a simple topic. Its study is itself  a huge, and still open, discipline. For a more detailed view there are many other sources where a much more thorough study can be found, for instance \cite{Nichuang}. 

\section{Quantum entanglement and entanglement measures}\label{entangsec}

Quantum entanglement is a feature of some multipartite quantum systems which is strongly related with non-locality. Basically, to describe entangled $k$-partite systems in quantum mechanics it is not enough with the description of the $k$ individual quantum states for each subsystem, even if the subsystems are spatially separated. This means that carrying out measurements on one of the subsystems we can gather information about the result of future measurements on any of the rest of the subsystems without directly acting on them beyond the limits imposed by classical physics \cite{Bellprime}.

 Quantum entanglement was central in the debate about the non-locality and completeness of quantum mechanics \cite{EPR0} which ended up with the banishing of local hidden-variable theories \cite{EPRtest}. More important, quantum entanglement is the principal resource for quantum information tasks such as quantum teleportation \cite{telep1} and quantum computing \cite{Nichuang} and, as we will discuss in this thesis, can be used to obtain information about quantum effects provoked by gravity.

In the case of pure states, if we have two quantum systems A and B and the Hilbert spaces for the states of these systems are $\mathcal{H}_\text{A}$ and $\mathcal{H}_\text{B}$ respectively, the Hilbert space of the composite system is the tensor product $\mathcal{H}_\text{A}\otimes\mathcal{H}_\text{B}$.  A bipartite state $\ket{\Psi}_{\text{AB}}$ is entangled when it is not possible to express $\ket{\Psi}_{\text{AR}}$ as the tensor product of states for the individual subsystems
\begin{equation}
\ket{\Psi}_{\text{AB}}\neq\ket{\phi}_\text{A}\otimes\ket\phi_{\text{B}}\Leftrightarrow \ket{\Psi}_{\text{AB}}\, \text{ Entangled.}
\end{equation}
In other words, if $\{\ket{j}_\text{A}\}$ and $\{\ket{k}_\text{B}\}$ are respectively bases of  $\mathcal{H}_\text{A}$ and $\mathcal{H}_\text{B}$ the most general bipartite state in  $\mathcal{H}_\text{A}\otimes\mathcal{H}_\text{B}$ has the form
\begin{equation}
\ket{\Psi}_{\text{AB}}=\sum_{j,k}c_{jk}\ket{j}_\text{A}\otimes\ket{k}_\text{B}.
\end{equation}
The state is separable when
 \begin{equation}\label{consep}
 c_{jk}=c_j^\text{A}c_k^\text{B},  
\end{equation}
yielding
\b
\left.
\begin{array}{l}
\displaystyle{\ket{\phi}_\text{A}=\sum_jc_j^\text{A}\ket{j}_\text{A}}\\[5mm]
\displaystyle{\ket{\phi}_\text{B}=\sum_kc_k^\text{B}\ket{k}_\text{B}}
\end{array}\right\}
\Rightarrow \ket{\Psi}_{\text{AB}}=\ket{\phi}_\text{A}\otimes\ket\phi_{\text{B}}.\e
If condition \eqref{consep} does not hold the state is entangled.

For mixed states the general definition is slightly more complicated. A general state is entangled if, and only if, it cannot be expressed as a probability distribution of the uncorrelated individual states. In other words, given a set of positive numbers $\{p_i\}$ such that $\sum_{i}p_i=1$ then
\begin{equation}
\rho_{AB}\neq \sum_{i}p_i\,\rho^A_i\otimes\rho^B_i  \Leftrightarrow \rho_{AR}\, \text{ Entangled.}
\end{equation}

Although determining if a state is entangled or not is conceptually simple, computationally speaking is a very hard problem for general states of arbitrary dimension. Actually there is no such thing as a unique measure of entanglement. Instead, a measure of entanglement is any positive function of the state $E(\rho)$ which satisfy the following axioms
\begin{itemize}
\item Must be maximum for maximally entangled states (Bell states)
\item Must be zero for separable states.
\item Must be non-zero for all non-separable states.
\item Must not grow under LOCC (Local Operations + Classical Communication)
\end{itemize}

For pure states the entanglement entropy (entropy of the reduced states of $A$ or $B$) is a natural measure of entanglement which have also a well understood physical interpretation, but it does not fulfill the previous axioms for non-pure states.

To account for the entanglement of general states let us introduce the partial transpose density matrix. For a general density matrix of a bipartite system $AB$
\begin{equation}
\rho_{AB}=\sum_{ijkl}\rho_{ijkl}\ket{i}_A\ket{j}_B\bra{k}_A\bra{l}_B,
\end{equation}
the partial transpose is defined as
\begin{equation}\label{ptranspdef}
\rho_{AB}^{p{T_B}}=\sum_{ijkl}\rho_{ijkl}\ket{i}_A\ket{l}_B\bra{k}_A\bra{j}_B
\end{equation}
or, equivalently for our purposes, as 
\begin{equation}
\rho_{AB}^{p{T_A}}=\sum_{ijkl}\rho_{ijkl}\ket{k}_A\ket{j}_B\bra{i}_A\bra{l}_B.
\end{equation}

There is a theorem for the lower dimensional cases, for bipartite systems of dimension $2\times 2$ (two-qubit states) and $3\times2$ (qutrit-qubit states) the well-known Peres criterion \cite{PeresCriterion} guarantees that a state is non-separable (and therefore, entangled) if, and only if, the partial transposed density matrix has, at least, one negative eigenvalue.

Unfortunately, for higher dimension the condition is no longer necessary and sufficient, but only sufficient due to the existence of bound entanglement: there are states which are entangled, but no pure entangled states can be obtained from them by means of local operations and classical communication (LOCC). Such states are called bound entangled states \cite{Bound} and its entanglement is of no utility to quantum information tasks. Peres criterion only accounts for the existence of entanglement that can be distilled and therefore useful to perform quantum information tasks. In this thesis we will only be interested in distillable entanglement so in principle we will not need to worry about the existence or not of bound entanglement.

Based on Peres criterion a number of entanglement measures have been introduced. In this thesis we have used negativity ($\mathcal{N}$) to account for the quantum correlations between the different bipartitions of the system \cite{Negat}. It is an entanglement monotone sensitive to distillable entanglement defined as the sum of the negative eigenvalues of the partial transpose density matrix, in other words, if $\sigma_i$ are the eigenvalues of any $\rho^{pT}_{AB}$ then
\begin{equation}\label{negativitydef}
\mathcal{N}_{AB}=\frac12\sum_{ i}(|\sigma_i|-\sigma_i)=-\sum_{\sigma_i<0}\sigma_i.
\end{equation}
The minimum value of negativity is zero (for states with no distillable entanglement) and its maximum (reached for maximally entangled states) depends on the dimension of the maximally entangled state. Specifically, for qubits $\mathcal{N}_{AB}^{\text{max}}=1/2$.

\section{Mutual information}\label{mutusec}

The mutual information of two random variables $(X,Y)$ is a function of these two variables that measures how much uncertainty about one of the variables is reduced by our knowledge about the other. It accounts for the correlations between the two variables.

Given two random variables $(X,Y)$ the mutual information $I_{XY}$ is defined as
\begin{equation}
I(X,Y)=H(X)+H(Y)-H(X,Y),
\end{equation}
where $H(X),H(Y)$ are the marginal Shanon entropies and $H(X,Y)$ the joint entropy defined as
\begin{align}
H(X,Y)&=-\sum_{x,y}P(x,y)\log_2\left[P(x,y)\right],\\*
H(X)&=-\sum_{x}P(x)\log_2\left[P(x)\right],\\*
H(Y)&=-\sum_{y}P(y)\log_2\left[P(y)\right],
\end{align}
where $P(x,y)$ is the joint probability distribution of the random variables $X,Y$ and
\begin{equation}
p(x)=\sum_yP(x,y),\qquad p(y)=\sum_xP(x,y)
\end{equation}
are the marginal probability distributions for $X$ and $Y$.

For a quantum bipartite system of density matrix $\rho_{\text{AB}}$ the quantum mutual information is expressed in terms of the Von Neumann Entropy
\begin{equation}
I_{\text{AB}}=S_\text{A}+S_\text{B}-S_{\text{AB}},
\end{equation}
where the Von Neumann entropies are\footnote{The $\log_2$ is chosen to be base 2 because in quantum information it is common to work with qubits, but any other basis can be chosen instead.}
\begin{align}
S_{\text{AB}}&=-\tr_{\text{AB}}\left(\rho_{\text{AB}}\log_2\rho_{\text{AB}}\right),\\*
S_{\text{A}}&=-\tr_{\text{A}}\left(\rho_{\text{A}}\log_2\rho_{\text{A}}\right),\\*
S_{\text{B}}&=-\tr_{\text{B}}\left(\rho_{\text{B}}\log_2\rho_{\text{B}}\right),
\end{align}
and the partial systems are $\rho_\text{A}=\tr_\text{B}\left(\rho_\text{AB}\right)$, $\rho_\text{B}=\tr_\text{A}\left(\rho_\text{AB}\right)$.

Mutual information accounts for both, classical and quantum correlations, so that it can be used together with an entanglement measure to distinguish the behaviour of classical correlations: in a system which has no quantum correlations, mutual information accounts exclusively for classical correlations.


\chapter[Introduction to QFT in curved spacetimes]{Introduction to quantum field theory in curved spacetimes}\label{INTROU}

This  section of the preliminars pursuits a double objective. First, it aims to give a brief introduction to the tools and the background of quantum field theory in curved spacetimes necessary to understand and to present the results obtained during the development of this thesis. Of course this is only a very brief introduction to a very complex and extense discipline, more thorough approaches to these topics can be found in many textbooks \cite{Wald2,Birrell,NavarroSalas}

Second, in section \ref{probexcitations} we analyse a problem present in most of the previous literature, the use of the `single mode approximation' introduced in \cite{Alsingtelep,AlsingMcmhMil} based on misleading assumptions. In this section we will introduce new material which will be necessary in order to discuss the new work presented in part \ref{part1} in the context of previous results in the literature, giving a correct interpretation for this approximation. However, it will be in chapter \ref{sma} when this topic will be thoroughly dealt with when we expose the new results obtained when going beyond such approximation. 

\section{Scalar field quantisation in Minkowski \mbox{spacetime}}\label{scalarfieldq}

In this section I present a brief review of the standard canonical quantisation procedure of a field in Minkowski spacetime. The aim of this section is to introduce the concepts and notation that are going to be used throughout the following sections. Therefore, for the sake of simplicity, we will focus on a real massless scalar field. 

Let us consider an inertial observer (Alice) of the flat spacetime whose proper coordinates are the Minkowskian coordinates $(t,x, y, z)$. She wants to build a quantum field theory for a free masless scalar field. 

The equation of motion for this field is the well-known Klein-Gordon equation in Minkowski coordinates
\begin{equation}\label{KG1}
(\Box -m^2)\Phi =0.
\end{equation}

As it is commonplace, we can expand an arbitrary solution $\Phi(\bm x,t)$ to this equation as a sum of `positive frequency' and `negative frequency' solutions. One could innocently ask what is the definition of positive and negative frequency solutions, but the answer is somewhat trivial if we work with the Minkowski spacetime as background. The Minkowski spacetime admits a global timelike Killing vector $\partial_t$. A positive frequency solution of \eqref{KG1} $u_k(\bm x,t)$ satisfies, therefore,
\begin{equation}\label{flatcriterion}
\partial_t u_k(\bm x,t)=-i\omega_ku_k(\bm x,t),
\end{equation}
and this criterion would be the same if instead of $t$ we use the proper time of any inertial observer. Hence, we will express  $\Phi(\bm x,t)$ as a combination of positive $u_i(\bm x,t)$ and negative  $u^*_i(\bm x,t)$ frequency solutions of \eqref{KG1} with respect to Alice's proper time\footnote{For notational convenience we are using the sum symbol $\sum_i$ meaning integration over frequencies. Note that in free space $\sum_i\rightarrow\int_{-\infty}^\infty \frac{d^Dk}{\sqrt{(2\pi)^d2\omega}}$ and the modes are normalised to Dirac's delta instead of Kronecker's delta} and her definition will agree  with the definition of any other inertial observer.
\begin{equation}\label{exp1}
\Phi(\bm x,t)=\sum_i \left[\alpha_i u_i(\bm x,t)+\alpha_i^* u^*_i(\bm x,t)\right].
\end{equation}

The solutions $u_i(\bm x,t)$ can be chosen to form an orthonormal basis of solutions with respect to the Klein-Gordon scalar product defined, through the continuity equation, as
\begin{equation}\label{KGsc}
(u_j,u_k)=-i\int \text{d}^3x\left(u_j\partial_tu_k^*-u_k^*\partial_tu_j\right),
\end{equation}
which on the space of positive energy solutions happens to be positive definite. Note that, obviously, the modes $u_j$ satisfy the orthonormality relations
\begin{equation}\label{orto}
(u_j,u_k)=\delta_{jk}=-(u_j^*,u_k^*),\qquad (u_j,u_k^*)=0.
\end{equation}
We can now construct a Fock space following the standard canonical field quantisation scheme.

First we promote the classical Klein-Gordon field to a quantum field operator satisfying the equal time commutation relations
\begin{equation}
\begin{array}{c}
\left[\Phi(\bm x,t),\Pi(\bm x',t)\right]=i\delta(\bm x-\bm x'),\\[3mm]
\left[\Phi(\bm x,t),\Phi(\bm x',t)\right]=\left[\Pi(\bm x,t),\Pi(\bm x',t)\right]=0,
\end{array}
\end{equation}
where $\Pi(\bm x,t)=\partial_t \Phi(\bm x,t)$ is the canonical conjugate momentum associated with the variable $\Phi$.

This promotion means that we have to replace the complex amplitudes $\alpha_i$ and $\alpha_i^*$ by annihilation and creation operators $a_i$ and $a_i^\dagger$ who inherit the following commutation relations
\begin{equation}
\begin{array}{c}
[a_i,a^\dagger_j]=(u_i,u_j)=\delta_{ij},\\[3mm]
[a_i,a_j]=[a^\dagger_i,a^\dagger_j]=0.
\end{array}
\end{equation}

Now we can construct the standard Fock space, first we characterise the vacuum state of the field (minimum energy state) as the state which is annihilated by all the operators $a_i$
\begin{equation}
a_i\ket{0}=0.
\end{equation}

Then we define the one-particle Hilbert space by applying the creation operators $a_i^\dagger$ on the vacuum state
\begin{equation}
\ket{1_i}=a^\dagger_i\ket{0},
\end{equation}
and so on and so forth we construct the complete Fock space
\begin{equation}
\ket{n^1_{i_1},n^2_{i_2},\dots,n^k_{i_k}}=\frac{1}{\sqrt{n^1!n^2!\dots n^k!}} (a^\dagger_{i_1})^{n^1}(a^\dagger_{i_2})^{n^2}\dots(a^\dagger_{i_k})^{n^k}\ket{0}.
\end{equation}

Note that this quantisation procedure is independent of the particular choice of the inertial observer Alice. Any other choice of time $t$ is related to this one via Poincar\'e transformations which do not modify what we would label as positive and negative frequency modes. As a consequence, the expansion \eqref{exp1} can be performed equivalently for any inertial reference frame and the splitting between positive and negative frequency modes is invariant. Hence, the vacuum state is also Poincar\'e invariant and the construction of the Fock space is equivalent for any inertial observer. This will not happen for a general spacetime, as we will see in the following sections.

\section{Field quantisation in curved spacetimes}

For general spacetimes we cannot assume that we have global Poincar\'e symmetry and we will run into many difficulties when trying to construct quantum fields.

Let us continue with the scalar field case for the sake of simplicity. First of all we generalise equation \eqref{KG1} by means of the covariant D'Alambert operator, obtained promoting the partial derivatives to the covariant derivatives $\Box =\partial_\mu\partial^\mu\rightarrow \nabla_\mu\nabla^\mu$ so that  equation\footnote{There are some subtleties that should not be overlooked: first of all, we are assuming that there is no coupling of the field with the scalar curvature (minimal coupling). Second of all, if the field had internal spin degrees of freedom one must be careful with the covariant derivative definition (See Appendix \ref{appB})} \eqref{KG1} now reads
\begin{equation}
\nabla_\mu\nabla^\mu \phi =0.
\end{equation}
 
 To extend the Klein-Gordon product \eqref{KGsc} to curved spacetime we need a complete set of initial data, in other words, a Cauchy hypersurface $\Sigma$ over which we have to extend the integral\footnote{It can be shown, using Gauss theorem, that the product is independent of the choice of the Cauchy hypersurface $\Sigma$}
 \begin{equation}\label{KGscc}
(u_j,u_k)=-i\int \text{d}\Sigma\, n^\mu\left(u_j\partial_\mu u_k^*-u_k^*\partial_\mu u_j\right),
 \end{equation}
where $d\Sigma$ is the volume element and $n^\mu$ is a future directed timelike unit vector which is orthogonal to $\Sigma$.

Whether the spacetime is stationary or not will be determinant in order to build a quantum field theory in it. For non-stationary spacetimes we run into difficulties to classify field modes as positive or negative frequency. These spacetimes do not have a global timelike Killing vector\footnote{A Killing vector field $\xi^\mu$ is an isometry of the metric tensor, which is to say, the Lie derivative of the metric tensor with respect to $\xi^\mu$ is zero: $\mathcal{L}_{\xi}g_{\mu\nu}=\nabla_\mu\xi_{\nu}+\nabla_\nu\xi_{\mu}=0$} and, therefore, there is no natural way to distinguish positive and negative frequency solutions of the Klein-Gordon equation.

In the absence of this metric symmetry there is an ambiguity when it comes to define particle states: without a natural way to split modes in positive and negative frequencies there is no objective way to construct a Fock space, starting form the fact that there is no unique notion of a vacuum state. However we will see in section \ref{nonstaint} that we can still find an `approximated' particle interpretation when the spacetime posses asymptotically stationary regions.

Conversely, if the spacetime has a timelike Killing vector field $\xi^\mu$ we have a natural way to define positive frequency modes ($u_j$) in an analogous way as we did for the flat spacetime in \eqref{flatcriterion} 
\begin{equation}\label{criterion2}
\xi^\mu\nabla_\mu u_j = -i \omega_j u_j,
\end{equation}
where $\omega_j>0$.

Of course we can construct a local set of coordinates whose timelike coordinate is the Killing time $\tau$ associated to the isometry $\xi^\mu$ such that it satisfies that $\xi^\mu\nabla_\mu \tau =1$. For the flat spacetime, a particular case of stationary spacetime, the role of $\tau$  is played by the coordinate $t$.

Therefore for stationary spacetimes we can readily generalise the field quantisation procedure explained in the previous section.

\section[Inertial and accelerated observers of quantum fields]{Inertial and accelerated observers of quantum fields in flat spacetime: Bogoliubov transformations}\label{Rindbogo}

Even for the simple case of the flat Minkowski spacetime, there are non-trivial differences between observers of a quantum field in different kinematic states. This is because the field quantisation procedure is different for different observers. Specifically, a completely new phenomenology appears when accelerated observers observe the inertial vacuum state of the field.

In this section we show this phenomenon in a spacetime as simple as the flat spacetime but for two different class of observers of a quantum field, inertial and constantly accelerated.

\subsection{Accelerated observers: Rindler coordinates}

To describe the point of view of an accelerated observer we introduce the so-called Rindler coordinates $(\tau,\xi)$ \cite{gravitation}, which are the proper coordinates of an accelerated observer moving with a fixed acceleration $a$. The correspondence between the Minkowskian coordinates $(t,x)$ and the accelerated frame ones $(\tau,\xi)$ is 
\begin{equation}\label{change}
ct=\xi \sinh\left(\frac{a\tau}{c}\right),\qquad x=\xi\cosh\left(\frac{a\tau}{c}\right),
\end{equation}
where, we have made $c$ explicit\footnote{Note that we are not using the conformal Rindler coordinates $t=c\,a^{-1}e^{a\xi/c^2}\sinh\left(\frac{a\tau}{c}\right)$ and $x=c^{2}a^{-1}e^{a\xi/c^2}\cosh\left(\frac{a\tau}{c}\right)$ (quite common in the literature) but the proper coordinates of an accelerated observer of acceleration $a$ such that the proper lengths and times measured in these units corresponds directly with physical distances and time intervals.}.  
\begin{figure}[H]
\begin{center}
\includegraphics[width=.70\textwidth]{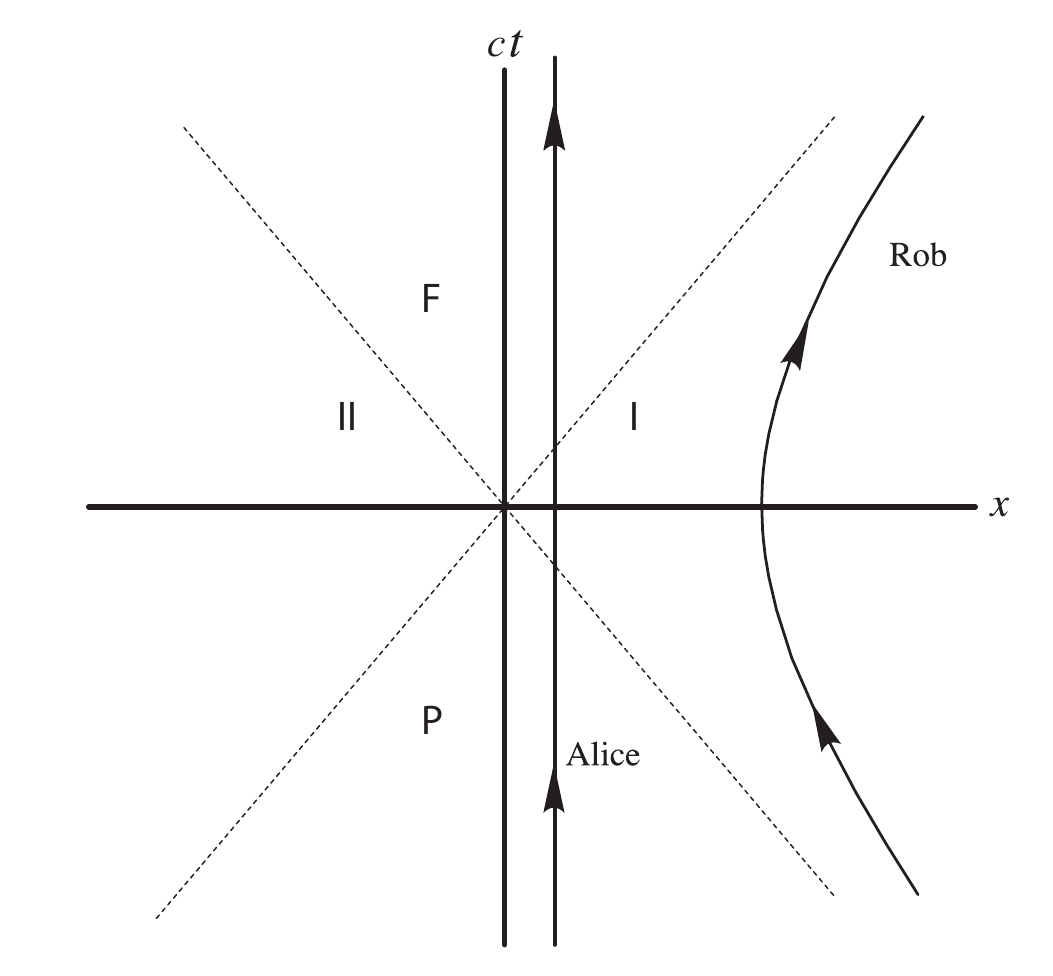}
\end{center}
\caption{Flat spacetime. Trajectories of an inertial (Alice) and accelerated (Rob) observer}
\label{rin2}
\end{figure}

Directly from \eqref{change} we see that for constant $\xi$ these coordinates describe hyperbolic trajectories in the spacetime whose asymptote is the light cone (as the observer accelerates his velocity tends to the speed of light, (i.e $ct\rightarrow x$). This means that a constantly accelerated observer follows trajectories such that $\xi=\text{const.}$ in the Rindler frame. However the observer for which these coordinates are his proper coordinates follows a particular trajectory:  To find its specific Rindler position we will use that all the Rindler observers are instantaneously at rest at time $t=0$ in the inertial frame, and at this time a Rindler observer with proper acceleration $a$ and, therefore, proper coordinates $(\xi,\tau)$ will be at Minkowskian position $x=c^2/a$. On the other hand, in this point $t=0\Rightarrow\xi=x$ instantaneously so, consequently, the constant Rindler position for this trajectory is $\xi= c^2/a$.

One can see that for accelerated observers an acceleration horizon appears: Any accelerated observer would be restricted to either region I or II of the spacetime. We will see in chapter \ref{blackhole1} that this horizon is locally very similar to an event horizon. The appearance of acceleration horizons is responsible for the Unruh effect \cite{Unruh}, as we will see in chapter \ref{tue}.  

A quick inspection reveals that the Rindler coodinates defined in \eqref{change} do not cover the whole Minkowski spacetime. Actually, these coordinates only cover the right wedge of the spacetime (Region I in Figure \ref{rin2}). This is so because an eternally  accelerated observer is always restricted to either region I or II depending if he is accelerating or decelerating with respect to the Minkowskian origin.

In fact to map the complete Minkowski spacetime we need three more sets of Rindler coordinates,
\begin{equation}\label{changeII}
ct=-\xi \sinh\left(\frac{a\tau}{c}\right),\qquad x=-\xi\cosh\left(\frac{a\tau}{c}\right),
\end{equation}
for region II, corresponding to an observer decelerating with respect to the Minkowskian origin, and
\begin{equation}\label{changeFP}
ct=\pm\xi \cosh\left(\frac{a\tau}{c}\right),\qquad x=\pm\xi\sinh\left(\frac{a\tau}{c}\right)
\end{equation}
for regions F and P.

Notice that for both relevant regions (I and II), the coordinates $(\xi,\tau)$ take values in the whole domain $(-\infty,+\infty)$. Therefore, they admit completely independent canonical field quantisation procedures.

\subsection{Field quantisation in Minkowski and Rindler coordinates}

For simplicity, imagine first an inertial observer (Alice) in a flat spacetime whose proper coordinates are the Minkowskian coordinates $(x,t)$. She wants to build a quantum field theory for a free massless scalar field. 

As explained in section \ref{scalarfieldq}, to build her Fock space she needs to find an orthonormal basis of solutions of the free massless Klein-Gordon equation in Minkowski coordinates. Of course, she can always use the positive energy plane wave solutions of the Klein-Gordon equation in her proper coordinates to build a complete set of solutions of this equation.

In this fashion the states $\ket{1_{\hat \omega}}_\text{M}=a^\dagger_{\hat
\omega,\text{M}}\ket{0}_\text{M}$ are free massless scalar field
modes, in other words, solutions of positive frequency $\hat\omega$
(with respect to the Minkowski timelike Killing vector $\partial_{
t}$) of the free Klein-Gordon equation:
\begin{align}\label{modmin}
\ket{1_{\hat\omega}}_\text{M}&\equiv u_{\hat\omega}^\text{M}\propto\frac{1}{\sqrt{2{\hat \omega}}}e^{-i {\hat \omega} \hat t},
\end{align}
where only the time dependence has been made explicit. The label $\text{M}$ just means that these states are expressed in the Minkowskian Fock space basis.

The field expanded in these modes takes the usual form \eqref{exp1}
\begin{equation}\label{minkexp12}
\phi=\sum_i \left(a_{\hat\omega_i,\text{M}}u_{\hat\omega_i}^\text{M}+a_{\hat\omega_i,\text{M}}^\dagger u_{\hat\omega_i}^{\text{M}*}\right),
\end{equation}
where we have eliminated redundant notation and M denotes that $u_{\hat\omega_i}^\text{M}$ and $a_{\hat\omega_i,\text{M}}$ are Minkowskian modes and operators.

An accelerated observer   can also define his vacuum and excited states
of the field. Actually, there are two natural vacuum states associated
with the positive frequency modes in regions $\text{I}$ and $\text{II}$
of Rindler spacetime. These are $\ket{0}_\text{I}$ and
$\ket{0}_{\text{II}}$, and subsequently we can define the field
excitations using Rindler coordinates $(\xi,\tau)$ as
\begin{align}\label{modrin}
\ket{1_\omega}_\text{I}&=a^\dagger_{\omega,\text{I}}\ket{0}_\text{I}
\equiv u_\omega^\text{I}\propto\frac{1}{\sqrt{2\omega}}e^{-i\omega  \tau},\nonumber\\
\ket{1_\omega}_{\text{II}}&=a^\dagger_{\omega,\text{II}}\ket{0}_\text{II}
\equiv u_\omega^\text{II}\propto\frac{1}{\sqrt{2\omega}}e^{i\omega  \tau}.
\end{align}
These modes are related by a spacetime reflection and only have support in regions I and II of the
Rindler spacetime respectively.

We can now expand the field \eqref{minkexp12} in terms of this complete set of solutions of the Klein-gordon equation in Rindler coordinates
\begin{equation}\label{rindexp1}
\phi=\sum_i \left(a_{\omega_i,\text{I}}u_{\omega_i}^\text{I}+a_{\omega_i,\text{I}}^\dagger u_{\omega_i}^{\text{I}*}+a_{\omega_i,\text{II}}u_{\omega_i}^\text{II}+a_{\omega_i,\text{II}}^\dagger u_{\omega_i}^{\text{II}*}\right).
\end{equation}

Expressions \eqref{minkexp12} and \eqref{rindexp1} are exactly equal and therefore Minkowskian modes can be expressed as function of Rindler modes by means of the Klein-Gordon scalar product \eqref{KGsc}
\begin{equation}
u_{\hat\omega_j}^\text{M}=\sum_i\left[(u_{\hat\omega_j}^\text{M},u_{\omega_i}^\text{I})u_{\omega_i}^\text{I}-
(u_{\hat\omega_j}^{\text{M}},u_{\omega_i}^{\text{II}*})u_{\omega_i}^\text{II*}+(u_{\hat\omega_j}^\text{M},u_{\omega_i}^\text{II})
u_{\omega_i}^\text{II}-(u_{\hat\omega_j}^{\text{M}},u_{\omega_i}^{\text{I}*})u_{\omega_i}^\text{I*}\right].
\end{equation}
Notice that we have taken into account the properties \eqref{orto}, and one has to be very careful with the signs given that $(u^*_i,u^*_j)=-\delta_{ij}$.

If we now define Bogoliubov coefficients as
\begin{equation}\label{bogo11}
\alpha^{\Sigma}_{ij}=\left(u_{\hat\omega_i}^{\text{M}},u_{\omega_j}^\Sigma\right),
\qquad\beta^{\Sigma}_{ij}=-\left(u_{\hat\omega_i}^{\text{M}},u_{\omega_j}^{\Sigma*}\right),
\end{equation}
where $\Sigma$ can take the values I and II, we have that
\begin{equation}\label{minkowskirindlerexp}
u_{\hat\omega_j}^\text{M}=\sum_i\left(\alpha^{\text{I}}_{ji}u_{\omega_i}^\text{I}+
\beta^{\text{II}}_{ji}u_{\omega_i}^\text{II*}+\alpha^{\text{II}}_{ji}
u_{\omega_i}^\text{II}+\beta^{\text{I}}_{ji}u_{\omega_i}^\text{I*}\right).
\end{equation}
We would like to know how the creation and annihilation operators in the Minkowski basis are related to operators in the Rindler bases. Since we know that $a_{\omega_i,\text{M}}=(\phi,u_{\omega_i}^\text{M})$, if we write $\phi$ in Rindler basis \eqref{rindexp1} we can readily obtain
\begin{equation}\label{temporary1}
a_{\hat\omega_i,\text{M}}=\sum_j\left[(u_{\omega_j}^\text{I},u_{\hat\omega_i}^\text{M})
a^{\phantom{\dagger}}_{\omega_j,\text{I}}+(u_{\omega_j}^{\text{I}*},u_{\hat\omega_i}^\text{M})
a^\dagger_{\omega_j,\text{I}}+(u_{\omega_j}^\text{II},u_{\hat\omega_i}^\text{M})
a^{\phantom{\dagger}}_{\omega_j,\text{II}}+(u_{\omega_j}^{\text{II}*},u_{\hat\omega_i}^\text{M})
a^\dagger_{\omega_j,\text{II}}\right).
\end{equation}
Using the properties of the KG product
\begin{equation}\label{KGproperties}(u_1,u_2)=(u_2,u_1)^*\qquad(u_1^*,u_2^*)=-(u_2,u_1)\end{equation}
we can write \eqref{temporary1} in terms of the Bogoliubov coefficients \eqref{bogo11} as 
\begin{equation}
a_{\hat\omega_i,\text{M}}=\sum_j\left(\alpha^{\text{I}*}_{ij}
a^{\phantom{\dagger}}_{\omega_j,\text{I}}-\beta^{\text{I}*}_{ij}
a^\dagger_{\omega_j,\text{I}}+\alpha^{\text{II}*}_{ij}
a^{\phantom{\dagger}}_{\omega_j,\text{II}}-\beta^{\text{II}*}_{ij}
a^\dagger_{\omega_j,\text{II}}\right).
\end{equation}

A completely analogous reasoning can be followed for the case of a Dirac field, with some differences that will be deeply analysed in chapter \ref{sma} of this thesis. Where we will also go through the computation of the coefficients \eqref{bogo11}.

For now and for the sake of this introduction let us say that the vacuum state in the Minkowskian basis can be expressed as a two mode squeezed state in the Rindler basis \cite{Takagi, Alicefalls,NavarroSalas}. Namely, for the scalar case considered in this introduction
\begin{equation}\label{scalarvacuum1}
\ket{0}_{\text{M}}=\frac{1}{\cosh r}\sum_{n=0}^\infty \tanh^n r_{\text{b},\omega} \ket{n}_{\text{I}}\ket{n}_{\text{II}}.
\end{equation}
where\footnote{The label b stands for `bosonic', this parameter has a different definition for fermionic and bosonic fields as we will see in section \ref{probexcitations}}
\begin{equation}\label{rbos1}
r_{\text{b},\omega}=\operatorname{atanh} \left[\exp\left(-\frac{\pi c\omega }{a}\right)\right].
\end{equation}

\section{The Unruh effect}\label{tue}

In the 70s Fulling, Davies and Unruh realised that the impossibility to map the whole Minkowski spacetime with only one set of Rindler coordinates has strong implications when accelerated and inertial observers describe states of a quantum field. Namely, the description of the vacuum state of the field in the inertial basis as seen by accelerated observers has a non-zero particle content. 

In very plain words, the Unruh effect is the fact that while inertial observers `see' the vacuum state of the field, an accelerated observer would `see' a thermal bath whose temperature is proportional to his acceleration.

Different approaches to this well-known effect can be found in multiple textbooks (let us cite \cite{Wald2,Birrell,NavarroSalas} as a token). However, in this section, we will provide a not so common but rather simple derivation of the effect in a way that will be useful in order to clearly present a feature of spacetime with horizons which turns out to be relevant when it comes to study entanglement.

Imagine that an inertial observer, Alice, is observing the vacuum state of a scalar field. Now imagine an accelerated observer, called Rob, who wants to describe the same quantum field state by means of his proper Fock basis. The first step we need to take is to change the vacuum state from the Fock basis build from solutions to the Klein-Gordon equation in Minkowskian coordinates \eqref{modmin} to the Fock basis build from solutions of the KG equations in Rindler coordinates \eqref{modrin}. This gives us equation \eqref{scalarvacuum1} which we presented in the section above.

The state \eqref{scalarvacuum1} is a pure state. However, the accelerated observer is restricted to either region I or II of the spacetime due to the appearance of an acceleration horizon (as shown in Figure \ref{rin2}), and, since both regions are causally disconnected, Rob has no access to the modes which have support in the opposite wedge of the spacetime. This is a key point to analyse information matters. 

This means that the quantum state accessible for Rob is no longer pure,
\begin{equation}
\rho_{\text{R}}=\tr_{\text{II}}\left(\proj{0}{0}\right)=\sum_{k}\bra{k}_{\text{II}}\ket{0}_\text{M}\bra{0}_\text{M}\ket{k}_{\text{II}}.
\end{equation}
Substituting $\ket{0}$ by its Rindler basis expression \eqref{scalarvacuum1} we have that
\begin{equation}
\rho_{\text{R}}=\frac{1}{\cosh^2 r_{\text{b},\omega}}\sum_k\sum_{n,m}\tanh^{m+n}r_{\text{b},\omega}\bra{k}_{\text{II}}\ket{n}_\text{I}\ket{n}_{\text{II}}\bra{m}_\text{I}\bra{m}_{\text{II}}\ket{k}_{\text{II}},
\end{equation}
which leads to
\begin{equation}\label{thermal}
\rho_{\text{R}}=\frac{1}{\cosh^2 r_{\text{b},\omega}}\sum_n\tanh^{2n}r\ket{n}_\text{I}\bra{n}_\text{I},
\end{equation}
which is a thermal state.

Indeed, if we compute the particle counting statistics that the accelerated observer would see we obtain
\begin{equation}
\langle N_{\omega,\text{R}}\rangle=\tr_{\text{I}}\left(\rho_\text{R}\,a^\dagger_{\omega,\text{I}}a_{\omega,\text{I}}\right)=\frac{1}{e^{2\pi c/\omega a}-1},
\end{equation}
which is a Bose-Einstein distribution with temperature
\begin{equation}
T_\text{U}=\frac{\hbar a}{2\pi K_\text{B}},
\end{equation}
which is nothing but the Unruh temperature.

Rob observes a thermal state\footnote{Note that thermal noise is only observed in the 1+1 dimensional case. In higher dimension Rob would observe a noisy distribution, very similar to a thermal one, but with different prefactors \cite{Takagi}. This is called the Rindler noise.} because he cannot see modes with support in region II due to the presence of an acceleration horizon. This is a very important point that will play a fundamental role in some of the results presented in this thesis. 

As we will see in chapters \ref{blackhole1} and \ref{stellarcollapse} this effect is closely related with the Hawking effect and the Hawking radiation emitted in a stellar collapse process.

\section[Bogoliubov transformations in non-stationary scenarios]{Bogoliubov transformations in non-stationary\\ scenarios}\label{nonstaint}

We have mentioned in previous sections the difficulty of carrying out the field quantisation when the spacetime is not stationary. However, there are some very interesting scenarios in which the spacetime is not stationary but posses stationary asymptotic regions. This is the case of some models of expansion of the Universe \cite{dun1} or the stellar collapse and formation of black holes \cite{NavarroSalas}. As part of the thesis deals with quantum information problems   in such scenarios, an introduction to field quantisation in this context is in order.

Consider a spacetime which has asymptotic stationary regions in the past and in the future. We will call them `in' and `out' respectively.

The existence of these regions can be used to give a particle interpretation to the solutions of the field equations. Namely, we can build two different complete sets of solutions of the Klein-Gordon equation, the first one $\{u_{\hat\omega_j}^{\text{in}}\}$ made of modes that have positive frequency $\hat\omega_j$ with respect to the inertial time in the asymptotic past. The second set $\{u_{\omega_j}^{\text{out}}\}$ would consist of modes that have positive frequency $\omega_j$ with respect to the inertial time in the future.

In this fashion we can now expand the quantum field in terms of either the first or the second complete set of modes
\begin{equation}\label{minkexp1}
\phi=\sum_i \left(a_{\hat\omega_i,\text{in}}u_{\hat\omega_i}^\text{in}+a_{\hat\omega_i,\text{in}}^\dagger u_{\hat\omega_i}^{\text{in}*}\right)=\sum_i \left(a_{\omega_i,\text{out}}u_{\omega_i}^\text{out}+a_{\omega_i,\text{out}}^\dagger u_{\omega_i}^{\text{out}*}\right).
\end{equation}

Moreover, since both set of modes are complete, one can also expand one set of modes in terms of the other by means of the Klein-Gordon scalar product
\begin{align}\label{modinout1}
u^{\text{in}}_{\hat\omega_j}&= \sum_i \left[(u_{\hat\omega_j}^{\text{in}},u_{\omega_i}^{\text{out}})u_{\omega_i}^{\text{out}}-(u_{\hat\omega_j}^{\text{in}},u_{\omega_j}^{\text{out}*})u_{\omega_j}^{\text{out}*}\right],\\
u^{\text{out}}_{\omega_j}&= \sum_i \left[(u_{\omega_j}^{\text{out}},u_{\hat\omega_i}^{\text{in}})u_{\hat\omega_i}^{\text{in}}-(u_{\omega_j}^{\text{out}},u_{\hat\omega_i}^{\text{in}*})u_{\hat\omega_i}^{\text{in}*}\right].
\end{align}
Let us define Bogoliubov coefficients as
\begin{equation}
\alpha_{ij}=(u_{\omega_i}^{\text{out}},u_{\hat\omega_j}^{\text{in}}),\qquad \beta_{ij}=-(u_{\omega_i}^{\text{out}},u_{\hat\omega_j}^{\text{in}*}),
\end{equation}
then, using the properties \eqref{KGproperties} we can rewrite \eqref{modinout1} as
\begin{align}\label{modinout2}
u^{\text{in}}_{\hat\omega_i}&= \sum_j \left[\alpha_{ji}^{*}u_{\omega_j}^{\text{out}}-\beta_{ji}u_{\omega_j}^{\text{out}*}\right],\\
u^{\text{out}}_{\omega_j}&= \sum_i \left[\alpha_{ji}u_{\hat\omega_i}^{\text{in}}+\beta_{ji}u_{\hat\omega_i}^{\text{in}*}\right].
\end{align}

Now, we can expand the particle operators associated with one basis in terms of operators of the other basis. To do this we use the fact that $a_{\hat\omega_i,\text{in}}=(\phi,u_{\hat\omega_j}^{\text{in}})$ and $a_{\omega_i,\text{out}}=(\phi,u_{\omega_j}^{\text{out}})$, after some trivial computations and use of the properties of the scalar product we obtain
\begin{align}
\label{uno1}a_{\hat\omega_i,\text{in}}&=\sum_j \left(\alpha_{ji} a_{\omega_j,\text{out}}+\beta_{ji}^* a_{\omega_j,\text{out}}^\dagger\right),\\
a_{\omega_i,\text{out}}&=\sum_j \left(\alpha^*_{ij} a_{\hat\omega_j,\text{in}}-\beta_{ij}^* a_{\hat\omega_j,\text{out}}^\dagger\right).
\end{align}

Now let us consider the vacuum state in the asymptotic past region $\ket0_{\text{in}}$, which fulfils $a_{\hat\omega_i,\text{in}}\ket0_{\text{in}}$ for all $\hat\omega_i$. One could ask how that state evolves subject exclusively to the gravitational interaction, in other words, we want to know the form of the state $\ket0_{\text{in}}$ in the basis of solutions of the KG equation in the asymptotic future.

To do this we will take advantage of the fact that  $a_{\hat\omega_i,\text{in}}\ket0_{\text{in}}=0$, if we substitute $a_{\hat\omega_i,\text{in}}$ in terms of out operators using equation \eqref{uno1} we obtain that
\begin{equation}\label{condit}
\sum_j \left(\alpha_{ji} a_{\omega_j,\text{out}}+\beta_{ji}^* a_{\omega_j,\text{out}}^\dagger\right)\ket{0}_\text{in}=0.
\end{equation}

We can assume a general form for the state $\ket0_{\text{in}}$ in terms of the out Fock basis as a sum of its n-particle amplitudes
\begin{equation}\label{ansatz1}
\ket{0}_{\text{in}}= C\ket{0}_{\text{out}}+C^{j_1}\ket{\Psi}_{j_1}+C^{j_1,j_2}\ket{\Psi}_{j_1,j_2}+\dots+C^{j_1,\dots,j_n}\ket{\Psi}_{j_1,\dots,j_n}+\dots
\end{equation}
where each component has the form
\begin{equation}
C^{j_1,\dots,j_n} \ket{\Psi_n}_{j_1,\dots,j_n}=\sum_{j_1,\dots,j_n} C^{j_1,\dots,j_n} a_{\omega_{j_1},\text{out}}^\dagger\dots a_{\omega_{j_n},\text{out}}^\dagger\ket{0}_\text{out}.
 \end{equation}

Substituting the ansatz \eqref{ansatz1} in \eqref{condit} we obtain an infinite set of conditions for the coefficients $C^{j_1,\dots,j_n}$. First of all, for $C^{j}$ we see that the zero particle component of equation \eqref{condit}  can only be obtained from the annihilator acting on the 1-particle component of \eqref{ansatz1} state this gives the condition
\begin{equation}\label{coefcond1}
\sum_j\alpha_{j i}C^j=0\Rightarrow C^j= 0.
\end{equation} 

Now, the $n$-particle component ($n>0$) of equation \eqref{condit} is obtained by the annihilator acting  on the $n+1$ particle component of \eqref{ansatz1} and the creator acting on the $n-1$ particle component of \eqref{ansatz1}, therefore we know that the coefficients $C^{j_1,\dots,j_{n+1}}$ can be written as functions of the coefficients $C^{j_1,\dots,j_{n-1}}$ allowing us to find a recurrence rule to write all the coefficients as funcions of $C$ (the vacuum coefficient).

This, together with \eqref{coefcond1}, means that the `in' vacuum evolves to a state in the asymptotic future that does not have components of an odd number of particles and that the coefficients of even components in \eqref{ansatz1} are related pairwise. 

To find the form of this coefficients we take advantage of the invertibility of the Bogoliubov coefficient matrix $\alpha_{ij}$ and therefore given arbitrary $C_j$ and $D_j$ the following identity holds
\begin{equation}\label{inve}
\sum_j\left(\alpha_{j i}C_j+\beta_{ji}^*D_j\right)=0\Rightarrow C_k=-\sum_{ij}\beta^*_{ji}\alpha_{ ik}^{-1}D_j.
\end{equation}

After some basic but lengthy algebra we obtain that the vacuum state in the asymptotic past is expressed in the basis of modes in the asymptotic future as
\begin{equation}\label{dynoresult}
\ket{0}_\text{in}=C\exp\left(-\frac12\sum_{ijk}\beta^*_{ik}\alpha_{ kj}^{-1}a_{\omega_{i},\text{out}}^\dagger a_{\omega_{j},\text{out}}^\dagger\right)\ket{0}_\text{out}.
\end{equation}
C is obtained imposing the normalisation of the state.

This proves that if $\beta_{ij}$ is different from zero one would observe particle production as a consequence of time evolution under the interaction with the gravitational field. 

\section{The problem of field excitations\footnote{E. Mart\'in-Mart\'inez, L.J. Garay, J. Le\'on. Phys. Rev. D 82, 064006 (2010)} \footnote{D. \!
Bruschi,\! J. Louko,\! E. Mart\'in-Mart\'inez,\! A. Dragan,\! I. Fuentes,\! Phys.\! Rev.\! A 82,\! 042332 (2010)}}\label{probexcitations}

Most of the work regarding field entanglement in non-inertial frames before this thesis has made use of what was called the ``single mode approximation'' \cite{Alsingtelep,AlsingMcmhMil} in order to build the excited states of the field and transform them to the Rindler basis. As part of the results presented in this thesis we will see that such approximation was flawed and cannot be used as it was first presented. However, this does not mean that the previous works are wrong and not useful, but that they need reinterpretation.

It is obvious from \eqref{minkowskirindlerexp} that a single frequency Minkowski mode does not correspond to a monochromatic mode when it is transformed into the Rindler basis. In other words, a monochromatic field excitation in the Minkowskian basis transforms to a non-monochromatic field excitation from the perspective of an accelerated observer.

However, until 2010, all the previous studies analysing entanglement in a general relativisitc context employed the so-called single mode approximation.  This approximation, which was introduced in \cite{Alsingtelep,AlsingMcmhMil}, has been extensively used in the literature not only in discussions concerning entanglement but also in other relativistic quantum information scenarios, for example, among many others \cite{Alicefalls,AlsingSchul,Bradler,highdim,chapucilla,chapucilla2,Shapoor,matsako,Ditta,Geneferm,DiracDiscord}. 

Such approximation invokes that a single Minkowski mode transforms to a peaked distribution of Rindler modes and, therefore, one can approximate a single  Minkowski mode by means of a single Rindler mode; but this assumption is wrong. As we have mentioned above, the distribution of Rindler modes that corresponds to a single frequency Minkowski mode is not peaked, but highly non-monochromatic.

But there are specific bases which have the property of having diagonal Bogoliubov coefficient matrices. We will see that if we work in one of these bases we obtain results that are are computationally equivalent to those made under the single mode approximation. Now we proceed to build such a basis.

There exist an infinite number of orthonormal bases
that define the same vacuum state, namely the Minkowski vacuum
$\ket{0}_\text{M}$, which can be used to expand the solutions of the
Klein-Gordon equation.

 More explicitly, since the modes
$u^\text{M}_{\hat\omega_i}$ have positive frequency, any complete set
made out of independent linear combinations of these modes only
(without including the negative frequency ones
$u^{\text{M}*}_{\hat\omega_i}$) will define the same vacuum
$\ket{0}_\text{M}$.

Specifically, as described in e.g. Refs. \cite{Unruh,Takagi,Birrell} and
explicitly constructed in chapter \ref{sma}, there exists an orthonormal basis
$\{\psi_{\omega_j}^\text{M}, \psi'^{\text{M}}_{\omega_j}\}$ determined by certain linear
combinations of monochromatic positive frequency modes 
$u^\text{M}_{\hat\omega_i}$:
\begin{equation}\label{modopsi}
\psi^\text{U}_{\omega_j}=\sum_i C_{ij}\, u_{\hat\omega_i}^\text{M},\qquad \psi'^{\text{U}}_{\omega_j}=\sum_i C'_{ij}\, u_{\hat\omega_i}^\text{M} 
\end{equation}
 such that the Bogoliubov coefficients that relate this basis
 $\{\psi_{\omega_j}^\text{U},\psi'^{\text{U}}_{\omega_j}\}$  and the Rindler basis
 $\{u^\text{I}_{\omega_i},u^\text{II}_{\omega_i}\}$ have the following form:
 \begin{align}\label{bogo1}
\hat\alpha^{\text{I}}_{ij}
&=\left(\psi_{\omega_j}^\text{U},u_{\omega_i}^\text{I}\right)=\cosh r_{\text{b},\omega_i}\,\delta_{ij},
\nonumber\\
\hat\alpha^{\text{II}}_{ij}
&=\left(\psi_{\omega_j}^\text{U},u_{\omega_i}^{\text{II}}\right)=0,\nonumber\\
\hat\beta^{\text{I}}_{ij}
&=-\left(\psi_{\omega_j}^{\text{U}},u_{\omega_i}^{\text{I}*}\right) = 0,
\nonumber\\
\hat\beta^{\text{II}}_{ij}&=-\left(\psi_{\omega_j}^{\text{U}},u_{\omega_i}^{\text{II}*}\right)= \sinh r_{\text{b},\omega_i}\,\delta_{ij},
\end{align}
and analogously for $\hat\alpha'^{\text{I,II}}_{ij}$ and $\hat\beta'^{\text{I,II}}_{ij}$ interchanging the labels $\text{I}$ and $\text{II}$ in the formulas above. In this expressions
\begin{equation}\label{defr1}
\tanh r_{\text{b},\omega_i}=\exp(-\pi c {\omega_i }/{a}),
\end{equation}
and the label $\text{s}$ in $r_{\text{b},\omega_i}$ has been introduced to
indicate that we are dealing with a scalar field. 

In this fashion a mode $\psi_{\omega_j}^\text{U}$ (or a mode $\psi'^{\text{U}}_{\omega_j}$) expands only in
terms of mode of frequency $\omega_j$ in Rindler regions
$\text{I}$ and $\text{II}$ and for this reason we have labeled $\psi_{\omega_j}^\text{U}$ and $\psi'^{\text{U}}_{\omega_j}$  with the frequency $\omega_j$ of the corresponding Rindler modes. In other words, we can express a given
monochromatic Rindler mode of frequency $\omega_j$   as a linear
superposition of the single Minkowski modes $\psi_{\omega_j}^{\text{U}}$ and $\psi'^{\text{U}*}_{\omega_j}$ or as a polychromatic combination of the
positive frequency Minkowski modes $u^\text{M}_{\hat\omega_i}$ and
their conjugates. These modes \eqref{modopsi} are nothing but a specific choice of the so-called Unruh modes \cite{Wald2,Birrell,NavarroSalas}. We will study this in detail in section \ref{sec3m7}.

Let us denote ${a_{\omega_j,\text{U}}}$ and $a_{\omega_j,\text{U}}^\dagger$ the
annihilation and creation operators associated with modes
$\psi^\text{U}_{\omega_j}$ (analogously we denote ${a'_{\omega_j,\text{U}}}$ and $a_{\omega_j,\text{U}}'^\dagger$ the ones associated with modes $\psi'^{\text{U}}_{\omega_i}$). The Minkowski vacuum
$\ket0_\text{M}$, which is annihilated by all the Minkowskian operators
$a_{\hat \omega_i,\text{M}}$, is also annihilated by all the operators
$a_{\omega_j,\text{U}}$ and $a_{\omega_j,\text{U}}'$, as we already mentioned. This comes out because any
combination of Minkowski annihilation operators annihilates the
Minkowskian vacuum.

Due to the Bogoliubov relationships \eqref{bogo1} being diagonal, each annihilation
operator $a_{\omega_i}$ can be expressed as a combination of Rindler
particle operators of only one Rindler frequency  $\omega_i$:
\begin{equation}\label{buenmod}
a_{\omega_i,\text{U}}=\cosh r_{\text{b},\omega_i}\, a^{\phantom{\dagger}}_{\omega_i,\text{I}}-
\sinh r_{\text{b},\omega_i}\,a^\dagger_{\omega_i,\text{II}},
\end{equation}
and analogously for $a'_{\omega_i}$ interchanging the labels I and II.

An analogous procedure can be carried out for fermionic fields (e.g. Dirac
fields). We can use linear combinations of monochromatic solutions of the
Dirac equation $\psi^\text{U}_{\omega_i,\sigma}$ and $\bar
\psi^\text{U}_{\omega_i,\sigma}$ (and their primed versions) built in the same fashion as for scalar
fields:
\begin{align}\label{modopsif}
\psi^\text{U}_{\omega_j,\sigma}=\sum_i D_{ij}\, u_{\hat\omega_i,\sigma,\text{M}}^{+},
&\qquad \bar \psi^\text{U}_{\omega_j,\sigma}=\sum_i E_{ij}\, u_{\hat\omega_i,\sigma,\text{M}}^{-},\nonumber\\
\psi'^{\text{U}}_{\omega_j,\sigma}=\sum_i D'_{ij}\, u_{\hat\omega_i,\sigma,\text{M}}^{+},
& \qquad\bar\psi'^{\text{U}}_{\omega_j,\sigma}=\sum_i E'_{ij}\, u_{\hat\omega_i,\sigma,\text{M}}^{-},
\end{align}
where $u^+_{\hat\omega_i,\sigma,\text{M}}$ and
$u^-_{\hat\omega_i,\sigma,\text{M}}$ are respectively monochromatic
solutions of positive (particle) and negative (antiparticle) frequency $\pm\hat\omega_i$ of the
massless Dirac equation with respect to the Minkowski Killing time. The
label $\sigma$ accounts for the possible spin degree of freedom of the
fermionic field\footnote{Throughout this work we will consider that the
spin of each mode is in the acceleration direction and, hence, spin will not
undergo Thomas precession due to instant Wigner rotations
\cite{AlsingSchul,Jauregui}.}.

The coefficients of these combinations are such that for the modes
$\psi_{\omega_i,\sigma}$ and $\bar \psi_{\omega_i,\sigma}$ the
annihilation operators are related with the Rindler ones by means of the
following Bogoliubov transformations:
\begin{eqnarray}\label{Bogoferm2}
\nonumber c_{{\omega_i},\sigma,\text{U}}&=&\cos{r_{\text{f},\omega_i}}\,c_{{\omega_i},
\sigma,\text{I}}-\sin r_{\text{f},\omega_i}\,d^\dagger_{{\omega_i},-\sigma,\text{II}},\\*
d_{{\omega_i},\sigma,\text{U}}^\dagger&=&\cos{r_{\text{f},\omega_i}}\,d^\dagger_{
{\omega_i},\sigma,\text{II}}+\sin r_{\text{f},\omega_i}\,c_{{\omega_i},-\sigma,\text{I}},
\end{eqnarray}
and analogously for $c'_{\omega_i,\text{U}}$ and $d'^\dagger_{\omega_i,\text{U}}$ interchanging the labels I and II, where \footnote{Although convenient, the notation for the parameters $r_{\text{f},\omega_i}$ and $r_{\text{b},\omega_i}$ can sometimes be excessive in the long and complex expressions that will appear throughout this thesis. Consequently, we will drop the labels b,f or $\omega_i$ to ease notation whenever there can be no ambiguity or confusion.}
\begin{equation}\label{defr}
\tan r_{\text{f},\omega_i}=\exp(-\pi c{{\omega_i}}/{a}).
\end{equation}
Here $c_{{\omega_i},\sigma}$, $d_{{\omega_i},\sigma}$ represent the
annihilation operators of modes $\psi^\text{U}_{\omega_i,\sigma}$ and
$\bar
\psi^\text{M}_{\omega_i,\sigma}$ for particles and antiparticles respectively. 
The label $\text{d}$ in $r_{\text{f},\omega_i}$ has been introduced to indicate
Dirac field. The specific form for $\psi^\text{U}_{\omega_i,\sigma}$
and $\bar
\psi^\text{U}_{\omega_i,\sigma}$ as a linear combination of monochromatic
solutions of the Dirac equation is given later in section \ref{secferme}, and can be also seen, for instance, in
\cite{Jauregui,chinada} among many other references. Notice again that,
although we are denotating $a_{\omega_i,\text{U}},
c_{\omega_i,\sigma,\text{U}}, d_{\omega_i,\sigma,\text{U}}$ the operators associated
with Minkowskian Unruh modes, those modes are not monochromatic, but a
linear combination of monochromatic modes given by \eqref{modopsi} and
\eqref{modopsif}.

To discuss fundamental issues and not an specific
experiment, there is no reason to adhere to a specific basis. Specifically,
if we work in the bases \eqref{modopsi} and
\eqref{modopsif} for Minkowskian modes we do
not need to carry out the single mode approximation.

We will work in this basis in  part \ref{part1} of this thesis. This will be useful to extend and compare some of the results obtained here with previous works that made use of the single mode approximation. In part \ref{part2} we will discuss the problems of the single mode approximations and how to relax it, presenting the most general way to work with Unruh modes and the new and interesting results that appear when we go beyond a specific choice of Unruh modes.

\cleardoublepage

\part[Entanglement, statistics and the Unruh-Hawking effect]{Entanglement, Statistics and the Unruh-Hawking Effect}
\label{part1}


\section*{Discussion: Previous results on field entanglement, dimension and statistics}
\markboth{Discussion}{Previous results on field entanglement, dimension and statistics}
\addcontentsline{toc}{chapter}{Discussion: Previous results on field entanglement} 
From the pioneering works that started the analysis of field entanglement behaviour in non-inertial frames it has been shown \cite{Alicefalls,AlsingSchul} that the Unruh effect degrades the entanglement between accelerated partners, affecting all the quantum information tasks that they could perform. 

Fundamental differences were found in these first works between field entanglement of a scalar field \cite{Alicefalls} and a spinless fermionic field (Grassmann scalar) \cite{AlsingSchul}.  Specifically, it was discovered that, as an observer of an entangled state of the field accelerates, entanglement is completely degraded for the scalar case and, conversely, some degree of entanglement survives (even in the limit of infinite acceleration) for the spinless fermionic field\footnote{In literature prior to this thesis only Grassmann scalar fields were considered, this is to say, formal spinless Dirac fields, see Appendix \ref{appB}}.

Let us highlight the main result in \cite{AlsingSchul}. This work presented two observers (Alice and Rob), one of them
inertial and the other one undergoing a constant acceleration $a$. Alice and Rob are the observers of a bipartite quantum state of a spinless fermion field which is maximally entangled for the inertial observer, namely a state of the form
\begin{equation}
\frac{1}{\sqrt{2}}\left(\ket{0_\text{A}0_\text{R}}+\ket{1_\text{A}1_\text{R}}\right).
\end{equation}
As Rob accelerates the Unruh effect
will introduce degradation in the state as seen by Rob, impeding all the
quantum information tasks between both observers.

If we quantify this entanglement by means of negativity $\mathcal{N}$ (see section \ref{entangsec}) following that work\footnote{In \cite{AlsingSchul}  the authors use logarithmic negativity (See \cite{logneg})  instead of negativity. Both entanglement measures are equally valid and are, in fact, very simply related: $L_{\mathcal{N}}=\log_2(2\mathcal N+1)$.} one obtains that the dependence of the negativity with the acceleration gets the simple expression 
\begin{equation}
\mathcal{N}=\frac12\cos^2 r_{\text{f},i},
\end{equation}
where
\begin{equation}
\tan r_{\text{f},i}=\exp(-\pi c{{\omega_i}}/{a}).
\end{equation}
Here $\omega_i$ is the frequency of the mode considered. Notice that in the limit $a\rightarrow\infty$ (which means infinite Unruh temperature), $\mathcal{N}\rightarrow\frac14$, implying that some fermionic entanglement survives the infinite acceleration limit. This was a very unexpected fact, above all taking into account that for the same scenario but considering a bosonic field entanglement is rapidly lost as the accelerated observer increases his acceleration, so that in the limit in which the accelerated observer would observe an infinite Unruh temperature thermal bath\footnote{When observing the inertial vacuum state of the field, see section \ref{tue}.} all entanglement vanishes \cite{Alicefalls}.
 
These previous studies stated that this different behaviour was caused for the difference in the Hilbert space dimension for the different fields. Every different frequency mode for a bosonic field can be excited to an unbounded number of levels, in other words, the dimension of the Fock space for each mode is not bounded. The Fock basis in the fermionic field is, however, limited due to Pauli exclusion principle. For instance, for a spinless fermion field (as the ones considered in these previous works), every frequency is either in the vacuum state or in the excited state, double excitations are forbidden.  For a spin $1/2$ we can have for each frequency one particle or two particles with opposite spin projection quantum numbers, so occupation number is bounded by 2.

As one can intuitively associate the Unruh effect with the observation of thermal noise, it was argued in works prior to this thesis that bosonic fields have a broader margin for the Unruh noise to stochastically excite more levels and degrade entanglement (similar to a decoherence process) while the smaller dimension of the fermionic fields protected them from these random excitations. 

In this first part of the thesis we will study the relationship between field statistics and entanglement behaviour. We will  gain knowledge about the strong relationship between statistics and entanglement degradation step by step. First we will study for the first time spin $1/2$ fermionic fields. This will allow us to study new kinds of entangled states. We will see  that the Dirac field entanglement behaves in exactly the same entanglement degradation that the Grassmann scalar case analysed in previous literature \cite{AlsingSchul}.

Then, by means of a multimode analysis we will prove this behaviour of fermionic fields is universal. Namely, it is independent of  i) the spin of the fermionic field,  ii) the kind of maximally entangled state from which we start, and iii) the dimension of the Fock space of the field for every mode.

We will discover more fundamental differences between fermions and bosons and we will disprove both ways of the statement: fermionic survival is not dependent on dimension and bosonic entanglement disappearance happens even in bosonic settings with a finite dimensional Fock space.

These results together banish the sensible, yet incorrect, argument that linked Fock space dimension with entanglement behaviour in non-inertial frames.

The final chapter of this part I of the thesis will deal with a related but very different topic. The only known work \cite{schacross} previous to this thesis that studies entanglement degradation in the background of a Schwarzschild black hole did not analyse the dependence with the distance to the black hole horizon. Instead the entanglement degradation for observers placed in the asymptotically flat region of spacetime was dealt with.

A more interesting scenario would be to consider observers very close to the event horizon where the gravitational field is strong and the loss of information due to the presence of the horizon is much more intense, so that we can analyse entanglement and correlations as a function of the distance to the horizon.

At the end of this part I, we will rigorously show that tools coming from the constantly accelerated case can be used to study a setting where correlated pairs are shared by observers free-falling into a Schwarzschild black hole and observers resisting the gravitational pull at a finite distance from the event horizon.

\chapter[Spin $1/2$ fields and fermionic entanglement in non-inertial frames]{Spin $1/2$ fields and fermionic entanglement in non-inertial frames\footnote{J. Le\'on, E. Mart\'in-Mart\'inez. Phys. Rev. A, 80, 012314 (2009).}}\label{onehalf}

\markboth{Chapter 3. Spin $1/2$ fields and non-inertial fermionic entanglement}{\rightmark}


Previous work \cite{AlsingSchul} on Unruh effect for Dirac field mode entanglement does not consider the spin of the parties. This work considered an effective `Grassmann field' where only two occupation numbers $n=(0,1)$ are allowed for each mode. Higher values of $n$ are forbidden by Pauli exclusion principle. However, addressing the effect of Unruh decoherence on spin entanglement can only be done by incorporating the spin of the parties in the framework from the very beginning. As a consequence, occupation number $n=2$ is also allowed. This fact will affect occupation number entanglement which has to be reconsidered in this new setting. For this purpose, we will study here the case of two parties (Alice and Rob) sharing a general superposition of Dirac vacuum and all the possible one particle spin states for both Alice and Rob. Alice is in an inertial frame while Rob undergoes a constant acceleration $a$.

We will show that Rob --when he is accelerated respect to an inertial observer of the Dirac vacuum-- would observe a thermal distribution of fermionic spin $1/2$ particles due to Unruh effect \cite{Unruh}. Next, we will consider that Alice and Rob share spin Bell states in a Minkowski frame. Then, the case in which Alice and Rob share a superposition of the Dirac vacuum and a specific one particle state in a maximally entangled combination. In both cases we analyse the entanglement and mutual information in terms of Rob's acceleration $a$.

Finally, we will study the case when the information about spin is erased from our setting. For this we implement a method to consistently erase such information, keeping only the occupation number information. Here, entanglement is more degraded than in \cite{AlsingSchul} as we are erasing part of the correlations from our system. 

We show that, even in the limit of $a\rightarrow\infty$, some degree of entanglement is preserved due to Pauli exclusion principle. Then we analyse Unruh effect on a completely different class of maximally entangled states (like $\ket{00}+\ket{ss'}$ where $s$ and $s'$ are $z$ component of spin labels) comparing it with the spin Bell states. In section \ref{sec6} we show that the erasure of spin information, in order to investigate occupation number entanglement alone, requires considering total spin states for the bipartite system. 

\section{The setting}\label{sec2}

We consider a free massless Dirac field in a Minkowski frame expanded in terms of the positive (particle) and the negative (antiparticle) energy solutions of Dirac equation in Minkowskian coordinates, denoted $u^+_{\hat\omega,s,\text{M}}$ and $u^-_{\hat\omega,s,\text{M}}$ respectively:
\begin{equation}\label{field}
\phi=\sum_{s}\int d^3k\, (c_{\hat\omega,s,\text{M}}u^+_{\hat\omega,s,\text{M}}+d_{\hat\omega,s,\text{M}}^\dagger u^-_{\hat\omega,s,\text{M}}).
\end{equation}
Here, the subscript $\hat\omega$ denotes Minkowskian frequency and labels the modes of the same energy and $s=\{\uparrow ,\downarrow\}$ is the spin label that indicates spin-up or spin-down along the quantisation axis. $c_{\hat\omega,s,\text{M}}$ and $d_{\hat\omega,s,\text{M}}$ are respectively the annihilation operators for particles and antiparticles, and satisfy the usual anticommutation relations.

For each mode of frequency $\hat\omega$ and spin $s$ the positive and negative energy modes have the form
\begin{equation}\label{eq2}
u^\pm_{\hat\omega,s,\text{M}} =\frac{1}{\sqrt{2\pi \hat\omega}}v^\pm_s(\bm k) e^{\pm i(\bm k\cdot\bm x- \hat\omega t)},\end{equation}
where $v^\pm_s(\hat\omega)$ is a spinor satisfying the usual normalisation relations.

The modes are classified as particle or antiparticle respect to $\partial_t$ (Minkowski Killing vector directed to the future). The Minkowski vacuum state is defined by the tensor product of each frequency mode vacuum
\begin{equation}\label{vacua}\ket0_{\text{M}}=\bigotimes_{\hat\omega}\ket{0_{\hat\omega}}^+_{\text{M}}\ket{0_{\hat\omega}}_{\text{M}}^-\end{equation}
such that it is annihilated by $c_{\hat\omega,s,\text{M}}$ and $d_{\hat\omega,s,\text{M}}$ for all values of $s$.

We will consider the spin structure for each mode, and hence, the maximum occupation number is two. This introduces the following notation
\begin{equation}c^\dagger_{\hat\omega,s,\text{M}}c^\dagger_{\hat\omega,s',\text{M}}\ket0=\ket{ss'_{\hat\omega}}_\text{M}\delta_{s,{-s'}}.\end{equation}
If $s=s'$ the two particles state is not allowed due to Pauli exclusion principle, so our allowed Minkowski states for each mode of particle/antiparticle are
\begin{equation}\{{\ket{0_{\hat\omega}}_\text{M}}^\pm,{\ket{\uparrow_{\hat\omega}}_\text{M}}^\pm,{\ket{\downarrow_{\hat\omega}}_\text{M}}^\pm,{\ket{\pa_{\hat\omega}}_\text{M}}^\pm\},\end{equation}
where $\ket{\pa_{\hat\omega}}^+_\text{M}=c^\dagger_{\hat\omega,\uparrow,\text{M}}c^\dagger_{\hat\omega,\downarrow,\text{M}}\ket{0}_\text{I}$ denotes the spin pair of frequency $\hat\omega$.

To build Rob's field excitations we will work  in the basis \eqref{modopsif} of solutions of the Dirac equation in Minkowski coordinates, whose particularities were explained in section \ref{probexcitations} and such that they correspond to a monochromatic mode when transformed into the Rindler basis. The reason of this is double, on the one hand as we are looking for fundamental behaviour and not the results of a specific experiment there is no reason to adhere to a specific basis, and on the other hand, it is in this basis when we recover the so-called single mode approximation, allowing us to compare our results with previous literature.

Particles and antiparticles will be classified with respect to the future-directed timelike Killing vector in each region. In region I the future-directed Killing vector is
\begin{equation}\label{KillingI}
\partial_\tau^I=\frac{\partial  t}{\partial \tau}\partial_{t}+\frac{\partial x}{\partial \tau}\partial_{x}=a(x\partial_{t}+t\partial_{x}),
\end{equation}
whereas in region II the future-directed Killing vector is $\partial_\tau^{II}=-\partial_\tau^{I}$.

Let us denote $(c_{\omega,s,\text{I}},c^{\dagger}_{\omega,s,\text{I}})$ the particle annihilation and creation operators in region I and $(d_{\omega,s,\text{I}},d^{\dagger}_{\omega,s,\text{I}})$ the corresponding antiparticle operators. Analogously we define $(c_{\omega,s,\text{II}},c^{\dagger}_{\omega,s,\text{II}}, d_{\omega,s,\text{II}},d_{\omega,s,\text{II}}^\dagger)$ the particle/antiparticle operators in region II.

These operators satisfy the usual anticommutation relations $\{c_{\omega,s,\Sigma},c^\dagger_{\omega',s',\Sigma'}\}=\delta_{\Sigma\Sigma'}\delta_{\omega\omega'}\delta_{ss'}$ where the label $\Sigma$ denotes the Rindler region of the operator $\Sigma=\{\text{I},\text{II}\}$. All other anticommutators are zero. That includes the anticommutators between operators in different regions of the Rindler spacetime.

We can relate Minkowskian Unruh modes $\Psi^\text{U}_{\omega,s}$ and Rindler creation and annihilation operators by taking appropriate inner products \cite{Takagi,Jauregui,Birrell,AlsingSchul} as explained in section \ref{probexcitations}. We recall the Bogoliubov coefficients expression \eqref{Bogoferm2}
\begin{eqnarray}\label{Bogoferm3}
\nonumber c_{{\omega_i},\sigma,\text{U}}&=&\cos{r_{\text{d},i}}\,c_{{\omega_i},
\sigma,\text{I}}-\sin r_{\text{d},i}\,d^\dagger_{{\omega_i},-\sigma,\text{II}},\\*
d_{{\omega_i},\sigma,\text{U}}^\dagger&=&\cos{r_{\text{d},i}}\,d^\dagger_{
{\omega_i},\sigma,\text{II}}+\sin r_{\text{d},i}\,c_{{\omega_i},-\sigma,\text{I}},
\end{eqnarray}
and analogously for $c'_{\omega_i}$ and $d'^\dagger_{\omega_i}$ interchanging the labels I and II, where
\begin{equation}\label{defr}
\tan r_{\text{d},i}=\exp(-\pi {c{\omega_i}}/{a}).
\end{equation}

Notice from Bogoliubov transformations \eqref{Bogoferm2} that the Unruh mode particle annihilator $c_{\omega_i,\sigma,\text{U}}$ transforms into a Rindler particle annihilator of frequency $\omega_i$ and spin $s$ in region I and an antiparticle creator of frequency $\omega_i$ and spin $-\sigma$ in region II.

\subsection{Unruh effect for fermion fields of spin $1/2$}\label{sec42m}

Now that we have the relationships between the creation and annihilation operators in the Minkowski and Rindler bases, we can obtain the expression of the Minkowski vacuum state for each mode $\ket{0_{\hat\omega}}_\text{M}$ in the Rindler Fock basis. For notation simplicity, we will drop the $\hat \omega$ and $\omega$ label in operators/states when it does not give any relevant information, but we will continue writing the spin label.

Let us introduce some notation for our states. We will denote with a subscript outside the kets if the mode state is referred to region I or II of the Rindler spacetime. The $\pm$ label of particle/antiparticle will be omitted throughout the chapter because, for the cases considered, a ket referred to Minkowski spacetime or Rindler's region I will always denote particle states and a ket referred to region II will always denote antiparticle states.

Note that here, and in the whole part I of this thesis we will only make use of the `unprimed' particle sector of the Hilbert space in the Minkowskian basis (see section \ref{probexcitations}), this means that Minkowski excitations will always be particles and Rindler excitations will be always particles in the Region I basis and antiparticles in the Region II basis (due to the time reversal anti-symmetry of both regions). From now on and for all this part we drop the label $\pm$ notating particles and antiparticles in the kets. The complete analysis of the antiparticle sector behaviour is dealt with in section \ref{secferme}.

Inside the ket we will write the spin state of the modes as follows
\begin{equation}\label{notation1}
\ket{s}_\text{I}=c^\dagger_{s,\text{I}}\ket{0}_\text{I},\qquad \ket{s}_{\text{II}}=d^\dagger_{s,\text{II}}\ket{0}_{\text{II}},
\end{equation}
which will denote a particle state in region I and an antiparticle state in region II respectively, both with spin $s$.

We will use the following definitions for the kets associated to particle pairs
\begin{eqnarray}\label{notation2}
\nonumber \ket{\pa}_\text{I}&=&c^\dagger_{\uparrow\text{I}}c^\dagger_{\downarrow\text{I}}\ket{0}_\text{I}=-c^\dagger_{\downarrow\text{I}}c^\dagger_{\uparrow\text{I}}\ket{0}_\text{I},\\*
\ket{\pa}_\text{II}&=&d^\dagger_{\uparrow\text{II}}d^\dagger_{\downarrow\text{II}}\ket{0}_\text{II}=-d^\dagger_{\downarrow\text{II}}d^\dagger_{\uparrow\text{II}}\ket{0}_\text{II},
\end{eqnarray}
and, being consistent with the anticommutation relations for different Rindler wedges,
\begin{equation}\label{notation3}
\nonumber \ket{s}_\text{I}\ket{s'}_\text{II}=c^\dagger_{s,\text{I}}d^\dagger_{s',\text{II}}\ket{0}_{I}\ket{0}_\text{II}=-d^\dagger_{s',\text{II}}c^\dagger_{s,\text{I}}\ket{0}_{I}\ket{0}_\text{II},
\end{equation}
\begin{equation}\label{notation4}
d^\dagger_{s',\text{II}}\biket{s}{0}=-\biket{s}{s'}.
\end{equation}

The Bogoliubov transformation \eqref{Bogoferm2} is a fermionic two-modes squeezing transformation for each $\omega_i$, indeed
\begin{equation}\label{upinthesky}\left(\!\begin{array}{c}
c_{\omega_i,s,\text{U}}\\
d^\dagger_{\omega_i,s,\text{U}}
\end{array}\!\right)=S\left(\!\begin{array}{c}
c_{\omega_i,s,\text{I}}\\
d^\dagger_{\omega_i,-s,\text{II}}
\end{array}\!\right)S^\dagger,\end{equation}
where
\begin{equation}\label{squeez}
S=\exp\left[r\left(c_{\omega_i,s,\text{I}}^\dagger\, d_{\omega_i,-s,\text{II}}+c_{\omega_i,s,\text{I}}\, d_{\omega_i,-s,\text{II}}^\dagger \right)\right].
\end{equation}
Note that, for notational convenience given the length and complexity of the expressions, we will drop the labels $\omega_i$ and d from $r$ when the distinction between different frequencies and fermions and bosons is not required to be explicitly shown.

Now, given \eqref{upinthesky} and analogously to \cite{Alicefalls,AlsingSchul}, it is reasonable to postulate that for every $\omega_i$ the Minkowski vacuum is a Rindler two-mode particles/antiparticles squeezed state with opposite spin and momentum states in I and II.

Considering that the modes have spin, occupation number is allowed to be 2 for each frequency, being higher occupation numbers forbidden by Pauli exclusion principle. We need to compute the coefficients of the squeezed state\footnote{One must be careful with the tensor product structure of the fermionic vacuum. Here we present only the `unprimed sector' of the vacuum state. While this is simpler and perfectly valid for the particular choice of inertial Unruh modes presented here, the whole vacuum must be considered in more general scenarios when we consider modes that combine `primed' and `unprimed' Unruh modes. This fact and its consequences are thoroughly analysed in section \ref{secferme}.}
\begin{equation}\label{vacuum0}
 \ket{0}=V\biket{0}{0}+A\biket{\uparrow}{\downarrow}+B\biket{\downarrow}{\uparrow}+ C\biket{\pa}{\pa}.
\end{equation}

To obtain the values of the coefficients $V,A,B,C$ we demand that the Minkowski vacuum is annihilated by the particle annihilator, $c_{\hat\omega, s,\text{M}}\ket0=0$ for all frequencies. As Unruh modes are purely positive frequency combination of Minkowskian modes, this is equivalent to demand that the vacuum is annihilated by all the particle annihilators of Unruh modes \eqref{modopsi}: $c_{\omega, s,\text{U}}\ket0=0$. Translating this into the Rindler basis we have
\begin{equation}\label{annihil2}
\left[\cos r\,c_{s,\text{I}}-\sin r\,d^\dagger_{-s,\text{II}}\right]\left[V\biket{0}{0}+A\biket{\uparrow}{\downarrow}+B\biket{\downarrow}{\uparrow}+ C\biket{\pa}{\pa}\right]=0.
\end{equation}
This equation gives 4 conditions (two for each value of $s$), although only 3 of them are independent:
\begin{equation}
\left.\begin{array}{lcr}
A\cos r - V\sin r&=&0\\
C\cos r - B\sin r&=&0\\
B\cos r - V\sin r&=&0\\
C\cos r - A\sin r&=&0\\
\end{array}\right\}\Rightarrow \begin{array}{ll}A=B=V \tan r\\
C=V\tan^2 r
\end{array}.
\end{equation}
To fix $V$ we impose the normalisation relation for each field mode $\braket{0}{0}=1\Rightarrow |V|^2=1-|A|^2-|B|^2-|C|^2$, this yields the values of the vacuum coefficients:
\begin{equation}\label{vaccoef}
\begin{array}{lcl}
V&=&\cos^2 r,\\
A&=&\sin r\,\cos r, \\
B&=&\sin r\,\cos r,\\
C&=&\sin^2 r.\\
\end{array}
\end{equation}

So, finally, the Minkowski vacuum state in the Rindler basis reads as follows
\begin{equation}\label{vacuum}
 \ket{0}=\cos^2 r\,\biket{0}{0}+\sin r\,\cos r\left(\biket{\uparrow}{\downarrow}+\biket{\downarrow}{\uparrow}\right)+\sin^2 r\,\biket{\pa}{\pa},
\end{equation}
where, to simplify notation, we drop the label M for the Minkowski vacuum which is annihilated by both Unruh and monochromatic Minkowskian modes particle and antiparticle annihilators.

Now we have to know how the excited states (of spin $s$) look like in the Rindler basis. This can be readily obtained by applying the Unruh particle creation operator to the vacuum state $\ket{s}_\text{U}=c^\dagger_{s,\text{U}}\ket0$, and translating it into the Rindler basis:
\begin{align}\label{onepart1}
 \nonumber\ket{s}_\text{U}&=\left[\cos r c^\dagger_{s,\text{I}}-\sin r d_{-s,\text{II}}\right]\left[\cos^2 r\biket{0}{0}+\sin r\cos r\left(\biket{\uparrow}{\downarrow}+\biket{\downarrow}{\uparrow}\right)\right.\\*
 &\left.+\sin^2 r\biket{\pa}{\pa}\right],
\end{align}
which gives
\begin{eqnarray}\label{onepart2a}
\nonumber\ket\uparrow_\text{U}&=&\cos r \biket{\uparrow}{0}+e^{i\phi}\sin r\biket{\pa}{\uparrow},\\*
\ket\downarrow_\text{U}&=&\cos r \biket{\downarrow}{0}-e^{i\phi}\sin r\biket{\pa}{\downarrow}.
\end{eqnarray}

Now, since Rob is experiencing a uniform acceleration, he will not be able to access to field modes in the causally disconnected region II, hence, Rob must trace over that unobservable region. Specifically, when Rob is in region I of Rindler spacetime and Alice observes the vacuum state, Rob can only observe a non-pure partial state given by $\rho_\text{R}=\tr_\text{II}\left(\ket{0}\bra{0}\right)$, which yields
\begin{equation}\label{partialvacuum}
\rho_\text{R}=\cos^4 r\ket{0}_\text{I}\!\!\bra{0}+\sin^2 r\,\cos^2 r\left(\ket{\uparrow}_\text{I}\!\!\bra{\uparrow}+\ket{\downarrow}_\text{I}\!\!\bra{\downarrow}\right)+\sin^4 r \ket{\pa}_\text{I}\!\bra{\pa}.
\end{equation}
Thus, while Alice would observe the vacuum state of mode $k$, Rob would observe a certain statistical distribution of particles. The expected value of Rob's number operator on the Minkowski vacuum state is given by
\begin{equation}
\langle N_\text{R}\rangle=\ematriz{0}{N_\text{R}}{0}=\tr_{I,II}\left(N_\text{R}\proj{0}{0}\right)=\tr_{I}\left(N_\text{R}\rho_\text{R}\right)=\tr_{I}\left[\left(c_{\uparrow\text{I}}^\dagger c_{\uparrow\text{I}}+c_{\downarrow\text{I}}^\dagger c_{\downarrow\text{I}}\right)\rho_\text{R}\right].
\end{equation}
Substituting the expression \eqref{partialvacuum} we obtain
\begin{equation}
\langle N_\text{R}\rangle=2\sin^2 r,
\end{equation}
which, using \eqref{defr}, reads 
\begin{equation}\label{Unruh}
\langle N \rangle=2\frac{1}{e^{2\pi\omega c/a}+1}=2\frac{1}{e^{\hslash\omega/K_BT}+1},
\end{equation}
where $k_B$ is the Boltzmann's constant and
\begin{equation}
T=\frac{\hslash\, a}{2\pi k_B c},
\end{equation}
is the Unruh temperature.

Equation \eqref{Unruh} is an alternative derivation of the well known Unruh effect \cite{DaviesUnr,Unruh} for a fermionic field. We have shown that an uniformly accelerated observer in region I detects a thermal Fermi-Dirac distribution when he observes the Minkowski vacuum. The factor 2 appears due to the degeneracy factor $2S+1$.

\section{Spin entanglement with an accelerated partner}\label{sec5}

In previous works \cite{Alicefalls,AlsingSchul} it was studied how Unruh decoherence affects occupation number entanglement in bipartite states of the type\footnote{Since Alice is an inertial observer, it is not relevant if Alice's excitations are monochromatic or `unprimed' Unruh modes. In either case all the results are exactly the same. To keep the notation as simple as possible we will label Alice's mode with the label `U' used for Unruh modes in this Part I of the thesis. See chapter \ref{sma} for further details.}
\begin{equation}\label{alsingst}
\ket{\Psi}=\frac{1}{\sqrt2}(\ket{0}\ket{0}+\ket{1}_\text{U}\ket{1}_\text{U})
\end{equation}of a  Grassmann scalar field, barring any reference to the spin of the field modes. The figures inside the kets represent occupation number of Alice and Rob modes respectively.

Here, where we have included the spin structure of each mode in our setting from the very beginning, it is possible to study the effects of acceleration in spin entanglement degradation, which is different from the mere occupation number entanglement.

Specifically, with this setting we are able to study how acceleration affects the entanglement of spin Bell states when Alice and Rob share a maximally entangled spin state and Rob accelerates. In the Minkowski basis, these 4 maximally entangled states take the form
\begin{equation}\label{Bellstates}\ket\psi^\pm=\frac{1}{\sqrt2}\left[\ket{\uparrow_\text{A}}_\text{U}\ket{\downarrow_\text{R}}_\text{U}\pm\ket{\downarrow_\text{A}}_\text{U},\ket{\uparrow_\text{R}}_\text{U}\right]\qquad\ket\phi^\pm=\frac{1}{\sqrt2}\left[\ket{\uparrow_\text{A}}_\text{U}\ket{\uparrow_\text{R}}_\text{U}\pm\ket{\downarrow_\text{A}}_\text{U},\ket{\downarrow_\text{R}}_\text{U}\right].\end{equation}

First of all, we build a general bipartite state that could be somehow analogous to state \eqref{alsingst} studied in \cite{AlsingSchul}. This state will be a superposition of the vacuum and all the possible 1-particle bipartite states for Alice and Rob.
\begin{equation}\label{genstate}
\ket{\Psi}=\mu\ket{0}\ket{0}+\alpha\ket{\uparrow}_\text{U}\ket{\uparrow}_\text{U}+\beta\ket{\uparrow}_\text{U}\ket{\downarrow}_\text{U}+ \gamma\ket{\downarrow}_\text{U}\ket{\uparrow}_\text{U}+\delta\ket{\downarrow}_\text{U}\ket{\downarrow}_\text{U},
\end{equation}
with $\mu=\sqrt{1-|\alpha|^2-|\beta|^2-|\gamma|^2-|\delta|^2}$.

This state generalises the Bell spin states (for instance, we have $\ket{\phi^+}$ choosing $\alpha=\delta=1/\sqrt2$) or a occupation number entangled state (for instance choosing $\alpha=\mu=1/\sqrt2$).  We will be able to deal with two different and interesting problems: 1. Studying the Unruh effect on  spin entangled states and 2. Investigating the impact of considering the spin structure of the fermion on the occupation number entanglement and its Unruh entanglement degradation.

The density matrix in the Minkowskian Unruh basis for the state \eqref{genstate} is
\begin{align}\label{Minkowdens}
\nonumber\rho^M&=\mu^2\proj{00}{00}+\mu\alpha^*\proj{00}{\uparrow\uparrow}+\mu\beta^*\proj{00}{\uparrow\downarrow}+\mu\gamma^*\proj{00}{\downarrow\uparrow}+\mu\delta^*\proj{00}{\downarrow\downarrow}\\*
\nonumber &+|\alpha|^2\proj{\uparrow\uparrow}{\uparrow\uparrow}+\alpha\beta^*\proj{\uparrow\uparrow}{\uparrow\downarrow}+\alpha\gamma^*\proj{\uparrow\uparrow}{\downarrow\uparrow}+\alpha\delta^*\proj{\uparrow\uparrow}{\downarrow\downarrow}+|\beta|^2\proj{\uparrow\downarrow}{\uparrow\downarrow}\\*
&+\beta\gamma^*\proj{\uparrow\downarrow}{\downarrow\uparrow}\!+\!\beta\delta^*\proj{\uparrow\downarrow}{\downarrow\downarrow}\!+\!|\gamma|^2\proj{\downarrow\uparrow}{\downarrow\uparrow}\!+\!\gamma\delta^*\proj{\downarrow\uparrow}{\downarrow\downarrow}\!+\!|\delta|^2\proj{\downarrow\downarrow}{\downarrow\downarrow}\!+\! (\text{H.c.})_{_{\substack{\text{non-}\\\text{diag.}}}},
\end{align}
where $(\text{H.c.})_{\text{non-diag.}}$ means non-diagonal Hermitian conjugate, and represents the Hermitian conjugate only for the non-diagonal elements.

Again, as Rob can only carry out measurements in his accessible Fock basis (made of only region I modes) we need to compute each term of \eqref{Minkowdens} in the Rindler basis and trace over the unobserved region II.

Computing the density matrix, taking into account that Rob is constrained to region I of Rindler spacetime, requires to rewrite Rob's mode in the Rindler basis and to trace over the unobservable Rindler's region II. It is useful to compute first the trace over II on all the operators that compose \eqref{Minkowdens}. Using equation \eqref{vacuum} we have
\begin{equation}
\proj{00}{00}\!=\!\ket{0_\text{A}}\left[\cos^2 r\biket{0}{0}\!+\!\sin r\cos r \left(\biket{\uparrow}{\downarrow}\!+\!\biket{\downarrow}{\uparrow}\right)\!+\!\sin^2 r\biket{\pa}{\pa}\right]\otimes \text{H.c.},
\end{equation}
where the label A indicates Alice's subsystem. Tracing over II:
\begin{equation}\label{trazapa1}
\tr_\text{II}\proj{00}{00}=\cos^4 r\proj{00}{00}+\sin^2r\,\cos^2r\left(\proj{0\uparrow}{0\uparrow}+\proj{0\downarrow}{0\downarrow}\right)+\sin^4r\proj{0\pa}{0\pa},
\end{equation}
where notation is different in each side of the equality: bras and kets in l.h.s. are in the Minkowskian basis $\ket{ss'}=\ket{s_\text{U}}\ket{s'_\text{U}}$ and in the r.h.s. they are referred to Alice's mode in the Minkowskian basis and Rob's mode in the Rindler's region I Fock basis $\ket{ss'}=\ket{s_\text{U}}\ket{s'_\text{R}}_\text{I}$.

In the same way, using expressions \eqref{vacuum}, \eqref{onepart2a} we have
\begin{align}
\nonumber\proj{00}{ss'}=&\ket{0_\text{A}}\left[\cos^2 r\,\biket{0}{0}+\sin r\,\cos r\left(\biket{\uparrow}{\downarrow}+\biket{\downarrow}{\uparrow}\right)+\sin^2 r\biket{\pa}{\pa}\right]\\*&\bra{s_\text{A}}\left(\cos r\,{\bra{s'}_\text{I}}\,{\bra{0}_\text{II}}+ \varepsilon\, \sin r\,{\bra{\pa}_\text{I}}\,{\bra{s'}_\text{II}}\right),
\end{align}
where $\varepsilon=1$ if $s=\uparrow$ and $\varepsilon=-1$ if $s=\downarrow$. Now, tracing over II
\begin{equation}\label{trazapa2}
\tr_\text{II}\proj{00}{ss'}=\cos^3 r\,\proj{00}{ss'}+\sin^2r\,\cos r\left(\delta_{s'\uparrow}\proj{0\downarrow}{s\pa}-\delta_{s'\downarrow}\proj{0\uparrow}{s\pa}\right),
\end{equation}
with the same notation used in \eqref{trazapa1}.

Similarly, using expression \eqref{onepart2a} we get
\begin{align}
\nonumber\proj{s_1s_2}{s_3s_4}&=\ket{{s_{1}}_\text{A}}\left[\cos r\,\biket{s_2}{0}+\varepsilon_2e^{-i\phi}\sin r\biket{\pa}{s_2}\right]\bra{{s_{3}}_\text{A}}\left[\cos r\,{\bra{s_4}_\text{I}}{\bra{0}_\text{II}}\right.\\*
&\left.+\,\varepsilon_4\sin r\,{\bra{\pa}_\text{I}}{\bra{s_4}_\text{II}}\right],
\end{align}
and tracing over II gives
\begin{equation}\label{trazapa3}
\tr_\text{II}\proj{s_1s_2}{s_3s_4}=\cos^2 r\,\proj{s_1s_2}{s_3s_4}+\delta_{s_2s_4}\sin^2r\proj{s_1\pa}{s_3\pa}.\\*
\end{equation}
Again, notation here is the same than in \eqref{trazapa1}.

 Using \eqref{trazapa1}, \eqref{trazapa2}, \eqref{trazapa3} we can easily compute the density matrix for Alice and Rob from \eqref{Minkowdens} since $\rho_{AR}=\tr_\text{II}\rho_M$, resulting in the long expression
\begin{align}\label{generaldensmat}
\nonumber\rho_{AR}&\!=\!\mu^2\Big[\!\cos^4r\proj{00}{00}\!+\!\sin^2r\cos^2r\left(\proj{0\!\uparrow}{0\!\uparrow}\!+\!\proj{0\!\downarrow}{0\!\downarrow}\right)\!+\!\sin^4r\proj{0\pa}{0\pa}\!\Big]\!+\!\mu\cos^3\!r\\*
\nonumber&\times\Big[\alpha^*\proj{00}{\uparrow\uparrow}\!+\!\beta^*\proj{00}{\uparrow\downarrow}\!+\!\gamma^*\proj{00}{\downarrow\uparrow}\!+\!\delta^*\proj{00}{\downarrow\downarrow}\Big]\!+\!\mu\sin^2r\cos r\Big[\alpha^*\proj{0\downarrow}{\uparrow\pa}\\*
\nonumber&-\beta^*\!\proj{0\!\uparrow}{\uparrow\!\pa}\!+\!\gamma^*\!\proj{0\!\downarrow}{\downarrow\!\pa}\!-\!\delta^*\!\proj{0\!\uparrow}{\downarrow\!\pa}\Big]\!+\!\cos^2 r\Big[|\alpha|^2\proj{\uparrow\uparrow}{\uparrow\uparrow}\!+\!\alpha\beta^*\proj{\uparrow\uparrow}{\uparrow\downarrow}\!+\!\alpha\gamma^*\\*
\nonumber&\times\proj{\uparrow\uparrow}{\downarrow\uparrow}\!+\!\alpha\delta^*\proj{\uparrow\uparrow}{\downarrow\downarrow}\!+\!|\beta|^2\proj{\uparrow\downarrow}{\uparrow\downarrow}\!+\!\beta\gamma^*\proj{\uparrow\downarrow}{\downarrow\uparrow}\!+\!\beta\delta^*\proj{\uparrow\downarrow}{\downarrow\downarrow}\!+\!|\gamma|^2\proj{\downarrow\uparrow}{\downarrow\uparrow}\\*
\nonumber&+\gamma\delta^*\proj{\downarrow\uparrow}{\downarrow\downarrow}+|\delta|^2\proj{\downarrow\downarrow}{\downarrow\downarrow}\Big]+\sin^2r\Big[\left(|\alpha|^2+|\beta|^2\right)\proj{\uparrow\pa}{\uparrow\pa}+\left(|\gamma|^2+|\delta|^2\right)\\*
&\times\proj{\downarrow\pa}{\downarrow\pa}+\left(\alpha\gamma^*+\beta\delta^*\right)\proj{\uparrow\pa}{\downarrow\pa}\Big]+(\text{H.c.})_{_{\substack{\text{non-}\\\text{diag.}}}}.
\end{align}
Here the notation is the same than in the r.h.s. of \eqref{trazapa1}. We see that the state, which in the Minkowski basis is pure, becomes mixed when the observer Rob is accelerated due to the causal disconnection of the accelerated observer with part of the spacetime.

Equation \eqref{generaldensmat} will be our starting point, from which we will study different entanglement settings and how Unruh effect affects them.

We will now compute how acceleration affects the entanglement of spin Bell states \eqref{Bellstates} when Alice and Rob share a maximally entangled spin state and Rob accelerates. A specific choice of coefficients in \eqref{genstate} gives us these states, namely
\begin{align}
\ket{\phi^\pm}\Rightarrow \alpha=\pm\delta=\frac{1}{\sqrt{2}},\\*
\ket{\psi^\pm}\Rightarrow \beta=\pm\gamma=\frac{1}{\sqrt{2}},
\end{align}
and all the other coefficients equal to zero. For such states in the Minkowski basis, the density matrix of Alice and Rob when he undergoes an acceleration $a$ is obtained from \eqref{generaldensmat}:
\begin{equation}\label{Phibell}
\rho^{\phi^\pm}_{AR}\!=\!\frac12\Big[\cos^2 r\Big(\proj{\uparrow\uparrow}{\uparrow\uparrow}\pm\proj{\uparrow\uparrow}{\downarrow\downarrow}\pm\proj{\downarrow\downarrow}{\uparrow\uparrow}\!+\!\proj{\downarrow\downarrow}{\downarrow\downarrow}\Big)+\sin^2 r\Big(\!\proj{\uparrow\!\pa}{\uparrow\!\pa}+\proj{\downarrow\!\pa}{\downarrow\!\pa}\!\Big)\Big],
\end{equation}
\begin{equation}\label{Psibell}
\rho^{\psi^\pm}_{AR}\!=\!\frac12\Big[\cos^2 r\Big(\proj{\uparrow\downarrow}{\uparrow\downarrow}\pm\proj{\uparrow\downarrow}{\downarrow\uparrow}\pm\proj{\downarrow\uparrow}{\uparrow\downarrow}\!+\!\proj{\downarrow\uparrow}{\downarrow\uparrow}\Big)+\sin^2 r\Big(\!\proj{\uparrow\!\pa}{\uparrow\!\pa}+\proj{\downarrow\!\pa}{\downarrow\!\pa}\!\Big)\Big].
\end{equation}

Notice that, in this case, Rob has a qutrit, since for his mode he could have three different possible orthogonal states: particle spin-up, particle spin-down and particle pair. This is an important difference with the spinless case: the Hilbert space dimension of the accelerated observer increases when he is accelerating, what was a qubit in the Minkowskian basis is now a qutrit in the Rindler basis for Rob.

To characterise its entanglement we will use the negativity \eqref{negativitydef}. To compute it we need the partial transpose density matrices (defined in \eqref{ptranspdef}) of the states \eqref{Phibell} and \eqref{Psibell}
\begin{equation}
\rho^{\phi^\pm pT}_{AR}\!\!\!=\!\frac12\Big[\cos^2 r\Big(\proj{\uparrow\uparrow}{\uparrow\uparrow}\pm\proj{\uparrow\downarrow}{\downarrow\uparrow}\pm\proj{\downarrow\uparrow}{\uparrow\downarrow}\!+\!\proj{\downarrow\downarrow}{\downarrow\downarrow}\Big)+\sin^2 r\Big(\!\proj{\uparrow\!\pa}{\uparrow\!\pa}+\proj{\downarrow\!\pa}{\downarrow\!\pa}\!\Big)\Big],
\end{equation}
\begin{equation}
\rho^{\psi^\pm pT}_{AR}\!\!\!=\!\frac12\Big[\cos^2 r\Big(\proj{\uparrow\downarrow}{\uparrow\downarrow}\pm\proj{\uparrow\uparrow}{\downarrow\downarrow}\pm\proj{\downarrow\downarrow}{\uparrow\uparrow}\!+\!\proj{\downarrow\uparrow}{\downarrow\uparrow}\Big)+\sin^2 r\Big(\!\proj{\uparrow\!\pa}{\uparrow\!\pa}+\proj{\downarrow\!\pa}{\downarrow\!\pa}\!\Big)\Big].
\end{equation}
We can write $\rho^{\phi^\pm pT}_{AR}$ matricially in the basis $\left\{\ket{\uparrow\uparrow},\ket{\uparrow\downarrow},\ket{\downarrow\uparrow},\ket{\downarrow\downarrow},\ket{\uparrow\pa},\ket{\downarrow\pa}\right\}$ and $\rho^{\psi^\pm pT}_{AR}$ in the basis $\left\{\ket{\uparrow\downarrow},\ket{\uparrow\uparrow},\ket{\downarrow\downarrow},\ket{\downarrow\uparrow},\ket{\uparrow\pa},\ket{\downarrow\pa}\right\}$. Both have the same expression
\begin{equation}
\frac12\left(\!\begin{array}{cccccc}
\cos^2 r & 0 &0 &0 & 0 & 0\\
0&0 &\pm\cos^2 r &0 & 0 & 0 \\
0  &\pm\cos^2 r  &0 &0 &0 &0 \\
0  & 0 &0 &\cos^2 r &0 &0 \\
0 & 0 &0&0& \sin^2 r &0 \\
0& 0& 0&0&0& \sin^2 r \\
\end{array}\!\right).
\end{equation}
Therefore the four Bell states will have the same eigenvalues. The only negative eigenvalue is
\begin{equation}
\lambda_-=-\frac12\cos^2r.
\end{equation}

Since $r=\arctan\left(e^{-\pi\frac{\omega c}{a}}\right)$ $a\rightarrow0\Rightarrow r\rightarrow 0$ and $a\rightarrow\infty\Rightarrow r\rightarrow \pi/4$ so that $\lambda_6$ is negative for all values of the acceleration. This implies, using Peres criterion \cite{PeresCriterion}, that the spin Bell states will be always entangled even in the limit of infinite acceleration.

We can readily evaluate the entanglement at the limits $a\rightarrow0$ and $a\rightarrow\infty$ if we compute the negativity for these states, which have the trivial form
\begin{equation}
\mathcal{N}(r)=\frac12\cos^2 r.
\end{equation}
In the limit $a\rightarrow0$ we obtain $\mathcal{N}=\dfrac12$ which is an expected result since $a\rightarrow0$ is the inertial limit.

However, in the limit $a\rightarrow\infty$ we obtain $\mathcal{N}=\frac{1}{4}$, which implies that spin entanglement degrades due to the Unruh effect up to a certain limit.

In \cite{AlsingSchul} it is discussed (for occupation number entangled states of a spinless fermion field) that Pauli exclusion principle protects the occupation number entanglement from degrading, and some degree of entanglement is preserved even at the limit $a\rightarrow\infty$. The striking result is that here we have obtained a mathematically identical result for the spin Bell states. We will discuss later how this result, along with many others presented in this thesis, will disprove the statement that the finite dimension of the Fock space is responsible for this entanglement survival in the $a\rightarrow\infty$ limit.

\section{Vacuum and 1-particle entangled superpositions}

Next, we will study the case in which Alice and Rob share a different class of maximally entangled state, we consider that in the Minkowski Unruh basis we have
\begin{equation}\label{minkstate2}
\ket{\Psi}=\frac{1}{\sqrt2}\left(\ket{0_\text{A}}\ket{0_\text{R}}+\ket{\uparrow_\text{A}}_\text{U}\ket{\downarrow_\text{R}}_\text{U}\right),
\end{equation}
which is a maximally entangled state that includes occupation number entanglement along with spin. We study this kind of states as a first analog to the state considered in previous literature \eqref{alsingst}. This state corresponds to the choice
\begin{align}\label{coef22}
\beta&=\mu=\frac{1}{\sqrt2},\\*
\alpha&=\gamma=\delta=0
\end{align}
in equation \eqref{generaldensmat}. The density matrix of such a state is
\begin{align}\label{minkstate2R}
\nonumber\rho&=\frac12\Big[\cos^4r\proj{00}{00}+\sin^2r\cos^2r\left(\proj{0\!\uparrow}{0\!\uparrow}+\proj{0\!\downarrow}{0\!\downarrow}\right)+\sin^4r\proj{0\pa}{0\pa}\\*
&+\cos^3r\left(\proj{00}{\uparrow\downarrow}+\proj{\uparrow\downarrow}{00}\right)-\sin^2r\cos r\left(\proj{0\!\uparrow}{\uparrow\!\pa}+\proj{\uparrow\!\pa}{0\!\uparrow}\right)+\cos^2r\proj{\uparrow\downarrow}{\uparrow\downarrow}\nonumber\\*
&+\sin^2r\proj{\uparrow\!\pa}{\uparrow\!\pa}\Big].
\end{align}
Notice the significant difference from the Bell spin states; considering that Rob accelerates means that, this time, Alice has a qubit and Rob has a four dimensional quantum system. Since in Rindler coordinates the state \eqref{minkstate2R} is qualitatively very different from the Minkowski Bell states \eqref{Phibell}, \eqref{Psibell}, it is therefore worthwhile to study its entanglement degradation as Rob accelerates.

The partial transpose of \eqref{minkstate2R} is
\begin{align}\label{minkstate2R2}
\nonumber\rho^{pT}&=\frac12\Big[\cos^4r\proj{00}{00}+\sin^2r\cos^2r\left(\proj{0\uparrow}{0\uparrow}+\proj{0\downarrow}{0\downarrow}\right)+\sin^4r\proj{0\pa}{0\pa}\\*
\nonumber&+\cos^3r\left(\proj{0\!\downarrow}{\uparrow\!0}+\proj{\uparrow\!0}{0\!\downarrow}\right)-\sin^2r\,\cos r\left(\proj{0\pa}{\uparrow\uparrow}+\proj{\uparrow\uparrow}{0\pa}\right)\!+\!\cos^2r\proj{\uparrow\downarrow}{\uparrow\downarrow}\\*
&+\sin^2r\proj{\uparrow\!\pa}{\uparrow\!\pa}\Big],
\end{align}
which is a $8\times8$ block-diagonal matrix. It has two non-positive eigenvalues
\begin{align}\label{eig2}
\nonumber\lambda_{1}&=\frac{1}{4}\left(\sin^2r\cos^2r-\sqrt{\sin^4r\cos^4r+4\cos^6r}\right)\\*
\lambda_{2}&=\frac14\left(\sin^4r-\sqrt{\sin^8r+4\sin^4r\,\cos^2r}\right).
\end{align}
As we can see, $\lambda_2$ is non-positive and $\lambda_1$ is negative for all values of $a$, therefore the state will always preserve some degree of distillable entanglement. If we calculate the negativity we will obtain
\begin{equation}
\mathcal{N}(r)=\frac12\cos^2r,
\end{equation}
which means that for this case, distillable entanglement behaves exactly as in the previous case. In other words, no matter if the entangled state is a spin Bell state or a superposition like \eqref{minkstate2}, entanglement degrades the same. To conclude this section we stress that the same results will be obtained if the state $\ket{\uparrow_\text{A}}_\text{U}\ket{\downarrow_\text{A}}_\text{U}$ in \eqref{minkstate2} is replaced by any other 1 particle bipartite spin state $\ket{ss'}_\text{U}$.

\section{Non-inertial occupation number entanglement for spin $1/2$ fermions}\label{sec6}

The previous work \cite{AlsingSchul} on occupation number entanglement between accelerated partners ignored the spin structure of the Dirac field modes. It is not possible to straightforwardly translate a state like \eqref{genstate} into mere occupation number states. This comes about because for a state like \eqref{genstate} the bipartite vacuum component does not have individual spin degrees of freedom as the other components do. In other words, by including the vacuum state in the superposition \eqref{genstate}, the Hilbert space ceases to be factorable in terms of individual spin times particle occupation number subspaces.

On the other hand, the bipartite vacuum is a well defined total spin singlet. Hence, the Hilbert space is factorable with respect to the total spin of the system A-R and the occupation number subspaces. Accordingly, to reduce the spin information in the general density matrix \eqref{generaldensmat} we will be forced to consider a factorisation of the Hilbert space as the product of the total spin and occupation number subspaces.

Taking this tensor product structure into account we can now formally consider that we are not able to access to the information of the total spin of the system $A-R$ and then, we should trace over total spin degree of freedom.

The equivalence between the standard basis (occupation number-individual spin) and the new basis (occupation number-total spin) is\footnote{The pair state in the same mode can only be a singlet of total spin due to anticommutation relations of fermionic fields.}
\begin{equation}\label{e1}
\begin{array}{cc}
\ket{00}=\ket{00}\ket{S}& \ket{0\pa}=\ket{02}\ket{S},\\*[1.5mm]
\ket{0\uparrow}=\ket{01}\ket{D_+}& \ket{0,\downarrow}=\ket{01}\ket{D_-},\\*[1.5mm]
\ket{\uparrow0}=\ket{10}\ket{D_+}&\ket{\downarrow0}=\ket{10}\ket{D_-},\\*[1.5mm]
\ket{\uparrow\pa}=\ket{12}\ket{D_+}&\ket{\downarrow\pa}=\ket{12}\ket{D_-},\\*[1.5mm]
\ket{\uparrow\uparrow}=\ket{11}\ket{T_+}&\ket{\downarrow\downarrow}=\ket{11}\ket{T_-},\\*[1.5mm]
\end{array}
\end{equation}
\begin{equation}\label{e2}
\begin{array}{c}
\ket{\uparrow\downarrow}=\frac{1}{\sqrt{2}}\ket{11}\left[\ket{T_0}+\ket{S}\right],\\*[1.5mm]
\ket{\downarrow\uparrow}=\frac{1}{\sqrt{2}}\ket{11}\left[\ket{T_0}-\ket{S}\right],
\end{array}
\end{equation}
where we are using the basis $\ket{n_\text{A}\,n_\text{R}}\ket{J,J_z}$ and the triplets, doublets and the singlet are denoted as
\begin{eqnarray}
\nonumber\ket{T_+}&=&\ket{J=1,J_z=1},\\*
\nonumber\ket{T_-}&=&\ket{J=1,J_z=-1},\\*
\nonumber\ket{T_0}&=&\ket{J=1,J_z=0},\\*
\nonumber\ket{D_+}&=&\ket{J=1/2,J_z=1/2},\\*
\nonumber\ket{D_-}&=&\ket{J=1/2,J_z=-1/2},\\*
\nonumber\ket{S}&=&\ket{J=0,J_z=0}.\\*
\end{eqnarray}
Rewriting the general state \eqref{genstate} in this basis we obtain
\begin{equation}\label{genstateb2}
\ket{\Psi}=\mu\ket{00}\ket{S}+\alpha\ket{11}\ket{T_+}+\frac{\beta+\gamma}{\sqrt2}\ket{11}\ket{T_0}+ \frac{\beta-\gamma}{\sqrt2}\ket{11}\ket{S}+\delta\ket{11}\ket{T_-}
\end{equation}
Considering the acceleration of Rob, the general state \eqref{generaldensmat} in terms of this new basis after reducing the information on the total spin by tracing over this degree of freedom is
\begin{equation}
\rho^n_{AR}=\sum_{J,J_z}\bra{J,J_z}\rho_{AR}\ket{J,J_z}.
\end{equation}
Which results in a state in the occupation number basis whose entanglement degradation incorporates the effect of the spin structure. This can be studied and compared with the results in reference \cite{AlsingSchul} in which spin existence is ignored.
\begin{align}\label{densityred}
\nonumber\rho^n_{AR}&=\mu^2\Big[\cos^4r\proj{00}{00}+2\sin^2r\,\cos^2r\proj{01}{01}+\sin^4r\proj{02}{02}\Big]+\mu\cos^3r\\*
&\!\!\times\left(\frac{\beta^*-\gamma^*}{\sqrt2}\proj{00}{11}\!+\!\frac{\beta-\gamma}{\sqrt2}\proj{11}{00}\right)\!+\!(1-\mu^2)\Big[\cos^2r\proj{11}{11}\!+\!\sin^2r\proj{12}{12}\Big].
\end{align}
We can readily compute the partial transpose 
\begin{align}\label{densityred}
\nonumber(\rho^n_{AR})^{pT}&=\mu^2\Big[\cos^4r\proj{00}{00}+2\sin^2r\,\cos^2r\proj{01}{01}+\sin^4r\proj{02}{02}\Big]+\mu\cos^3r\\*
&\!\!\!\!\!\!\!\!\!\!\!\!\!\!\!\times\left(\frac{\beta^*-\gamma^*}{\sqrt2}\proj{01}{10}+\frac{\beta-\gamma}{\sqrt2}\proj{10}{01}\right)\!+\!(1-\mu^2)\Big[\cos^2r\proj{11}{11}\!+\!\sin^2r\proj{12}{12}\Big],
\end{align}
which only has a negative egigenvalue. The negativity is, in this case,
\begin{equation}\label{negativitymode}
\mathcal{N}=\cos^2r\left|\mu^2\sin^2r-\mu\sqrt{\mu^2\sin^4r+\cos^2r\frac{|\beta-\gamma|^2}{2}}\right|,
\end{equation}
which depends on the proportion of singlet \mbox{$|\beta-\gamma|/\sqrt2$} of the $\ket{11}$ component in the state \eqref{genstate}. When there is no singlet component ($\beta=\gamma$) the negativity is zero. Indeed in the limit $a\rightarrow0$ (Minkowskian limit)
\begin{equation}\label{negativitymode0}
\mathcal{N}_0=\frac{1}{\sqrt2}\left|\mu\right|\left|\beta-\gamma\right|,
\end{equation}
which shows that the maximally entangled Minkowski occupation number state (Negativity $=1/2$) arises tracing over total spin when the starting state is
\begin{equation}\label{modesmaxen}
\ket{\Psi}=\frac{1}{\sqrt2}\ket{00}\pm\frac12\big[\ket{\uparrow\downarrow}-\ket{\downarrow\uparrow}\big],
\end{equation}
or, in the occupation number-total spin basis
\begin{equation}\label{modesmaxen2}
\ket{\Psi}=\frac{1}{\sqrt2}\Big[\ket{00}\ket{S}\pm\ket{11}\ket{S}\big].
\end{equation}
This means that, for occupation number entanglement, the only way to have an entangled state of the bipartite vacuum $\ket{00}$ and the one particle state $\ket{11}$ of a Dirac field is through the singlet component of total spin for the $\ket{11}$ component.

On the contrary, the state
\begin{equation}
\ket{\Psi}=\frac{1}{\sqrt2}\Big[\ket{00}\ket{S}\pm\ket{11}\ket{T_{0,\pm}}\big]
\end{equation}
will become separable after tracing over total spin due to the orthonormality of the basis \eqref{e1}, \eqref{e2}.

We have established that the Minkowski maximally entangled state for occupation number arises after tracing over total spin in a state as \eqref{modesmaxen}. Now we will compute the limit of the negativity when the acceleration goes to $\infty$ in order to see its Unruh decoherence and to compare it with the results for occupation number entanglement from \cite{AlsingSchul}.

Taking $a\rightarrow\infty \Rightarrow r\rightarrow\pi/4$ in \eqref{negativitymode}
\begin{equation}\label{negativitymodeinfty}
\mathcal{N}_\infty=\frac14\left|\mu^2-\sqrt{\mu^4+\mu^2|\beta-\gamma|^2}\right|.
\end{equation}
Therefore, for the maximally Minkowski entangled state we have $\mu=1/\sqrt{2}$, $|\beta-\gamma|=1$ and the negativity in the limit is
\begin{equation}\label{negalimi}
\mathcal{N}_\infty=\frac{\sqrt3-1}{8}.
\end{equation}

\begin{figure}[H]
\begin{center}
\includegraphics[width=.85\textwidth]{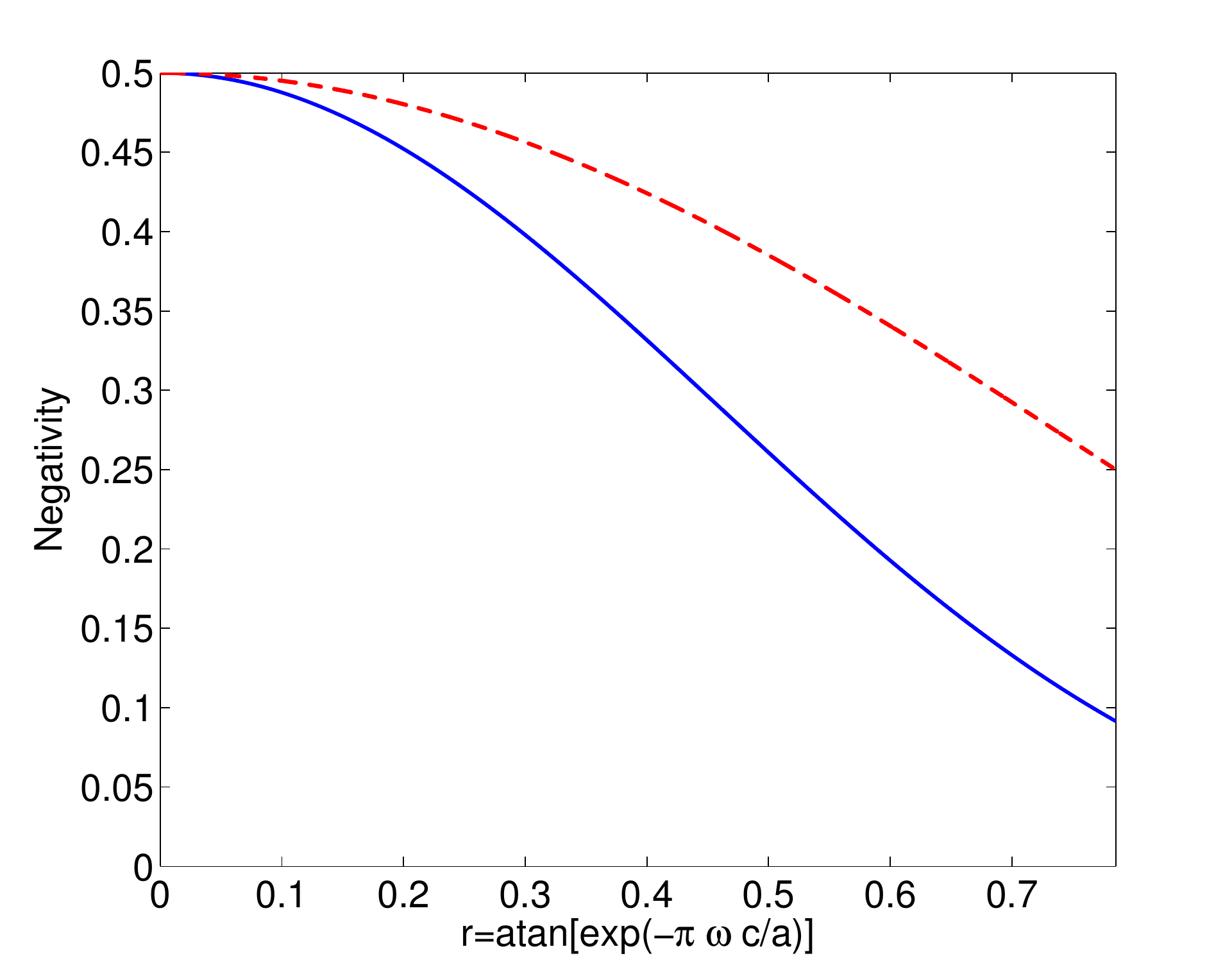}
\end{center}
\caption{(Blue solid) Negativity in the occupation number degree of freedom as a function of the acceleration of Rob when A and R share an occupation number maximally entangled state \eqref{modesmaxen} in the Minkowski basis after tracing over total spin. (Red dashed) Negativity a as a function of the acceleration of Rob when A and R share spin Bell states and the state $\frac{1}{\sqrt2}\left(\ket{00}+\ket{ss'}\right)$ (both cases coincide).}
\label{fig5}
\end{figure}

This result shows that when we reduce the total spin information, looking at the occupation number entanglement alone, we see that it is more degraded by the Unruh effect than when we considered spin Bell states in the previous section. More importantly, the occupation number entanglement is more degraded than in \cite{AlsingSchul}, where the spin structure of the modes was considered nonexistent. This happens because we are removing part of the correlations when we erase total spin information. The negativity dependence on the acceleration is shown in Fig. \ref{fig5} compared with the negativity for the maximally entangled states \eqref{Bellstates} and \eqref{minkstate2}.

\section{Discussion}\label{sec7}

It is known \cite{Alicefalls,AlsingSchul} that Unruh decoherence degrades entanglement of occupation number states of fields. Here we have shown a richer casuistic that appears when we take into account that each Dirac mode has spin structure. This fact enables us to study interesting effects (such as entanglement degradation for spin Bell states) and develop new procedures to erase spin information from the system in order to study occupation number entanglement.

Along this chapter we have analysed how a maximally entangled spin Bell state losses entanglement when one of the partners accelerates. We have seen that, while in the Minkowski basis Alice and Rob have qubits, when Rob accelerates the system becomes a non-pure state of a qubit for Alice and a qutrit for Rob. In this case spin entanglement for a Dirac field is degraded when Rob accelerates. However some degree of entanglement survives even at the limit $a\rightarrow\infty$.

A first difference with previous literature where spin was not taken into account is that in this case the dimension of the Hilbert space changes when we consider the state in Rob's Fock basis. A first analog to the well studied state $(1/\sqrt{2})(\ket{00}+\ket{11})$ but including spin, $(1/\sqrt{2})(\ket{00}+\ket{\uparrow\downarrow})$, has been studied. This state,  becomes $2\times 4$ dimensional when Rob accelerates. On the other hand,  Spin Bell states becomes a $2\times3$  dimension system. Nevertheless, we have demonstrated distillable entanglement degrades exactly in the same way as for spin Bell states and for other maximally entangled superposition. This fact will be deeply analysed in following chapters.

The Fock space for every mode of a Dirac field has higher dimension than for the spinless fermionic field as the one analysed in \cite{AlsingSchul} (the basis for every mode changes from $\ket{0},\ket{1}$ to $\{\ket{0},\ket{\uparrow},\ket{\downarrow},\ket{\pa}$). Indeed, we have seen how for a Dirac field, the dimension of the Hilbert space is increased when Rob accelerates due to the excitation of spin pairs.

 However, we have seen that in spite of the evident differences between the spinless fermionic field analysed in the literature and the spin $1/2$ field analysed here, the entanglement degradation due to the Unruh effect is exactly the same in both cases. More striking: even spin Bell states (which had no analog in a Grassmann scalar field) have the same entanglement behaviour as the states described above. This starts to suggest some sort of universal behaviour present in fermionic fields. This is a first argument against  the idea that the Hilbert space dimension is connected with the entanglement survival in the infinite acceleration limit for fermions. We will explore this in the following chapters. 

We have also introduced a procedure to consistently erase spin information from our setting preserving the occupation number information. We have done it by tracing over total spin. The maximally entangled occupation number state is obtained from the total spin singlet \eqref{modesmaxen} after tracing over total spin. Finally we have shown that its entanglement is more degraded than in \cite{AlsingSchul} where the spin structure of Dirac modes was neglected.

\cleardoublepage


\chapter{Multimode analysis of fermionic non-inertial entanglement\footnote{E. Mart\'in-Mart\'inez, J. Le\'on, Phys. Rev. A, 80, 042318 (2009)}}\label{multimode}

We have seen in the previous chapter that fermionic maximally entangled states seem to degrade the same no matter the kind of fermionic field considered (Dirac or Grassmann) and the state analysed (Spin Bell or occupation number). This prompts suspicions about the relationship between dimension of the Hilbert space and the entanglement survival in the limit of infinite acceleration for fermions. 

In this chapter we will present a manifestly multimode formalism equivalent to that presented in the chapters before but that will reveal useful to study multimode entangled states. We will show that when we consider states that mix different frequency modes a larger number of modes can become excited by the Unruh effect even for fermion fields, and so, the argument about the Hilbert space dimension playing a role in the degradation phenomenon looks lees plausible. Here a fundamental question arises; does fermionic statistics protect the entanglement in these different frequencies entangled states? In this chapter we shall show that such entanglement survival is fundamentally inherent in the Fermi-Dirac statistics, and that it is independent of the number of modes considered, of the maximally entangled state we start from, and even of the spin of the fermion field studied.

We will begin revisiting the derivation of the entanglement for the states computed in chapter \ref{onehalf} and in \cite{AlsingSchul} with this multimode formalism that will prove to be useful to handle multimode scenarios. After that we will study a state that has no analog with any state studied before: an entangled state of two different spins and frequencies. This state dwells in a Hilbert space of higher dimension than the previous ones and, in principle, there is no guarantee that entanglement is going to behave in the same way as the others.

In section \ref{s2} we will revisit the field vacuum and one particle state in the basis of an accelerated observer and for two different kinds of fermionic fields (a Dirac field and a Grassmann\footnote{See appendix \ref{appB}} ``spinless'' fermion field) in the context of this new formalism. After that, we will analyse entanglement degradation for two different kinds of maximally entangled states that were already analysed with the single mode formalism: the case of vacuum entangled with one particle state studied in chapter \ref{onehalf} and a maximally entangled state of a ``spinless fermion'' field (considered with a single mode formalism in \cite{AlsingSchul}). Studying these known cases where we will learn about the multimode formalism, we    analyse the case of a maximally entangled state of two different frequency modes. We will see that even for the radically different final states obtained in each case, after non-trivial computations entanglement degradation ends up being the same for all of them and the dependence of entanglement on $a$ turns out to be exactly the same as in the other fermionic cases analysed.

\section{Vacuum and 1-Particle states in the multimode scenario}\label{s2}

In this section we shall build the vacuum state and the 1-particle excited state for two very different kinds of fermionic fields in the explicitly multimode formalism: First a Dirac field and then a spinless fermion field. Both kinds of fields were analysed before in a pure single mode scenario (the spinless case in \cite{AlsingSchul} and the Dirac field in chapter \ref{onehalf}).

To begin with, let us consider that a discrete number $n$ of different modes of a Dirac field $\omega_1,\dots,\omega_n$ is relevant.  We label with $s_i$ the spin degree of freedom of each mode. We will rederive a expression for the Minkowski vacuum in a way that takes explicitly into account all the relevant frequency modes that will be useful for further considerations. 

As seen before, the Minkowski multimode vacuum should be expressed in the Rindler basis as a squeezed state, which is an arbitrary superposition of spins and frequencies as it is discussed in chapter \ref{onehalf}
\begin{equation}
\label{vacuumCOMP}\ket{0}=\sum_{m=0}^{2n}\sum_{\substack{s_1,\dots,s_{m}\\\omega_1,\dots,\omega_{m}}}\!\!\!\!C^{m}_{s_1,\dots,s_{m},\omega_1,\dots,\omega_{m}}
\xi_{s_1,\dots,s_{m}}^{\omega_1,\dots,\omega_{m}} \biket{\tilde{m}}{\tilde{m}},
\end{equation}
where, the notation is
\begin{equation}\label{notationmod}
|{\tilde i}\rangle_\text{I}|{\tilde i}\rangle_\text{II}=\biket{s_1,\!\omega_1;\dots;\!s_i,\!\omega_i}{-s_1,\omega_1;\dots;-s_i,\omega_i},
\end{equation}
with
\begin{equation}
\ket{\omega_1,s_1;\dots;\omega_m,s_m}_\text{I}=c^\dagger_{I,\omega_m,s_m}\dots c^\dagger_{I,\omega_1,s_1}\ket{0}_\text{I}.
\end{equation}
The label outside the kets notates Rindler spacetime region, and the symbol $\xi$ is 0 if $\{\omega_i,s_i\}=\{\omega_j,s_j\}$ for any $i\neq j$, and it is $1$ otherwise, imposing Pauli exclusion principle constraints on the state (quantum numbers of fermions cannot coincide).

Due to the anticommutation relations of the fermionic operators, terms with different orderings are not independent.
So, without loss of generality, we could choose not to write all the possible orderings in \eqref{vacuumCOMP}, selecting one of them instead. In this fashion we will write the elements \eqref{notationmod} with the following ordering criterion:
\begin{eqnarray}\label{ordering}
 \nonumber & \omega_i\le \omega_{i+1}, &\\
 &\omega_i=\omega_{i+1}\Rightarrow s_i=\uparrow,s_{i+1}=\downarrow.&
 \end{eqnarray}
The coefficients $C^m$ are constrained because the Minkowski vacuum should satisfy 
$a_{\omega,s}\ket0=0$, $\forall \omega,s$. 
As the elements \eqref{notationmod} form an orthogonal set, this implies that all the terms proportional to different elements of the set should be zero simultaneously, which gives the following conditions on the coefficients
\begin{itemize}
\item $C^1_{s,\omega}$ as a function of $C^0$\\[-9mm]
\end{itemize}
\begin{eqnarray}
\label{01} C^1_{\uparrow,\omega}\cos r-C^0\sin r&=&0,\\*
C^1_{\downarrow,\omega}\cos r-C^0\sin r&=&0. \label{02}
\end{eqnarray}
Since equations \eqref{01},\eqref{02} should be satisfied $\forall \omega$, we obtain that $C^1_{\uparrow,\omega}=C^1_{\downarrow,\omega}=\text{const.}$ because $C^0$ does not depend on $\omega$ or $s$. We will denote $C^1_{s,\omega}\equiv C^1$.
\begin{itemize}
\item $C^2_{s_1,s_2,\omega_1,\omega_2}$ as a function of $C^1$\\[-9mm]
\end{itemize}
\begin{eqnarray}
\label{03}  C^1\sin r- C^2_{ss',\omega_1,\omega_2}\cos r&=&0,\\*
\label{04}  C^1\sin r- C^2_{ss',\omega_1,\omega_2}\cos r&=&0,
\end{eqnarray}
where we consider\footnote{Our interest here is to show that the increase on the Hilbert space dimension does not play a role in entanglement behaviour. With this motivation, we consider that all the $N$ frequencies $\{\omega^j_\text{R}\}_{j=1}^N$ are close enough to roughly approximate $r^1\approx r^2\approx\cdots r^N\equiv r$. This will show the result we want to prove more transparently. } $r_1\approx r_2\equiv r$. Since equations \eqref{03}, \eqref{04} should be satisfied $\forall \omega_0$, we obtain that $C^2_{s_1,s_2,\omega_1,\omega_2}=C^2$ where $C^2$ does not depend on spins or frequencies since $C^1$ does not depend on $\omega$ or $s$, the only dependence of the coefficients \eqref{vacuumCOMP} with $\omega_i$ and $s_i$ is given by Pauli exclusion principle. This dependence comes through the symbol $\xi$.

In fact it is very easy to show inductively that all the coefficients are independent of $s_i$ and $\omega_i$ --apart from the Pauli exclusion principle constraint--. Using the fact that $C^0$ does not depend on $s_i$ and $\omega_i$ and noticing that by applying the annihilator on the vacuum state and equalling it to zero we will always obtain the linear relationship between $C^{n}$ and $C^{n-1}$ given below.
\begin{itemize}
\item $C^m$ as a function of $C^{m-1}$\\[-9mm]
\end{itemize}
\begin{eqnarray}
\label{05}  C^{m-1}\sin r- C^m\cos r&=&0,\\*
\label{06}  C^{m-1}\sin r- C^m\cos r&=&0.
\end{eqnarray}
We finally obtain that $C^m$ is a constant which can be expressed as a function of $C^0$ as
\begin{equation}\label{coeff2}
C^m=C^0 \tan^m r,
\end{equation}
where $\tan r=\exp\left(-\pi \omega c/a\right)$. $C^m$ is independent of $s_i$ and $\omega$. Therefore, we obtain the vacuum state by substituting \eqref{coeff2} in \eqref{vacuumCOMP} and factoring the coefficients out of the summation.
\begin{equation}\label{vacuumCOMP2}
\ket{0}=C^0\sum_{m=0}^{2n}\tan^m r\sum_{\substack{s_1,\dots,s_m\\\omega_1,\dots,\omega_m}}\!\!\!\!\xi_{s_1,\dots,s_m}^{\omega_1,\dots,\omega_m} \biket{\tilde m}{\tilde m}.\\
\end{equation}
The only parameter not yet fixed is $C^0$. To derive $C^0$ except for a global phase, we impose the normalisation of the vacuum state in the Rindler basis $\braket00=1$, from \eqref{vacuumCOMP2}, we see that this means that
\begin{equation}\label{normalispinap}C^0=\left[\sum_{m=0}^n\Upsilon_m\tan^{2m}r+\sum_{m=n+1}^{2n}\Upsilon_{2n-m}\tan^{2m}r\right]^{-1/2},\end{equation}
where
\begin{equation}\label{upsilon}\Upsilon_m=\sum_{\substack{s_1,\dots,s_m\\\omega_1,\dots,\omega_m}}\!\!\!\!\xi_{s_1,\dots,s_m}^{\omega_1,\dots,\omega_m}.
\end{equation}
Now, we are going to show that \eqref{upsilon}, has the form
\begin{equation}\label{upsilonap}\Upsilon_m=\sum_{p=0}^{\lfloor \frac{m}
{2}\rfloor}\binom{n-p}{m-2p}\binom{n}{p}2^{m-2p}.
\end{equation}
To see how this expression comes from Pauli exclusion principle, we have to read $p$ as an index that represents the number of possible spin pairs ($\omega_i=\omega_{i+1},s_i=\uparrow,s_{i+1}=\downarrow$) which can be formed, and goes from $0$ to the integer part of $m/2$, and then
\begin{itemize}
\item The combinatory number $\binom{n-p}{m-2p}$ represents the possible number of combinations of modes that can be formed taking into account that $p$ different frequencies $\omega_i$ are not available since they are already occupied by the $p$ pairs. Hence, it is given by the combinations of the $n-p$ available frequencies taken $m-2p$ at time, since $m-2p$ is the number of free momentum `slots' (the total number of different frequencies $m$ minus the number of positions taken by pairs $2p$).
\item The combinatory factor $\binom{n}{p}$ represents the different possible combinations for the configuration of the $p$ pairs, which have $n$ possible different frequencies to be combined among them without repetition and in a particular order.
\item The factor $2^{m-2p}$ represents the possible combination for the spin degree of freedom of each mode. As a spin pair only admits one spin configuration, only the unpaired modes will give different spin contributions, so the factor is $(2S+1)^{m-2p}$ giving the formula \eqref{upsilon}
\end{itemize}

After some lengthy but elementary algebra we can see that
\begin{equation}\label{upsilon22}\Upsilon_m=\binom{2n}{m},
\end{equation}
and using the property $\binom{a}{a-b}=\binom{a}{b}$, 
we can express \eqref{normalispinap} as
\begin{equation}\label{normalispin}
C^0=\left[\sum_{m=0}^{2n}\binom{2n}{m}\tan^{2m}r\right]^{-1/2}=\cos^{2n}r 
\end{equation}
and, therefore, rewrite the vacuum \eqref{vacuumCOMP} as
\begin{equation}\label{vacuumCOMP2b}
\ket{0}=\cos^{2n}r\sum_{m=0}^{2n}\tan^m r\sum_{\substack{s_1,\dots,s_m\\\omega_1,\dots,\omega_m}}\!\!\!\!\xi_{s_1,\dots,s_m}^{\omega_1,\dots,\omega_m} \biket{\tilde m}{\tilde m}.\\
\end{equation}

Next, the 1-particle state can be worked out translating the Minkowski one particle Unruh state $\ket{\omega,s}=a^\dagger_{\omega,s}\ket0$ into the Rindler basis
\begin{equation}\label{onepart23m}
\ket{k,s}_\text{U}=\sum_{m=0}^{2n-1}A^m\sum_{\substack{s_1,\dots,s_m\\\omega_1,\dots,\omega_m}}\!\!\!\!\xi_{s_1,\dots,s_m,s}^{\omega_1,\dots,\omega_m,\omega} \biket{\tilde m;\omega,s}{\tilde m},
\end{equation}
where
\begin{equation}\label{Am}
A^m=(C^m\cos r+C^{m+1}\sin r)
\end{equation}
and the notation $\ket{\tilde m;\omega,s}_\text{I}$, consequently with \eqref{notationmod}, means the ordered version of 
$\ket{s_1,\!\omega_1;\dots;\!s_n,\!\omega_n;\omega,s}_\text{I}$. 

Another different kind of field that we are going to consider appears by neglecting spin while keeping the fermionic statistics (Grassmann scalar 
fields). We will analyse Unruh decoherence in this multimode formalism. The Minkowski multimode vacuum state would be expressed as
\begin{equation}\label{vacuzero}
\ket{0}=\sum_{m=0}^n\sum_{\omega_1,\dots,\omega_m}\xi_{\omega_1,\dots,\omega_m}\hat C^m_{\omega_1,\dots,\omega_m}\ket{\tilde m}_\text{I}\ket{\tilde m}_\text{II},
\end{equation}
where, in this occasion $\ket{\tilde m}_\text{I}\ket{\tilde m}_\text{II}=\ket{\omega_1,\dots,\omega_m}_\text{I}$ $\ket{\omega_1,\dots,\omega_m}_\text{II}$. Using the same procedures as for the spin $1/2$ 
case \eqref{vacuumCOMP} we can prove that all the coefficients are independent of $\omega_i$ and can be related to $\hat C^0$ as in \eqref{coeff2}, $\hat C^m=\hat C^0 \tan^m r$. We can now fix $\hat C^0$ imposing the normalisation relation $\braket00=1$ giving
\begin{equation}\label{normalizeropre}
\hat C^0=\left[\sum_{m=0}^n\chi_m\tan^{2m}r\right]^{-1/2}.
\end{equation}
For the spinless fermion field we have
\begin{equation}\label{XIIap}
\chi_m\equiv\sum_{\omega_1,\dots,\omega_m}\xi_{\omega_1,\dots,\omega_m}=\binom{n}{m},
\end{equation}
corresponding to the possible combinations of m values of $\omega_i$ imposing that $\omega_i\neq \omega_j$ if $j\neq i$ (which is the translation of Pauli exclusion principle to spinless modes). This expression can be readily obtained taking into account that the $n$ possible values of $\omega_i$ should be combined without repetition in a particular ordering of the $m$ modes, so the possible combinations are simply the combinatory number $\binom{n}{m}$.

Therefore, \eqref{normalizeropre} can be simplified to
\begin{equation}\label{normalizero}
\hat C^0=\left[\sum_{m=0}^n\binom{n}{m}\tan^{2m}r\right]^{-1/2}=\cos^n r.
\end{equation}
 
Finally, the Grassmann one particle state $a_\omega^\dagger\ket0$ is
\begin{equation}\label{vacuuno}
\ket{\omega}_\text{U}=\sum_{m=0}^{n-1}\hat A^m\!\!\sum_{\omega_1,\dots,\omega_m}\!\!\xi_{\omega_1,\dots,\omega_m,k}\biket{\tilde m,k}{\tilde m},
\end{equation}
where $\hat A^m$ has the expression \eqref{Am} but substituting $C^m$ by $\hat C^m$.

\section{Entanglement degradation for a Dirac field}\label{caso1}

In the following we will analyse Unruh entanglement degradation in various \mbox{settings} corresponding to different maximally entangled states of fermion fields. First we consider a state that was already analysed in chapter \ref{onehalf} but computing entanglement with this new formalism. This will be useful as a pedagogical example of operation of this explicitly multimode formalism and to compare with the results obtained when we go beyond the cases studied in previous chapters. Let us consider the state 
\begin{equation}\label{minkowstate1}\ket\Psi=\frac{1}{\sqrt2}\big(\ket{0}\ket{0}+\ket{\omega_A,s_A}_\text{U}\ket{\omega_\text{R},s_\text{R}}_\text{U}\big).\end{equation}
The density matrix for the accelerated observer Rob is obtained after expressing Rob's state in the Rindler basis --which means using \eqref{vacuumCOMP} and \eqref{onepart23m} in Rob's part of \eqref{minkowstate1}-- and then, tracing over Rindler's region II since Rob is causally disconnected from it and he is not to extract any information from beyond the horizon. Following this procedure we obtain the density matrix
\begin{align}\label{densmat1}
\nonumber\rho&=\frac{1}{2}\Big[\sum_{m=0}^{2n}\Big(D_{0}^m\!\!\!\sum_{\substack{s_1,\dots,s_m\\\omega_1,\dots,\omega_m}}\!\!\!
\xi_{s_1,\dots,s_m}^{\omega_1,\dots,\omega_m}\ket{0}_{\text{U}}\ket{\tilde m}_\text{I}\bra{0}_{\text{U}}\bra{\tilde m}_\text{I}\Big)\\*
\nonumber&+\sum_{m=0}^{2n-1}\Big(D_{1}^m\!\!\!\sum_{\substack{s_1,\dots,s_m\\\omega_1,\dots,\omega_m}}\!\xi_{s_1,\dots,s_m,s_\text{R}}^{\omega_1,\dots,\omega_m,\omega_\text{R}}
\ket{0}_{\text{U}}\ket{\tilde m}_\text{I}\bra{\omega_A,s_A}_{\text{U}}\bra{\tilde m;\omega_\text{R},s_\text{R}}_\text{I}\!\!\Big)\\*
&+\sum_{m=0}^{2n-1}\Big(D_{2}^m\!\!\!\!\sum_{\substack{s_1,\dots,s_m\\\omega_1,\dots,\omega_m}}\!\!\!\!
\xi_{s_1,\dots,s_m,s_\text{R}}^{\omega_1,\dots,\omega_m,\omega_\text{R}}\ket{\omega_A,s_A}_{\text{U}}\ket{\tilde m;\omega_\text{R},s_\text{R}}_\text{I}\bra{\omega_A,s_A}_{\text{U}}\bra{\tilde m;\omega_\text{R},s_\text{R}}_\text{I}\Big)\Big]+(\text{H.c.})_{_{\substack{\text{non-}\\\text{diag.}}}},
\end{align}
where $(\text{H.c.})_{\text{non-diag.}}$ means Hermitian conjugate of only the non-diagonal terms and
\begin{equation}\label{Des}
D_i^m=|C^0|^2\frac{\tan^{2m}r}{\cos^i r}
\end{equation}
with $i=0,1,2$. The derivation of \eqref{densmat1} can be found in the appendix to this chapter (section \ref{ap2}). 

Notice that as Rob accelerates, the state becomes mixed, showing all the available modes $(\omega_1,\dots,\omega_n)$ excitations explicitly. 

As in the previous chapter we will compute the negativity as a function of $a$ as a measure of the state entanglement.

The partial transpose of \eqref{densmat1} has a $2\times2$ and $1\times1$ blocks structure. Each eigenvalue in the $1\times1$ blocks is non-negative (since 
$D_i^m\ge0$), so we are interested in the $2\times2$ which are the ones that may have negative eigenvalues. These $2\times2$ blocks expressed in the basis
\begin{equation}\label{blocksbasis}
\Big\{\ket0_{A}\ket{\tilde m;\omega_\text{R},s_\text{R}}_\text{I},\ket{s_A,\omega_A}_\text{U}\ket{\tilde m}_\text{I}\Big\}_{m=0}^{2n-1}
\end{equation}
are of the form
\begin{equation}\label{blocks3m}
\frac12
\left(\begin{array}{cc}
D^{m+1}_0 & \pm D_1^m\\
\pm D_1^m & 0
\end{array}\right).
\end{equation}
There is no matrix element proportional to $D_2^m$ because it would correspond to $\ket{\omega_A,s_A}_{\text{U}}\ket{\tilde m;\omega_\text{R},s_\text{R}}_\text{I}\bra{\omega_A,s_A}_{\text{U}}\bra{\tilde m;\omega_\text{R},s_\text{R}}_\text{I}$ 
which cannot have any element within this block as Pauli exclusion principle imposes $\omega_\text{R},s_\text{R}\not\in\left\{\omega_i,s_i\right\}_{i=1,\dots,m}$.

Each $2\times2$ block of \eqref{blocks3m} appears a number of times $B_m$. Taking a look at the basis in which those blocks are expressed \eqref{blocksbasis}, we can see that the expression for $B_m$ is given by two terms:
\begin{itemize}
\item  The number of possible combinations of $m$ modes with $n$ possible different frequencies $\omega_i$ and two possible spins $s_i$ according to Pauli exclusion principle as in \eqref{upsilonap}.
\item A negative contribution which comes from excluding those combinations in which $\{\omega_\text{R},s_\text{R}\}$ coincides with any $\{\omega_i,s_i\}$, which means excluding the number of combinations in \eqref{upsilonap} which have one of their values fixed to $\{\omega_i,s_i\}=\{\omega_\text{R},s_\text{R}\}$. This number is given by the combinatory number $\binom{2n-1}{m-1}$ provided that $m>0$ and it is zero if $m=0$.
\end{itemize}

To see where this negative contribution comes from let us assume that $\{\omega_i,s_i\}$ is the mode which coincides with $\{\omega_\text{R},s_\text{R}\}$ we will have $2n-1$ possible choices for each $\{\omega_{j\neq i},s_{j\neq i}\}$ ($2$ values for $s$ and $n$ for $\omega$ excepting $\omega_i,s_i$ due to Pauli exclusion principle). This happens for all the combinations of all the possible values $\{\omega_j,s_j\}$ with $j\neq i$. Hence, as there are $m$ modes and one of them is fixed $\omega_i=\omega_\text{R}, s_i=s_\text{R}$, we have to consider the combinations of $2n-1$ elements taken $m-1$ at time.

If $m>n$ the situation is equivalent to having $m'=2n-m$. Since having more modes $m$ than possible values of $\omega_i$ we are forced to have $n-m$ pairs and we lose freedom to combine the available modes.

Now if we compute
\begin{equation}\label{Nblocksap}B_m=\Upsilon_m-\binom{2n-1}{m-1}=\binom{2n}{m}-\binom{2n-1}{m-1},
\end{equation}
after some basic algebra we obtain
\begin{equation}\label{Nblocks3m}B_m=\binom{2n-1}{m}\end{equation}

Using \eqref{Des}, the negative eigenvalue of each block can be expressed
\begin{equation}\label{neigenb}
|\lambda^-_m|=\frac12|C^0|^2\,\tan^{2m} r,
\end{equation}
where $C^0$ is given by \eqref{normalispin}. Therefore, the negativity is expressed as the sum of the negative eigenvalue of each block $|\lambda_m^-|$ multiplied by the number of times $B_m$ that that block appears in the partially transposed density matrix. The 
summation of the series is
\begin{equation}\label{negativitypre1}
\mathcal{N}=\sum_{m=0}^{2n-1} B_m|\lambda^-_m|=\frac{\cos^{4n}r}{2}\sum_{m=0}^{2n-1}\binom{2n-1}{m}\tan^{2m}r ,
\end{equation}
but this result can be easily simplified to
\begin{equation}\label{negativitypre}
\mathcal{N}=\frac12\cos^2 r,
\end{equation}
which is independent of the number of modes that we have considered. This is the same result obtained in chapter \ref{onehalf}. This expected result shows that this multimode formalism is valid to analyse the entanglement degradation due to Unruh effect. This also emphasises a somewhat non trivial result: despite the fact that all the available modes are excited when Rob accelerates \eqref{densmat1}, the quantum correlations behave as if we were considering only one possible mode for the field. This is nothing but a consequence of the tensor product structure of the Hilbert space showed in the previous chapter.

\section{Entanglement degradation for a spinless fermion field}\label{caso3}

We can also revisit the results on the literature \cite{AlsingSchul} and consider a spinless field on which we have imposed the fermionic statistics. We will re-obtain with the explicitly multimode formalism the entanglement degradation for the maximally entangled state with vacuum and one particle components
\begin{equation}\label{minkowstate3}
\ket\Psi=\frac{1}{\sqrt{2}}\Big(\ket{0}_{\text{U}}\ket{0}_\text{R}+\ket{\omega_A}_{\text{U}}\ket{\omega_\text{R}}_\text{R}\Big).
\end{equation}
As it is shown in appendix \ref{ap2}, this leads to the following density matrix for the accelerated observer Rob after using expressions \eqref{vacuzero} and \eqref{vacuuno} and after tracing over Rindler's region II 
\begin{align}\label{densmat3}
\nonumber\rho&=\frac{1}{2}\Big[\sum_{m=0}^{n}\!\hat D_{0}^m\!\!\!\sum_{\omega_1,\dots,\omega_m}\!\!\!\xi_{\omega_1,\dots,\omega_m}\ket{0}_{\text{U}}\ket{\tilde m}_\text{I}\bra{0}_{\text{U}}\bra{\tilde m}_\text{I}\\*
\nonumber&+
\sum_{m=0}^{n-1}\Big(\hat D_{1}^m\!\!\!\sum_{\omega_1,\dots,\omega_m}\!\xi_{\omega_1,\dots,\omega_m,\omega_\text{R}}\ket{0}_{\text{U}}\ket{\tilde m}_\text{I}\bra{\omega_A}_{\text{U}}\bra{\tilde m;\omega_\text{R}}_\text{I}\\
 &+\hat D_{2}^m\!\!\!\sum_{\omega_1,\dots,\omega_m}\xi_{\omega_1,\dots,\omega_m,\omega_\text{R}}\ket{\omega_A}_{\text{U}}\ket{\tilde m;\omega_\text{R}}_\text{I}\bra{\omega_A}_{\text{U}}\bra{\tilde m;\omega_\text{R}}_\text{I}\Big)\Big]+(\text{H.c.})_{_{\substack{\text{non-}\\\text{diag.}}}},
\end{align}
where $\hat D^,_i$ is given by the expression \eqref{Des} but substituting $C^0$ by $\hat C^0$ (given in equation \eqref{normalizero}). 

Analogously to \eqref{densmat1}, the partial transpose of \eqref{densmat3} has a $2\times2$ and $1\times1$ blocks structure. The 
$2\times2$ blocks expressed in the basis
\begin{equation}\label{blocksbasis3}
\Big\{\ket{0}_\text{U}\ket{\tilde m;\omega_\text{R}}_\text{I},\ket{\omega_A}_\text{U}\ket{\tilde m}_\text{I}\Big\}_{m=0}^{n-1}
\end{equation}
would have the form 
\begin{equation}\label{blockszero}
\frac12
\left(\begin{array}{cc}
\hat D^{m+1}_0 & \pm \hat D_1^m\\
\pm \hat D_1^m & 0
\end{array}\right).
\end{equation}
The main difference with \eqref{blocks3m} is that $\hat C^0$ is given by \eqref{normalizero} (instead of $C^0$ given by \eqref{normalispin}). Here, $\hat D_2^m$ does not appear because Pauli exclusion principle imposes that $\omega_\text{R}\not\in\left\{\omega_i\right\}_{i=1,\dots,m}$. Now, each $2\times2$ block multiplicity is $W_m$.

$W_m$ can be easily obtained taking into account that the number of $2\times2$ blocks \eqref{blocksbasis3} is given by the number of mode combinations allowed by Pauli principle \eqref{XIIap}, subtracting the terms having $\omega_i=\omega_\text{R}$. The number of possible $\omega_j$ values allowed for the rest $m-1$ modes having fixed $\omega_i=\omega_\text{R}$ is $n-1$, so the number of combinations we must subtract is the combinatory number $\binom{n-1}{m-1}$, obtaining
 \begin{equation}\label{Nblocks3ap}W_m=\binom{n}{m}-\binom{n-1}{m-1}=\binom{n-1}{m}.\end{equation}

The negative eigenvalue of each block is given by the same expression \eqref{neigenb} but $C^0$ is now given by 
\eqref{normalizero}, which is to say
\begin{equation}\label{neigenb0}
|\lambda^-_m|=\frac12|\hat C^0|^2\,\tan^{2n} r=\frac12\cos^{2m}\,\tan^{2m} r.
\end{equation}
 The negativity yields
\begin{equation}\label{negazeropre}
\mathcal{N}=\sum_{m=0}^{n-1} W_m|\lambda^-_m|=\frac12\cos^{2n} r\sum_{m=0}^{n-1}\binom{n-1}{m}\tan^{2m}r.
\end{equation}
At this point, the reader might not be surprised by the resulting negativity after straightforward simplification
\begin{equation}\label{negazero}
\mathcal{N}=\frac{1}{2}\cos^2 r
\end{equation}
which is the same result as in the cases \eqref{minkowstate1} and 
\eqref{minkowstate2}. Again, entanglement degradation due to Unruh effect is 
the same as considering one mode of a Dirac field.

Although we have seen that the derivation here does not add anything new from the standard mode-by-mode expressions for the vacuum and one-particle states, it will be useful to study entangled states of a discrete number of different frequency modes.

\section[Entanglement degradation between different modes]{Entanglement degradation between different frequency modes}\label{caso2}

In this section we will go beyond the states analysed in previous sections of this thesis and the published literature. We will analyse a state that in the Minkowskian basis is a maximally entangled superposition of different (but very close) frequency modes with arbitrary spin components. In principle each mode would suffer its own decoherence induced by the Unruh noise and the naive expectation would be that, even though the frequencies are close, the state presents a qualitative different entanglement behaviour than the other single-mode states analysed previously.

If instead of \eqref{minkowstate1} we start from a Bell momentum-spin state in the Minkowskian basis
\begin{equation}\label{minkowstate2}\ket\Psi=\frac{1}{\sqrt2}\Big(\ket{\omega_A^1,s_A^1}_{\text{U}}\ket{\omega^1_\text{R},s^1_\text{R}}_\text{U}+\ket{\omega^2_A,s^2_A}_{\text{U}}\ket{\omega^2_\text{R},s^2_\text{R}}_\text{U}\Big).
\end{equation}
As it can be seen in the appendix \ref{ap2} the density matrix for Rob takes the form 
\begin{align}\label{densmat2}
\rho&=\nonumber\sum_{m=0}^{2n-1}\frac{D_{2}^m}{2}\!\!\!\sum_{\substack{s_1,\dots,s_m\\\omega_1,\dots,\omega_m}}\!\!\!\!
\Big(\xi_{s_1,\dots,s_m,s_\text{R}^1}^{\omega_1,\dots,\omega_m,\omega_\text{R}^1}\!\ket{\omega^1_A,s^1_A}_\text{U}\!\ket{\tilde m;\omega^1_\text{R},s^1_\text{R}}_\text{I}\bra{\omega^1_A,s^1_A}_{\text{U}}\bra{\tilde m,\omega^1_\text{R},s^1_\text{R}}_\text{I}\\*
&\nonumber+\xi_{s_1,\dots,s_m,s_\text{R}^2}^{\omega_1,\dots,\omega_m,\omega_\text{R}^2}\ket{\omega^2_A,s^2_A}_{\text{U}}\ket{\tilde m;\omega^2_\text{R},s^2_\text{R}}\bra{\omega_A^2,s_A^2}_{\text{U}}\bra{\tilde m;\omega^2_\text{R},s_\text{R}^2}_\text{I}\\*
&+\xi_{s_1,\dots,s_m,s_\text{R}^1,s_\text{R}^2}^{\omega_1,\dots,\omega_m,\omega_\text{R}^1,\omega_\text{R}^2}\ket{\omega^1_A,s^1_A}_{\text{U}}\ket{\tilde m;\omega_\text{R}^1,s_\text{R}^1}_\text{I}\bra{\omega^2_A,s^2_A}_{\text{U}}\bra{\tilde m;\omega^2_\text{R},s^2_\text{R}}_\text{I}\Big)+(\text{H.c.})_{_{\substack{\text{non-}\\\text{diag.}}}}.
\end{align}
Analogously to \eqref{densmat1}, the partial transpose of \eqref{densmat2} has a $2\times2$ and $1\times1$ blocks structure. Again, we are interested in the 
$2\times2$ blocks --the ones that may have negative eigenvalues--. These blocks expressed in the basis
\begin{equation}\label{blocksbasis2}
\Big\{\ket{\omega^1_A,s^1_A}_\text{U}\ket{\tilde m;\omega^2_\text{R},s^2_\text{R}}_\text{I},\ket{s^2_A,\omega^2_A}_\text{U}\ket{\tilde m,\omega^1_\text{R},s^1_\text{R}}_\text{I}\Big\}_{m=0}^{2n-2}
\end{equation}
are of the form
\begin{equation}\label{blocks2}
\frac12
\left(\begin{array}{cc}
0 & \pm D_2^m\\
\pm D_2^m & 0
\end{array}\right).
\end{equation}
Notice that there is no diagonal elements in the block because the terms that would go in the diagonal are forbidden by Pauli exclusion principle,  which imposes 
that $\omega^1_\text{R},s^1_\text{R};\omega^2_\text{R},s^2_\text{R}\not\in\left\{\omega_i,s_i\right\}_{i=1,\dots,m}$. This time, each $2\times2$ block of the form \eqref{blocks2} appears a number of  times $B'_m$. The derivation of $B'_m$ is quite straightforward considering the derivation of $B_m$. Looking at the basis of the $2\times2$ blocks \eqref{blocksbasis2} we can see that this case would be exactly the same as the previous one but now $\{\omega_i,s_i\}$ cannot coincide neither with $\{\omega_\text{R}^1,s_\text{R}^1\}$ nor $\{\omega_\text{R}^2,s_\text{R}^2\}$. Repeating the same reasoning as before we have to do three operations as follows
\begin{itemize}
\item Discounting the combinations which have a coincidence $\{\omega_i,s_i\}=\{\omega_\text{R}^1,s_\text{R}^1\}$ from the total number \eqref{upsilonap} and obtain the expression \eqref{Nblocksap}
\item Subtracting the combinations with coincidences $\{\omega_j,s_j\}=\{\omega_\text{R}^2,s_\text{R}^2\}$
\item Taking into account that we have subtracted twice the cases in which we have double coincidences, we need to add the number of double coincidences once to compensate it.
\end{itemize}
 The number of cases with double coincidences (which require $m>1$) is the combinatory number $\binom{2n-2}{m-2}$, as we have $2n$ possible spins and frequencies minus the two fixed possibilities ($\{\omega_i,s_i\}=\{\omega_\text{R}^1,s_\text{R}^1\}$ and $\{\omega_j,s_j\}=\{\omega_\text{R}^2,s_\text{R}^2\}$) and $m$ modes being 2 of them fixed. Taking this into account
\begin{equation}\label{apend2}
B'_m=\Upsilon_m -\binom{2n-1}{m-1}-\binom{2n-1}{m-1} + \binom{2n-2}{m-2}.
\end{equation}
This expression can be simplified to
\begin{equation}\label{Nblocks2ap}B'_m=B_m-\binom{2n-2}{m-1}=\binom{2n-2}{m}.
\end{equation}

The negative eigenvalue of each block is
\begin{equation}\label{neigen2}
|\lambda^-_m|=\frac{D_2^m}{2}=\frac{\cos^{4n-2}r}{2}\tan^{2m}r,
\end{equation}
where $C^0$ has been substituted by \eqref{normalispin}. Therefore, the negativity results
\begin{equation}\label{negativity2pre}
\mathcal{N}=\sum_{m=0}^{2n-2} B'_m|\lambda^-_m| =\frac{\cos^{4n-2}r}{2}\sum_{m=0}^{2n-2}\binom{2n-2}{m}  \tan^{2m}r.
\end{equation}
This can be readily simplified to
\begin{equation}\label{negativity2pre}
\mathcal{N}= \frac12\cos^2 r.
\end{equation}
Strikingly we run into the same simple result as above\footnote{It can be proved that if we relax the approximation $r_1\approx r_2$ the negativity is the geometric mean of each mode negativity $\mathcal{N}=\frac12\cos r_1\cos r_2$.} \eqref{negativitypre}. Even starting from a spin Bell state, the entanglement is degraded by Unruh effect in the same way as in the previous case.

\section{Discussion}

Let us summarise our results so far. We have studied entanglement degradation by Unruh effect due to Rob's acceleration for  three different Minkowskian maximally entangled states: 1) Vacuum-vacuum plus one-particle-one-particle maximally entangled state of a Dirac field, 2) Vacuum-vacuum plus one-particle-one-particle maximally entangled state of a spinless fermion field 3) Multimode Bell state for a Dirac field. In spite of the essential differences among these states and the very different dimension of the Hilbert space for the three cases, the negativity degrades in exactly the same way for any acceleration. This result may look surprising but this is an outcome of fermionic statistics.

In the bosonic case acceleration excites an infinite number of modes being the Hilbert subspace that contains the state of higher dimension, and this completely degrades the entanglement in the limit $a\rightarrow\infty$. If the Hilbert space dimension were determinant for this phenomenon one could expect a similar behaviour here when we increase the number of modes involved in a fermionic entangled state, but our results show that  when two different frequency modes are involved, entanglement behaves in the same way as for the single mode entangled state. 

This striking result can be traced back to the fanciful block structure of Rob density matrix, which produces the same negativity even when the characteristics of the entangled states (and even the field) change. The culprit of this structure is fermionic statistics, (as we have discussed after \eqref{blocks3m}, \eqref{blocks2}, \eqref{blockszero}) which is responsible for the identical, and somewhat unforeseen, negativity behaviour. This is a global feature of maximally entangled states of fermion fields and not a consequence of the specific cases chosen and the number of modes considered. 

So, $\mathcal{N}\rightarrow 
1/4$ when $a\rightarrow\infty$, and this happens independently of the number of modes of the field that we are considering, of the starting maximally entangled state, and even of the spin of the field which we study. What all the cases have in common is the fermionic statistics itself, so, widening the margin for Unruh degradation for fermionic fields will not affect entanglement degradation.

Notice that a very different scenario would come from a setting in which we erase partial information for the state as Rob accelerates (e.g. angular momentum). In that case, it was shown in the previous chapter that entanglement degradation is greater than in the cases where all the information is taken into account, but this has more to do with this erasure of information than which the fermionic nature of the states.

One question immediately arises from these results; are the remaining correlations purely statistical? As all the states undergo the same degradation,  the quantum correlations which survive the infinite acceleration limit may only contain the information about the fermionic nature of the system and nothing else. We will understand better this question when we analyse fermionic entanglement beyond the single mode approximation in chapter \ref{parantpar}.

\section[Appendix to the Chapter: Density matrix construction]{Appendix to Chapter \ref{multimode}: Density matrix construction}\label{ap2}

In this appendix we will derive expressions \eqref{densmat1}, \eqref{densmat2}, \eqref{densmat3} for the density matrix of the system Alice-Rob.

Using expression \eqref{vacuumCOMP2} we see that the Alice-Rob Minkowskian operator $P_{00}\equiv\proj{0_\text{A}0_\text{R}}{0_\text{A}0_\text{R}}$ when Rob is accelerating translates into
\begin{equation}\label{0000}
 P_{00}=\sum_{m=0}^{2n}\sum_{l=0}^{2n}C^m (C^{l})^*\!\!\!\sum_{\substack{s_1,\dots,s_m\\\omega_1,\dots,\omega_m}}\!\xi_{s_1,\dots,s_m}^{\omega_1,\dots,\omega_m}\sum_{\substack{s'_1,\dots,s'_{l}\\\omega_1,\dots,\omega'_{l}}}\!\xi_{s'_1,\dots,s'_l}^{\omega'_1,\dots,\omega'_l}\ket{0}_{\text{U}}\biket{\tilde m}{\tilde m}\langle{\tilde l}|_{\text{II}}\langle{\tilde l}|_{\text{I}}\bra{0}_{\text{U}},
\end{equation}
where
\begin{align}
\nonumber\langle{\tilde l}|_{\text{I}}&=\bra{\omega'_1,s'_1;\dots; \omega'_l,s'_l}_\text{I}=\bra0_I c_{I,\omega'_1,s'_1}\dots c_{I,\omega'_m,s'_m}\\*
\langle{\tilde l}|_{\text{II}}&=\bra{\omega'_1,-s'_1;\dots; \omega'_l,-s'_l}_\text{I}=\bra0_{\text{II}} c_{II,\omega'_1,-s'_1}\dots c_{II,\omega'_m,-s'_m}.
\end{align}

Using expression \eqref{onepart23m} we can compute $P_{11}^{ij}\equiv | \omega^i_\text{A},s^i_\text{A};\omega^i_\text{R},s^i_\text{R}\rangle_\text{U}\langle \omega^j_\text{A},s^j_\text{A};\omega^j_\text{R},s^j_\text{R}|_\text{U}$ in the Rindler basis for Rob
\begin{align}\label{1111}
 \nonumber P_{11}^{ij}&=\sum_{m=0}^{2n-1}\sum_{l=0}^{2n-1}A^m(A^l)^*\sum_{\substack{s_1,\dots,s_m\\\omega_1,\dots,\omega_m}}\!\xi_{s_1,\dots,s_m,s^i_\text{R}}^{\omega_1,\dots,\omega_m,\omega^i_\text{R}}\sum_{\substack{s'_1,\dots,s'_{l}\\\omega_1,\dots,\omega'_{l}}}\!\xi_{s'_1,\dots,s'_l,s^j_\text{R}}^{\omega'_1,\dots,\omega'_l,\omega^j_\text{R}} \ket{\omega^i_\text{A},s^i_\text{A}}_{\text{U}}\biket{\tilde m;\omega^i_\text{R},s^i_\text{R}}{\tilde m}\\*
 &\times\langle \tilde l|_{\text{II}}\langle \tilde l;\omega^j_\text{R},s^j_\text{R}|_\text{I}\langle{\omega^j_\text{A},s^j_\text{A}}|_\text{A},
\end{align}
where $A^m$ is given by \eqref{Am}.

Notice that the objects $\ket{\tilde m;\omega^i_\text{R},s^i_\text{R}}_\text{I}$ represent the appropriate ordering of the elements inside with its sign, taking the criterion \eqref{ordering} into account.

Now we can use expressions \eqref{vacuumCOMP2} and \eqref{onepart23m} to obtain the operator $P_{01}\equiv\proj{00}{\omega_A,s_A;\omega_\text{R},s_\text{R}}_\text{U}$ in the Rindler basis for Rob.
\begin{equation}\label{0011}
 P_{01}=\sum_{m=0}^{2n}\sum_{l=0}^{2n-1}C^m (A^{l})^*\!\!\!\sum_{\substack{s_1,\dots,s_m\\\omega_1,\dots,\omega_m}}\!\xi_{s_1,\dots,s_m}^{\omega_1,\dots,\omega_m}\sum_{\substack{s'_1,\dots,s'_{l}\\\omega_1,\dots,\omega'_{l}}}\!\xi_{s'_1,\dots,s'_l,s_\text{R}}^{\omega'_1,\dots,\omega'_l,\omega_\text{R}}\ket{0}_{\text{U}}\biket{\tilde m}{\tilde m}\langle{\tilde l}|_{\text{II}}\langle{\tilde l;\omega_\text{R},s_\text{R}}|_{\text{I}}\bra{0}_{\text{U}}.
\end{equation}

After obtaining the expressions for the operators $P_{00},P_{11},P_{01}$ we can write the density matrix associated with the state \eqref{minkowstate1} in the Rindler basis for Rob, 
\begin{equation}\label{roimp1}
\rho=\frac12\left(P_{00}+P_{01}+P_{01}^\dagger+P^{ii}_{11}\right)
\end{equation}
Where for $P_{11}^{ii}$ we are considering $\{\omega^i_\text{R},s^i_\text{R}\}=\{\omega^j_\text{R},s^j_\text{R}\}\equiv \{\omega_\text{R},s_\text{R}\}$ and $\{\omega^i_\text{A},s^i_\text{A}\}=\{\omega^j_\text{A},s^j_\text{A}\}\equiv \{\omega_A,s_A\}$.

We can do the same to obtain the density matrix associated with \eqref{minkowstate2} in the Rindler basis for Rob
\begin{equation}\label{roimp2}
\rho=\frac12\left(P^{11}_{11}+P^{22}_{11}+P^{12}_{11}+(P^{12}_{11})^\dagger\right).
\end{equation}

Now, we must consider that, as Rob is causally disconnected from Ridler's region II, we should trace over that region to obtain Rob's density matrix. Hence,  we need to compute the trace over II for each of the previous operators \eqref{0000}, \eqref{1111}, \eqref{0011}.

Taking this trace is actually quite straightforward taking into account the orthonormality of our basis once we have chosen one particular ordering criterion \eqref{ordering}, 
\begin{equation}\label{products}
\braket{\tilde m}{\tilde m'}_\text{II}=\delta_{mm'}\left(\delta_{s_1,s'_1}\delta_{\omega_1,\omega'_1}\dots\delta_{s_m,s'_m}\delta_{\omega_m,\omega'_m}\right).
\end{equation} 
Hence,
\begin{equation}\label{traza00}
\tr_{\text{II}} P_{00}=\sum_{m'=0}^{2n}\bra{\tilde m'}_\text{II}P_{00}\ket{\tilde m'}_\text{II}.
\end{equation}
Using \eqref{products} only the diagonal elements in region II survive and \eqref{traza00} turns out to be
\begin{equation}\label{t00pre}
\tr_{\text{II}} P_{00}=\sum_{m=0}^{2n}|C^m|^2 \!\!\!\sum_{\substack{s_1,\dots,s_m\\\omega_1,\dots,\omega_m}}\!\!\!\!\xi_{s_1,\dots,s_m}^{\omega_1,\dots,\omega_m}\ket{0}_{\text{U}}\ket{\tilde m}_{\text{I}}\langle{\tilde m}|_{\text{I}}\bra{0}_{\text{U}},
\end{equation}
which, substituting $C^m$ as a function of $C^0$ using \eqref{coeff2} and then \eqref{Des}, is expressed as
\begin{equation}\label{t00}
\tr_{\text{II}} P_{00}=\sum_{m=0}^{2n}D_0^m \!\!\!\sum_{\substack{s_1,\dots,s_m\\\omega_1,\dots,\omega_m}}\!\!\!\!\xi_{s_1,\dots,s_m}^{\omega_1,\dots,\omega_m}\ket{0}_{\text{U}}\ket{\tilde m}_{\text{I}}\langle{\tilde m}|_{\text{I}}\bra{0}_{\text{U}}.
\end{equation}

Now, let us compute the trace of $P^{ij}_{11}$ over region II:
\begin{equation}\label{traza11}
\tr_{\text{II}} P^{ij}_{11}=\sum_{m'=0}^{2n}\bra{\tilde m'}_\text{II}P^{ij}_{11}\ket{\tilde m'}_\text{II},
\end{equation}
\begin{equation}\label{t11pre}
 \tr_{\text{II}} P_{11}^{ij}=\sum_{m=0}^{2n-1}|A^m|^2 \!\!\!\sum_{\substack{s_1,\dots,s_m\\\omega_1,\dots,\omega_m}}\xi_{s_1,\dots,s_m,s^j_\text{R}}^{\omega_1,\dots,\omega_m,\omega^j_\text{R}}\ket{\omega_A,s_A}_{\text{U}}\ket{\tilde m;\omega_\text{R},s_\text{R}}_{\text{I}}\langle{\tilde m;\omega_\text{R},s_\text{R}}|_{\text{I}}\bra{\omega_A,s_A}_{\text{U}}.
\end{equation}
Substituting $C^m$ as a function of $C^0$ (combining \eqref{Am} and \eqref{coeff2}) we can express
\begin{equation}
|A^m|^2=|C^0|^2 \tan^{2m}r\left(\cos r+\frac{\sin^2 r}{\cos r}\right)^2=|C^0|^2 \frac{\tan^{2m}r}{\cos^2 r}=D^m_2
\end{equation}
such that we obtain
\begin{equation}\label{t11}
\tr_{\text{II}} P^{ii}_{11}=\sum_{m=0}^{2n-1}D^m_2 \!\!\!\sum_{\substack{s_1,\dots,s_m\\\omega_1,\dots,\omega_m}}\!\xi_{s_1,\dots,s_m,s_\text{R}}^{\omega_1,\dots,\omega_m,\omega_\text{R}}\ket{\omega^i_\text{A},s^i_\text{A}}_{\text{U}}\ket{\tilde m;\omega^i_\text{R},s^i_\text{R}}_{\text{I}}\langle{\tilde m;\omega^j_\text{R},s^j_\text{R}}|_{\text{I}}\langle{\omega^j_\text{A},s^j_\text{A}}|_{\text{U}}.
\end{equation}
When $\{\omega^i_\text{R},s^i_\text{R}\}=\{\omega^j_\text{R},s^j_\text{R}\}$ $\equiv$ $\{\omega_\text{R},s_\text{R}\}$, $\{\omega^i_\text{A},s^i_\text{A}\}=\{\omega^j_\text{A},s^j_\text{A}\}\equiv \{\omega_A,s_A\}$, and
\begin{equation}\label{t112}
\tr_{\text{II}} P^{ij}_{11}=\sum_{m=0}^{2n-1}D^m_2 \!\!\!\sum_{\substack{s_1,\dots,s_m\\\omega_1,\dots,\omega_m}}\!\xi_{s_1,\dots,s_m,s^i_\text{R},s^j_\text{R}}^{\omega_1,\dots,\omega_m,\omega^i_\text{R},\omega^j_\text{R}}\ket{\omega^i_\text{A},s^i_\text{A}}_{\text{U}}\ket{\tilde m;\omega^i_\text{R},s^i_\text{R}}_{\text{I}}\langle{\tilde m;\omega^j_\text{R},s^j_\text{R}}|_{\text{I}}\langle{\omega^j_\text{A},s^j_\text{A}}|_{\text{U}}
\end{equation}
for $i\neq j$.

Now, let us compute the trace
\begin{equation}\label{traza01pre}
\tr_{\text{II}} P_{01}=\sum_{m'=0}^{2n}\bra{\tilde m'}_\text{II}P_{01}\ket{\tilde m'}_\text{II},
\end{equation}
\begin{equation}\label{traza01pre2}
\tr_{\text{II}} P_{01}=\sum_{m=0}^{2n-1}C^m (A^{m})^*\!\!\!\sum_{\substack{s_1,\dots,s_{m}\\\omega_1,\dots,\omega_{m}}}\!\xi_{s_1,\dots,s_m,s_\text{R}}^{\omega_1,\dots,\omega_m,\omega_\text{R}}\ket{0}_{\text{U}}\ket{\tilde m}_\text{I}\langle{\tilde l;\omega_\text{R},s_\text{R}}|_{\text{I}}\bra{0}_{\text{U}}.
\end{equation}
From \eqref{Am} and \eqref{coeff2} we see that the product $C^m(A^m)^*$ is real and has the expression
\begin{equation}
C^m(A^m)^*=|C^0|^2\tan^{2m}r\left(\cos r+\frac{\sin^2 r}{\cos r}\right)=|C^0|^2\frac{\tan^{2m}r}{\cos r}=D^m_1
\end{equation}
so that
\begin{equation}\label{traza01}
 \tr_{\text{II}} P_{01}=\sum_{m=0}^{2n-1}D^m_1\!\!\!\sum_{\substack{s_1,\dots,s_{m}\\\omega_1,\dots,\omega_{m}}}\!\xi_{s_1,\dots,s_m,s_\text{R}}^{\omega_1,\dots,\omega_m,\omega_\text{R}}\ket{0}_{\text{U}}\ket{\tilde m}_\text{I}\langle{\tilde l;\omega_\text{R},s_\text{R}}|_{\text{I}}\bra{0}_{\text{U}}.
\end{equation}

Now we can compute Rob's density matrices for each case tracing over II in expressions \eqref{roimp1} and \eqref{roimp2}. First the matrix \eqref{roimp1} is, after tracing over II, 
\begin{equation}
\tr_{\text{II}}\rho=\frac12\left(\tr_{\text{II}}P_{00}+\tr_{\text{II}}P_{01}+\tr_{\text{II}}P_{01}^\dagger+\tr_{\text{II}}P^{ii}_{11}\right).
\end{equation}
Substituting expressions \eqref{t00}, \eqref{t112}, \eqref{traza01} we get expression \eqref{densmat1}.

Now, concerning \eqref{roimp2}
\begin{equation}
\tr_{\text{II}} \rho=\frac12\tr_{\text{II}}\left( P^{11}_{11}+P^{22}_{11}+P^{12}_{11}+(P^{12}_{11})^\dagger\right).
\end{equation}
Substituting expressions \eqref{t11} and \eqref{t112} we obtain expression \eqref{densmat2}.

The derivation of \eqref{densmat3} is completely analogous to \eqref{densmat1}, taking now into account that we have $\hat C^m$ and $\hat D^m$ instead of $C^m$ and $D^m$ and that we have no spin degree of freedom. Notice that, even though the structure of \eqref{densmat3} is completely analogous to the structure of \eqref{densmat1}, and therefore, repeating the derivation will add nothing to this appendix,  these density matrices are completely different due to the different dimensions, the different values of $\hat C^0$ and $C^0$ and the number of $2\times 2$ blocks which give negative eigenvalues.

\cleardoublepage


\chapter{Entanglement through the acceleration horizon\footnote{E. Mart\'in-Mart\'inez, J. Le\'on, Phys. Rev. A, 81, 032320 (2010)}}\label{etanthrough}

As thoroughly discussed in previous chapters, the Unruh effect degrades the entanglement between the two partners affecting all the quantum information tasks that they could perform. Specifically, it has been shown that, as Rob accelerates, entanglement is completely degraded for a scalar field \cite{Alicefalls} and, conversely, some degree of entanglement is preserved for fermionic fields. This behaviour of fermionic fields has been shown to be universal.  Namely, it is independent of  i) the spin of the fermionic field,  ii) the kind of maximally entangled state from which we start, and iii) the number of participating modes when studying a non-single mode state.

When Rob accelerates, the description of his partial state must be done by means of a basis built from solutions to the field equation in Rindler coordinates \cite{gravitation,Takagi}. As it will be shown below, the description of the system splits in three different subsystems; Alice's Minkowskian system, a subsystem in region I of Rindler spacetime (which we assign to Rob) and another subsystem, called AntiRob, constituted by the modes of the field in region $\text{II}$ of Rindler space time.

\begin{figure}[h]
\begin{center}
\includegraphics[width=.80\textwidth]{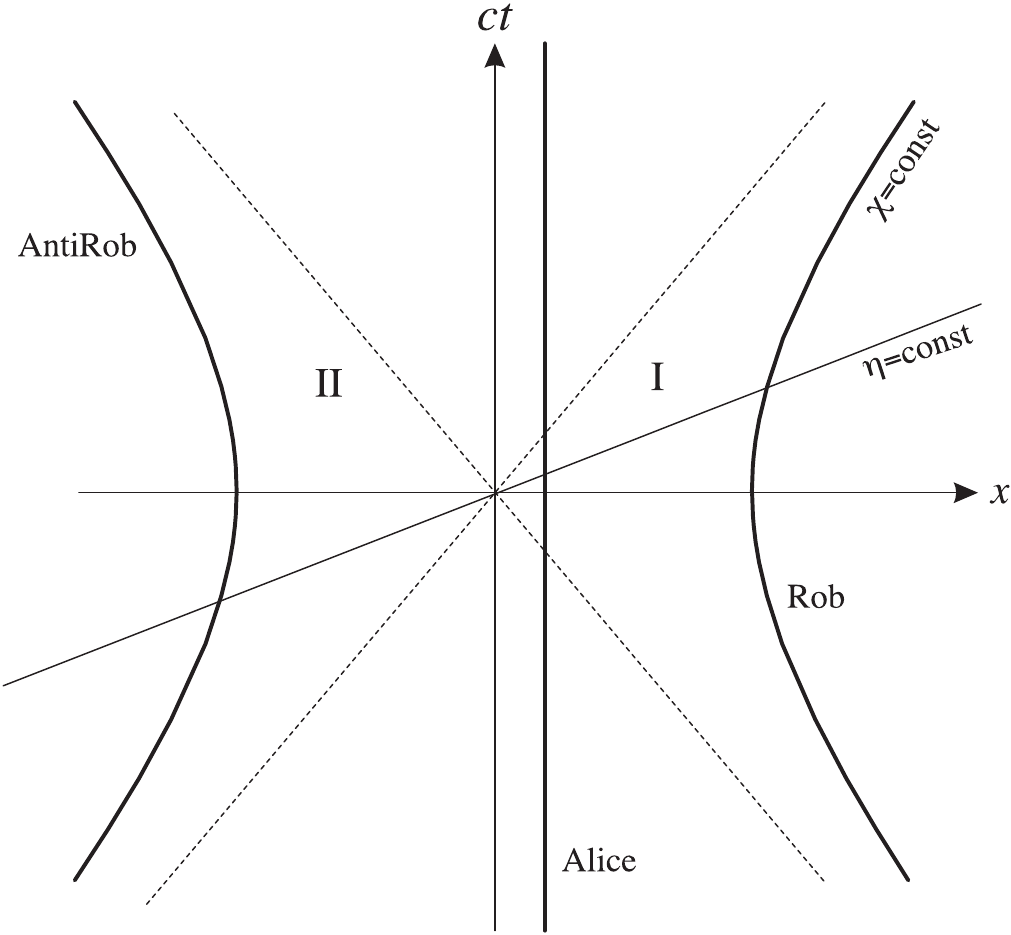}
\end{center}
\caption{Flat spacetime. Trajectories of an inertial (Alice) and causally disconnected accelerated observers (Rob and AntiRob)}
\label{rar}
\end{figure}

Any accelerated observer is constrained to either region I or $\text{II}$ of Rindler spacetime. If we select region I coordinates to account for the accelerated observer Rob, he would remain causally disconnected from region $\text{II}$, and therefore, Rob would be unable to communicate with the hypothetical observer AntiRob (who is accelerating with the same proper acceleration than Rob but decelerates with respect to the origin) in region $\text{II}$ as shown in Figure \ref{rar}.

To gain a deeper understanding of the entanglement degradation mechanisms it is useful to study how entanglement is lost as one traces over regions of the Rindler spacetime. Although the system is obviously bipartite (Alice and Rob), we saw that shifting to the Rindler basis for Rob the mathematical description of the system (\cite{Alicefalls,AlsingSchul} and previous chapters) admits a straightforward tripartition: Minkowskian modes (Alice), Rindler region I modes (Rob), and Rindler region $\text{II}$  modes (AntiRob). 

 Let us revisit the physical meaning  of each of
these three `observers'. Alice represents an observer in an inertial frame.
For Alice the states
\eqref{Min1} and \eqref{Minf} are maximally entangled.
Rob represents an accelerated observer moving in a $x=a^{-1}$
trajectory in Region I of Rindler spacetime (as seen in Fig. \ref{rar}) who shares a bipartite entangled state \eqref{Min1} or \eqref{Minf} with Alice.
AntiRob represents an observer  moving in a $x=a^{-1}$ trajectory in
Region II with access to the information to which Rob is not able to
access due to the presence of the Rindler horizon.

In this chapter, instead of considering only the Alice-Rob bipartition, we deal with all the different bipartitions of the system to study the correlation tradeoff among them. These three bipartitions are\begin{enumerate}
\item Alice-Rob $(\text{AR})$
\item Alice-AntiRob $(\text{A}{\bar{\text{R}}})$
\item Rob-AntiRob $(\text{R}{\bar{\text{R}}})$
\end{enumerate}

The first bipartition is the most commonly considered in the literature. It represents the system formed by an inertial observer and the modes of the field which an accelerated observer is able to access.

The second bipartition represents the subsystem formed by the inertial observer Alice and the modes of the field which Rob is not able to access due to the presence of an horizon as he accelerates.

The third bipartition lacks physical meaning in terms of information theory because communication between Rob and AntiRob is not allowed. Anyway, studying this bipartition is still useful to account for the correlations which are created between the spacetime regions separated by horizons and, therefore, its study is necessary and complementary to the previous ones in order to give a complete description of the information behaviour in non-inertial settings.

In \cite{AlsingSchul} the existence of these three possible bipartitions was considered only for spinless fermion fields. Bosonic fields were analysed in a completely different formalism (covariance matrices) in \cite{Adeschul} finding relationships between the correlations in both sides of the horizon. In this chapter we will go beyond previous analysis and we will compare the correlation tradeoff among different bipartitions for bosonic and fermionic fields in the standard formalism introduced before, showing the leading role of statistics in the behaviour of information in non-inertial frames. 

The work presented in this chapter will be useful to take a step in the discussion and refutation of the argument that the dimension of the Hilbert space is responsible for the difference between fermionic and bosonic entanglement behaviour in the presence of horizons. Indeed, we will present here an entanglement tradeoff between the bipartitions Alice-Rob and AliceAntiRob that only occurs for the fermionic case and that will reveal to be deeply connected with the fermionic entanglement survival in the limit $a\rightarrow\infty$.

\section{Scalar and Dirac fields from constantly accelerated frames}\label{sec2}

Following the standard conventions, let us denote the particle annihilation and creation operators in region I as $(a_{\omega,\text{I}}^{\phantom{\dagger}},a^{\dagger}_{\omega,\text{I}})$ for the scalar field  and $(c^{\phantom{\dagger}}_{\omega,s,\text{I}},c^{\dagger}_{\omega,s,\text{I}})$ for the Dirac field. $(d^{\phantom{\dagger}}_{\omega,s,\text{I}},d^{\dagger}_{\omega,s,\text{I}})$ are the corresponding antiparticle Dirac field operators. Analogously we define $(a^{\phantom{\dagger}}_{\omega\text{II}},a^{\dagger}_{\text{II},k})$, $(c^{\phantom{\dagger}}_{\omega,s,\text{II}},c^{\dagger}_{\omega,s,\text{II}})$, $(d^{\phantom{\dagger}}_{\omega,s,\text{II}},d_{\omega,s,\text{II}}^\dagger)$ as the particle/antiparticle operators in region \text{II}.

The bosonic operators satisfy the commutation relations $[a^{\phantom{\dagger}}_{\omega,\Sigma},a^\dagger_{\omega,\Sigma'}]=\delta_{\Sigma\Sigma'}\delta_{\omega\omega'}$. The fermionic ones satisfy  anticommutation relations $\{d^{\phantom{\dagger}}_{\omega,s,\Sigma},d^\dagger_{\omega',s',\Sigma'}\}=\delta_{\Sigma\Sigma'}\delta_{\omega\omega'}\delta_{ss'}$. The label $\Sigma$ notates the Rindler region of the operator  $\Sigma=\{\text{I},\text{II}\}$.  All other commutators and anticommutators are zero. This includes the anticommutators between operators in different Rindler regions.

We can relate Minkowski Unruh operators and Rindler creation and annihilation operators recalling the Bogoliubov relationships \eqref{buenmod}, \eqref{Bogoferm2}.

For a scalar field, the Bogoliubov relationships for the annihilation operator of modes with positive frequency are
\begin{equation}\label{bogoboson}
 a_{\omega,\text{U}}=\cosh r_\text{b}\, a_{\omega,\text{I}} - \sinh r_\text{b}\, a^\dagger_{\omega,\text{II}},
\end{equation}
where
\begin{equation}\label{defr1}
\tanh r_\text{b}=e^{-\pi \frac{\omega c}{a}}.
\end{equation}

For a Dirac field, the Bogoliubov relationships take the form
\begin{eqnarray}\label{bogodirac}
\nonumber c_{\omega,s,\text{U}}&=&\cos{r_\text{f}}\,c_{\omega,s,\text{I}}-\sin r_\text{f}\,d^\dagger_{\omega,-s,\text{II}}\\*
d^\dagger_{\omega,s,\text{U}}&=&\cos{r_\text{f}}\,d^\dagger_{\omega,s,\text{II}}+\sin r_\text{f}\,c_{\omega,-s,\text{I}},
\end{eqnarray}
where
\begin{equation}\label{defr2}
\tan r_\text{f}=e^{-\pi \frac{\omega c}{a}}.
\end{equation}

\section{Vacuum and one particle states}\label{sec3}

As it is shown in \cite{Alicefalls} and in previous chapters, the vacuum state of a scalar field as seen from the perspective of an accelerated observer is
\begin{equation}\label{scavac}
\ket{0}=\prod_{\omega}\frac{1}{\cosh r_\text{b}}\sum_{n=0}^\infty \tanh^n r_\text{b} \ket{n_\omega}_\text{I}\ket{n_{\omega}}_{\text{II}},
\end{equation}
and as it was discussed in previous chapters, the unprimed sector of the vacuum state for a Dirac field as seen from the accelerated frame is
\begin{equation}\label{diravac}
 \ket{0_\omega}=\cos^2 r_\text{f}\biketn{0}{0}+\sin r_\text{f}\cos r_\text{f}\left(\biketn{\uparrow}{\downarrow}+\biketn{\downarrow}{\uparrow}\right)+\sin^2 r_\text{f}\biketn{\pa}{\pa},
\end{equation}
where $\ket{\pa_\omega}^\pm$ represents the pair of particles/antiparticles of frequency $\omega$.

From now on we will drop the sign $\pm$ as as reasoned in section \ref{sec42m} a mode in region I will always be a particle mode and a mode in region $\text{II}$ will always represent an antiparticle mode. To simplify notation we will also drop the $\omega$ label as we focus on a single mode state.

As usual we will also need the Minkowskian Unruh one particle state in the Rindler basis which is obtained by applying $a^\dagger_{\text{U}}$ to the vacuum state. i.e. 
\begin{equation}
\ket{1}_\text{U}=\frac{1}{\cosh^2 r_\text{b}}\sum_{n=0}^\infty \tanh^n r_\text{b} \,\sqrt{n+1}\ket{n+1}_\text{I}\ket{n}_{\text{II}}
\end{equation} 	
for the scalar field \cite{Alicefalls} and 
\begin{eqnarray}\label{onepart2}
\nonumber\ket\uparrow_\text{U}&=&\cos r_\text{f} \biket{\uparrow}{0}+\sin r_\text{f}\biket{\pa}{\uparrow}\\*
\ket\downarrow_\text{U}&=&\cos r_\text{f} \biket{\downarrow}{0}-\sin r_\text{f}\biket{\pa}{\downarrow}
\end{eqnarray}
for the Dirac field (See chapter \ref{onehalf}).

Now we need to consider the following maximally entangled states in the Minkowski Unruh basis
\begin{eqnarray}
\label{entangledsca}\ket{\Psi_\text{b}}&=&\frac{1}{\sqrt{2}}\left(\ket{0}\ket{0}+\ket{1}_\text{U}\ket{1}_\text{U}\right),\\*
\label{entangleddir}\ket{\Psi_\text{f}}&=&\frac{1}{\sqrt{2}}\left(\ket{0}\ket{0}+\ket{\uparrow}_\text{U}\ket{\downarrow}_\text{U}\right).
\end{eqnarray}
These two maximally entangled states are analogous, both are bipartite qubit states superpositions of the vacuum and the one particle state. The difference is that in \eqref{entangleddir} we have a Dirac field state and hence, the one particle states have spin and follow fermionic statistics.

For $\ket{\Psi_\text{f}}$ we have selected one amongst the possible values for the spin of the terms with one particle for Alice and Rob, but it can be shown (see chapter \ref{onehalf}) that the choice of a specific value for these spins is not relevant when considering the behaviour of correlations. Then, the results presented here are independent of the particular choice of a spin state for the superposition \eqref{entangleddir}.

\section{Correlations for the Dirac field}\label{sec4m4}

The density matrix for the whole tripartite state, which includes modes on both sides of the Rindler horizon along with Minkowskian modes, is built from \eqref{entangleddir}
\begin{equation}\label{tripadir}
\rho^{A\text{R}{\bar{\text{R}}}}_\text{f}=\proj{\Psi_\text{f}}{\Psi_\text{f}}.
\end{equation}

The three different bipartite partial density matrices are obtained by partial tracing:
\begin{eqnarray}
\label{AR2}\rho^\text{AR}_\text{f}&\!\!=\!&\tr_{\text{II}}\rho^{A\text{R}{\bar{\text{R}}}}_\text{f}=\!\!\!\!\sum_{s\in\{0,\uparrow,\downarrow,\pa\}} \bra{s}_{\text{II}}\rho_\text{f}^{A\text{R}{\bar{\text{R}}}}\ket{s}_{\text{II}},\\*
\label{AAR2}\rho^{\text{A}{\bar{\text{R}}}}_\text{f}&\!\!=\!&\tr_\text{I}\rho^{A\text{R}{\bar{\text{R}}}}_\text{f}=\!\!\!\!\sum_{s\in\{0,\uparrow,\downarrow,\pa\}} \bra{s}_\text{I}\rho_\text{f}^{A\text{R}{\bar{\text{R}}}}\ket{s}_\text{I},\\*
\label{RAR2}\rho^{\text{R}{\bar{\text{R}}}}_\text{f}&\!\!=\!&\tr_\text{U}\rho^{A\text{R}{\bar{\text{R}}}}_\text{f}=\!\!\!\!\sum_{s\in\{0,\uparrow,\downarrow,\pa\}} \bra{s}_\text{U}\rho_\text{f}^{A\text{R}{\bar{\text{R}}}}\ket{s}_\text{U},
\end{eqnarray}
and the density matrix for each individual subsystem is obtained by tracing over the other subsystems,
 \begin{eqnarray}
\label{A2}\rho^{\text{A}}_\text{f}&=&\tr_\text{I}\rho^\text{AR}_\text{f}=\tr_{\text{II}}\rho^{\text{A}{\bar{\text{R}}}}_\text{f},\\*
\label{R2}\rho^{\text{R}}_\text{f}&=&\tr_{\text{II}}\rho^{\text{R}{\bar{\text{R}}}}_\text{f}=\tr_\text{U}\rho^\text{AR}_\text{f},\\*
\label{aR2}\rho^{{\bar{\text{R}}}}_\text{f}&=&\tr_\text{I}\rho^{\text{R}{\bar{\text{R}}}}_\text{f}=\tr_\text{U}\rho^{\text{A}{\bar{\text{R}}}}_\text{f}.
\end{eqnarray}
In the cases AR and $\text{A}\bar{\text{R}}$, there are physical
arguments to justify the need for the partial trace beyond mere
quantum information considerations. Namely, Rob will never be  able to
access region II of the spacetime due to the presence of the Rindler
horizon so that $\bar{\text{R}}$ (Region $\text{II}$) must be traced
out. Likewise, AntiRob is not able to access region I  because of the
horizon and hence R (Region I) must be traced out. For the subsystem
${\text{R}\bar{\text{R}}}$ taking the partial trace over subsystem A corresponds
to the standard procedure for analysing correlations between two parts
of a multipartite system.

The different bipartitions are characterised by the following density matrices
\begin{align}\label{rhoars2}
\nonumber \rho^\text{AR}_\text{f}&=\frac12\Big[\cos^4 r_\text{f} \proj{00}{00}+\sin^2 r_\text{f}\,\cos^2 r_\text{f}\Big(\proj{0\uparrow}{0\uparrow}+\proj{0\downarrow}{0\downarrow}\Big)+\sin^4 r_\text{f}\proj{0\pa}{0\pa}\\
&+\cos^3r_\text{f}\Big(\ket{00}\bra{\uparrow\downarrow}  +\proj{\uparrow\downarrow}{00}\Big)-\sin^2 r_\text{f}\cos r_\text{f}\Big(\proj{0\uparrow}{\uparrow\pa}+\proj{\uparrow\pa}{0\uparrow}\Big)\nonumber\\*
& +\cos^2 r_\text{f} \proj{\uparrow\downarrow}{\uparrow\downarrow}+\sin^2 r_\text{f}\proj{\uparrow\pa}{\uparrow\pa}\Big],
\end{align}
\begin{align}\label{rhoa-rs2}
\nonumber \rho^{\text{A}{\bar{\text{R}}}}_\text{f}&=\frac12\Big[\cos^4 r_\text{f} \proj{00}{00}+\sin^2 r_\text{f}\,\cos^2 r_\text{f}\Big(\proj{0\downarrow}{0\downarrow}+\proj{0\uparrow}{0\uparrow}\Big)+\sin^4 r_\text{f}\proj{0\pa}{0\pa} \nonumber\\*
& -\sin^3r_\text{f}\Big(\proj{0\pa}{\uparrow\downarrow}+\proj{\uparrow\downarrow}{0\pa}\Big)+\sin r_\text{f}\cos^2 r_\text{f}\Big(\ket{0\uparrow}\bra{\uparrow0}+\proj{\uparrow0}{0\uparrow}\Big)\nonumber\\*
&+\cos^2 r_\text{f} \proj{\uparrow0}{\uparrow0}+\sin^2 r_\text{f}\proj{\uparrow\downarrow}{\uparrow\downarrow}\Big],
\end{align}
\begin{align}\label{rhor-rs2}
\nonumber \rho^{\text{R}{\bar{\text{R}}}}_\text{f}&=\frac{1}{2}\Big[\cos^4 r_\text{f} \proj{00}{00}+\sin r_\text{f}\, \cos^3 r_\text{f}\Big(\proj{00}{\uparrow\downarrow}+\ket{00}\bra{\downarrow\uparrow}+\proj{\uparrow\downarrow}{00}+\proj{\downarrow\uparrow}{00}\Big)\\*
&\nonumber +\sin^2 r_\text{f}\cos^2 r_\text{f}\Big(\ket{00}\bra{\pa\pa}\!+\!\proj{\uparrow\downarrow}{\uparrow\downarrow}\!+\!\proj{\uparrow\downarrow}{\downarrow\uparrow}\!+\!\proj{\downarrow\uparrow}{\uparrow\downarrow}\!+\!\proj{\downarrow\uparrow}{\downarrow\uparrow}+\proj{\pa\pa}{00}\Big)\\*
&\nonumber + \sin^3 r_\text{f}\,\cos r_\text{f}\Big(\proj{\uparrow\downarrow}{\pa\pa}+\proj{\pa\pa}{\uparrow\downarrow}+\ket{\downarrow\uparrow}\bra{\pa\pa}\!+\!\proj{\pa\pa}{\downarrow\uparrow}\Big)+\cos^2 r_\text{f}\proj{\downarrow 0}{\downarrow 0}\\*
&\nonumber+\sin^2 r_\text{f}\proj{\pa\downarrow}{\pa\downarrow}-\cos r_\text{f}\,\sin r_\text{f}\Big(\proj{\downarrow0}{\pa\downarrow}+\ket{\pa\downarrow}\bra{\downarrow0}\Big)+\sin^4r_\text{f}\proj{\pa\pa}{\pa\pa}\Big],\\*
\end{align}
where the bases are
\begin{eqnarray}\label{barbolbasis2}
 \ket{nm}&=&\ket{n^\text{A}}_\text{U}\ket{m^\text{R}}_\text{I},\\*
\ket{nm}&=&\ket{n^\text{A}}_\text{U}|m^{{\bar{\text{R}}}}\rangle_{\text{II}},\\*
\ket{nm}&=&\ket{n^\text{R}}_\text{I}|m^{{\bar{\text{R}}}}\rangle_{\text{II}}
\end{eqnarray}
respectively for \eqref{rhoars2}, \eqref{rhoa-rs2} and \eqref{rhor-rs2}.

On the other hand, the density matrices for the individual subsystems \eqref{A2}, \eqref{R2},\eqref{aR2} are
\begin{align}\label{robfpartialstate}
\nonumber \rho^{\text{R}}_\text{f}&=\frac12\Big[\sin^2r_\text{f}(1+\sin^2 r_\text{f})\proj{\pa}{\pa}+\sin^2 r_\text{f}\cos^2r_\text{f}\proj{\uparrow}{\uparrow}+\cos^2 r_\text{f}(1+\sin^2 r_\text{f})\proj{\downarrow}{\downarrow}\\*
&+\cos^4r_\text{f}\proj{0}{0}\Big],
\end{align}
\begin{align}\label{arobfpartialstate}
\nonumber \rho^{{\bar{\text{R}}}}_\text{f}&=\frac12\Big[\cos^2r_\text{f}(1+\cos^2r_\text{f})\proj{0}{0}+\sin^2 r_\text{f}\cos^2r_\text{f}\proj{\uparrow}{\uparrow}+\sin^2 r_\text{f}(1+\cos^2 r_\text{f})\proj{\downarrow}{\downarrow}\\*&+\sin^4r_\text{f}\proj{\pa}{\pa}\Big],
\end{align}
\begin{equation}\label{alicefpartialstate}
 \rho^{\text{A}}_\text{f}=\frac12\left(\proj{0}{0}+\proj{\uparrow}{\uparrow}\right).
\end{equation}

\subsection{Mutual Information: creation, exchange and conservation}

In this section we will compute mutual information (see section \ref{mutusec}) which accounts for correlations (both quantum and classical) between two different parts of a system. It is defined as
\begin{equation}\label{mutualdef}
I_{AB}=S_A+S_B-S_{AB},
\end{equation}
where $S_A$, $S_B$ and $S_{AB}$ are respectively the Von Neuman entropies for the individual subsystems $A$ and $B$ and for the joint system $AB$.

To compute the mutual information  for each bipartition we will need the eigenvalues of the corresponding density matrices. We shall go through all the process step by step in the lines below.

\subsubsection{Bipartition Alice-Rob}

The eigenvalues of the matrix for the system Alice-Rob \eqref{rhoars2} are
\begin{eqnarray}\label{eigAR4m1}
\nonumber \lambda_1&=&\lambda_2=0,\\*
\nonumber \lambda_3&=&\frac12\sin^2r_\text{f}\cos^2r_\text{f},\\*
\nonumber \lambda_4&=&\frac12\sin^4r_\text{f},\\*
\nonumber \lambda_5&=&\frac12\cos^2r_\text{f}\left(1+\cos^2r_\text{f}\right),\\*
\lambda_6&=&\frac12\sin^2r_\text{f}\left(1+\cos^2r_\text{f}\right).
\end{eqnarray}

\subsubsection{Bipartition Alice-AntiRob}

The eigenvalues of the matrix for the system Alice-AntiRob \eqref{rhoa-rs2} are
\begin{eqnarray}\label{eigAaR4m1}
\nonumber \lambda_1&=&\lambda_2=0,\\*
\nonumber \lambda_3&=&\frac12\sin^2r_\text{f}\cos^2 r_\text{f},\\*
\nonumber \lambda_4&=&\frac12\cos^4 r_\text{f},\\*
\nonumber \lambda_5&=&\frac12\sin^2r_\text{f} \left(1+\sin^2 r_\text{f}\right),\\*
 \lambda_6&=&\frac12\cos^2 r_\text{f}\left(1+\sin^2 r_\text{f}\right).
\end{eqnarray}

\subsubsection{Bipartition Rob-AntiRob}

All the eigenvalues of the matrix for the system Rob-AntiRob \eqref{rhor-rs2} are zero excepting two of them
\begin{equation}\label{eigRaR4m1}
\lambda_1=\lambda_2=\frac12,\qquad \lambda_{i>2}=0.
\end{equation}

\subsubsection{Von Neumann entropies for each subsystem and mutual information}

To compute the Von Neumann entropies we need the eigenvalues of every bipartition and the individual density matrices. The eigenvalues of $\rho^\text{AR}_\text{f}$, $\rho^{\text{A}{\bar{\text{R}}}}_\text{f}$, $\rho^{\text{R}{\bar{\text{R}}}}_\text{f}$ are respectively \eqref{eigAR4m1}, \eqref{eigAaR4m1} and \eqref{eigRaR4m1}.

The eigevalues of the individual systems density matrices can be directly read from \eqref{robfpartialstate}, \eqref{arobfpartialstate} and \eqref{alicefpartialstate} since $\rho^\text{R}_\text{b}$, $\rho^{{\bar{\text{R}}}}_\text{b}$ and $\rho^\text{A}_\text{b}$ have diagonal forms in the given basis.

The Von Neumann entropy for a partition $B$ of the system is
\begin{equation}\label{Vonneu}
S_B=-\tr(\rho\log_2 \rho)=-\sum \lambda_\text{I}\log_2\lambda_\text{I}.
\end{equation}

At this point, computing the entropies is quite straightforward. The Von Neumann entropies for all the partial systems are
\begin{align}
\nonumber S_R&=1-\sin^2r_\text{f}\log_2(\sin^2r_\text{f})-\frac32\cos^2r_\text{f}\log_2(\cos^2r_\text{f})-\frac{1+\sin^2r_\text{f}}{2}\log_2(1+\sin^2r_\text{f}),\\*
\nonumber S_{\bar R}&=1-\cos^2r_\text{f}\log_2(\cos^2r_\text{f})-\frac32\sin^2r_\text{f}\log_2(\sin^2r_\text{f})-\frac{1+\cos^2r_\text{f}}{2}\log_2(1+\cos^2r_\text{f}),
\end{align}
\begin{equation}
 S_{AR}=S_{\bar R}; \qquad S_{\text{A}{\bar{\text{R}}}}=S_{R}; \qquad S_{R\bar R}=S_{A}=1.
 \end{equation}

And then, the mutual information for all the possible bipartitions of the system will be
\begin{eqnarray}
\nonumber I_{AR}&=&S_A+S_R-S_{A R}=1+S_R -S_{\bar R},\\*
\nonumber I_{\text{A}{\bar{\text{R}}}}&=&S_A+S_{\bar R}-S_{\text{A}{\bar{\text{R}}}}=1+ S_{\bar R}-S_R,\\*
\nonumber I_{R\bar R}&=& S_R+S_{\bar R}-S_{R\bar R}=S_R+S_{\bar R}-1.
\end{eqnarray}  
At first glance we see a conservation law for the mutual information for the system Alice-Rob and Alice-AntiRob
\begin{equation}\label{conservation1}
 I_{AR} + I_{\text{A}{\bar{\text{R}}}}=2,
\end{equation}
which suggests a correlation transfer from the system Alice-Rob to Alice-AntiRob as the acceleration increases.

\begin{figure}[h]
\begin{center}
\includegraphics[width=.85\textwidth]{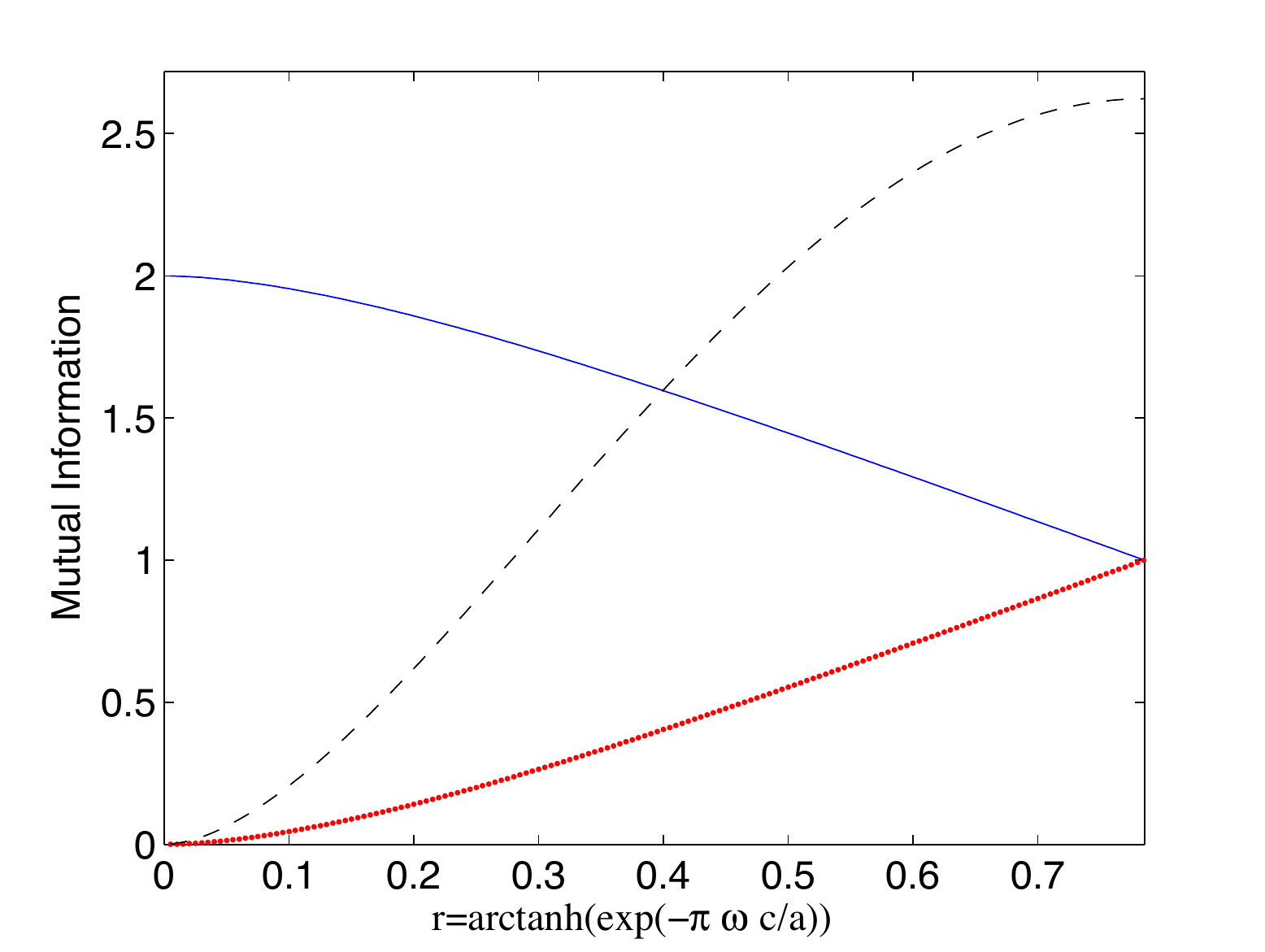}
\end{center}
\caption{ Dirac field: Mutual information tradeoff and conservation law between the systems Alice-Rob and Alice-AntiRob as acceleration varies. It is also shown the behaviour of the mutual information for the system Rob-AntiRob. Blue continuous line: Mutual information $AR$, red dotted line: Mutual information $\text{A}{\bar{\text{R}}}$, black dashed line: Mutual information $R\bar R$.}
\label{mutuferm}
\end{figure}

Fig. \ref{mutuferm} shows the behaviour of the mutual information for the three bipartitions. It also shows how the correlations across the horizon (Rob and AntiRob) increase, up to certain finite limit, as Rob accelerates. 

If we recall the results on spinless fermion fields \cite{AlsingSchul} we see that the conservation law obtained here is also valid for that spinless fermion case. This result was expected according to the universality argument stated  in the previous chapter.

However, something different occurs with the system Rob-AntiRob. The creation of correlations between modes on both sides of the horizon is greater in the Dirac field case. One have to be careful interpreting this result since the negativity entropy upper bound is a function of the dimension of the Hilbert space. The  fermionic field has a finite upper limit. For bosons the unbounded dimension of the Hilbert space implies that negativity can grow unboundedly. In principle this does not guarantee that one can extract more information from bosons than from fermions, even more when we are concerned about correlations between Rob and Anti-Rob that cannot communicate with each other.

\subsection{Entanglement conservation and behaviour across the Rindler horizon}\label{conservanet}

To compute the negativity, we will need the partial transpose of the bipartite density matrices \eqref{rhoars2}, \eqref{rhoa-rs2} and \eqref{rhor-rs2}, which we will notate as $\eta^\text{AR}_\text{f}$, $\eta^{\text{A}{\bar{\text{R}}}}_\text{f}$ and $\eta^{\text{R}{\bar{\text{R}}}}_\text{f}$  respectively.

\begin{align}\label{etaARd}
\nonumber\eta^\text{AR}_\text{f}&=\frac12\Big[\cos^4r_\text{f}\proj{00}{00}+\sin^2r_\text{f}\cos^2r_\text{f}\left(\proj{0\uparrow}{0\uparrow}+\proj{0\downarrow}{0\downarrow}\right)+\sin^4r_\text{f}\proj{0\pa}{0\pa}\\*
\nonumber&+\cos^3r_\text{f}\left(\proj{0\downarrow}{\uparrow0}+\proj{\uparrow0}{0\downarrow}\right)-\sin^2r_\text{f}\,\cos r_\text{f}\left(\proj{0\pa}{\uparrow\uparrow}+\proj{\uparrow\uparrow}{0\pa}\right)\\*
&+\cos^2r_\text{f}\proj{\uparrow\downarrow}{\uparrow\downarrow}+\sin^2r_\text{f}\proj{\uparrow\pa}{\uparrow\pa}\Big],
\end{align}
\begin{align}\label{etaAaRd}
\nonumber \eta^{\text{A}{\bar{\text{R}}}}_\text{f}&=\frac12\Big[\cos^4 r_\text{f} \proj{00}{00}+\sin^2 r_\text{f}\cos^2 r_\text{f}\Big(\proj{0\downarrow}{0\downarrow}+\proj{0\uparrow}{0\uparrow}\Big)+\sin^4 r_\text{f}\proj{0\pa}{0\pa} \nonumber\\*
& -\sin^3r_\text{f}\Big(\proj{0\downarrow}{\uparrow\pa}+\proj{\uparrow\pa}{0\downarrow}\Big)+\sin r_\text{f}\cos^2 r_\text{f}\nonumber\Big(\ket{00}\bra{\uparrow\uparrow}+\proj{\uparrow\uparrow}{00}\Big)\nonumber\\*
&+\cos^2 r_\text{f} \proj{\uparrow0}{\uparrow0}+\sin^2 r_\text{f}\proj{\uparrow\downarrow}{\uparrow\downarrow}\Big],
\end{align}
\begin{align}\label{etaRaRd}
\nonumber \eta^{\text{R}{\bar{\text{R}}}}_\text{f}&\!=\!\frac{1}{2}\Big[\cos^4 r_\text{f} \proj{00}{00}\!+\!\sin r_\text{f} \cos^3 r_\text{f}\Big(\proj{0\!\downarrow}{\uparrow\! 0}\!+\!\ket{0\!\uparrow} \bra{\downarrow\! 0}\!+\!\proj{\uparrow 0}{0 \downarrow}\!+\!\proj{\downarrow\! 0}{0\!\uparrow}\Big)\\*
&\nonumber+\sin^2 r_\text{f} \cos^2 r_\text{f}\Big(\ket{0\pa}\bra{\pa0}\!+\!\proj{\uparrow\downarrow}{\uparrow\downarrow}\!+\!\proj{\uparrow\uparrow}{\downarrow\downarrow} +\proj{\downarrow\downarrow}{\uparrow\uparrow}+\proj{\downarrow\uparrow}{\downarrow\uparrow}+\proj{\pa0}{0\pa}\Big)\\*
&\nonumber\!+\! \sin^3 r_\text{f}\cos r_\text{f}\Big(\proj{\uparrow\!\pa}{\pa\!\downarrow}\!+\!\proj{\pa\!\downarrow}{\uparrow\!\pa}\!+\!\ket{\downarrow\!\pa}\bra{\pa\!\uparrow}\!+\!\proj{\pa\!\uparrow}{\downarrow\!\pa}\Big)\!+\!\cos^2 r_\text{f}\proj{\downarrow\! 0}{\downarrow\! 0}\\*
&+\sin^2 r_\text{f} \ket{\pa\downarrow} \bra{\pa\downarrow}-\cos r_\text{f}\,\sin r_\text{f}\Big(\proj{\downarrow\downarrow}{\pa0}+\ket{\pa0}\bra{\downarrow\downarrow}\Big)+\sin^4r_\text{f}\proj{\pa\pa}{\pa\pa}\Big].
\end{align}

In the following paragraphs we shall compute the negativity for each bipartition of the system.
 
\subsubsection{Bipartition Alice-Rob}

The non-positive eigenvalues of the partial transpose density matrix  for the bipartition Alice-Rob \eqref{etaARd} are
\begin{align}\label{eigetaARf}
\nonumber \lambda_{1}&=\frac14\sin^2 r_\text{f}\cos^2 r_\text{f}\left(1-\sqrt{1+\frac{4\cos^2 r_\text{f}}{\sin^4 r}}\right),\\*
 \lambda_{2}&=\frac14\sin^4 r_\text{f}\left(1-\sqrt{1+\frac{4\cos^2 r_\text{f}}{\sin^4 r_\text{f}}}\right).
\end{align}
The negativity, after some basic algebra turns out to be
\begin{equation}
\mathcal{N}_\text{f}^\text{AR}=\frac12\cos^2r_\text{f}.
\end{equation}

\subsubsection{Bipartition Alice-AntiRob}

The non-positive eigenvalues of the partial transpose density matrix for the bipartition Alice-AntiRob \eqref{etaAaRd} are
\begin{eqnarray}\label{eigetaAaRf}
\nonumber \lambda_{1}&=&\frac14\sin^2 r_\text{f}\cos^2 r_\text{f}\left(1-\sqrt{1+\frac{4\tan^2 r_\text{f}}{\cos^2 r_\text{f}}}\right),\\*
\lambda_{2}&=&\frac14\cos^4 r_\text{f}\left(1-\sqrt{1+\frac{4\tan^2 r_\text{f}}{\cos^2 r_\text{f}}}\right).
\end{eqnarray}
The negativity yields in this case
\begin{equation}
\mathcal{N}^{\text{A}{\bar{\text{R}}}}_\text{f}=\frac12\sin^2 r_\text{f}.\end{equation}

It is remarkable --and constitutes one of the most suggestive results of this chapter-- that we have obtained here a conservation law for the entanglement Alice-Rob and Alice-AntiRob, since the sum of both negativities is independent of the acceleration
\begin{equation}\label{conservationN}
\mathcal{N}^{\text{A}{\bar{\text{R}}}}_\text{f} +\mathcal{N}_\text{f}^\text{AR} =\frac12.
\end{equation}
This is similar to the result \eqref{conservation1} for mutual information. Again, one could check that for spinless fermion fields the same  conservation law \eqref{conservationN} obtained here applies. This was again expected due to the universality principle exposed in previous chapters.

As we will see below, this is a genuine statistical effect: this conservation of quantum correlations is exclusive of fermionic fields and nothing of the sort will be found for bosonic fields.

\begin{figure}[h]
\begin{center}
\includegraphics[width=.85\textwidth]{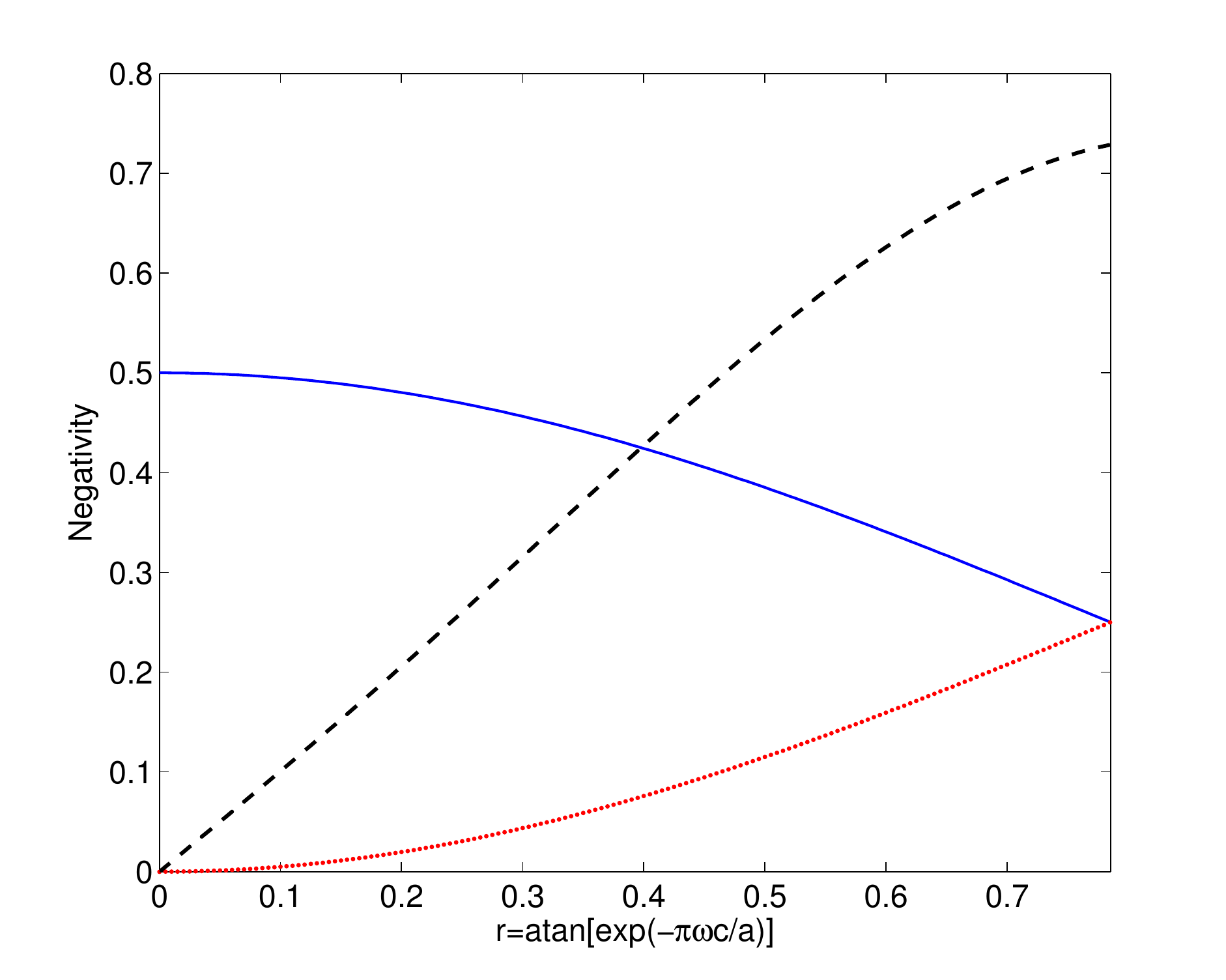}
\end{center}
\caption{ Dirac field: Negativity tradeoff and conservation law between the systems Alice-Rob and Alice-AntiRob  as acceleration varies. It is also shown the behaviour of the quantum correlations for the system Rob-AntiRob. Blue continuous line: Negativity $AR$, red dotted line: Negativity $\text{A}{\bar{\text{R}}}$, black dashed line: Negativity $R\bar R$ }
\label{negaferm}
\end{figure}

For the bipartition Rob-AntiRob the expression for the negativity is not as simple as it was for the previous cases, this negativity is plotted in Fig. \ref{negaferm}. We can see that the entanglement between Rob and AntiRob, created as Rob accelerates, grows up to a finite value. Although this entanglement is useless for quantum information tasks because of the impossibility of classical communication between both sides of an event horizon, the result obtained here may be a useful hint in order to understand how information behaves in the proximity of horizons.

Comparing again this result with spinless fermions \cite{AlsingSchul}, we see that for Dirac fields, the maximum value of the negativity is greater. Again this is strongly related with the dimension of the Hilbert space\footnote{For the Dirac case correlations between the the spin-up mode for Rob and the spin-down mode for AntiRob show up as well, even though the spin-up mode was not excited in \eqref{entangleddir}. This phenomenon cannot happen for the Grassmann scalar case where there is no spin.} that imposes a bound in negativity as mentioned before.

\section{Correlations for the scalar field}\label{sec5}

The density matrix for the whole tripartite state, which includes modes in both sides of the horizon along with Minkowskian modes, is built from \eqref{entangledsca}
\begin{equation}\label{tripasca}
\rho^{A\text{R}{\bar{\text{R}}}}_\text{b}=\proj{\Psi_\text{b}}{\Psi_\text{b}}.
\end{equation}

As in the fermion case, the three different bipartitions for the scalar field case are obtained as follows
\begin{eqnarray}
\label{AR1}\rho^\text{AR}_\text{b}&=&\tr_{\text{II}}\rho^{A\text{R}{\bar{\text{R}}}}_\text{b},\\*
\label{AAR1}\rho^{\text{A}{\bar{\text{R}}}}_\text{b}&=&\tr_\text{I}\rho^{A\text{R}{\bar{\text{R}}}}_\text{b},\\*
\label{RAR1}\rho^{\text{R}{\bar{\text{R}}}}_\text{b}&=&\tr_\text{U}\rho^{A\text{R}{\bar{\text{R}}}}_\text{b}.
\end{eqnarray}
and the density matrix for each individual subsystem
 \begin{eqnarray}
\label{A1}\rho^{\text{A}}_\text{b}&=&\tr_\text{I}\rho^\text{AR}_\text{b}=\tr_{\text{II}}\rho^{\text{A}{\bar{\text{R}}}}_\text{b},\\*
\label{R1}\rho^{\text{R}}_\text{b}&=&\tr_{\text{II}}\rho^{\text{R}{\bar{\text{R}}}}_\text{b}=\tr_\text{U}\rho^\text{AR}_\text{b},\\*
\label{aR1}\rho^{{\bar{\text{R}}}}_\text{b}&=&\tr_\text{I}\rho^{\text{R}{\bar{\text{R}}}}_\text{b}=\tr_\text{U}\rho^{\text{A}{\bar{\text{R}}}}_\text{b}.
\end{eqnarray}

The bipartite systems are characterised by the following density matrices
\begin{align}\label{rhoars14m}
\nonumber\rho^\text{AR}_\text{b}&=\sum_{n=0}^{\infty}\frac{\tanh^{2n}r_\text{b}}{2\cosh^2 r_\text{b}}\Big[\proj{0n}{0n}+\frac{\sqrt{n+1}}{\cosh r_\text{b}}\Big(\proj{0n}{1\, n+1}+\proj{1\, n+1}{0n}\Big)\\*
&+\frac{n+1}{\cosh^2 r_\text{b}}\proj{1\,n+1}{1\,n+1}\Big],
\end{align}
\begin{align}\label{rhoa-rs14m}
\nonumber\rho^{\text{A}{\bar{\text{R}}}}_\text{b}&=\sum_{n=0}^{\infty}\frac{\tanh^{2n}r_\text{b}}{2\cosh^2 r_\text{b}}\Big[\!\proj{0n}{0n}\!+\!\frac{\sqrt{n+1}}{\cosh r_\text{b}}\tanh r_\text{b}\Big(\!\ket{0\,n+1}\times\bra{1 n}+\proj{1 n}{0\,n+1}\Big),\\*
&+\frac{n+1}{\cosh^2 r_\text{b}}\proj{1n}{1n}\Big].
\end{align}
\begin{align}\label{rhor-rs14m}
\rho^{\text{R}{\bar{\text{R}}}}_\text{b}&=\sum_{\substack{n=0\\m=0}}^{\infty}\frac{\tanh^{n+m}r_\text{b}}{2\cosh^2 r_\text{b}}\Big(\proj{nn}{mm}+\frac{\sqrt{n+1}\sqrt{m+1}}{\cosh^2 r_\text{b}}\times\proj{n+1\,n}{m+1\,m}\!\Big),
\end{align}
where the bases are respectively
\begin{eqnarray}\label{barbolbasis}
 \ket{nm}&=&\ket{n^\text{A}}_\text{U}\ket{m^\text{R}}_\text{I},\\*
\ket{nm}&=&\ket{n^\text{A}}_\text{U}|m^{{\bar{\text{R}}}}\rangle_{\text{II}},\\*
\ket{nm}&=&\ket{n^\text{R}}_\text{I}|m^{{\bar{\text{R}}}}\rangle_{\text{II}}
\end{eqnarray}
for \eqref{rhoars14m}, \eqref{rhoa-rs14m} and \eqref{rhor-rs14m}.

On the other hand, the density matrices for the individual subsystems \eqref{A1}, \eqref{R1},\eqref{aR1} are
\begin{equation}\label{Robpartialm4}
\rho^{\text{R}}_\text{b}=\sum_{n=0}^\infty\frac{\tanh^{2(n-1)}r_\text{b}}{2\cosh^2 r_\text{b}}\left[\tanh^2 r_\text{b}+\frac{n}{\cosh^2r_\text{b}}\right]\proj{n}{n},
\end{equation}
\begin{equation}\label{ARobpartialm4}
\rho^{{\bar{\text{R}}}}_\text{b}=\sum_{n=0}^\infty\frac{\tanh^{2n}r_\text{b}}{2\cosh^2 r_\text{b}}\left[1+\frac{n+1}{\cosh^2r_\text{b}}\right]\proj{n}{n},
\end{equation}
\begin{equation}\label{AlicedeAliceRobm4}
\rho^{\text{A}}_\text{b}=\frac12\left(\proj{0}{0}+\proj{1}{1}\right).
\end{equation}

\subsection{Mutual Information: creation, exchange and conservation}

To compute the mutual information  for each bipartition we need the eigenvalues of the corresponding density matrices. We shall go through all the process in detail in the lines below.

\subsubsection{Bipartition Alice-Rob}

The density matrix for the system Alice-Rob \eqref{rhoars14m}  consists on an infinite number of $2\times2$ blocks in the basis $\{\ket{0 n},\ket{1\, n+1}\}_{n=0}^\infty$ which have the form
\begin{equation}
\frac{\tanh^{2n}r_\text{b}}{2\cosh^2 r_\text{b}}
\left(\!\begin{array}{cc}
1 & \dfrac{\sqrt{n+1}}{\cosh r_\text{b}}\\
\dfrac{\sqrt{n+1}}{\cosh r_\text{b}} & \dfrac{n+1}{\cosh^2r_\text{b}}
\end{array}\!\right),
\end{equation}
whose eigenvalues are
\begin{eqnarray}\label{eigAR4m}
\nonumber\lambda^1_n&=&0,\\*
\lambda^2_n&=&\frac{\tanh^{2n}r_\text{b}}{2\cosh^2 r_\text{b}}\left(1+\frac{n+1}{\cosh^2 r_\text{b}}\right).
\end{eqnarray}

\subsubsection{Bipartition Alice-AntiRob}

Excepting the diagonal element corresponding to $\proj{00}{00}$ (which forms a $1\times1$ block itself) the density matrix for the system Alice-AntiRob \eqref{rhoa-rs14m} consists on an infinite number of $2\times2$ blocks in the basis $\{\ket{0 n},\ket{1\, n-1}\}_{n=1}^\infty$ which have the form
\begin{equation}
\frac{\tanh^{2n}r_\text{b}}{2\cosh^2 r_\text{b}}
\left(\!\begin{array}{cc}
1 & \dfrac{\sqrt{n}}{\sinh r_\text{b}} \\
\dfrac{\sqrt{n}}{\sinh r_\text{b}} & \dfrac{n}{\sinh^2 r_\text{b}}\\
\end{array}\!\right).
\end{equation}
We can gather all the eigenvalues in the expressions
\begin{eqnarray}\label{eigAaR4m}
\nonumber\lambda^1_n&=&\frac{\tanh^{2n}r_\text{b}}{2\cosh^2 r_\text{b}}\left(1+\frac{n}{\sinh^2r_\text{b}}\right),\\*
\nonumber \lambda^2_n&=&0.\\*
\end{eqnarray}

\subsubsection{Bipartition Rob-AntiRob}

It is easy to see that the density matrix for Rob-AntiRob  \eqref{rhor-rs14m} --which basically consists in the direct sum of two blocks of infinite dimension-- only has rank $\operatorname{rank}(\rho^{\text{R}{\bar{\text{R}}}}_\text{b})=2$. Therefore all its eigenvalues are zero except for two of them, which are
\begin{eqnarray}\label{eigRaR4m}
\nonumber\lambda^{\text{R}{\bar{\text{R}}}}_1&=&\sum_{n=0}^{\infty}\frac{\tanh^{2n}r_\text{b}}{2\cosh^2r_\text{b}}=\frac12\label{lambda1RaR},\\*\label{lambda2RaR}
\lambda^{\text{R}{\bar{\text{R}}}}_2&=&\sum_{n=0}^{\infty}\frac{(n+1)\tanh^{2n}r_\text{b}}{2\cosh^4r_\text{b}}=\frac12,
\end{eqnarray}
so that the Von Neumann entropy for $\rho^{\text{R}{\bar{\text{R}}}}$ is
\begin{equation}\label{entrop}
S^{\text{R}{\bar{\text{R}}}}=1.
\end{equation}

\subsubsection{Von Neumann entropies for each subsystem and mutual information}

To compute the Von Neumann entropies we need the eigenvalues of every bipartition and the individual density matrices. The eigenvalues of $\rho^\text{AR}_\text{b}$, $\rho^{\text{A}{\bar{\text{R}}}}_\text{b}$, $\rho^{\text{R}{\bar{\text{R}}}}_\text{b}$ are respectively \eqref{eigAR4m}, \eqref{eigAaR4m} and \eqref{eigRaR4m}.

The eigevalues of the individual systems density matrices can be directly read from \eqref{Robpartialm4}, \eqref{ARobpartialm4} and \eqref{AlicedeAliceRobm4} since $\rho^\text{R}_\text{b}$, $\rho^{{\bar{\text{R}}}}_\text{b}$ and $\rho^\text{A}_\text{b}$ have diagonal forms in the Fock basis. The Von Neumann entropy for a partition $B$ of the system is \eqref{Vonneu}.

At this point, computing the entropies is quite straightforward. Von Neumann entropy for Rob's partial system is
\begin{equation}\label{entropyref}
S_R\!=\!-\sum_{n=0}^\infty\frac{\tanh^{2(n-1)}r_\text{b}}{2\cosh^2 r_\text{b}}\Big(\tanh^2 r_\text{b}+  \frac{n}{\cosh^2 r_\text{b}}\Big)\log_2\!\left[\frac{\tanh^{2(n-1)}r_\text{b}}{2\cosh^2 r_\text{b}}\Big(\tanh^2 r_\text{b} \!+\! \frac{n}{\cosh^2 r_\text{b}}\Big)\!\right].
\end{equation}
The rest of the partial matrices have a similar mathematical structure and, as a consequence we can express the non-trivial entropies for the all the possible partitions as a function of the entropy \eqref{entropyref} for Rob's partial system
\begin{eqnarray}\label{entropiesbos}
\nonumber &S_{\bar R}=\dfrac{S_R}{\tanh^2 r_\text{b}}-\dfrac{1}{2\sinh^2 r_\text{b}}\log_2\left(\dfrac{1}{2\cosh^2 r_\text{b}}\right)+\log_2\Big(\tanh^2 r_\text{b}\Big),&\\[2mm]
 &S_{AR}=S_{\bar R},\qquad \!\! S_{\text{A}{\bar{\text{R}}}}=S_{R}, \qquad \!\! S_{R\bar R}=S_{A}=1.&
\end{eqnarray} 
Notice that the expression for $S_{\bar R}$ may appear to blow up as $r_\text{b}\rightarrow0$, however this is not the case and it can be checked analytically using \eqref{entropyref} that $\lim_{r\rightarrow 0} S_{\bar R} = 0$.

Using \eqref{entropiesbos}, the mutual information for all the possible bipartitions of the system can be written as
\begin{eqnarray}
\nonumber I_{AR}&=&S_A+S_R-S_{A R}=1+S_R -S_{\bar R},\\*
\nonumber I_{\text{A}{\bar{\text{R}}}}&=&S_A+S_R-S_{\text{A}{\bar{\text{R}}}}=1+ S_{\bar R}-S_R,\\*
\nonumber I_{R\bar R}&=& S_R+S_{\bar R}-S_{R\bar R}=S_R+S_{\bar R}-1.
\end{eqnarray}  

Again we obtain a conservation law of the mutual information for the system Alice-Rob and Alice-AntiRob
\begin{equation}\label{conservationbos}
 I_{AR} + I_{\text{A}{\bar{\text{R}}}}=2.
\end{equation}
which again suggests a correlation transfer from the system Alice-Rob to Alice-AntiRob as the acceleration increases.

\begin{figure}[h]
\begin{center}
\includegraphics[width=.85\textwidth]{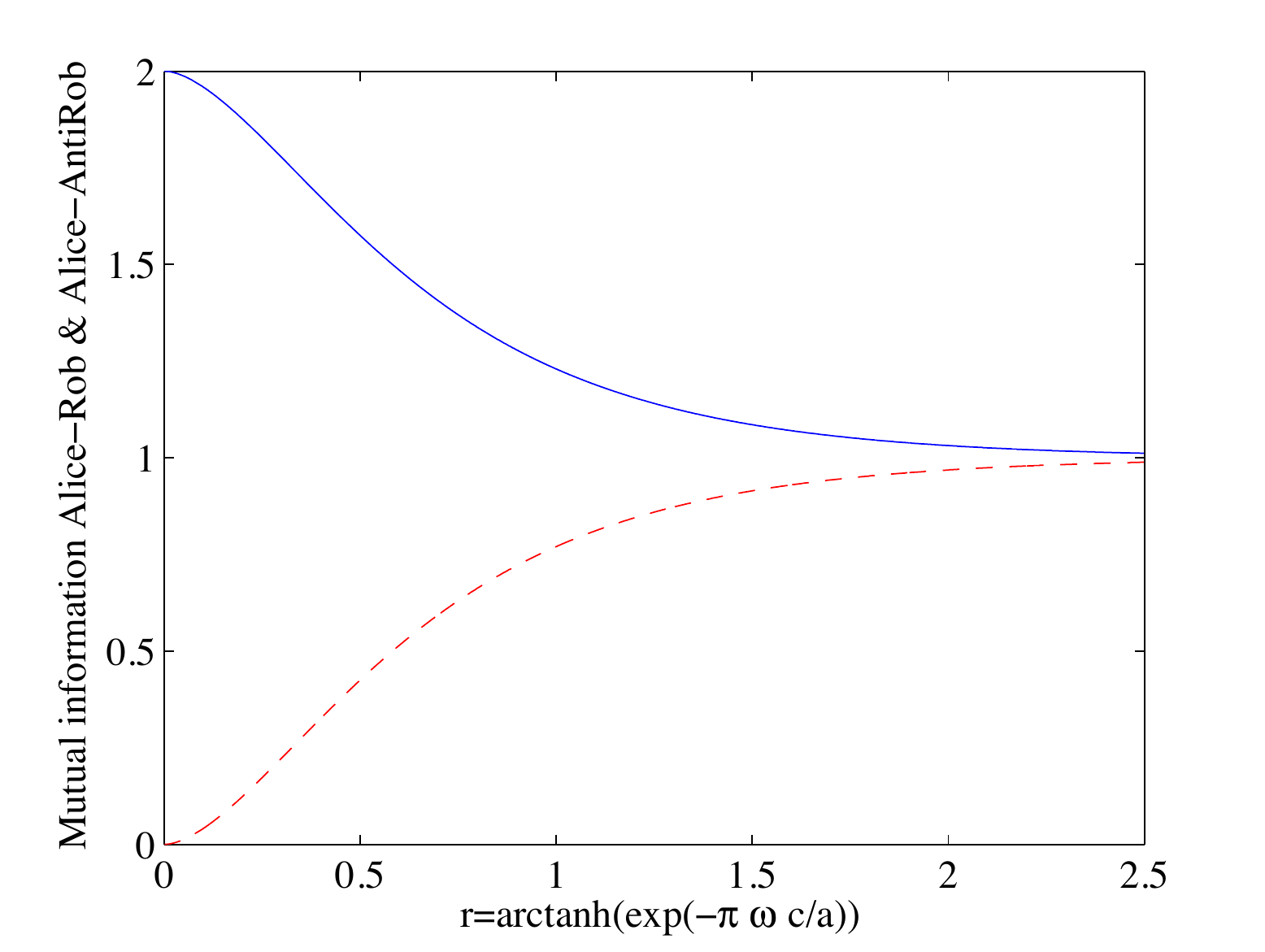}
\end{center}
\caption{ Scalar field: Mutual information conservation law for Alice-Rob and Alice-AntiRob. Blue continuous line: Mutual information $AR$, red dashed line: Mutual information $\text{A}{\bar{\text{R}}}$.}
\label{mututradeoffbos}
\end{figure}

Although the conservation law is the same as for fermion fields \eqref{conservation1}, the specific dependance of the mutual information with the acceleration is different, as it can be seen in Fig. \ref{mututradeoffbos}. Later, when we analyse the negativity for all the bipartitions, we will see that, even though mutual information fulfills this conservation law, we must wait for the analysis of quantum correlations to appreciate the striking differences between fermions and bosons.  

\begin{figure}[h]
\begin{center}
\includegraphics[width=.85\textwidth]{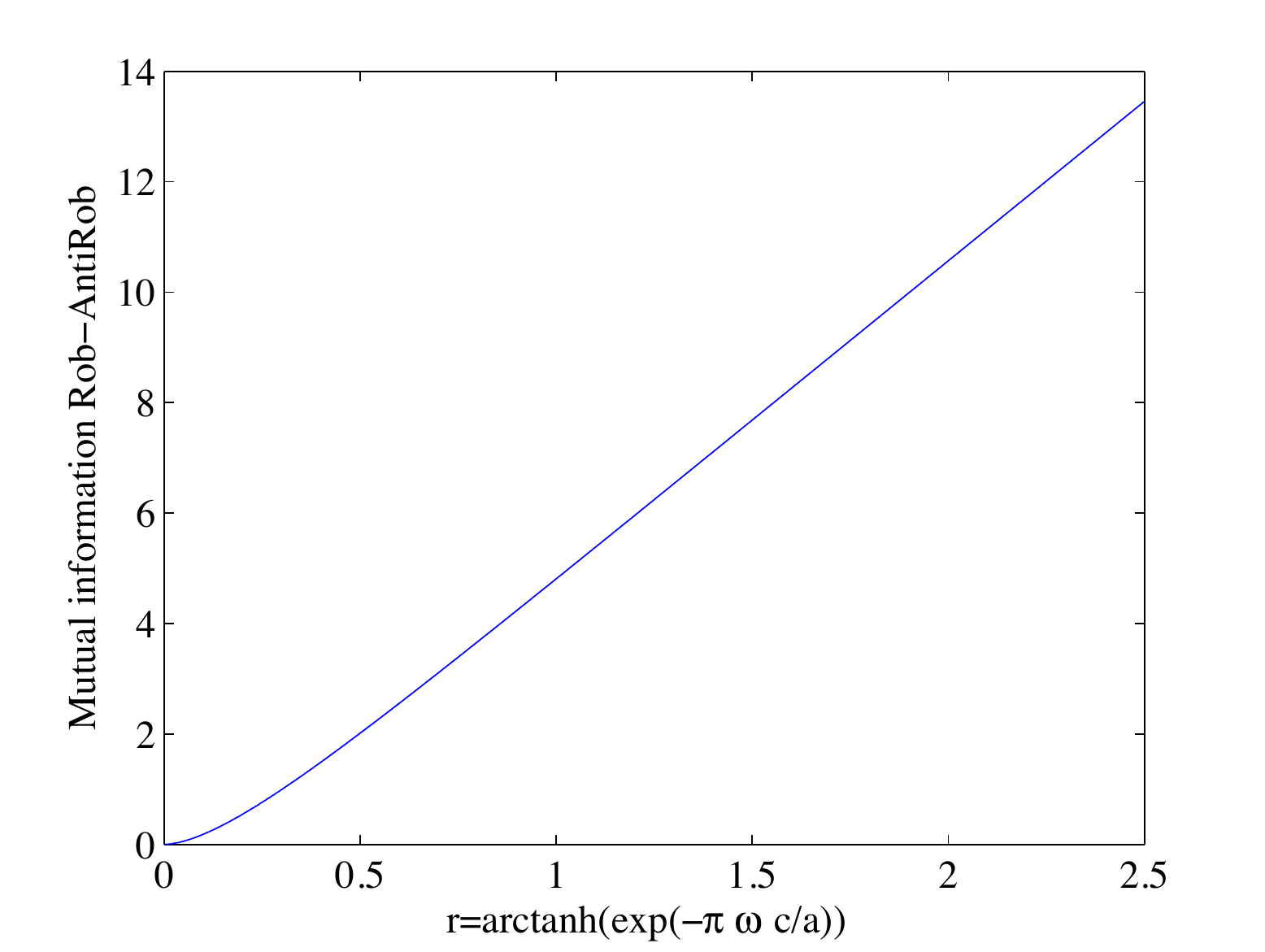}
\end{center}
\caption{Scalar field: Mutual information for the system Rob-AntiRob as acceleration varies.}
\label{mutuRARbos4m}
\end{figure}

Fig. \ref{mutuRARbos4m} shows how the correlations across the horizon (Rob and AntiRob) increase with no bound as Rob accelerates showing that (unusable) correlations are created between observers in the causally disconnected regions. 

\subsection{Entanglement behaviour}\label{negatsecm4}

 As we did for fermionic fields, we will compute the negativity for the scalar case.  To do so, we need the partial transpose of the bipartite density matrices \eqref{rhoars14m}, \eqref{rhoa-rs14m} and \eqref{rhor-rs14m}, which we will notate as $\eta^\text{AR}_\text{b}$, $\eta^{\text{A}{\bar{\text{R}}}}_\text{b}$ and $\eta^{\text{R}{\bar{\text{R}}}}_\text{b}$  respectively.
\begin{align}\label{etaARs4m}
\eta^\text{AR}_\text{b}&=\sum_{n=0}^{\infty}\frac{\tanh^{2n}r_\text{b}}{2\cosh^2 r_\text{b}}\Big[\proj{0n}{0n}+\frac{\sqrt{n+1}}{\cosh r_\text{b}}\Big(\proj{0\, n+1}{1n}\nonumber\\*
&+\proj{1 n}{0\,n+1}\Big)+\frac{n+1}{\cosh^2 r_\text{b}}\proj{1\,n+1}{1\,n+1}\Big],
\end{align}
\begin{align}\label{etaAaRs4m}
\nonumber\eta^{\text{A}{\bar{\text{R}}}}_\text{b}&=\sum_{n=0}^{\infty}\frac{\tanh^{2n}r_\text{b}}{2\cosh^2 r_\text{b}}\Big[\proj{0n}{0n}+\frac{\sqrt{n+1}}{\cosh r_\text{b}}\tanh r_\text{b}\Big(\ket{0 n}\bra{1\, n+1}\\*
&+\proj{1\, n+1}{0n}\!\Big)\!+\!\frac{n+1}{\cosh^2 r_\text{b}}\proj{1n}{1n}\!\Big],
\end{align}
\begin{equation}\label{etaRaRs4m}
\eta^{\text{R}{\bar{\text{R}}}}_\text{b}=\sum_{\substack{n=0\\m=0}}^{\infty}\frac{\tanh^{n+m}r_\text{b}}{2\cosh^2 r_\text{b}}\Big(\proj{nm}{mn}+\frac{\sqrt{n+1}\sqrt{m+1}}{\cosh^2 r_\text{b}}\proj{n+1\,m}{m+1\,n}\!\Big).
\end{equation}

In the following paragraphs we shall compute the negativity of each bipartition of the system.
 
\subsubsection{Bipartition Alice-Rob}

Excepting the diagonal element corresponding to $\proj{00}{00}$ (which forms a $1\times1$ block itself), the partial transpose of the density matrix $\rho^{A R}_\text{b} $ \eqref{etaARs4m} has a $2\times2$ block structure in the basis $\{ \ket{0\, n+1},\ket{1 n}\}$
\begin{equation}\label{blocks}
\frac{\tanh^{2n}r_\text{b}}{2\cosh^2 r_\text{b}}
\left(\!\begin{array}{cc}
\tanh^2 r_\text{b} & \dfrac{\sqrt{n+1}}{\cosh r_\text{b}}\\
\dfrac{\sqrt{n+1}}{\cosh r_\text{b}} & \dfrac{n}{\sinh^2r_\text{b}}
\end{array}\!\right).
\end{equation}
Hence, the eigenvalues of \eqref{etaARs4m} are
\begin{align}
\nonumber\lambda^1&=\frac{1}{2\cosh^2r_\text{b}},\\*
\lambda^2_n&=\frac{\tanh^{2n} r_\text{b}}{4\cosh^2 r_\text{b}}\left[\left(\frac{n}{\sinh^2r_\text{b}}+\tanh^2 r_\text{b}\right)\pm\sqrt{\left(\frac{n}{\sinh^2r_\text{b}}+\tanh^2 r_\text{b}\right)^2+\frac{4}{\cosh^2 r_\text{b}}}\right].
\end{align}
And then the negativity for this bipartition is
\begin{equation}
\mathcal{N}^\text{AR}_\text{b}\!=\!\sum_{n=0}^\infty\frac{\tanh^{2n} r_\text{b}}{4\cosh^2 r_\text{b}}\left|\!\left(\frac{n}{\sinh^2r_\text{b}}+\tanh^2 r_\text{b}\right)\!-\sqrt{\left(\frac{n}{\sinh^2r_\text{b}}+\tanh^2 r_\text{b}\right)^2\!\!+\frac{4}{\cosh^2 r_\text{b}}}\right|.
\end{equation}

\begin{figure}[h]
\begin{center}
\includegraphics[width=.85\textwidth]{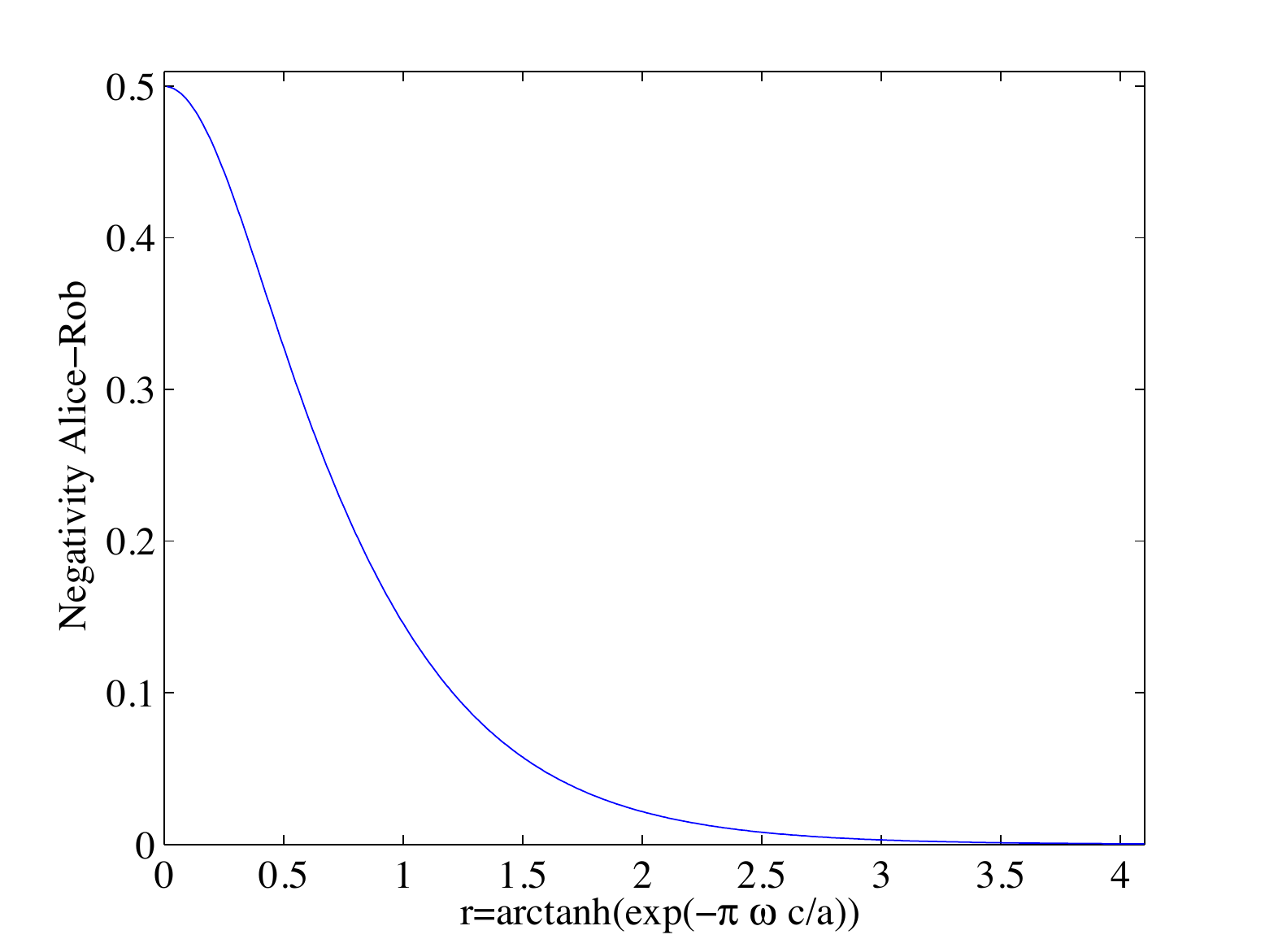}
\end{center}
\caption{ Scalar field: Behaviour of the negativity for the bipartition Alice-Rob as Rob accelerates}
\label{negARbosfig}
\end{figure}

Fig. \ref{negARbosfig} shows $\mathcal{N}^\text{AR}_\text{b}$ as a function of $r_\text{b}$.

\subsubsection{Bipartition Alice-AntiRob}

Excepting the diagonal element corresponding to $\proj{10}{10}$ (which forms a $1\times1$ block itself), the partial transpose of the density matrix $\rho_\text{b}^{\text{A}{\bar{\text{R}}}}$ \eqref{etaAaRs4m} has a $2\times2$ block structure in the basis $\{ \ket{0 n},\ket{1\, n+1}\}$ 
\begin{equation}\label{blocksbosAaR}
\frac{\tanh^{2n}r_\text{b}}{2\cosh^2 r_\text{b}}
\left(\!\begin{array}{cc}
1 & \dfrac{\tanh r_\text{b}}{\cosh r_\text{b}}\sqrt{n+1}\\
\dfrac{\tanh r_\text{b}}{\cosh r_\text{b}}\sqrt{n+1} & \dfrac{\tanh^2 r_\text{b}}{\cosh^2r_\text{b}}(n+2)
\end{array}\!\right).
\end{equation}
Hence, the eigenvalues of \eqref{etaAaRs4m} are
\begin{align}
\nonumber\lambda^1&=\frac{1}{2\cosh^4 r_\text{b}},\\*
\lambda^2_n&=\frac{\tanh^{2n}r_\text{b}}{4\cosh^2 r_\text{b}}\left[\left(1+(n+2)\frac{\tanh^2 r_\text{b}}{\cosh^2 r_\text{b}}\right)\pm\sqrt{\left(1+(n+2)\frac{\tanh^2 r_\text{b}}{\cosh^2 r_\text{b}}\right)^2-\frac{4\tanh^2 r_\text{b}}{\cosh^2 r_\text{b}}}\right].
\end{align}
Therefore, the negtivity for this bipartition is always $0$, independently of the value of Rob's acceleration. This is a striking difference with the fermionic case: In the fermionic case there is an entanglement tradeoff between the partitions Alice-Rob and Alice-AntiRob but in the bosonic case all the entanglement in the first bipartition is lost while no entanglement at all is created in the A${\bar{\text{R}}}$ bipartition. We will see in chapter \ref{boundedpop} that this happens even if we consider limited occupation number bosons, therefore this is a purely statistical effect.

\subsubsection{Bipartition Rob-AntiRob}

The partial transpose of the density matrix $\rho^{\text{R}{\bar{\text{R}}}}_\text{b}$  \eqref{etaRaRs4m} has a block structure, but the blocks themselves are of different dimensions  which grow up to infinity. Because of this, negativity is not as easily computable as for the other cases, not being possible to write it in a closed form.

However it is still possible to compute the eigenvalues of \eqref{etaRaRs4m} numerically taking into account that the blocks which form the matrix are  endomorphisms which act in the subspace expanded by the basis $B_{D}=\{\ket{mn}\}$ in which $m+n=D-1=\text{constant}$, which is to say, the fisrt block acts within the subspace expanded by the basis $B_1=\{\ket{00}\}$, the second $B_2=\{\ket{01},\ket{10}\}$, the third $B_3=\{\ket{02},\ket{20},\ket{11}\}$, the fourth $B_4=\{\ket{03},\ket{30},\ket{12},\ket{21}\}$ and so forth. In this fashion, the whole matrix is an endomorphism within the subspace $\bigoplus_{i=1}^\infty S_i$ being $S_i$ the subspace (of dimension $D=i$) expanded by the basis $B_i$.

Let us denote $M_D$ the blocks which form the matrix \eqref{etaRaRs4m}, being $D$ the dimension of each block. Then its structure is
\begin{equation}\label{blockss}
M_D=\left(\!
\begin{array}{cccccccc}
0  & a_1  & 0 & 0 & \cdots & \cdots& \cdots& 0 \\
a_1 & 0 & a_2 & 0 & \cdots & \cdots& \cdots & 0\\
0 & a_2 & 0 & a_3 & \cdots & \cdots& \cdots& 0\\
0 & 0 & a_3 &0 & a_4 & \cdots& \cdots& 0\\
0 & 0 & 0 &  \ddots &\ddots &  \ddots &\cdots& 0\\
\vdots  & \vdots  & \vdots  & \vdots  & \ddots  & \ddots  & \ddots  & \vdots \\
0 & 0 & 0 &  0 &\cdots&  \ddots &0& a_{D-1}\\
0 & 0 & 0 &  0 &0&  \dots &a_{D-1}& a_{D}\\
\end{array}\!\right),
\end{equation}
which is to say, the diagonal terms are zero except for the last one, and the rest of the matrix elements are zero excepting the two diagonals on top and underneath the principal diagonal. The elements $a_n$ are defined as follows
\begin{equation}
a_{2l+1}=\frac{(\tanh r_\text{b})^{D-1}}{2\cosh^2 r_\text{b}},
\end{equation}
\begin{equation}
a_{2l}=\sqrt{D-l}\,\sqrt{l}\frac{(\tanh r_\text{b})^{D-2}}{2\cosh^4r_\text{b}}.
\end{equation}
Notice that the elements are completely different when the value of the label $n$ is odd or even.

\begin{figure}[h]
\begin{center}
\includegraphics[width=.85\textwidth]{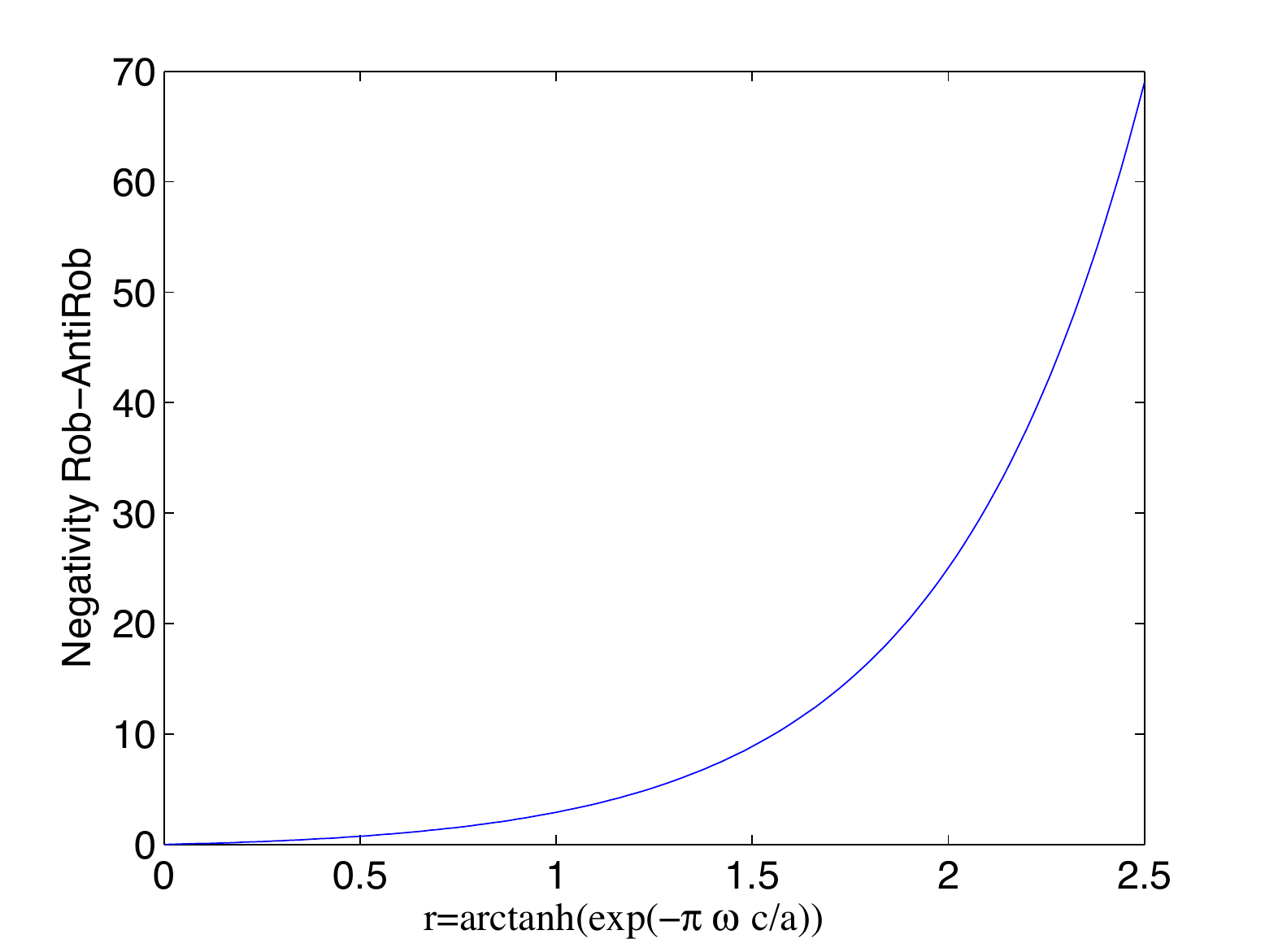}
\caption{ Scalar field: Behaviour of the negativity for the bipartition Rob-AntiRob as Rob accelerates.}
\label{negaRARbos}
\end{center}
\end{figure}

As the whole matrix is the direct sum of the blocks
\begin{equation}
\eta_\text{b}^{\text{R}{\bar{\text{R}}}}=\bigoplus_{D=1}^\infty M_D,
\end{equation}
the eigenvalues and, specifically, the negative eigenvalues of $\eta_\text{b}^{\text{R}{\bar{\text{R}}}}$ would be the negative eigenvalues of all the blocks $M_D$ gathered togheter. It can be shown that the absolute value of the negative eigenvalues of the blocks decreases quickly as the dimension increases. Thus, the negativity $\mathcal{N}^{\text{R}{\bar{\text{R}}}}_\text{b}$ promptly converges to a finite value for a given value of $r_\text{b}$.  Fig  \ref{negaRARbos} shows the behaviour of $\mathcal{N}^{\text{R}{\bar{\text{R}}}}_\text{b}$ with $r_\text{b}$, showing that the entanglement increases unboundedly between Rob and AntiRob.

Let us compare these results with the fermion case. First, as it was shown in \cite{AlsingSchul}, the negativity of the system Alice-Rob decreases as Rob accelerates, vanishing in the limit $a\rightarrow\infty$, instead of remaining finite as in the fermionic cases (\cite{AlsingSchul} and previous chapters of this thesis) .

What may be more surprising is the behaviour of quantum correlations of the system Alice-AntiRob. In the fermion case negativity grows monotonically from zero (for $a=0$) to a finite value (for $a=\infty$). Nevertheless, for scalars, Alice-AntiRob negativity is identically zero for all acceleration. Hence, there is no transfer of entanglement from Alice-Rob to Alice-AntiRob as it was the case for fermions. Still, correlations (classical) are not lost as it can be concluded from \eqref{conservationbos}.

Why do we obtain such loss of entanglement for the bosonic case and, conversely, this does not happen in the fermionic case? The answer is, once again, statistics. 

One could think of the infinite dimensionality of the Hilbert space for scalars (compared to the finite dimension for fermions) as the cause of this different behaviour. However we shall prove that it has to do with the bosonic nature of the field rather than with the infinite dimensionality of the Hilbert space. We will see this when we consider limited dimension bosons instead of scalars, which is to say, limiting the occupation number for the bosonic modes to a certain finite limit $N$ instead of taking $N\rightarrow\infty$. By doing so, we transform the infinite dimension Hilbert space for bosons into a finite dimension one. That is what we will do in the next chapter.

\begin{figure}[h]
\begin{center}
\includegraphics[width=.85\textwidth]{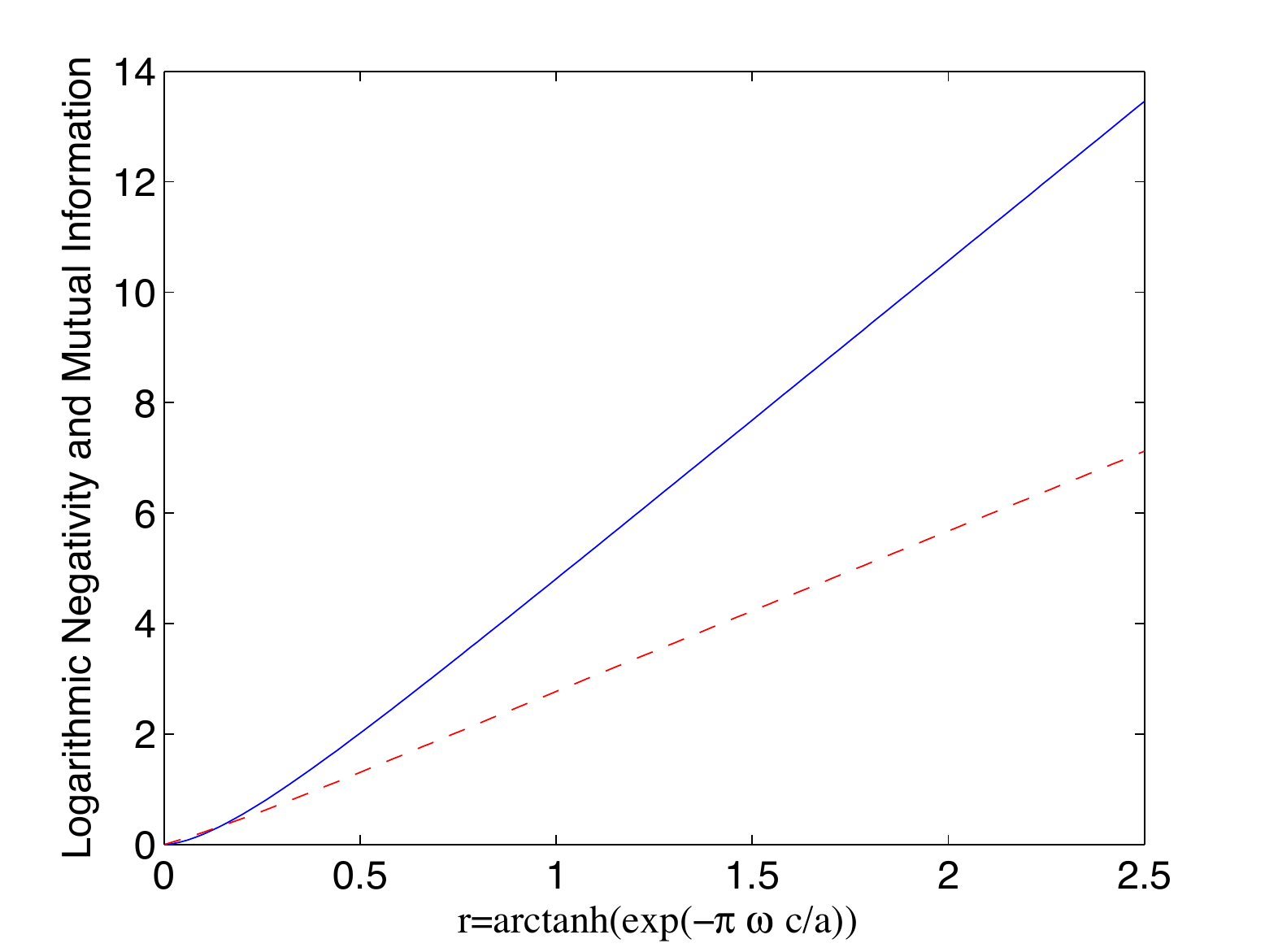}
\caption{Scalar field: Comparison of growth of quantum and all (quantum+classical) correlations for the system Rob-AntiRob as acceleration increases. Quantum correlations are accounted for by logarithmic negativity (Red dashed line). This figure compares this entanglement measurement with mutual information (Blue continuous line).}
\label{logneg2}
\end{center}
\end{figure}

As for the bipartition Rob-AntiRob, we observe that entanglement grows unboundedly for this bipartition, conversely to the fermion case in which negativity increases up to a certain finite limit as Rob accelerates.  At first glance at Fig. \ref{mutuRARbos4m} and \ref{negaRARbos} one could think that there might be some inconsistency between the behaviour of entanglement and mutual information, as the latter grows linearly while negativity seems to grow exponentially. Since mutual information accounts for all the correlations (quantum and classical) between Rob and AntiRob, the result may appear paradoxical. However this apparently inconsistent results are due to the fact that negativity cannot be identified as the entanglement itself, but as a monotone which grows as the degree of entanglement does.  The specific functional form chosen for the monotone is not imposed by physical motivations. Actually, we could have chosen logarithmic negativity  --instead of negativity-- as our entanglement monotone since it is in fact better to be compared with mutual information due to its additivity properties \cite{logneg}. The result obtained in this case, shown in Fig. \ref{logneg2}, is that when acceleration grows both growths become linear.

\section{Discussion}\label{conclusions}

This chapter focused on the bipartite correlations between different spacetime domains in the presence of an acceleration horizon. Specifically, we analyse all the possible bipartitions of an entangled system composed by an inertial observer and an accelerated one, who sees an acceleration horizon.

First of all, we have studied the relation between the entanglement behaviour of Alice-Rob and Alice-AntiRob bipartitions, which are the ones where communication is allowed. 

Here we have disclosed a great difference in the behaviour of quantum correlations for fermions and bosons. In the fermionic case we showed that, at the same time as Unruh decoherence destroys the entanglement of the system Alice-Rob, entanglement is created between Alice and AntiRob. This means that the quantum entanglement lost between Alice and the field modes in region I is gained between Alice and the modes in region II. This is expressed through the entanglement conservation law \eqref{conservationN}, which we have deduced for fermions.

Nevertheless, for bosonic states it was shown that, as acceleration increases, entanglement is quickly and completely lost between Alice and Rob while no quantum correlations are created between Alice and AntiRob. Moreover, no entanglement of any kind survives among any physical bipartition of the system in the limit $a\rightarrow\infty$ for the bosonic case. This contrasts with the fermionic case where the amount of entanglement among all the physical bipartitions of the system remains always constant.

Another remarkable result is the conservation law for mutual information for fermions \eqref{conservation1} and bosons \eqref{conservationbos} shown in  in Fig. \ref{mutuferm} and Fig. \ref{mututradeoffbos}. The detailed  behaviour is different for both cases but the same conservation law is obtained for the mutual information of the bipartitions Alice-Rob, and Alice AntiRob. Mutual information accounts for both classical and quantum correlations (despite the fact that in general there is no direct relation between negativity and mutual information). However in the bosonic case, mutual information distributes more rapidly between the  Alice-Rob and Alice-AntiRob than in the fermion case.

This result for mutual information means that correlations are always conserved for the systems Alice-Rob and Alice-AntiRob, despite the fact that quantum entanglement vanishes for the bosonic case and it is preserved (this is the effect of statistics) in the fermionic case. The fact that classical correlations behave in a similar way for fermions and bosons while quantum correlations behave so differently suggests again that the quantum entanglement which survives the infinite acceleration limit has a statistical origin, as if the information of `being fermion' cannot be killed by the presence of the horizon. Something to reflect upon: this scenario has some resemblance with the results found in quantum mechanical fermionic systems \cite{sta1} in which it was demonstrated that identical fermions systems has some degree of entanglement that is `built in' in their wavefunction.

Another difference between fermion and bosons appears when analysing the correlations between the wedges I and II. It is interesting to notice that, as the non-inertial partner accelerates, correlations between these two regions  are created. We have found that for Dirac fields these correlations, quantum and classical, grow as Rob accelerates up to a finite value at the limit $a\rightarrow\infty$. This limit is greater than the analogous limit  obtained for spinless fermions in \cite{AlsingSchul} whose Hilbert space for each mode is smaller. For the bosonic case, on the contrary, those correlations grow unboundedly, diverging when $a\rightarrow\infty$.


\chapter{Population bound effects on non-inertial bosonic correlations\footnote{E. Mart\'in-Mart\'inez, J. Le\'on. Phys. Rev. A, 81, 052305 (2010}}\label{boundedpop}
\markboth{Chapter 6. Non-inertial bosonic correlations: Population bound effects  }{\rightmark}

We have seen before, studying different fermionic fields and for different kinds of states, that the Fock space dimension does not play a role in fermionic entanglement behaviour in non-inertial frames. In this chapter we will analyse the effects of artificially bounding the occupation number of the modes of a bosonic field while conserving Bose-Einstein-like statistics. Doing that we will be able to study from the bosonic perspective if the Hilbert space dimension has any relevance in non-inertial entanglement behaviour.

As we did in previous chapters, we will consider once again a bipartite system (Alice-Rob), in which one of the partners (Rob) is undergoing a uniform acceleration and therefore describing the world which Rindler coordinates. As pointed out in \cite{AlsingSchul} and the previous chapter, there are 3 possible bipartitions that can be considered when analysing entanglement in this setting; 1) The entanglement of the inertial observer with field modes in Rindler's region I (Alice-Rob, AR), 2) The entanglement of the inertial observer with field modes in Rindler's region II (Alice-AntiRob, $\text{A}{\bar{\text{R}}}$) and 3) The entanglement between modes in regions I and II of the Rindler spacetime (Rob-AntiRob $\text{R}{\bar{\text{R}}}$). Partitions AR and $\text{A}{\bar{\text{R}}}$ are especially important as these are the partitions in which classical communication is allowed (we will refer to them as CCA bipartitions from now on).

In chapter \ref{etanthrough} we have explored the radical differences between fermionic and bosonic entanglement behaviour in the presence of Rindler and event horizons, showing that the real cause of these differences is fermionic/bosonic statistics. This contradicts the naive argument that the differences come from the finite dimensional nature of the fermionic Hilbert space for each frequency as opposed to the built-in infinite dimension of the Fock space for bosons.

In an attempt to settle down the discussion about what effects are influenced by the dimension of the Hilbert space and which ones are not, we are going to study a finite dimensional analog to bosonic fields with a limited dimension of each frequency Fock space. This means engineering a method to impose a maximum occupation number $N$ in each scalar field frequency mode. The construction of a finite dimensional scalar field state for a non-inertial observer can be problematic, thus it is an issue which will need to be tackled in order to conduct the proposed analysis.

We will present results about entanglement of the CCA bipartitions that will strengthen the argument discussed in previous chapters on the capital importance of statistics in the phenomenon of entanglement degradation due to the Unruh effect. Specifically we will prove that the behaviour is fundamentally independent of the Fock space dimension. However, bosonic entanglement for AR is slightly sensitive to Fock space dimension variation, in opposition to what happens with fermions. 

We shall point out that those variations strongly oppose once again what is said in previous literature in which it is argued that the Unruh decoherence degrades the entanglement quicker as the dimension of the Hilbert space is higher. Instead, we will show that quantum correlations can be more or less quickly degraded for different dimensions depending on the value of the acceleration. This completely banishes the former argument. 

Furthermore, we will show remarkable results concerning correlations between modes in Rindler regions I and II. We will see how they are ruled by both statistics and Hilbert space dimension. There are differences and similarities between fermions and bosons concerning correlations $\text{R}{\bar{\text{R}}}$. We will analyse the different bosonic cases comparing them with their Fock space dimension fermionic analogs, in order to comprehend the relative importance of dimensionality and statistics in the behaviour of such correlations.

We will also show how classical correlations between AR and $\text{A}{\bar{\text{R}}}$ are affected by the bound on the occupation number. Specifically we will show that the effect of imposing a finite dimensional Fock space affects the conservation law for mutual information found in previous works. We will compare this with the fermionic cases and will prove some results about mutual information to be completely universal.

\section{Limiting the occupation number}\label{sec3}

The Minkowski vacuum state for a $\omega$-frequency mode of a scalar field seen from the perspective of an accelerated observer is
\begin{equation}\label{scavacinfm5}
\ket{0_\omega}=\frac{1}{\cosh r}\sum_{n=0}^\infty \tanh^n r \ket{n}_{\text{I}}\ket{n}_{\text{II}}.
\end{equation}
We will drop the frequency label as we will study single Unurh mode states as we did in previous chapters.

The Minkowskian Unruh one particle state results from applying the creation operator to the vacuum state. Its translation to the Rindler basis is
\begin{equation}\label{unoinfm5}
\ket{1}_\text{U}=\frac{1}{\cosh^2 r}\sum_{n=0}^{\infty} \tanh^n r \,\sqrt{n+1}\ket{n+1}_{\text{I}}\ket{n}_{\text{II}}.
\end{equation} 	
Now we will consider the following maximally entangled state in the Minkowski basis
\begin{equation}
\label{entangledscainf}\ket{\Psi}=\frac{1}{\sqrt{2}}\left(\ket{0}\ket{0}+\ket{1}_\text{U}\ket{1}_\text{U}\right).
\end{equation}
This is a qubit state which is a superposition of the bipartite vacuum and the bipartite one particle state completely equivalent to that studied in chapters before. 

For our purposes we need to limit the dimension of the Hilbert space. To do so we are going to limit the maximum occupation number for the Rindler modes up to $N$.

If we want to go beyond qualitative effects and do a completely rigorous analysis we would run into important problems. Namely, we would not be able to normalise both states  \eqref{scavacinfm5}, \eqref{unoinfm5} simultaneously. In other words, the translation into the Rindler basis of the Minkowskian creation operator applied on the vacuum state would not preserve normalisation and it would be no longer true that applying the annihilation operator to the one particle state we recover the vacuum of the theory. Therefore the canonical quantisation rules of bosonic fields would be ill-defined (for instance, problems would appear when applying the commutator to the one particle state). As statistics is fundamental to explain the Unruh decoherence mechanism, a rigorous analysis would require that we consider the vacuum of our theory as expressed in \eqref{scavacinfm5} with an unbounded occupation number.

Alternately, we will define finite dimension analogs to the vacuum and one particle states
\begin{equation}\label{scavac}
\ket{0_N}_\text{U}=\frac{1}{\cosh r}\sum_{n=0}^N \tanh^n r \ket{n}_{\text{I}}\ket{n}_{\text{II}},
\end{equation}
\begin{equation}\label{uno}
\ket{1_N}_\text{U}=\frac{1}{\cosh^2 r}\sum_{n=0}^{N-1} \tanh^n r \,\sqrt{n+1}\ket{n+1}_{\text{I}}\ket{n}_{\text{II}},
\end{equation} 	
in which we have cut off the higher occupation numbers and thus these two states are not exactly the vacuum of our theory and the first excitation. Instead, they could be understood as approximations in which Rob is not able to notice occupation numbers larger than $N$. Indeed this is a consistent approximation as the coefficients of higher $n$ become more and more smaller as $n$ grows. This simple construct allows us to consider a bounded occupation number along with bosonic statistics. Therefore we can now disentangle the statistical effects from the ones derived from the dimensionality of the Hilbert space. 

We will then consider the following entangled state in Minkowski coordinates
\begin{equation}
\label{entangledsca}\ket{\Psi}=\frac{1}{C_N(r)}\left(\ket{0}_\text{U}\ket{0_N}_\text{U}+\ket{1}_\text{U}\ket{1_N}_\text{U}\right),
\end{equation}
in which the one particle state and the vacuum for Rob are substituted by the bounded occupation number approximations\footnote{Alternatively, we could have considered a scalar field which is quantised with the following commutation rules: $[a,a^\dagger]=1+(N-1)\proj{N}{N}$. This field would share the same Rindler-Minkowski Bogoliubov coefficients than the standard scalar field and a maximum occupation number $N$ would be naturally imposed. Although the state normalisation of \eqref{scavac} and \eqref{uno} would have not been the same, \eqref{entangledsca} would have exactly the same form once it is normalised. The two approaches are equivalent for our purposes and produce the same results.}.

Notice that a factor $1/C_N(r)$ must now be included as our occupation number cutoff implies that $\ket{0_N}_\text{U}$ and $\ket{1_N}_\text{U}$ are not normalised. Its value is
\begin{equation}
C_N(r)=\sqrt{\braket{0_N}{0_N}_\text{U}+\braket{1_N}{1_N}_\text{U}},
\end{equation}
or, explicitly,
\begin{equation}
C_N(r)=\sqrt{2-\tanh^{2N}r\left(\tanh^2 r + 1 + \frac{N}{\cosh^2r}\right)}.
\end{equation}
In the limit $N\rightarrow\infty$, $C_{N}(r)\rightarrow\sqrt2$ recovering the standard scalar maximally entangled state \eqref{entangledscainf}.

Since we have restricted our whole Hilbert space to the sector of $N$ particles and no operation takes us out from it we can guarantee that Unruh decoherence will not affect higher occupation number modes.

The density matrix for the whole tripartite state, which includes modes in both sides of the horizon along with Minkowskian modes, is built from \eqref{entangledsca} changing to the Rindler basis for Rob
\begin{equation}\label{tripasca}
\rho^{A\text{R}{\bar{\text{R}}}}=\proj{\Psi}{\Psi}.
\end{equation}

The partial subsystems are obtained as usual
\begin{align}
\label{AR1}\rho^{AR}&=\tr_{\text{II}}\rho^{A\text{R}{\bar{\text{R}}}},\\*
\label{AAR1}\rho^{A\bar R}&=\tr_{\text{I}}\rho^{A\text{R}{\bar{\text{R}}}},\\*
\label{RAR1}\rho^{\text{R}{\bar{\text{R}}}}&=\tr_{\text{U}}\rho^{A\text{R}{\bar{\text{R}}}},
\end{align}
and the density matrix for each individual subsystem
 \begin{align}
\label{A1}\rho^{A}&=\tr_{\text{I}}\rho^{AR}=\tr_{\text{II}}\rho^{A\bar R},\\*
\label{R1}\rho^{R}&=\tr_{\text{II}}\rho^{\text{R}{\bar{\text{R}}}}=\tr_{\text{U}}\rho^{AR},\\*
\label{aR1}\rho^{\bar R}&=\tr_{\text{I}}\rho^{\text{R}{\bar{\text{R}}}}=\tr_{\text{U}}\rho^{\text{A}{\bar{\text{R}}}}.
\end{align}

The bipartite systems are characterized by the following density matrices
\begin{align}\label{rhoars1}
\rho^{AR}&=\!\left\{\sum_{n=0}^{N-1}\frac{\tanh^{2n}r}{\cosh^2r}\left[\proj{0n}{0n}+\frac{\sqrt{n+1}}{\cosh r}\Big(\proj{0n}{1\, n+1}\right.\nonumber+\proj{1\, n+1}{0n}\Big)\right.\nonumber\\*
&\left.\left.+\frac{n+1}{\cosh^2 r}\proj{1\,n+1}{1\,n+1}\right]+\frac{\tanh^{2N}r}{\cosh^2r}\proj{0N}{0N}\right\}\frac{1}{C_N(r)^2},
\end{align}
\begin{align}\label{rhoa-rs1}
\nonumber\rho^{A\bar R}&=\left\{\sum_{n=0}^{N-1}\frac{\tanh^{2n}r}{\cosh^2r}\left[\proj{0n}{0n}\!+\!\frac{\sqrt{n+1}}{\cosh r}\tanh r\Big(\!\ket{0\,n+1}\right.\bra{1 n}+\proj{1 n}{0\,n+1}\Big)\right.\nonumber\\*
&\left.\left.+\frac{n+1}{\cosh^2 r}\proj{1n}{1n}\right]+\frac{\tanh^{2N}r}{\cosh^2r}\proj{0N}{0N}\right\}\frac{1}{C_N(r)^2},
\end{align}
\begin{align}\label{rhor-rs1}
\nonumber\rho^{\text{R}{\bar{\text{R}}}}&=\frac{1}{C_N(r)^2}\Bigg\{\sum_{\substack{n=0\\m=0}}^{N}\frac{\tanh^{n+m}r}{\cosh^2r}\proj{nn}{mm}\\
&+\sum_{\substack{n=0\\m=0}}^{N-1}\frac{\tanh^{n+m}r}{\cosh^4r}\sqrt{n+1}\sqrt{m+1}\proj{n+1\,n}{m+1\,m}\Bigg\},
\end{align}
where the bases are respectively
\begin{eqnarray}\label{barbolbasis}
 \ket{nm}&=&\ket{n^A}_{\text{U}}\ket{m^R}_{\text{I}},\\*
\ket{nm}&=&\ket{n^A}_{\text{U}}|m^{\bar R}\rangle_{\text{II}},\\*
\ket{nm}&=&\ket{n^R}_{\text{I}}|m^{\bar R}\rangle_{\text{II}}
\end{eqnarray}
for \eqref{rhoars1}, \eqref{rhoa-rs1} and \eqref{rhor-rs1}.

On the other hand, the density matrices for the individual subsystems \eqref{A1}, \eqref{R1},\eqref{aR1} are
\begin{equation}\label{Robpartial}
\rho^{R}=\frac{1}{C_N(r)^2}\sum_{n=0}^N\frac{\tanh^{2n}r}{\cosh^2r}\left[1+\frac{n}{\sinh^2r}\right]\proj{n}{n},
\end{equation}
\begin{equation}\label{ARobpartial}
\rho^{\bar R}=\frac{1}{C_N(r)^2}\left[\sum_{n=0}^{N-1}\frac{\tanh^{2n}r}{\cosh^2r}\left(1+\frac{n+1}{\cosh^2r}\right)\proj{n}{n}+\frac{\tanh^{2N}r}{\cosh^2r}\proj{N}{N}\right],
\end{equation}
\begin{equation}\label{AlicedeAliceRob}
\rho^{A}=\frac{1}{C_N(r)^2}\left(D^0_N(r)\proj{0}{0}+D^1_N(r)\proj{1}{1}\right),
\end{equation}
where
\begin{equation}\label{S1}
D^0_N(r)=\sum_{n=0}^N\frac{\tanh^{2n}r}{\cosh^2 r}=1-(\tanh r)^{2(N+1)},
\end{equation}
\begin{equation}\label{S2}
D^1_N(r)\!=\!\! \sum_{n=0}^{N-1}(n+1)\frac{\tanh^{2n}r}{\cosh^2 r} =1-\left(1+\dfrac{N}{\cosh^2 r}\right) \tanh^{2 N}\!r.
\end{equation}
Notice that $D^0_N(r)+D^1_N(r)=C_N^2(r)$ and consequently all the density matrix traces are 1 as it must be. As all the probability is within the modes that we are considering, all the possible `decoherence' is confined to the finite occupation number Hilbert space we are studying. We are not just taking part of the complete vacuum and one particle states losing probability in our approximation, instead we have artificially imposed that the Unruh effect will only excite every mode up to a maximum occupation number $N$.

As an effect of the imposition of the finite dimension $\rho_A\rightarrow\proj00$ as $a\rightarrow\infty$ for any finite $N$, but it tends to $\frac12(\proj00+\proj11)$ when $N\rightarrow\infty$ for all $a$. It is important to notice that both limits do not commute. The limit $N\rightarrow\infty$ should be taken first in order to recover the standard scalar field result.

\section{Analysis of correlations}\label{sec4}

In this section we will analyse the correlations tradeoff among all the possible bipartitions of the system. We will account for the entanglement by means of the negativity, and we will study the total correlations by means of the mutual information, which accounts for both classical and quantum.

\subsection{Quantum entanglement}\label{negatsec}

We will study quantum entanglement for the three bipartitions in this settings. For this, we will use negativity as an entanglement measure. To compute  negativity we need the partial transpose of the bipartite density matrices \eqref{rhoars1}, \eqref{rhoa-rs1} and \eqref{rhor-rs1}, which we will notate as $\eta^{AR}$, $\eta^{A\bar R}$ and $\eta^{R \bar R}$  respectively.
\begin{align}\label{etaARs}
\eta^{AR}&=\!\left\{\sum_{n=0}^{N-1}\frac{\tanh^{2n}r}{\cosh^2r}\left[\proj{0n}{0n}+\frac{\sqrt{n+1}}{\cosh r}\Big(\proj{0\,n+1}{1\, n}\right.\nonumber+\proj{1 n}{0\,n+1}\Big)\right.\nonumber\\*
&\left.\left.+\frac{n+1}{\cosh^2 r}\proj{1\,n+1}{1\,n+1}\right]+\frac{\tanh^{2N}r}{\cosh^2r}\proj{0N}{0N}\right\}\frac{1}{C_N(r)^2},
\end{align}
\begin{align}\label{etaAaRs}
\nonumber\eta^{A\bar R}&=\left\{\sum_{n=0}^{N-1}\frac{\tanh^{2n}r}{\cosh^2r}\left[\proj{0n}{0n}\!+\!\frac{\sqrt{n+1}}{\cosh r}\tanh r\Big(\!\ket{0n}\right.\bra{1\, n+1}+\proj{1\, n+1}{0\,n}\Big)\right.\nonumber\\*
&\left.\left.+\frac{n+1}{\cosh^2 r}\proj{1n}{1n}\right]+\frac{\tanh^{2N}r}{\cosh^2r}\proj{0N}{0N}\right\}\frac{1}{C_N(r)^2},
\end{align}
\begin{align}\label{etaRaRs}
\nonumber\eta^{\text{R}{\bar{\text{R}}}}&=\frac{1}{C_N(r)^2}\Bigg\{\sum_{\substack{n=0\\m=0}}^{N}\frac{\tanh^{n+m}r}{\cosh^2r}\proj{nm}{mn}\\
&+\sum_{\substack{n=0\\m=0}}^{N-1}\frac{\tanh^{n+m}r}{\cosh^4r}\sqrt{n+1}\sqrt{m+1}\proj{n+1\,m}{m+1\,n}\Bigg\}.
\end{align}

In the following subsections we shall compute the negativity of each bipartition of the system.
 
\subsubsection{Alice-Rob Bipartition}

Apart from the diagonal elements corresponding to $\proj{00}{00}$ and $\proj{1N}{1N}$ (which form two $1\times1$ blocks themselves), the partial transpose of the density matrix $\rho^{A R} $ \eqref{etaARs} has a $2\times2$ block structure in the basis $\{ \ket{0\, n+1},\ket{1 n}\}_{n=0}^{N-1}$
\begin{equation}\label{blocks}
\frac{\tanh^{2n}r}{C(r)^2\cosh^2 r}
\left(\!\begin{array}{cc}
\tanh^2 r & \dfrac{\sqrt{n+1}}{\cosh r}\\
\dfrac{\sqrt{n+1}}{\cosh r} & \dfrac{n}{\sinh^2r}
\end{array}\!\right).
\end{equation}
Hence, the eigenvalues of \eqref{etaARs} are
\begin{align}
\nonumber\lambda^\pm_n&\!=\!\frac{\tanh^{2n} r}{2C_N(r)^2\cosh^2 r}\Bigg[\!\left(\frac{n}{\sinh^2r}+\tanh^2 r\right)\left.\!\pm\sqrt{\left(\frac{n}{\sinh^2r}+\tanh^2 r\right)^2\!+\frac{4}{\cosh^2 r}}\right]_{n=0}^{N-1}\!\!\!\!,\\*
\lambda_N&=\frac{1}{C_N(r)^2\cosh^2r};\qquad\qquad \lambda_{N+1}=\frac{N(\tanh r)^{2N-2}}{C_N(r)^2\cosh^4 r}.
\end{align}
Here, the notation $\displaystyle{|_{n=a_1}^{a_N}}$ means that $n$ takes all the integer values from $a_1$ to $a_N$.

Therefore the negativity for this bipartition is
\begin{equation}
\mathcal{N}^{AR}\!\!=\!\!\sum_{n=0}^{N-1}\frac{\tanh^{2n} r}{2C_N(r)^2\cosh^2 r}\left|\!\left(\frac{n}{\sinh^2r}+\tanh^2 r\right)\!-\sqrt{\left(\frac{n}{\sinh^2r}+\tanh^2 r\right)^2\!\!\!\!+\!\frac{4}{\cosh^2 r}}\right|\!.
\end{equation}

\begin{figure}[h]
\begin{center}
\includegraphics[width=.85\textwidth]{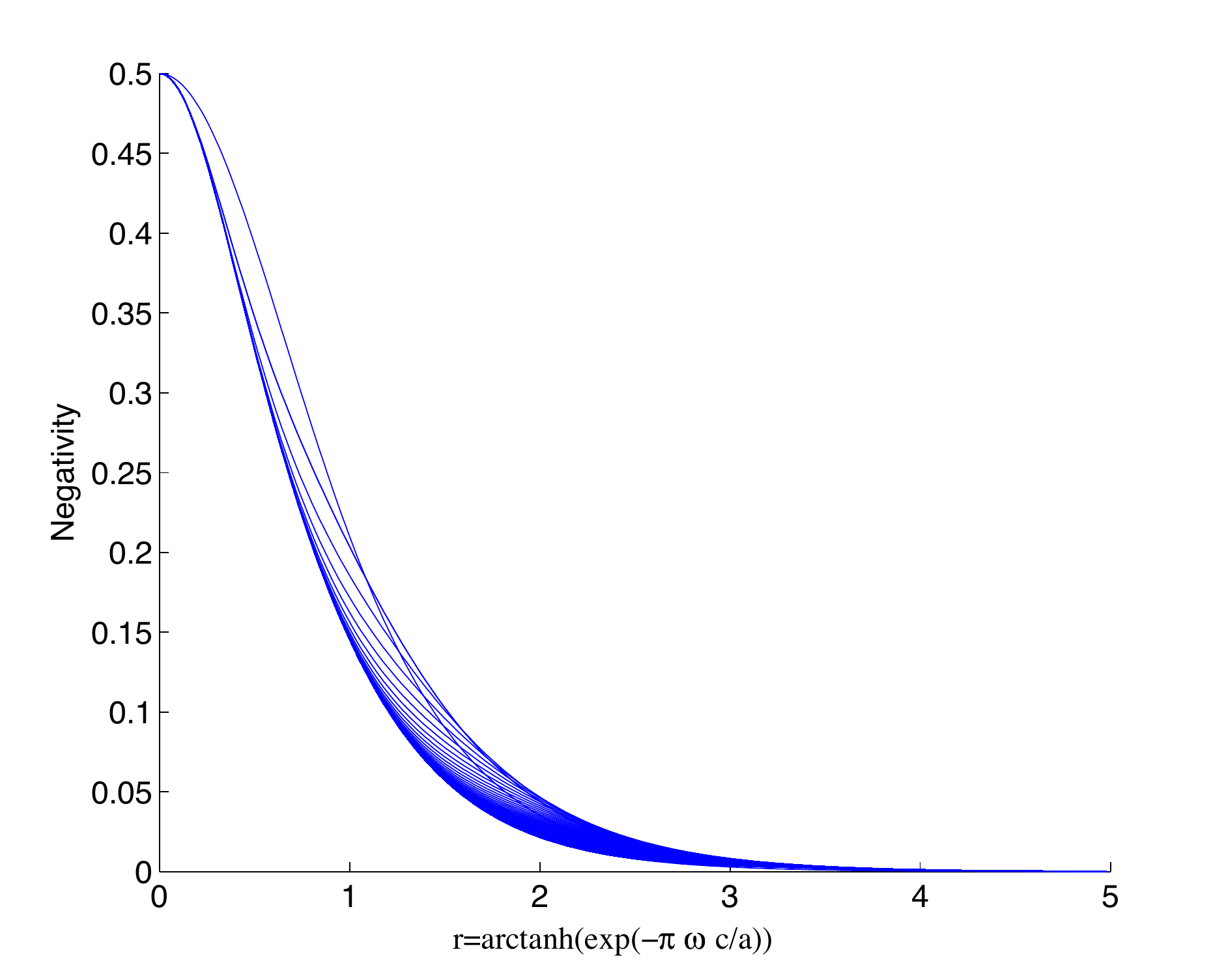}
\end{center}
\caption{Bundle of curves displaying the negativity for the bipartition AR for all values of $N$ as a function of the acceleration.}
\label{bundle}
\end{figure}

Figure \ref{bundle} shows the behaviour of negativity for all values of $N$, which is clearly similar for all cases no matter how many dimensions we are allowing for each mode. Despite this, negativity AR is slightly sensitive to dimension variations.

\begin{figure}[h]
\begin{center}
\includegraphics[width=.85\textwidth]{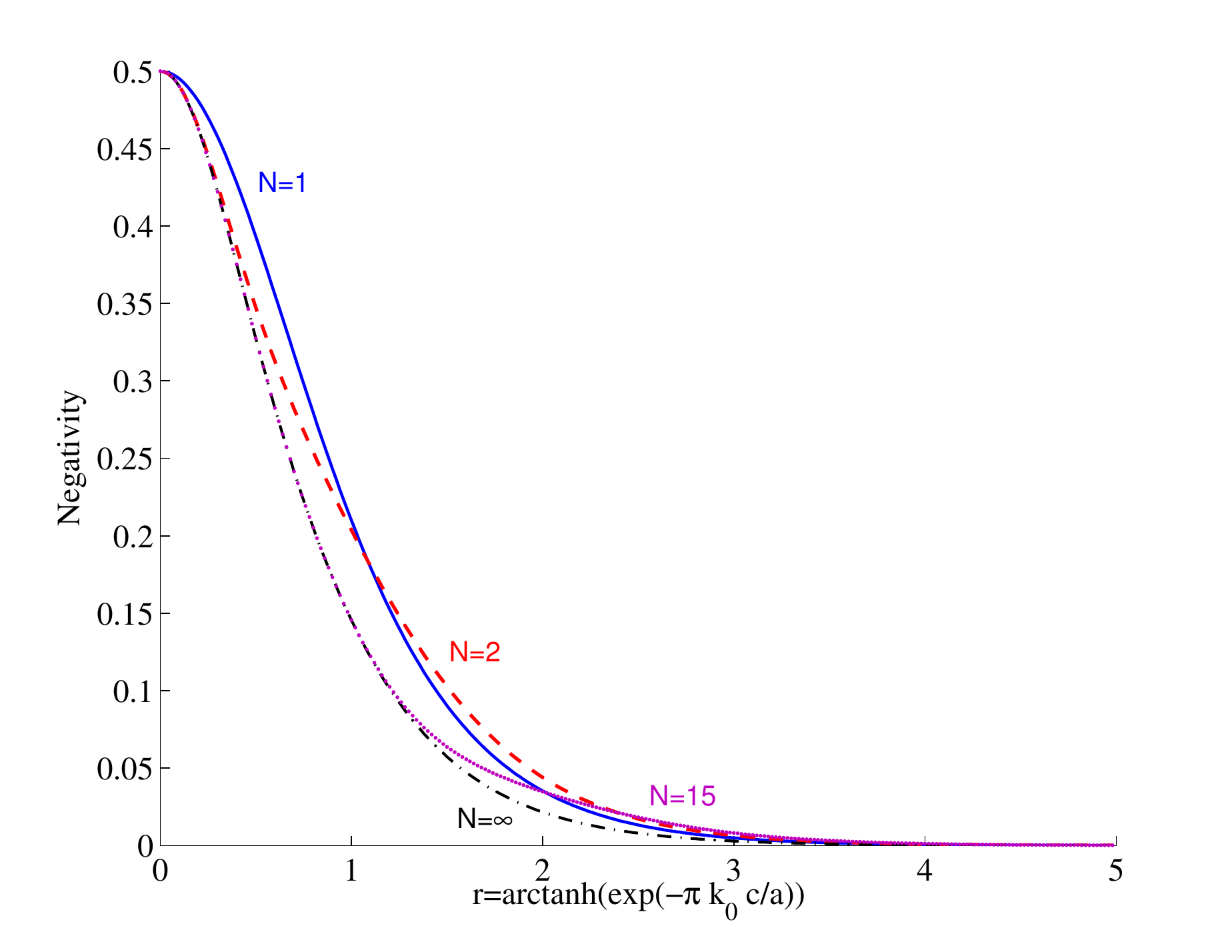}
\end{center}
\caption{Same as fig. \ref{bundle} particularised for different values of the occupation number bound $N$ showing the existence of crossing points. To the right (left) of these points entanglement degrades less (more) for lesser $N$. Solid blue line $N=1$, dashed red line $N=2$, dotted purple line $N=15$, black dash-dotted line $N=\infty$.}
\label{negaARcomp}
\end{figure}

$\mathcal{N}^{AR}$ is shown in figure \ref{negaARcomp} as a function of $r$ for different values of $N$, comparing them with the case $N=1$. 
A very interesting result emerges here, for any pair of values for the maximum occupation number $N_1<N_2$ both negativity curves cross in a point $a=a_c(N_1,N_2)$. This means that for any finite value of the Hilbert space dimension there is a region $a<a_c(N_1,N_2)$ (low accelerations) in which entanglement is more degraded for higher dimension, and another region $a>a_c(N_1,N_2)$ (high accelerations)  in which entanglement is more degraded for lower dimension.

\begin{figure}[h]
\begin{center}
\includegraphics[width=.85\textwidth]{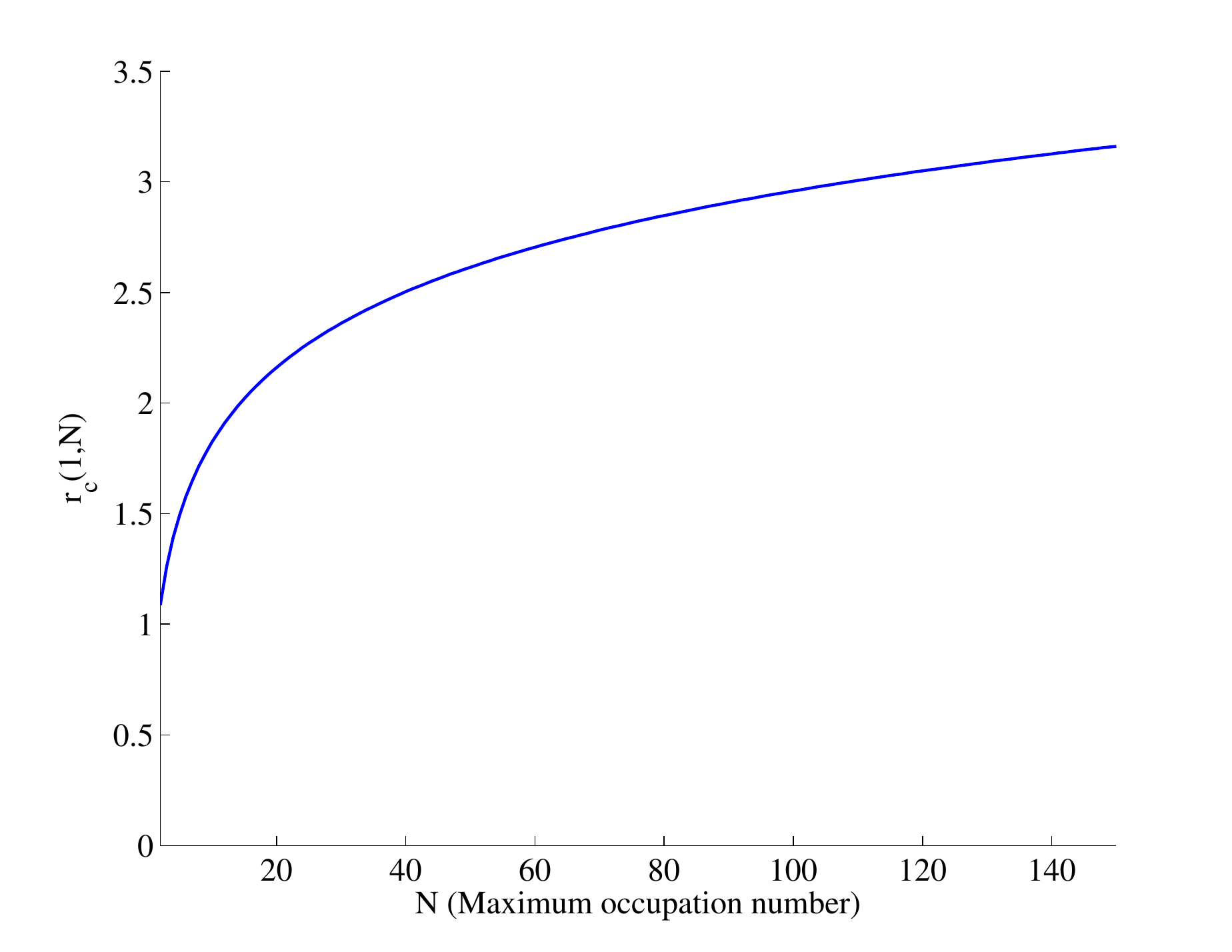}
\end{center}
\caption{$r$ for which negativity curves for a bound $N$ cross the negativity curve for $N=1$. In the region above the curve entanglement degrades faster for $N=1$ than for $N>1$.}
\label{rcritical}
\end{figure}

This disagrees the naive argument that higher dimension would lead to higher Unruh decoherence which is not necessarily true. Figure \ref{rcritical} shows the behaviour of $r_c(1,N)$ as $N$ grows. The crossing point with the negativity curve for $N=1$ grows as we consider larger $N$ curves. $r_c(N_1,N_2)$ is related with $a_c(N_1,N_2)$ by means of the relationship \eqref{defr1}.

For the limit $N_2\rightarrow\infty$,  $a_c(N_1,N_2)\rightarrow\infty$, which means that infinite dimension negativity is below all the finite dimensional curves.

\subsubsection{Bipartition Alice-AntiRob}

Excepting the diagonal elements corresponding to $\proj{10}{10}$ and $\proj{0N}{0N}$  (which form two $1\times1$ blocks themselves), the partial transpose of the density matrix $\rho^{\text{A}{\bar{\text{R}}}}$ \eqref{etaAaRs} has a $2\times2$ block structure in the basis $\{ \ket{0 n},\ket{1\, n+1}\}_{n=0}^{N-1}$ 
\begin{equation}\label{blocksbosAaR}
\frac{\tanh^{2n}r}{C(r)^2\cosh^2 r}
\left(\!\begin{array}{cc}
1 & \dfrac{\tanh r}{\cosh r}\sqrt{n+1}\\[3mm]
\dfrac{\tanh r}{\cosh r}\sqrt{n+1} & \dfrac{\tanh^2 r}{\cosh^2r}(n+2)
\end{array}\!\right).
\end{equation}
Hence, the eigenvalues of \eqref{etaAaRs} are
\begin{align}
\nonumber\lambda^{\pm}_n&\!=\!\frac{\tanh^{2n}r}{2C(r)^2\cosh^2 r}\Bigg[\!\left(1\!+\!(n\!+\!2)\frac{\tanh^2 r}{\cosh^2 r}\right)\left.\!\!\pm\sqrt{\!\left(1+(n+2)\frac{\tanh^2 r}{\cosh^2 r}\right)^2\!\!\!-\frac{4\tanh^2 r}{\cosh^2 r}}\!\right]_{n=0}^{N-1}\!\!\!\!,\nonumber\\*
\lambda_N&=\frac{1}{C(r)^2\cosh^4 r};\qquad\qquad \lambda_{N+1}=\frac{\tanh^{2N}r}{C(r)^2\cosh^2 r}.
\end{align}
Therefore, the negativity for this bipartition is always $0$, independently of the value of the acceleration parameter and the occupation number bound $N$.

From this results can be concluded that limiting the dimension has no effect in the creation or not of quantum correlations between Alice and AntiRob. As far as the field is bosonic, no entanglement is created in the CCA bipartitions of the system no matter how we limit the dimension of the Hilbert space.

\subsubsection{Bipartition Rob-AntiRob}

The partial transpose of the density matrix $\rho^{\text{R}{\bar{\text{R}}}}$  \eqref{etaRaRs} has a block structure. Namely, it is formed by $2N+1$ blocks whose dimension varies. In the following we will detailedly analyse the blocks.

\begin{enumerate}
\item First of all, we have $N+1$ blocks $\left\{M_D\right\}_{D=1}^{N+1}$  which are  endomorphisms that act in the subspace (of dimension $D$) expanded by the basis $B_{D}=\{\ket{mn}\}$ in which $m+n=D-1\le N$. 
\item Then we have $N$ more blocks $\left\{M'_D\right\}_{D=1}^N$ that  act in the subspace (of dimension $D$) expanded by the basis $B'_{D}=\{\ket{m'n'}\}$ in which $m'+n'=2N-D+1>N$. Notice that not all the possible $m'$ and $n'$ are allowed due to the limitation to the occupation number $m',n'\le N$.
\end{enumerate}

As an example which will perfectly clarify this construction, if $N=4$ there will be 9 blocks, $M_1,M_2,M_3,M_4,M_5, M'_4,M'_3,M'_2,M'_1$ each one is an endomorphism which acts in the subspace expanded by the bases
\begin{eqnarray}
\nonumber B_1&=&\left\{\ket{00}\right\},\\*
\nonumber B_2&=&\left\{\ket{01},\ket{10}\right\},\\*
\nonumber B_3&=&\left\{\ket{02},\ket{20},\ket{11}\right\},\\*
\nonumber B_4&=&\left\{\ket{03},\ket{30},\ket{12},\ket{21}\right\},\\*
\nonumber B_5&=&\left\{\ket{04},\ket{40},\ket{13},\ket{31},\ket{22}\right\},\\*
\nonumber B'_4&=&\left\{\ket{14},\ket{41},\ket{23},\ket{32}\right\},\\*
\nonumber B'_3&=&\left\{\ket{24},\ket{42},\ket{33}\right\},\\*
\nonumber B'_2&=&\left\{\ket{34},\ket{43}\right\},\\*
\nonumber B'_1&=&\left\{\ket{44}\right\}\\*
\end{eqnarray}
respectively.

In this fashion, the whole matrix is an endomorphism within the subspace R=$\bigoplus_{i=1}^{N+1} S_{i}\oplus\bigoplus_{j=1}^N S'_j$, being $S_{i}$ the subspace (of dimension $D=i$) expanded by the basis $B_{i}$ and $S'_j$ the subspace (of dimension $D=j$) expanded by the basis $B'_j$ .

The blocks $M_1,\dots,M_{N+1}$ and $M'_1,\dots,M'_{N}$ which form the matrix \eqref{etaRaRs} have the following form
\begin{equation}\label{blockss}
M_D=\left(\!
\begin{array}{cccccccc}
0  & a_1  & 0 & 0 & \cdots & \cdots& \cdots& 0 \\
a_1 & 0 & a_2 & 0 & \cdots & \cdots& \cdots & 0\\
0 & a_2 & 0 & a_3 & \cdots & \cdots& \cdots& 0\\
0 & 0 & a_3 &0 & a_4 & \cdots& \cdots& 0\\
0 & 0 & 0 &  \ddots &\ddots &  \ddots &\cdots& 0\\
\vdots  & \vdots  & \vdots  & \vdots  & \ddots  & \ddots  & \ddots  & \vdots \\
0 & 0 & 0 &  0 &\cdots&  \ddots &0& a_{D-1}\\
0 & 0 & 0 &  0 &0&  \dots &a_{D-1}& a_{D}\\
\end{array}\!\right).
\end{equation}

\begin{equation}
M'_D=\left(\!
\begin{array}{cccccccc}
0  & b_{1}  & 0 & 0 & \cdots & \cdots& \cdots& 0 \\
b_{1} & 0 & b_{2} & 0 & \cdots & \cdots& \cdots & 0\\
0 & b_{2} & 0 & b_{3} & \cdots & \cdots& \cdots& 0\\
0 & 0 & b_{3} &0 & b_{4} & \cdots& \cdots& 0\\
0 & 0 & 0 &  \ddots &\ddots &  \ddots &\cdots& 0\\
\vdots  & \vdots  & \vdots  & \vdots  & \ddots  & \ddots  & \ddots  & \vdots \\
0 & 0 & 0 &  0 &\cdots&  \ddots &0& b_{D-1}\\
0 & 0 & 0 &  0 &0&  \dots &b_{D-1}& b_{D}\\
\end{array}\!\right).
\end{equation}

The matrix elements $a_n$ and $b_n$ are defined as follows
\begin{align}
\nonumber a_{2l+1}&=\frac{(\tanh r)^{D-1}}{C(r)^2\cosh^2 r},\\*
\nonumber a_{2l}&=\sqrt{D-l}\,\sqrt{l}\frac{(\tanh r)^{D-2}}{C(r)^2\cosh^4r},\\*
\nonumber b_{2l+1}&=\frac{(\tanh r)^{2N-D+1}}{C(r)^2\cosh^2 r},\\*
 b_{2l}&=\sqrt{N+1-l}\,\sqrt{l+N-D+1}\frac{(\tanh r)^{2N-D}}{C(r)^2\cosh^4r}.
\end{align}
Notice that the elements are completely different when the value of the label $n$ is odd or even.

\begin{figure}[h]
\begin{center}
\includegraphics[width=.85\textwidth]{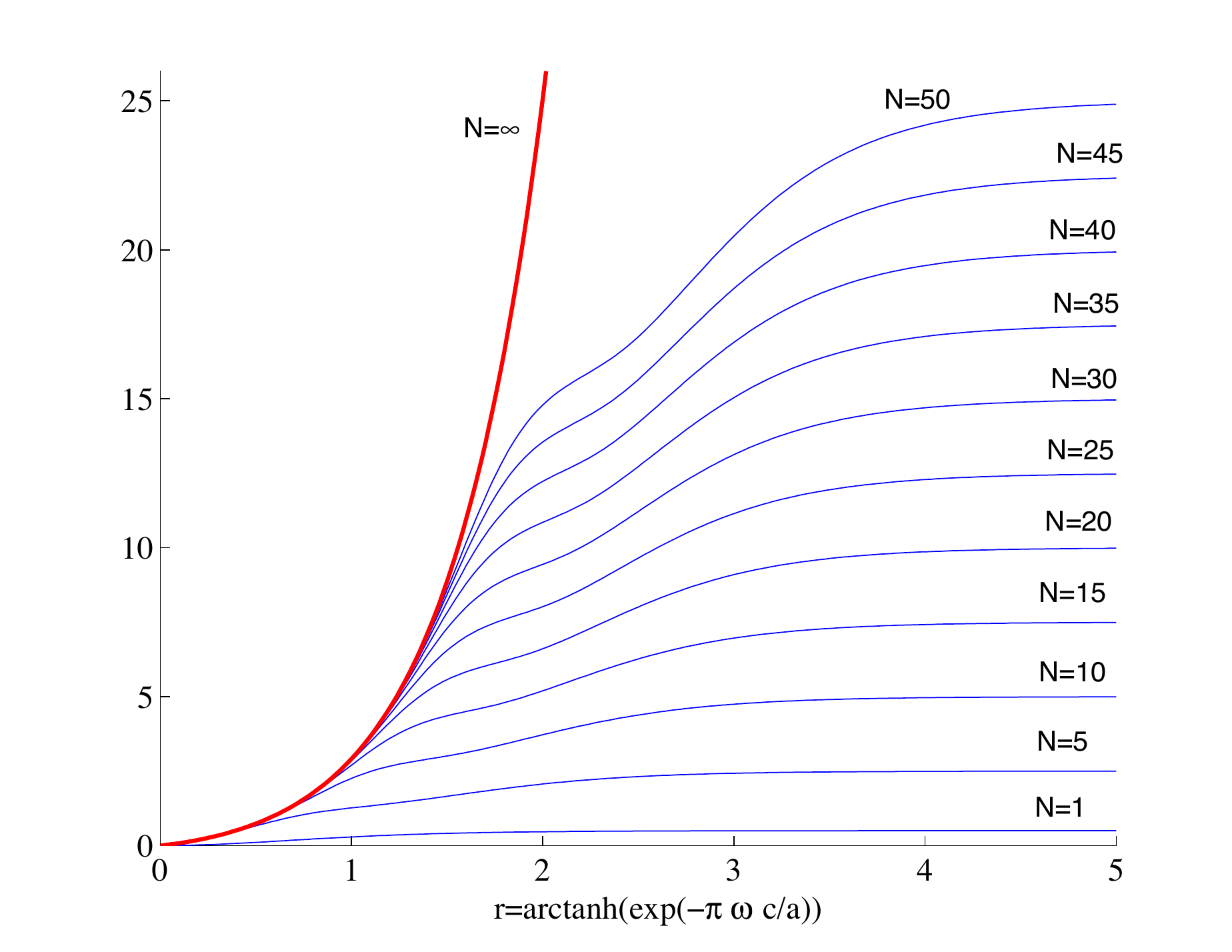}
\end{center}
\caption{ Negativity $\text{R}{\bar{\text{R}}}$ for different values of $N$, showing the upper bound reached when $a\rightarrow\infty$. Negativity diverges when $a\rightarrow\infty$ only for $N\rightarrow\infty$.}
\label{negaRAR}
\end{figure}

As the whole matrix is the direct sum of the blocks
\begin{equation}
\eta^{R \bar R}=\left(\bigoplus_{D=1}^{N+1} M_D\right) \oplus \left(\bigoplus_{D=1}^{N} M'_D\right),
\end{equation}
the eigenvalues and, specifically, the negative eigenvalues of $\eta^{R \bar R}$ would be the negative eigenvalues of all the blocks $M_D$ and $M'_D$  gathered togheter, which can be easily computed numerically.  Figure \ref{negaRAR} shows the behaviour of $\mathcal{N}^{\text{R}{\bar{\text{R}}}}$ with $r$ and for different values of $N$.

We can now compare the finite $N$ bosonic case with their same dimension analog for fermions. Namely, a Grassmann scalar field (spinless fermion) has the same Hilbert space dimension as the scalar case with $N=1$, the relevant difference is the anticommutation of the field operators instead of the commutation which applies for bosons. On the other hand, scalars limited to $N=3$ and $N=2$ can be considered as two different analogs to the Dirac field as the former has the same Hilbert space dimension as Dirac modes and the latter would share the same possible maximum occupation number. 

\begin{figure}[h]
\begin{center}
\includegraphics[width=.85\textwidth]{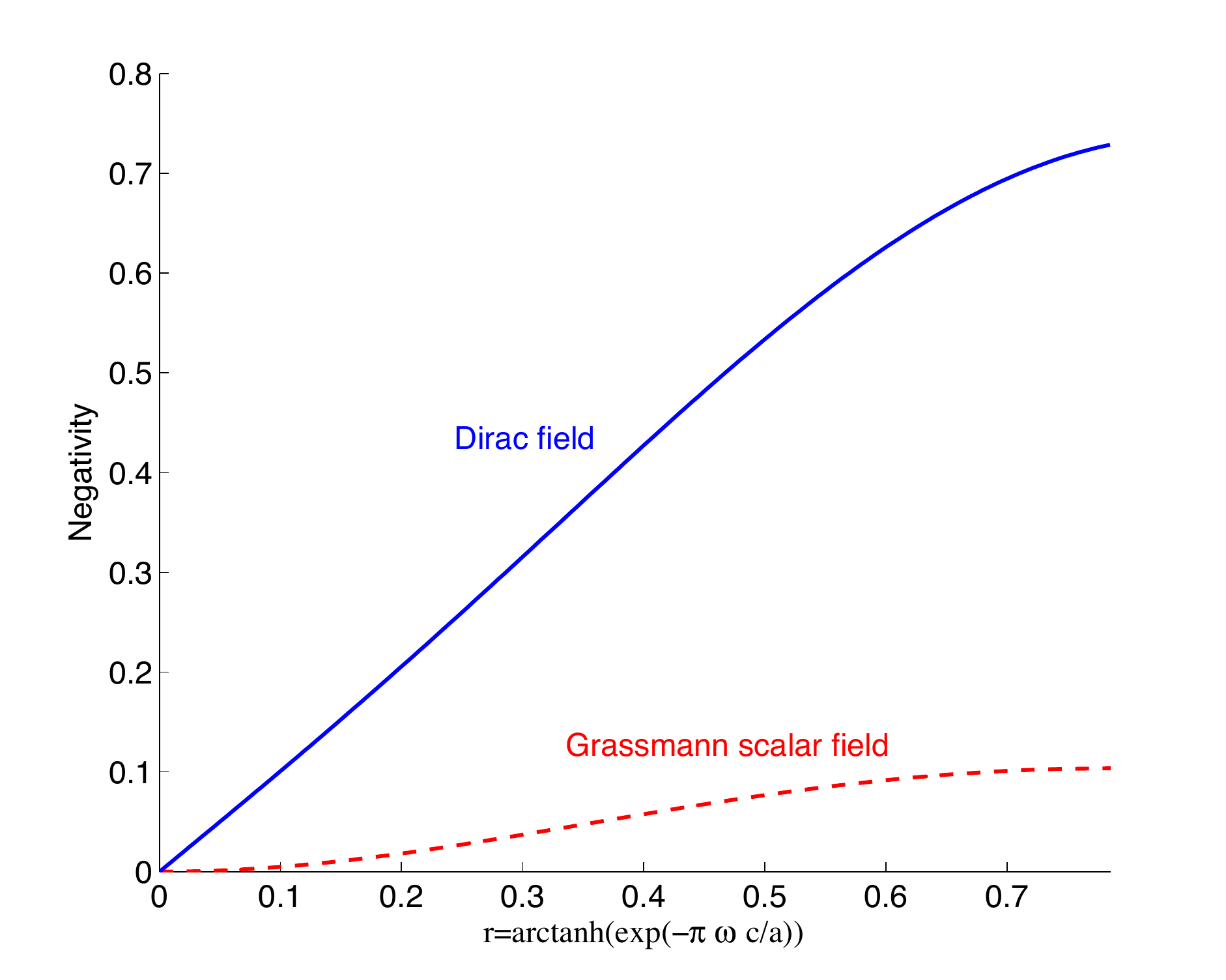}
\end{center}
\caption{ Negativity of the bipartition $\text{R}{\bar{\text{R}}}$ for fermion fields. Grassmann scalar (red dashed line) and Dirac (blue solid line). Negativity upper bound is greater for the Dirac case as $\dim(\mathcal{H}_{\text{Dirac}})>\dim(\mathcal{H}_{\text{Grassmann}})$. Notice that here $r=\operatorname{atan}(e^{-\pi k_0/a})$ instead of the hyperbolic tangent and therefore $r\rightarrow\pi/4\Rightarrow a\rightarrow\infty$. }
\label{comparison1}
\end{figure}

\begin{figure}[h]
\begin{center}
\includegraphics[width=.85\textwidth]{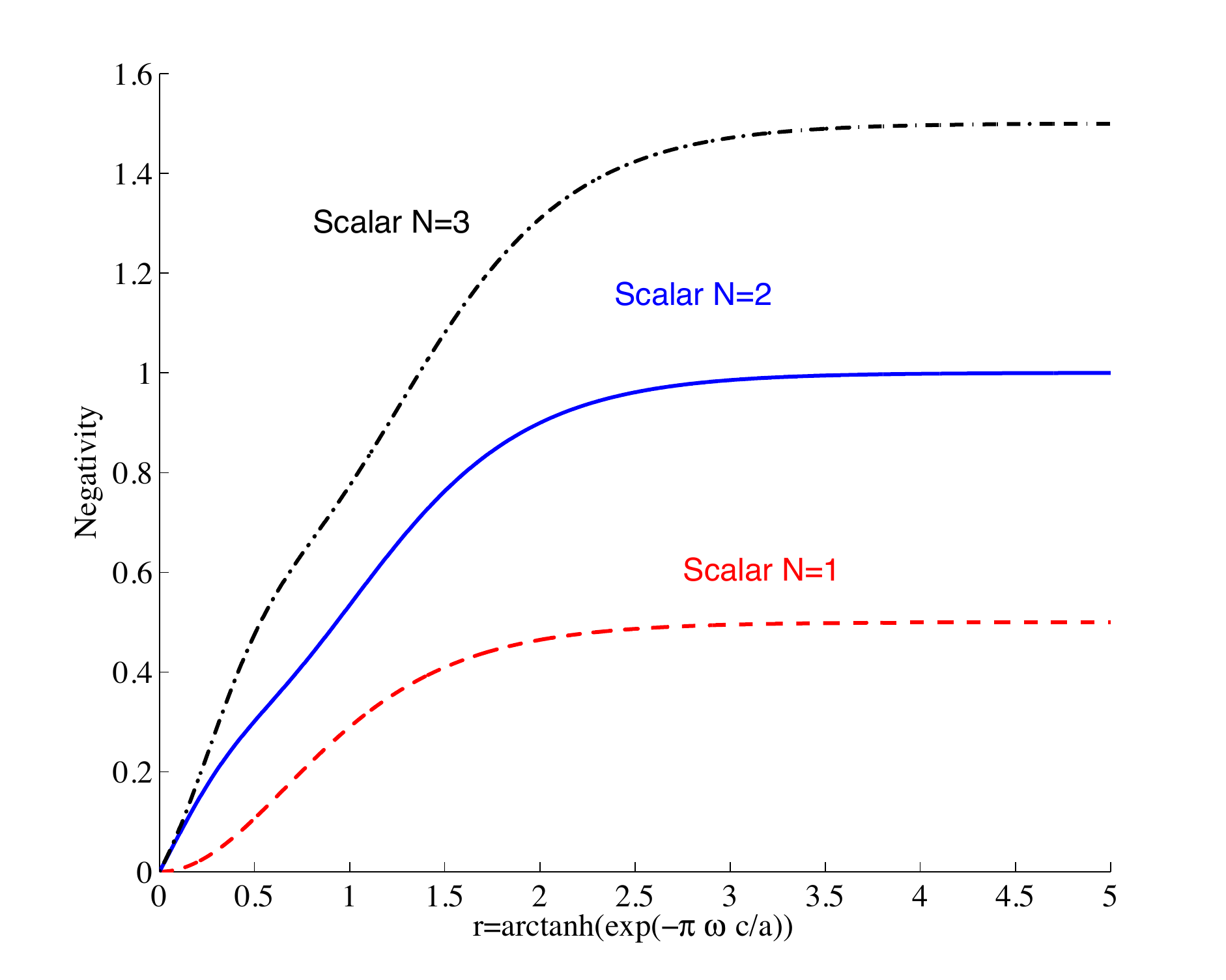}
\end{center}
\caption{ Same as fig. \ref{comparison1} but for bounded occupation number scalar fields. $N=1$ (red dashed line) is dimensionally analogous to the Grassmann scalar case. $N=3$ (black dash-dotted line) is dimensionally analogous to the Dirac case. $N=2$ (blue solid line) is analogous to the Dirac field in maximum occupation number.}
\label{comparison2}
\end{figure}

This comparison can be seen in figures \ref{comparison1}, \ref{comparison2}. We see that the behaviour is similar (monotonic growth from zero to a finite limit for $a\rightarrow\infty$) but the functional dependence is still very different in both cases. Specifically, as $a$ increases the bosonic cases grow a higher entanglement between the modes of the field on both sides of the horizon than the same dimension fermionic analogs. 


This clearly shows another important difference between fermionic and bosonic fields. Pauli exclusion principle prevents the total degradation of fermionic  entanglement in CCA bipartitions, whereas, conversely, impedes entanglement creation between $\text{R}{\bar{\text{R}}}$.

\subsection{Mutual information}\label{mutualsec}

Mutual information accounts for correlations (both quantum and classical) between two different partitions of a system (see section \ref{mutusec}). It is defined as
\begin{equation}\label{mutualdef}
I_{AB}=S_A+S_B-S_{AB},
\end{equation}
where $S_A$, $S_B$ and $S_{AB}$ are respectively the Von Neumann entropies for the individual subsystems $A$ and $B$ and for the joint system $AB$.

To compute the mutual information  for each bipartition we will need the eigenvalues of the corresponding density matrices. We shall go through all the process detailedly in the lines below.

\subsubsection{Alice-Rob Bipartition}

Excepting the element $\proj{0N}{0N}$ (which forms a $1\times1$ block itself) the density matrix for the system Alice-Rob \eqref{rhoars1} consists on $N$ $2\times2$ blocks in the basis $\{\ket{0 n},\ket{1\, n+1}\}_{n=0}^{N-1}$ which have the form
\begin{equation}
\frac{\tanh^{2n}r}{C_N(r)^2\cosh^2r }
\left(\!\begin{array}{cc}
1 & \dfrac{\sqrt{n+1}}{\cosh r}\\
\dfrac{\sqrt{n+1}}{\cosh r} & \dfrac{n+1}{\cosh^2r}
\end{array}\!\right).
\end{equation}
Hence, the eigenvalues of \eqref{rhoars1} are
\begin{eqnarray}\label{eigAR}
\nonumber\lambda_n&=&\left.\frac{\tanh^{2n}r}{C_N(r)^2\cosh^2r}\left(1+\frac{n+1}{\cosh^2 r}\right)\right|_{n=0}^{N-1}\\*
\lambda_N&=&\frac{\tanh^{2N} r}{C_N(r)^2\cosh^2r},
\end{eqnarray}
along with $N$ identically zero eigenvalues.

\subsubsection{Alice-AntiRob Bipartition}

Except from the diagonal element corresponding to $\proj{00}{00}$  (which forms one $1\times1$ block itself) the density matrix for the system Alice-AntiRob \eqref{rhoa-rs1} consists on $(N-1)$ $2\times2$ blocks in the basis $\{\ket{0 n},\ket{1\, n-1}\}_{n=1}^N$ which have the form
\begin{equation}
\frac{\tanh^{2n}r}{C_N(r)^2\cosh^2r}
\left(\!\begin{array}{cc}
1 & \dfrac{\sqrt{n}}{\sinh r} \\
\dfrac{\sqrt{n}}{\sinh r} & \dfrac{n}{\sinh^2 r}\\
\end{array}\!\right).
\end{equation}
Therefore the eigenvalues of \eqref{rhoa-rs1} are
\begin{eqnarray}\label{eigAaR}
\lambda_n&=&\left.\frac{\tanh^{2n}r}{C_N(r)^2\cosh^2r}\left(1+\frac{n}{\sinh^2r}\right)\right|_{n=0}^{N},
\end{eqnarray}
along with $N$ identically zero eigenvalues.

\subsubsection{Rob-AntiRob Bipartition}

The density matrix for Rob-AntiRob  \eqref{rhor-rs1} consists in the direct sum of two blocks 
\begin{equation}\label{newdoe}
\rho^{\text{R}{\bar{\text{R}}}}=X\oplus Y
\end{equation}
of dimensions $\dim(X)=N+1$, $\dim(Y)=N$.
The matrix elements of $X$ and $Y$ are
\begin{equation}
X_{ij}=\frac{(\tanh r)^{i+j-2}}{C_N(r)^2\cosh^2r }\qquad Y_{ij}=\sqrt{i}\sqrt{j} \frac{(\tanh r)^{i+j-2}}{C_N(r)^2\cosh^4r}
\end{equation}
in the bases $\left\{\ket{nn}\right\}_{n=0}^{N}$ and $\left\{\ket{n+1\,n}\right\}_{n=0}^{N-1}$ respectively.

It is easy to see that $\operatorname{rank}(X)=\operatorname{rank}(Y)=1$. This means that all the eigenvalues of  \eqref{rhor-rs1} are zero  except for two of them, which we can readily compute
\begin{equation}\label{eigRaR}
\lambda^{\text{R}{\bar{\text{R}}}}_{X}=\frac{D_N^0(r)}{C(r)^2}\qquad\lambda^{\text{R}{\bar{\text{R}}}}_{Y}= \frac{D_N^1(r)}{C(r)^2},
\end{equation}
where $D_N^0(r)$ and $D_N^1(r)$ are given by \eqref{S1} and \eqref{S2}.

\subsubsection{Von Neumann entropies for each subsystem and mutual information}

To compute the Von Neumann entropies we need the eigenvalues of every bipartition and the individual density matrices. The eigenvalues of $\rho^{AR}$, $\rho^{A\bar R}$, $\rho^{\text{R}{\bar{\text{R}}}}$ are respectively \eqref{eigAR}, \eqref{eigAaR} and \eqref{eigRaR}.

The eigenvalues of the individual systems density matrices can be directly read from \eqref{Robpartial}, \eqref{ARobpartial} and \eqref{AlicedeAliceRob} since $\rho^R$, $\rho^{\bar R}$ and $\rho^A$ have diagonal forms in the Fock basis. The Von Neumann entropy for a partition $B$ of the system is $S=-\tr(\rho\log_2\rho)$. Therefore their entropies are
\begin{align}\label{entropARaR}
\nonumber S_{R}&=-\sum_{n=0}^{N}\frac{\tanh^{2n}r}{C_N(r)^2\cosh^2r}\left[1+\frac{n}{\sinh^2r}\right]\log_2\left[\frac{\tanh^{2n}r}{C_N(r)^2\cosh^2r }\left(1+\frac{n}{\sinh^2r}\right)\right],\\*
\nonumber S_{\bar R}&=-\sum_{n=0}^{N-1}\frac{\tanh^{2n}r}{C_N(r)^2\cosh^2r}\left[1+\frac{n+1}{\cosh^2r}\right]\log_2\left[\frac{\tanh^{2n}r}{C_N(r)^2\cosh^2r }\left(1+\frac{n+1}{\cosh^2r}\right)\right],\\*
\nonumber&-\frac{\tanh^{2N}r}{C_N(r)^2\cosh^2r }\log_2\left(\frac{\tanh^{2N}r}{C_N(r)^2\cosh^2r}\right)\\*
S_A&=2\log_2\left[C_N(r)\right]-\frac{1}{C_N(r)^2}\sum_{i=0,1}D^i_N(r)\log_2\left[D^i_N(r)\right].
\end{align}

We obtain a universal result which relates the entropies of the different bipartitions of the system, 
\begin{equation}
S_{AR}=S_{\bar R},\qquad S_{A\bar R}=S_R,\qquad S_{\text{R}{\bar{\text{R}}}}= S_A.
\end{equation}  
These results can be summarised in the expression
\begin{equation}
S_{IJ}=S_K,
\end{equation}
where $I,J$ and $K$ labels represent different subsystems. Whichever values $I\neq J\neq K$ will satisfy the identity. This is also true for standard scalar fields, Grassmann scalar fields (spinless fermions) and Dirac fields. These relationships are completely universal, being independent of statistics and dimension, and reflect a fundamental aspect of Unruh decoherence in terms of the entropy of the partial systems, namely,  the way in which the entropy of the bipartitions behaves as acceleration increases  is not independent from the way the individual entropies do. 

The mutual information for all the possible bipartitions of the system will be
\begin{eqnarray}
\nonumber I_{AR}&=&S_A+S_R-S_{A R}=S_A+S_R -S_{\bar R},\\*
\nonumber I_{A\bar R}&=&S_A+S_R-S_{A\bar R}=S_A+ S_{\bar R}-S_R,\\*
\nonumber I_{\text{R}{\bar{\text{R}}}}&=& S_R+S_{\bar R}-S_{\text{R}{\bar{\text{R}}}}=S_A+ S_{\bar R}+S_R.
\end{eqnarray}  

The first notable difference from the standard bosonic field is that we do not obtain here the conservation law of the mutual information for the system Alice-Rob and Alice-AntiRob, instead
\begin{equation}\label{noconservationbos}
 I_{AR} + I_{A\bar R}=2S_{A},
\end{equation}
where for any finite $N$, $S_A$ goes to zero when $a\rightarrow\infty$. Only in the limit of $N\rightarrow\infty$ where $S_{A}\rightarrow 1$  $\forall a$ the conservation law is restored. 

\begin{figure}[h]
\begin{center}
\includegraphics[width=.85\textwidth]{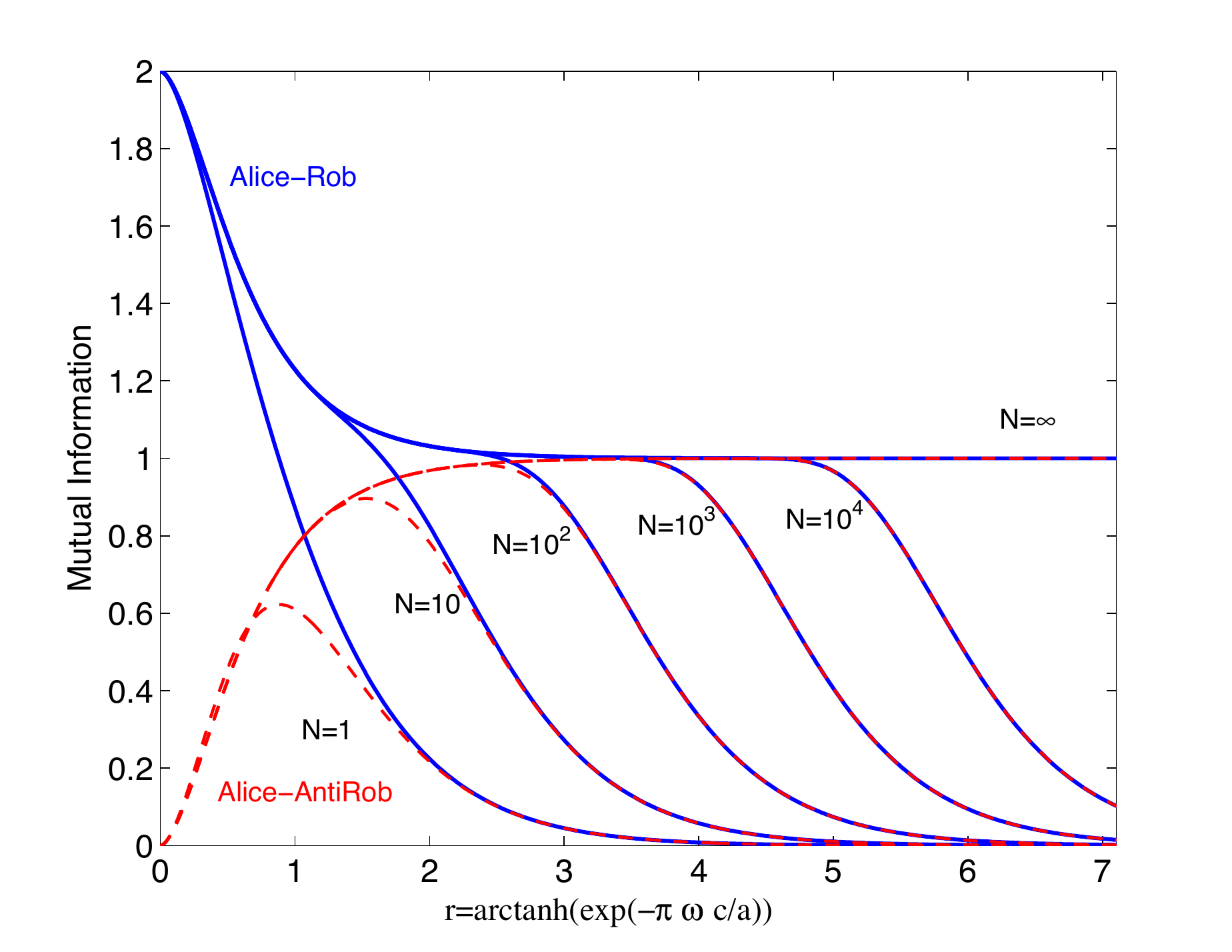}
\end{center}
\caption{ Mutual information for the systems Alice-Rob (blue continuous lines) and Alice-AntiRob (red dashed lines) as the acceleration parameter varies. Several values of $N$ are plotted along with the $N=\infty$ case. A conservation law is satisfied until acceleration reaches a critical value which is displaced to the right logarithmically as $N$ increases.}
\label{MutuARAAR1}
\end{figure}

Figure \ref{MutuARAAR1} shows  the mutual information tradeoff of the systems AR and $\text{A}{\bar{\text{R}}}$ from $N=1$ to $N=10^4$, along with the limit $N\rightarrow\infty$ in which the conservation law is fulfilled for all values of $a$. 

\begin{figure}[h]
\begin{center}
\includegraphics[width=.85\textwidth]{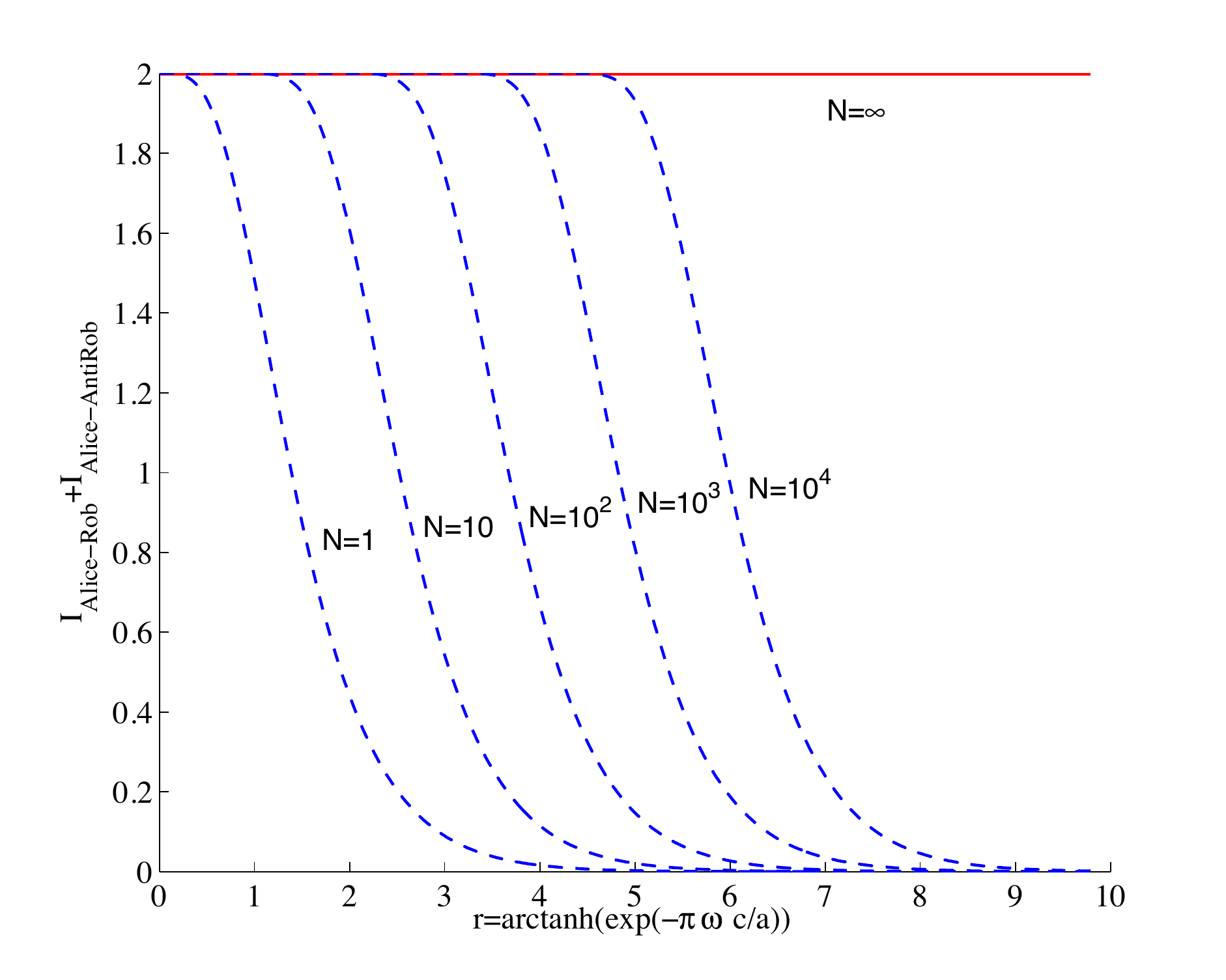}
\end{center}
\caption{ Violation of mutual information conservation law for the systems AR and $\text{A}{\bar{\text{R}}}$. The conservation law is fulfilled until $a$ reaches a critical value which is logarithmically displaced to the right as $N$ increases. The conservation law is completely restored when $N\rightarrow\infty$.}
\label{conserva}
\end{figure}

As it can be seen in Figure \ref{conserva} the largest deviation from the conservation law is obtained for $N=1$. As it is shown in the Figure, for a given $N$ the conservation law is fulfilled until the acceleration reaches a critical value $a=a_{l}$, then correlations go rapidly to zero. This critical value increases logarithmically with $N$. 

We showed above that quantum correlations between AR are quickly lost as $a$ increases for all $N$  and no entanglement is created between $\text{A}{\bar{\text{R}}}$. This means that for the high acceleration regime (where quantum entanglement vanishes) classical correlations dominate mutual information. Therefore, what we learn from mutual information in this regime is the behaviour of purely classical\footnote{There might be also quantum discord \cite{Discord}, but it is irrelevant in our analysis since our interest here is to distinguish entanglement from the rest of correlations.} correlations which are usually very difficult to be studied separately from entanglment.

For the scalar field we have seen that, conversely to fermionic fields,  limiting the dimension produces boundary effects which make classical correlations go to zero. Here, conservation of these correlations for all values of $a$ requires infinite dimension. In other words, finite dimensions schemes kill classical correlations as Rob accelerates in the bosonic case.

One would expect something similar for fermions since their states are naturally of finite dimension. Hence, similar `border effects' in classical correlations should appear in the same fashion as for bosons. Despite this fact, mutual information for fermions does not vanish. The explanation for this difference between bosons and fermions comes from fermionic quantum correlations. As shown in chapter \ref{etanthrough}, there is a conservation law for fermionic quantum entanglement 
\begin{equation}\label{connetfer}
\mathcal{N}^{AR}_{\text{fermions}}+\mathcal{N}^{\text{A}{\bar{\text{R}}}}_{\text{fermions}}=\frac12.
\end{equation}
Since classical correlations for the finite dimensional case eventually go to zero, quantum correlations rule mutual information behaviour for fermions. Therefore, it can be concluded that  the conservation law for the mutual information for fermions must be strongly related with the conservation of the fermionic entanglement which has its origin in statistics. 

We can conclude then that the origin of the universal mutual information conservation law is different for fermions and bosons. On one hand, for bosons, it appears as a classical correlations conservation law. On the other hand, for fermions, this law reflects a quantum correlations conservation. This can also explain why mutual information behaves so similarly to negativity for fermions as it was obtained in chapter \ref{etanthrough}.  

\begin{figure}[H]
\begin{center}
\includegraphics[width=.85\textwidth]{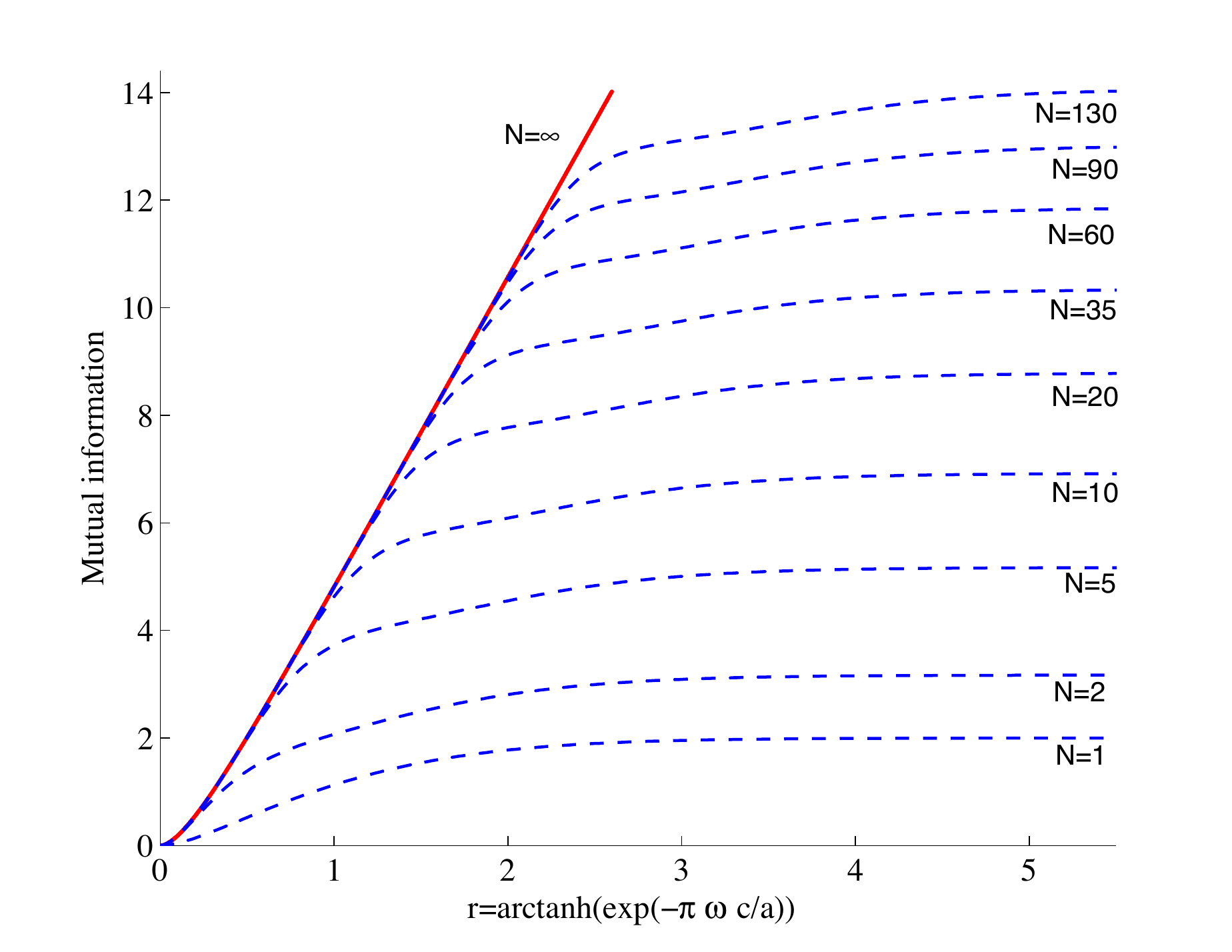}
\end{center}
\caption{Mutual information for the system R-${\bar{\text{R}}}$ as acceleration varies for different values of $N$. Only for $N\rightarrow\infty$ Mutual information diverges.}
\label{GOREBADA}
\end{figure}

To finish this work and complete the analysis of mutual information let us show in Figure \ref{GOREBADA}  how the behaviour of $I_{\text{R}{\bar{\text{R}}}}$ changes as $N$ is increased and how the divergent limit is obtained when $N\rightarrow\infty$. The results about mutual information here are coherent with the thorough analysis of the correlations for the $\text{R}{\bar{\text{R}}}$ bipartition performed above when we analysed negativity.

\section{Discussion}\label{conclusions}

In this chapter we answer the question of the actual impact of Fock space dimensionality on the Unruh entanglement degradation phenomena. To do so, we have studied the dimensional dependence of scalar field correlations when one observer is non-inertial. With this end in sight we have built a scalar field entangled state in which we have imposed a maximum occupation number $N$ for Rindler modes.

We have shown that the entanglement for AR and $\text{A}{\bar{\text{R}}}$ is only slightly influenced by $N$. In other words, the qualitative behaviour (quick loss of entanglement for AR as shown in Figure \ref{bundle} and no entanglement creation for the system $\text{A}{\bar{\text{R}}}$) is the same for finite and infinite $N$. This again points to the argument given in previous chapters that it is statistics and not dimensionality that conditions the behaviour of correlations in the presence of horizons.

However, we have shown that AR entanglement is sensitive to variations of $N$. This opposes what we found for fermions, whose correlations are completely insensitive to Hilbert space dimension variations (for example going from Grassmann scalars to Dirac fields).

In previous works we found a universal behaviour for the fermionic entanglement of the bipartition AR. Specifically the functional form of the negativity was exactly the same; independent of the maximally entangled state selected, the spin of the field, and the number of modes considered going beyond SMA. In all the cases Unruh decoherence degrades fermionic entanglement exactly in the same way. Here we see that for bosons this universality principle does exist but it is not as strong due to the sensitivity of AR to dimension changes. 

We have also seen that lesser $N$ does not necessarily imply faster entanglement degradation. Instead, we have shown that for two different finite values of $N$, namely $N_1<N_2$ there is a region $a<a_c$ in which entanglement is more degraded for $N_2$ and another region $a>a_c$ in which entanglement is more degraded for $N_1$. In other words, for high accelerations, higher dimension means less entanglement degradation by Unruh effect. This result clashes again with the extended idea that lesser dimension would protect correlations better than higher dimension, one misconception that after all this research should be banished from the explanation of these phenomena.

We have also showed that, since $a_c$ shifts to the right as $N$ is increased, in the limit $N\rightarrow\infty$, $a_c\rightarrow\infty$, so that entanglement is more degraded for the infinite dimensional case than for any finite $N$ whatever the value of the acceleration.

It is remarkable that there is no entanglement tradeoff in the CCA bipartitions even for finite dimension; no entanglement appears in the bipartition $\text{A}{\bar{\text{R}}}$ whatever the dimension limit $N$. This reflects again that the differences between fermions and bosons have nothing to do with the finite dimensionality, but with the different statistics. 

Concerning mutual information, we have shown that the conservation law found for scalar fields in chapter \ref{etanthrough} for the systems AR and $\text{A}{\bar{\text{R}}}$ is violated for finite values of $N$. We have obtained that for a finite $N$ the conservation law is fulfilled until acceleration reaches a critical value, in which correlations quickly drop. This critical value grows logarithmically with $N$, which means that the conservation law is satisfied for all $a$ when $N\rightarrow\infty$. Therefore, the violation of this conservation law can be associated with the boundary effects of imposing a dimensional limit. It is important to observe that in the bosonic case, mutual information is mainly accounting for classical correlations since quantum entanglement in the bosonic case are quickly lost as $a$ increases.

One could expect something similar in the fermionic case since fermions have a limited dimension Hilbert space for each mode. However, they lack these boundary effects. This big difference between fermions and bosons is related with the conservation of quantum entanglment in the $a\rightarrow\infty$ for fermions: in the high acceleration regime fermionic entanglement does not die (unlike the bosonic case) and therefore, for fermions, mutual information is accounting for quantum entanglement. This entanglement satisfies itself a conservation law \eqref{connetfer} which is `inherited' by mutual information. This also explains the similitude between mutual information and negativity behaviour for fermions found in chapter \ref{etanthrough}.

The conclusion here is that the universal conservation law for mutual information is found for both fermions and bosons, however the nature of this conservation is completely different in each case. For bosons it is due to classical correlations conservation, whereas for fermions it is due to quantum entanglement conservation. This illustrates once again that statistics is a paramount feature in order to explain how correlations behave in the presence of horizons.

The dimension of the Fock space has the largest impact in the behaviour of correlations between regions I and II separated by the horizon. Comparing the limited dimension scalar fields with their fermionic analogs we have found that the behaviour of correlations $\text{R}{\bar{\text{R}}}$ is somewhat similar for bosons and fermions and mainly ruled by the dimensionality of the Hilbert space. For both fermionic and bosonic fields, entanglement is always created between $\text{R}{\bar{\text{R}}}$ reaching a maximum value when $a\rightarrow\infty$ for any finite dimension.

The scalar cases of finite $N$ present, important differences compared with their fermionic Fock space dimensional analogs (namely $N=1$ is analogous to the Grassmann scalar case and $N=2,3$ to the Dirac field case). In the infinite acceleration limit, scalar states entanglement created between I and II modes is greater than in the fermionic case.

This implies that, regarding correlations between Rob and AntiRob modes, the effect of statistics is the opposite to what happened for the CCA bipartitions. Here bosonic correlations reach higher values than their corresponding fermionic analogs.


\chapter{Quantum entanglement near a black hole event horizon\footnote{E. Mart\'in-Mart\'inez, L. J. Garay, J. Le\'on. Phys. Rev. D, 82, 064006 (2010)}}\label{blackhole1}

In previous chapters we studied the entanglement degradation
phenomenon produced when one of the partners of an entangled bipartite
system undergoes a constant acceleration; this phenomenon, sometimes
called Unruh decoherence, is strongly related to the Unruh effect. Its
study revealed that there are very strong differences between fermionic
and bosonic field entanglement
(\!\!\cite{Alicefalls,AlsingSchul,Ditta,DiracDiscord} and previous chapters). The reason for these
differences  was traced back to fermionic/bosonic statistics and not to
the difference between bosonic and fermionic mode population as
previously thought (see previous chapters). In these earlier studies some
conclusions were drawn about the infinite acceleration limit, in which the
situation is similar to being arbitrarily close to an event horizon of a
Schwarzschild black hole.

However there are many subtleties and differences between Rindler and
Schwarzschild spacetimes. For example Schwarzschild spacetime
presents a real curvature singularity while Rindler metric is nothing but
the usual Minkowski metric represented in different coordinates and,
therefore, has no singularities. The Rindler horizon is also of very
different nature from the Schwarzschild's event horizon. Namely, the
Rindler horizon is an acceleration horizon experienced only by
accelerated observers (at rest in Rindler coordinates). On the other hand,
a Schwarzschild horizon is an event horizon, which affects the global
causal structure of the whole spacetime, independently of the observer.
Also, for the Rindler spacetime there are two well defined families of 
timelike Killing vectors with respect to which modes can be classified
according to the criterion of being of positive or negative frequency.
Contrarily, Schwarzschild spacetime has only one timelike Killing vector
(outside the horizon).

Therefore, to analyse the entanglement degradation produced due to the
Hawking effect near a Schwarzschild black hole we must be careful,
above all if we want to do a deeper study than simply taking the limit in
which the Rindler acceleration parameter becomes infinite. In this chapter
we will see how we can use the tools coming from the study of the Unruh
degradation in uniformly accelerated scenarios without restricting only
to the exact infinite acceleration limit and controlling to what extent
such tools are valid.

Consequently, we will be able to compute the entanglement degradation
introduced by the Hawking effect as a precise function of three physical
parameters, the distance of Rob to the event horizon, the mass of the
black hole, and the frequency of the mode that Rob has entangled with
Alice's field state. As a result of this study we will obtain not only the
explicit form of the quantum correlations as a function of the physical
parameters mentioned above,  but also a  quantitative control on what
distances from the horizon can be still analysed using the mathematical
toolbox coming from the Rindler results.

The setting consists in two observers (Alice and Rob), one of them
free-falling into a Schwarzschild black hole  close to the horizon (Alice)
and the other one standing at a small distance  from the event horizon
(Rob). Alice and Rob are the observers of a bipartite quantum state which
is maximally entangled for the observer in free fall. The Hawking effect
will introduce degradation in the state as seen by Rob, impeding all the
quantum information tasks between both observers.

In this context, we will analyse not only the classical and quantum
correlations between Alice and Rob, but also those that both observers would acquire with the mode fields on
the part of the spacetime that is classically unaccessible due to the
presence of the event horizon.

By means of this study we will show that all the interesting entanglement
behaviour occurs in the vicinity of the event horizon. What is more, we will
argue that as the entangled partners go 
 away from the horizon the effects on entanglement become unnoticeably small and, as a consequence, quantum information
tasks  in universes that contain event horizons are not jeopardised.

We will also see that the phenomenon of Hawking degradation is
universal for every Schwarzschild black hole, which is to say, it is ruled by
the presence of the event horizon and is not fundamentally influenced by
the specific value of the black hole parameters when the analysis is
performed using natural units to the black hole.  Furthermore, we will discuss the validity of the results obtained when instead of the usual plane wave basis we work in a base of wave packets, for which the states of Alice and Rob can be spatially localised.



\section{Revisiting entanglement degradation due to acceleration}

To introduce the new notation that we will need to use in this chapter we will now summarise the results that have been obtained concerning
the effects of an uniform acceleration on quantum correlations. 

As seen in previous chapters, the Minkowski vacuum state of the field $\ket{0}$ is
annihilated by the annihilation operators $a_{\hat\omega_i,\text{M}}$ as well as
by the operators $a_{\omega_i,\text{U}}$ and $a'_{\omega_i,\text{U}}$. For the excited states of the field,
we will work with the orthonormal basis $\{\psi^\text{U}_{\omega_i},\psi'^{\text{U}}_{\omega_i}\}$
defined in \eqref{modopsi} such that
\begin{equation}\label{onepartgoodb}
\ket{1_{\omega_i}}_\text{U}=a_{\omega_i,\text{U}}^\dagger \ket{0},\qquad |1'_{\omega_i}\rangle_\text{U}=a'^\dagger_{\omega_i\text{U}} \ket{0}
\end{equation}
are solutions of the free Klein-Gordon equation which are not
monochromatic, but linear superpositions of plane waves of positive
frequency $\hat\omega_j$.

We have learnt that we can express the Minkowski vacuum
state and the first Unruh excitation in terms of the Rindler Fock space basis,
\begin{equation}\label{scavacinf1}
\ket{0_{\omega_i}}_\text{M}=\frac{1}{\cosh r_{\text{b},\omega_i}}\sum_{n=0}^\infty
(\tanh r_{\text{b},\omega_i})^n \ket{n_{\omega_i}}_\text{I}\ket{n_{\omega_i}}_{\text{II}}.
\end{equation}
and
\begin{equation}\label{unoinf1}
\ket{1_{\omega_i}}_\text{U}=\frac{1}{(\cosh r_{\text{b},\omega_i})^2}
\sum_{n=0}^{\infty}  (\tanh  r_{\text{b},\omega_i})^n
\sqrt{n+1}\ket{n+1_{\omega_i}}_\text{I} \ket{n_{\omega_i}}_{\text{II}}.
\end{equation}
where\footnote{In this chapter we employ (for convenience) the natural system of units $\hbar=c=G=1$}
\begin{equation}
\tanh r_{\text{b},\omega_i}=\exp(-\pi \omega_i/a).
\end{equation}
The mode $ |1'_{\omega_i}\rangle_\text{U}$ is analogous but swapping the labels I and II.

Analogously, same states can be obtained for a Dirac field.  As for the scalar case, the vacuum state of the field
$\ket{0}$ is annihilated by the annihilation operators
$c_{\hat\omega_{i},\sigma,\text{M}}$ and $d_{\hat\omega_{i},\sigma,\text{M}}$ for all
$\hat\omega_i,\sigma$ as well as by the operators
$c_{\omega_i,\sigma,\text{U}}$ and $d_{\omega_i,\sigma,\text{U}}$ for all
$\omega_i,\sigma$.

For the excited states of the field, we will work with the orthonormal
basis \eqref{modopsif} such that
\begin{equation}\label{onepartgood}
\ket{\sigma_{\omega_i}}_\text{U}=c_{\omega_i,\sigma,\text{U}}^\dagger \ket{0}
\end{equation}
are positive frequency solutions of the free Dirac equation which are not
monochromatic, but linear superpositions of plane waves of positive
frequency $\hat\omega_i$.

As it can be seen in chapter \ref{onehalf}, the projection onto the unprimed sector of the basis \eqref{modopsif} of the Minkowski vacuum state written in the
Rindler basis, is as follows
\begin{align}\label{vacuumf1}
 \ket{0_{{\omega_i}}}&=(\cos r_{\text{f},\omega_i})^2\biket{0}{0}+
 \sin r_{\text{f},\omega_i} \cos r_{\text{f},\omega_i}
 \left(\biket{\uparrow_{{\omega_i}}}{\downarrow_{{\omega_i}}}
+ \biket{\downarrow_{{\omega_i}}}{\uparrow_{{\omega_i}}}\right)\nonumber\\
&+
(\sin r_{\text{f},\omega_i})^2\biket{\pa_{{\omega_i}}}{\pa_{{\omega_i}}},
\end{align}
where
\begin{equation}
\tan r_{\text{f},\omega_i}=\exp(-\pi\omega_i/a).
\end{equation}
It is straightforward to check that the vacuum is annihilated by
$c_{{\omega_i},\sigma,\text{U}}$ and $d_{{\omega_i},\sigma,\text{U}}$ simply using
\eqref{Bogoferm2} and applying both operators to \eqref{vacuumf1}.

The one particle state (projected onto the sector $\psi^\text{M}_{\omega_i}$ of \eqref{modopsif}) in the Rindler basis can be readily obtained by
applying the particle creation operator $c^\dagger_{{\omega_i},\sigma}$ to \eqref{vacuumf1}:
\begin{align}\label{onepartf1}
\ket{\uparrow_{{\omega_i}}}_\text{M}&=\cos r_{\text{f},\omega_i}
\biket{\uparrow_{{\omega_i}}}{0}+
\sin r_{\text{f},\omega_i}\biket{\pa_{{\omega_i}}}{\uparrow_{{\omega_i}}},\nonumber\\
\ket{\downarrow_{{\omega_i}}}_\text{M}&=\cos r_{\text{f},\omega_i}
\biket{\downarrow_{{\omega_i}}}{0}-
\sin r_{\text{f},\omega_i}\biket{\pa_{{\omega_i}}}{\downarrow_{{\omega_i}}}.
\end{align}

Let us first consider the following maximally entangled state for a scalar
field:
\begin{equation}\label{Min1}
\ket{\Psi}_\text{s}=\frac{1}{\sqrt2}\left(\ket{0}\ket{0}
+\ket{1_\text{A}}\ket{1_{\omegar}}_\text{U}\right),
\end{equation}
where the label A denotes Alice's subsystem and R denotes Rob's
subsystem. In this expression, $\ket{0}_{\text{A,R}}$ represents the
Minkowski vacuum for Alice and Rob, $\ket{1}_\text{A}$ is an arbitrary
one particle state excited from the Minkowski vacuum for Alice, and the
one particle state for Rob is expressed in the basis~\eqref{modopsi} and
characterised by the frequency $\omegar$ observed by Rob.
\begin{figure}[h]
\begin{center}
\includegraphics[width=.85\textwidth]{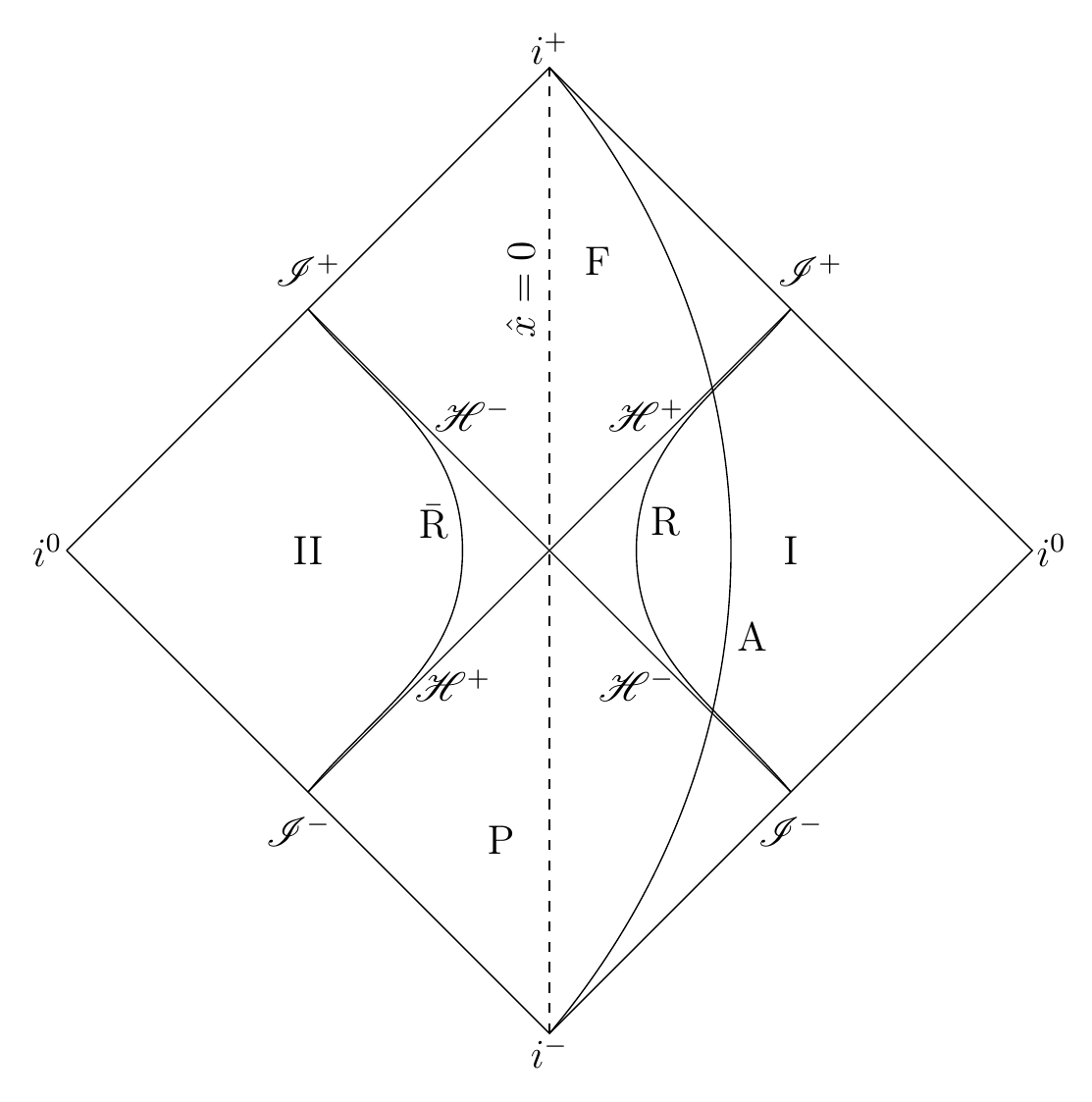}
\caption{Flat spacetime conformal diagram showing Alice, Rob and AntiRob
trajectories. $i^0$ denotes the spatial infinities, $i^-$, $i^+$ are
respectively the timelike  past and future infinities,
$\mathscr{I}^-$ and $\mathscr{I}^+$ are the null past and future infinities
respectively, and $\mathscr{H}^\pm$ are the Rindler horizons.}
\end{center}
\label{Rindler}
\end{figure}
As usual, since the second partner (Rob) ---who  observes the bipartite state
\eqref{Min1}--- is accelerated, it is convenient to map the second partition
of this state  into the Rindler Fock space basis, which can be computed
using equations
\eqref{scavacinf1} and \eqref{unoinf1}.
\begin{equation}\label{Minscalar}
\ket\Psi_\text{s}=\sum_{n=0}^\infty \frac{(\tanh r_{\text{b}})^n}{\sqrt{2}
\cosh r_{\text{b}}} \bigg(\ket{0}_\text{A}
\ket{n_{\omegar}}_\text{I}\ket{n_{\omegar}}_{{\text{II}}}+\left.\frac{\sqrt{n+1}}{\cosh r_{\text{b}}}\ket{1}_\text{A}
 \ket{n+1_{\omegar}}_\text{I}\ket{n_{\omegar}}_{\text{II}}\right).
 \end{equation}
 
The same can be done in the case of a Dirac field. Let us now consider the
following maximally entangled state for a Dirac field in the Minkowskian
basis
\begin{equation}\label{Minf}
\ket{\Psi}_\text{d}=\frac{1}{\sqrt2}\left(\ket{0}\ket{0}+
\ket{\uparrow}_\text{A}\ket{\downarrow_\omegar}_\text{U}\right).
\end{equation}
As for the bosonic case, if Rob, who observes this bipartite state, is
accelerated, it is convenient to map the second partition of this state into
the Rindler Fock space basis, which can be computed using Eqns.
\eqref{vacuumf1} and \eqref{onepartf1}. The explicit form of such state
can be seen in chapter \ref{onehalf}.

Notice that we have chosen a specific maximally entangled state
\eqref{Minf} of all the possible choices. This election has no relevance
since in previous chapters it was shown the universality of the degradation of
fermionic entanglement. All fermionic maximally entangled states are
equally degraded by the Unruh effect, no matter what kind of maximally
entangled state is (either occupation number or spin Bell state), or even if
we work with a Grassmann scalar field instead of a Dirac field.

Let us denote
\begin{equation}\label{dens}
\rho^\text{s}_{\text{AR}\bar{\text{R}}}=\proj{\Psi_\text{s}}{\Psi_\text{s}},\qquad \rho^\text{d}_{\text{AR}\bar{\text{R}}}=\proj{\Psi_\text{d}}{\Psi_\text{d}},
\end{equation}
the tripartite density matrices for the bosonic and fermionic cases,  in
which we use the Minkowski basis for Alice and the Rindler basis for
Rob-AntiRob.

In the standard Unruh entanglement degradation scenario we must trace over AntiRob degrees of freedom when accounting
for the quantum state shared by Alice and Rob. This provokes, for
instance, the observation of a thermal bath by Rob while Alice observes
the Minkowski vacuum as it can be seen elsewhere
(\cite{Birrell,AlsingSchul} and sections \ref{tue} and \ref{sec42m}). As a consequence the state becomes
mixed, which can cause some degree of correlation loss in the system AR as
we increase the value of the acceleration $a$. In references
\cite{Alicefalls,AlsingSchul,Ditta,DiracDiscord} and in previous chapters it was
studied how this phenomenon affects the entanglement for different
fields.

The correlation
trade-off among the all possible bipartitions of the system, 
 Alice-Rob (AR),  Alice-AntiRob $(\text{A}\bar{\text{R}})$, and
  Rob-AntiRob $(\text{R} \bar{\text{R}})$ It has been also studied in chapter \ref{etanthrough}.
  
 Classical communication between the two partners is only
allowed for the bipartitions AR and $\text{A}\bar{\text{R}}$. We refer to these bipartitions as  `Classical communication allowed'. These bipartitions are the only ones in which quantum information
tasks are possible.

On the other hand, no quantum information tasks can be performed using
$\text{R}\bar{\text{R}}$ correlations since classical communication
between Rob and AntiRob is not allowed. Anyway, studying this
bipartition is still necessary to give a complete description of the
behaviour of the correlation created between the spacetime regions
separated by the horizon.

The partial quantum states
for each bipartition are obtained by tracing over the third subsystem as usual
\begin{equation}\label{traza1}
\rho^{\text{AR}} = \tr_{{\text{II}}}\rho^{\text{AR}\bar{\text{R}}},\qquad \rho^{\text{A}\bar{\text{R}}} = \tr_{\text{I}}\rho^{\text{AR}\bar{\text{R}}},\qquad \rho^{\text{R}\bar{\text{R}}}=\tr_{\text{A}}\rho^{\text{AR}\bar{\text{R}}}.
\end{equation}

The properties of the correlations among these
subsystems have been analysed in previous chapters, showing a completely
different behaviour of quantum correlations for the CCA bipartitions
depending on whether the system is fermionic or bosonic.

For fermionic fields we saw that quantum correlations are conserved as Rob
accelerates (see subsection \ref{conservanet}). Specifically, as entanglement in the
bipartition AR is reduced, entanglement in the system
$\text{A}\bar{\text{R}}$ is increased. In the limit of
$a\rightarrow\infty$ some entanglement survives in all the bipartitions of
the system.

For the scalar field the situation is radically different, namely, no
entanglement is created in the CCA bipartitions (see subsection \ref{negatsecm4}). Moreover, the
entanglement in the AR bipartition is very quickly lost as Rob accelerates,
even if we artificially limit the dimension of the Hilbert space
(see chapter \ref{boundedpop}).

We have seen that this different behaviour is only ruled by statistics, which
plays a crucial role in the phenomenon of Unruh entanglement
degradation. The role of statistics is so important that, for fermions, the
behaviour of quantum correlations has been proven to be universal. Also, the survival of entanglement for the fermionic case, is
arguably related to statistical correlations as we will see later. All these
aspects will be discussed in depth later on, when we present the results
for the Schwarzschild black hole.

\section{The ``Black Hole Limit'': from Rindler to Kruskal}\label{sec2m6}

In this section we will study a completely new setting using the tools
learned from previous chapters. We will prove in a
constructive way that the entanglement degradation in the vicinity of an
eternal black hole can be studied in detail with these well-known tools. By
means of the construction shown below we will be able to deal with new
problems such as computing entanglement loss between a free-falling
observer and another one placed at fixed distance to the event horizon as
a function of the distance, studying the behaviour of quantum
correlations in the presence of black holes.

To begin this section let us work a little bit with the Schwarzschild
metric
 \begin{equation}
 \diff s^2=-\left(1-\frac{2m}{r}\right)\diff t^2+\left(1-\frac{2m}{r}\right)^{-1}
 \diff r^2 +r^2\diff \Omega^2,
 \end{equation}
 where $m$ is the black hole mass and $\diff\Omega^2$ is the line element in the unit sphere.
Due to the symmetry of the problem we are going to restrict the analysis
to the radial coordinate. To shorten notation let us write the radial part
of metric as
 \begin{equation}\label{propsch}
 \diff s^2=-f\diff t^2+f^{-1}\diff r^2,
 \end{equation}
 where $f=1-2m/r$.

 We can choose to write the metric in terms of the proper time $t_0$ of an observer placed in $r=r_0$ as follows,
  \begin{equation}\label{propsch2}
 \diff s^2=-\frac{f}{f_0}\diff t_0^2+f^{-1}\diff r^2,
 \end{equation}
where $f_0=1-2m/r_0$.
The relationship between $t_0$ and $t$ is given by the norm of the
timelike Killing vector $\xi=\partial_t$ in $r=r_0$, namely
 $t_0=\sqrt{f_0}\,t$.

We can now change the spatial coordinate such that the new coordinate
vanishes at the Schwarzschild radius $r=R_{\text{b}}=2m$. Let us define
$z$ in the following way
 \begin{equation}\label{defz}
 r-2m=\frac{z^2}{8m}\quad\Rightarrow\quad
  f=\frac{(\kappa z)^2}{1+(\kappa z)^2},
 \end{equation}
 with $\kappa =1/(4m)$ being the surface gravity of the black hole. Then the metric \eqref{propsch2} results
 \begin{equation}\label{change1}
 \diff s^2=-\frac{1}{f_0}\frac{(\kappa z)^2}{1+(\kappa z)^2}\diff t_0^2+
 \left[1+(\kappa z)^2\right]\diff z^2.
 \end{equation}
Near the event horizon ($z\approx0$), we can expand this metric to
lowest order in $z$ and approximate it by
\begin{equation}\label{misli}
\diff s^2=-\left(\frac{\kappa z}{\sqrt{f_0}}\right)^2\diff t_0^2+\diff z^2,
\end{equation}
which is a Rindler metric with acceleration parameter
$\kappa/\sqrt{f_0}$.

  On the other hand, Eq. \eqref{misli} represents
the metric near the event horizon in terms of the proper time of an
observer placed at $r=r_0$. The next step is giving a physical meaning to
this Rindler-like acceleration parameter. For this, we need to compute the
proper acceleration of a Schwarzschild observer placed at $r=r_0$,
which is, indeed, different from $\kappa$ (as $\kappa$ would be the
acceleration of an observer arbitrarily close to the horizon as seen from
a free-falling frame).

To compute $a$ for this observer as seen by himself (proper
acceleration) we must start from the Schwarzschild metric. The value of
the proper acceleration for an accelerated observer at arbitrary fixed
position $r$ is $a=\sqrt{a_\mu a^\mu}$ where $a^\mu=v^\nu\nabla_\nu
v^\mu$ is the observer 4-acceleration at such position, whereas $v^\mu$
is his 4-velocity.

The 4-velocity for a Schwarzschild observer in an arbitrary position $r$
is
\begin{equation}
v^\mu= {\xi^\mu}/{|\xi|},
\end{equation}
where $\xi\equiv\partial_t$ is the Schwarzschild timelike Killing vector.
As $\xi^\mu=(1,0,0,0)$ in Schwarzschild coordinates, 
$|\xi|=\sqrt{|g_{00}|}=\sqrt{f}$, and therefore
$v^\mu=\xi^\mu/\sqrt{f}$. Thus, we can compute the acceleration
4-vector
\begin{equation}
a^\mu=v^\nu\nabla_\nu v^\mu=\frac{1}{|\xi|}\xi^\nu\nabla_\nu
 \frac{\xi^\mu}{|\xi|} .
\end{equation}
Taking into account that $\xi^\mu$ is a Killing vector and, therefore, it
satisfies $\nabla_\mu\xi_\nu+\nabla_\nu\xi_\mu=0$, we easily obtain
\begin{equation}
a_\mu=\frac12\frac{\partial_\mu|\xi|^2}{|\xi|^2} =
 \frac{\partial_\mu f}{2f}=\frac{1}{2f}\left(0,\partial_r f ,0,0\right).
\end{equation}
Hence, since $g^{rr}=f$, the   proper acceleration for this observer is
\begin{equation}\label{properpropera}
a=\sqrt{g^{\mu\nu}a_\mu a_\nu}=\sqrt{\frac{(\partial_r f)^2}{4f} } .
\end{equation}
For an observer placed at $r=r_0$,
\begin{equation}\label{properpropera3}
a_0=\frac{\kappa}{\sqrt{f_0}}(1-f_0)^2.
\end{equation}

We know from \eqref{defz} that $1-f_0=[1+(\kappa z_0)^2]^{-1}$. So,
if the observer in $r=r_0$ is close to the event horizon ($r_0\approx
R_{\text{b}}$), then, to lowest order, $1+(\kappa z_0)^2\approx 1$ and
\begin{equation}\label{properproperaf}
a\approx {\kappa}/{\sqrt{f_0}}.
\end{equation}
Therefore, under this approximation, we can re-write \eqref{misli} as
\begin{equation}\label{Rindleradapt}
\diff s^2=-\left(a_0 z\right)^2\diff t_0^2+\diff z^2.
\end{equation}
This shows that the Schwarzschild metric can be approximated, in the
proximities of the event horizon, by a Rindler metric whose acceleration
parameter is the proper acceleration of an observer resisting in a position
$r_0$ close enough to the event horizon.

This approximation holds if
\begin{equation}\label{conditioaproximatio}
\left(\frac{z_0}{2R_{\text{b}}}\right)^2\ll1
\end{equation}
or, in other words, if
\begin{equation}\label{conditioaproximatio2}
\frac{\Delta_0}{R_\text{S}}\ll1,
\end{equation}
where $\Delta_0\equiv r_0-R_\text{S}$ is the coordinate distance from $r_0$ to
the event horizon. In the limit $r_0\rightarrow R_{\text{b}}$ we obtain
that $f_0\rightarrow0$ and, from \eqref{properproperaf},
$a_0\rightarrow\infty$. This shows rigorously that being very close to
the event horizon of a Schwarzschild black hole can be very well
approximated by the infinite acceleration Rindler case, as it was
suggested in \cite{Alicefalls,AlsingSchul}. This also enables us
to study what would happen  with the entanglement between observers
placed at different distances of the event horizon as far as the Rindler
approximation holds.

\begin{figure}[h]
\begin{center}
\includegraphics[width=.85\textwidth]{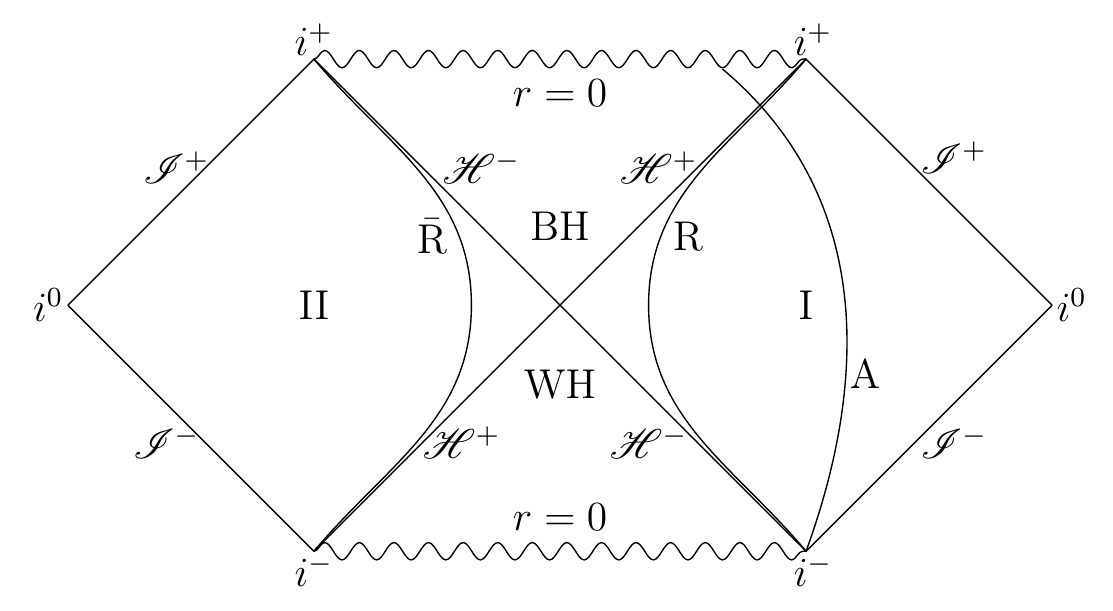}
\caption{Kruskal spacetime  conformal diagram showing trajectories for Alice,
Rob and AntiRob. $i^0$ denotes the spatial infinities, $i^-$, $i^+$
are respectively the timelike past and future infinities,
$\mathscr{I}^-$ and $\mathscr{I}^+$ are the null past and future infinities
respectively, and $\mathscr{H}^\pm$ are the event horizons.}
\label{Kruskal}
\end{center}
\end{figure}

Now let us identify again who is who in this new scenario. For this, we
introduce the null Kruskal-Szeckeres coordinates
\begin{equation}\label{eq:szeckeres}
u=-\kappa^{-1}\exp[-\kappa (t-r^*)],\quad v=\kappa^{-1}\exp[\kappa (t+r^*)],
\end{equation}
where $r^*=r+2m\log|1-r/2m|$. In terms of these coordinates the radial
part of the Schwarzschild metric is
\begin{equation}
\diff s^2=\frac{-1}{2\kappa r}e^{-2\kappa r}\diff u\diff v,
\end{equation}
where $r$ is implicitly defined by (\ref{eq:szeckeres}). The Penrose
diagram for the maximal analytic extension of Schwarzschild spacetime obtained from these coordinates is shown in fig.~\ref{Kruskal}. In this coordinates, 
near the horizon the metric can be written to  lowest order as
\begin{equation}
\diff s^2=-e^{-1}\diff u \diff v
\end{equation}
and $uv=-(\kappa z)^2$.

Hence, there are three regions in which we can clearly define physical
timelike vectors respect to which we can classify positive and negative
frequencies:
\begin{itemize}
\item  $\partial_{\hat t}\propto (\partial_{u}+\partial_{v})$.
The parameter $\hat t$ for this timelike vector corresponds to the
proper time of a free-falling observer close to the horizon, and it is
analogous to the Minkowskian   timelike Killing vector. Positive
frequency modes associated to this timelike vector define a vacuum
state known as the Hartle-Hawking vacuum $\ket{0}_\text{H}$,
which is analogous to $\ket{0}_\text{M}$ in the Rindler case.

\item $\partial_{t}\propto (u\partial_{u}-v\partial_{v})$.
 This is the Schwarzschild timelike Killing vector, which (when properly
normalised) corresponds to an observer whose acceleration at the
horizon equals the surface gravity $\kappa$ of the black hole with
respect to a Minkowskian observer, or, in other words, with proper
acceleration $a_0\approx\kappa/\sqrt{f_0}$ close to the horizon.
The vacuum state corresponding to positive frequencies associated to
this timelike Killing vector is called the Boulware vacuum
$\ket{0}_\text{B}$. This state is analogous to the Rindler vacuum
$\ket{0}_\text{I}$.

\item There is another timelike Killing vector $-\partial_{t}$
(as in Rindler) for region II that will allow us to define another
Boulware vacuum  in region II. We will call it AntiBoulware vacuum
$\ket0_{\bar{\text{B}}}$, analogous to $\ket{0}_\text{II}$ in the
Rindler case.
\end{itemize}

Now, in this scenario,
$\ket{1_{\hat\omega}}_\text{H}=a^\dagger_{\hat\omega,\text{H}}\ket{0}_\text{H}$
are free scalar field modes, in other words, solutions of positive
frequency ${\hat\omega}$ with respect to $\partial_{\hat t}$  of the
free Klein-Gordon equation close to the horizon
\begin{equation}
\ket{1_{\hat\omega}}_\text{H}\equiv u_{\hat\omega}^\text{H}\propto
\frac{1}{\sqrt{2\hat\omega}}e^{-i\hat\omega \hat t}.
\end{equation}
The label H just means that those states are expressed in the
Hartle-Hawking Fock space basis.

An observer located at a fixed distance from the black hole can also define his own vacuum and
excited states of frequency $\omega$ respect to the Killing vector
$\partial_{t}$. Actually, there are two natural vacuum states associated
with the positive frequency modes in both sides of the horizon these are
$\ket{0}_\text{B}$ and $\ket{0}_{\bar{\text{B}}}$, vacua for the
positive frequency modes in regions I and II respectively (fig.
\ref{Kruskal}). Subsequently, for a scalar field, we can define the field
excitations as
\begin{align}
\ket{1_\omega}_{\text{B}}&=a^\dagger_{\omega,\text{B}}\ket{0}_\text{B}\equiv
u_\omega^{\text{B}}\propto\frac{1}{\sqrt{2\omega}}e^{-i\omega  t},\nonumber\\
\ket{1_\omega}_{\bar{\text{B}}}&=a^\dagger_{\omega,\bar{\text{B}}}
\ket{0}_{\bar{\text{B}}}\equiv u_\omega^{\bar{\text{B}}}\propto\frac{1}{\sqrt{2\omega}}e^{i\omega  t}.
\end{align}

Then, the analogy between the Rindler-Minkowski and the Boulware-Hartle-Hawking states,  and their relation with 
the standard Alice-Rob-AntiRob notation is as follows:
\begin{equation}\label{identif}
\begin{array}{lclcl}
\ket{0}_\text{R}&\leftrightarrow&\ket{0}_\text{I}&\leftrightarrow&\ket{0}_\text{B},\\
\ket{0}_{\bar{\text{R}}}&\leftrightarrow&\ket{0}_{\text{II}}&
\leftrightarrow&\ket{0}_{\bar{\text{B}}},\\
\ket{0}_{\text{A}}&\leftrightarrow&\ket{0}_\text{M}&\leftrightarrow&
\ket{0}_{\text{H}}.
\end{array}
\end{equation}
The change of basis between Hartle-Hawking modes
and Boulware modes is completely analogous to the change of basis
between Minkowskian modes and Rindler modes with an acceleration
parameter $a_0=\kappa/\sqrt{f_0}$.

In the same fashion as for Rindler we define an orthonormal basis 
of Hartle-Hawking scalar field modes $\{\psi^\text{H}_{\omega_j},\psi'^{\text{H}}_{\omega_j}\}$ whose elements are superpositions of  positive-frequency solutions
$u_{\hat\omega_i}^\text{H}$ of the Klein-Gordon equation with
respect to the Kruskal time $\hat t$
 such that each element  corresponds to Boulware modes of one single frequency in the Kruskal regions I and II
($u^\text{B}_{\omega_j}$ and $u^{\bar{\text{B}}*}_{\omega_j}$). The same can be
done for the Dirac field.

We can express the Hartle-Hawking vacuum state in terms of the
Boulware Fock space basis.  To do so, we use what we learned from the
Rindler case. Taking into account that $\ket{0}_\text{H}=\bigotimes_{
i}\ket{0_{\omega_i}}_\text{H}$, we have that
\begin{equation}\label{scavacinf}
\ket{0_{\omega_i}}_\text{H}=\frac{1}{\cosh q_{\text{b},\omega_i}}
\sum_{n=0}^\infty (\tanh q_{\text{b},\omega_i})^n  \ket{n_{\omega_i}}_\text{B}
\ket{n_{\omega_i}}_{{\bar{\text{B}}}},
\end{equation}
where
\begin{equation}\label{defr3}
\tanh q_{\text{b},\omega_i}=\exp\left({-\pi \sqrt{f_0}\,\omega_i/{\kappa}}\right)
.
\end{equation}
The unprimed Hartle-Hawking one particle state in the basis $\{\psi^\text{H}_{\omega_j},\psi'^{\text{H}}_{\omega_j}\}$ results
from applying the corresponding creation operator to the vacuum state.
We can also translate this state to the Boulware basis:
  \begin{equation}\label{unoinf}
\ket{1_{\omega_i}}_\text{H} = \frac{1}{(\cosh q_{\text{b},\omega_i})^2}\sum_{n=0}^{\infty} (\tanh  q_{\text{b},\omega_i})^n\sqrt{n+1}
\ket{n+1_{\omega_i}}_\text{B}\!\ket{n_{ \omega_i}}_{{\bar{\text{B}}}}.
\end{equation} 

The Hartle-Hawking  vacuum (projected onto the unprimed sector) for the Dirac case is expressed in the
Boulware basis as follows
\begin{align}\label{vacuumf}
 \ket{0_{\omega_i}}_{\text{H}} &= (\cos q_{\text{f},\omega_i})^2
 \bikete{0_{\omega_i}}{0_{\omega_i}}
 +\sin q_{\text{f},\omega_i}\cos q_{\text{f},\omega_i}
 \left(\ket{\uparrow_{\omega_i}}_{{\text{B}}}
  \ket{\downarrow_{\omega_i}}_{\bar{\text{B}}}+
 \bikete{\downarrow_{\omega_i}}{\uparrow_{\omega_i}}\right)
 \nonumber\\
 &+(\sin q_{\text{f},\omega_i})^2\bikete{\pa_{\omega_i}}{\pa_{\omega_i}},
\end{align}
whereas the projected Hartle-Hawking one particle state 
is expressed in the Boulware basis as
\begin{align}\label{onepartf}
 \ket{\uparrow_{\omega_i}}_\text{H}&= \cos q_{\text{f},\omega_i}
 \bikete{\uparrow_{\omega_i}}{0_{\omega_i}}+
 \sin q_{\text{f},\omega_i}\bikete{\pa_{\omega_i}}{\uparrow_{\omega_i}},\nonumber\\
\ket{\downarrow_{\omega_i}}_\text{H}&=
\cos q_{\text{f},\omega_i} \bikete{\downarrow_{\omega_i}}{0_{\omega_i}}-
\sin q_{\text{f},\omega_i}\bikete{\pa_{\omega_i}}{\downarrow_{\omega_i}},
\end{align}
where this time
\begin{equation}\label{defr4}
\tan q_{\text{f},\omega_i}=\exp\left(-\pi \sqrt{f_0}\,\omega_i /{\kappa}\right)
.
\end{equation}

Thus, in this new scenario, we can consider a bipartite states for fermions and bosons analogous to
the states \eqref{Min1} and \eqref{Minf} for the Rindler scenario, which have the following form in the basis of a free-falling observer (Alice)
\begin{align}\label{Haw1}
\ket\Psi_\text{s}&=\frac{1}{\sqrt{2}}
\left(\ket{0}\ket{0}+
\ket{1}_\text{A}\ket{1_{\omegar}}_\text{H}\right),\\*
\label{Hawf}
\ket{\Psi}_\text{d}&=\frac{1}{\sqrt2}\left(\ket{0}
\ket{0}+\ket{\uparrow}_\text{A}
\ket{\downarrow_{\omegar}}_\text{H}\right).
\end{align}
This bipartite system consists in two subsystems, the first one is going to
be observed by Alice, who is free-falling into the black hole and close to
the event horizon, and the second one will be observed by Rob, who is near
the event horizon at $r=r_0\approx R_\text{S}$. Therefore, the second
partner who observes the bipartite states
\eqref{Haw1} and \eqref{Hawf}  describes \eqref{Haw1} and  \eqref{Hawf} using the Boulware
basis, so that it is convenient to map the second partition of these states
into the Boulware Fock space basis.

Following the notation \eqref{identif}, to analyse the correlations among
the bipartite subsystems we need to trace out the third subsystem
analogously to what we did in \eqref{traza1}:
\begin{eqnarray}\label{traza2}
\nonumber\rho^{\text{AR}}&\!\!=\!&\tr_{{\text{II}}}\rho^{\text{AR}\bar{\text{R}}},\\*
\nonumber\rho^{\text{A}\bar{\text{R}}}&\!\!=\!&\tr_{\text{R}}\rho^{\text{AR}\bar{\text{R}}},\\*
\rho^{\text{R}\bar{\text{R}}}&\!\!=\!&\tr_{\text{A}}\rho^{\text{AR}\bar{\text{R}}}.
\end{eqnarray}

It can be seen in fig. \ref{Kruskal} that all the information beyond the
event horizon cannot be accessed by Rob. Actually, what happens beyond
the horizon is determined by the information that Rob can access along
with the information that AntiRob can access. In this context it makes
sense to say that studying the system $\rho^{\text{R}\bar{\text{R}}}$
gives an idea of the correlations across the horizon.

\section{Correlations behaviour}\label{sec4}

In this section we will use the machinery we already have from the Rindler
set-ups to compute the entanglement degradation as a function of the
position of Rob.

First we will consider that Rob's frequency $\omegar$ is measured in
natural units adapted to each black hole. This will show how modes of
different frequencies  suffer different correlation degradation. It will
also show how less massive black holes produce a higher degradation than
the heavier ones. Furthermore, this analysis will show the universality of
the phenomenon of the Hawking entanglement degradation for
Schwarzschild black holes.

After that, we will  analyse the different degree of entanglement
degradation experimented by an observer of fixed Boulware frequency
$\omegar$ standing at fixed distances from the event horizon for
different black hole masses.

In the following subsections we will see that all the interesting behaviour
happens in regions in which the Rindler approximation
\eqref{Rindleradapt} is valid. Specifically, we will see in the plots below that the
values of the distance to the horizon where the interesting
entanglement behaviour appears are in the regime $\Delta_0\lesssim 0.05R_\text{S}$
in all the cases considered in this section for which, consequently, the
approximation \eqref{Rindleradapt} holds.

\subsection{Adapted frequency}

In terms of the mode frequency measured by Rob  (written in units
natural to the black hole, i.e. in terms of the surface gravity $\kappa$) and his position measured in Schwarzschild
radii,
\begin{align}
\Omega&=2\pi\omegar
/\kappa=8\pi m\omegar,\\
R_0&=r_0/R_\text{S}=r_0/(2m),
\end{align}
 Eqns. \eqref{defr3} and
\eqref{defr4} can be written as
\begin{align}\label{defr32}
\tanh q_\text{s}&=\exp\left({-\frac{\Omega}{2} \sqrt{1-\frac{1}{R_0}}}\right),
\\
\label{defr42}
\tan q_\text{d}&=\exp\left({-\frac{\Omega}{2} \sqrt{1-\frac{1}{R_0}}}\right),
\end{align}
showing that the phenomenon of Hawking entanglement degradation
presents universality, which is to say, if the frequency is measured in natural
units, every Schwarzschild black hole behaves in the same way, as
expected.

\subsubsection{Quantum correlations}

\begin{figure}[h]
\begin{center}
\includegraphics[width=.85\textwidth]{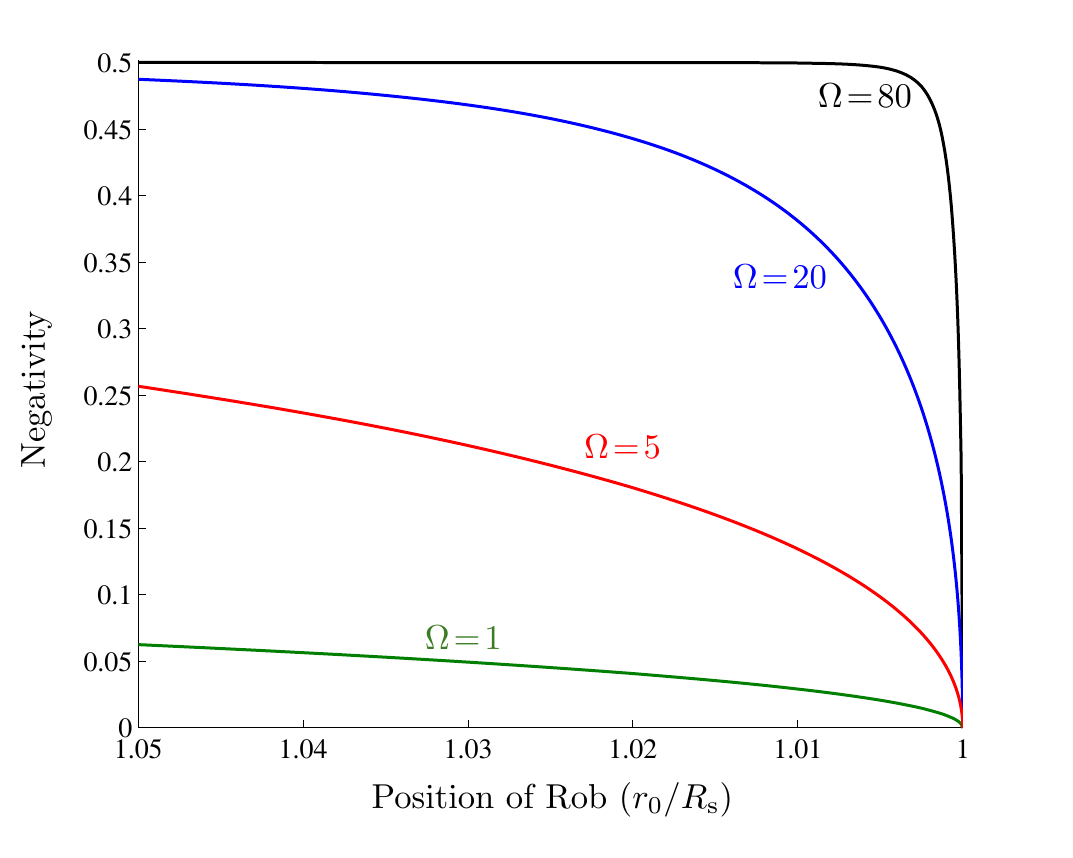}
\caption{Scalar field: Entanglement of the system Alice-Rob as a function of the position of Rob for different values of $\Omega$. Entanglement vanishes as Rob approaches
the Schwarzschild radius while no entanglement is created between Alice and AntiRob. The smaller the value of $\Omega$ the more degradation is produced by the black hole.}
\label{sncca}
\end{center}
\end{figure}

\begin{figure}[h]
\begin{center}
\includegraphics[width=.85\textwidth]{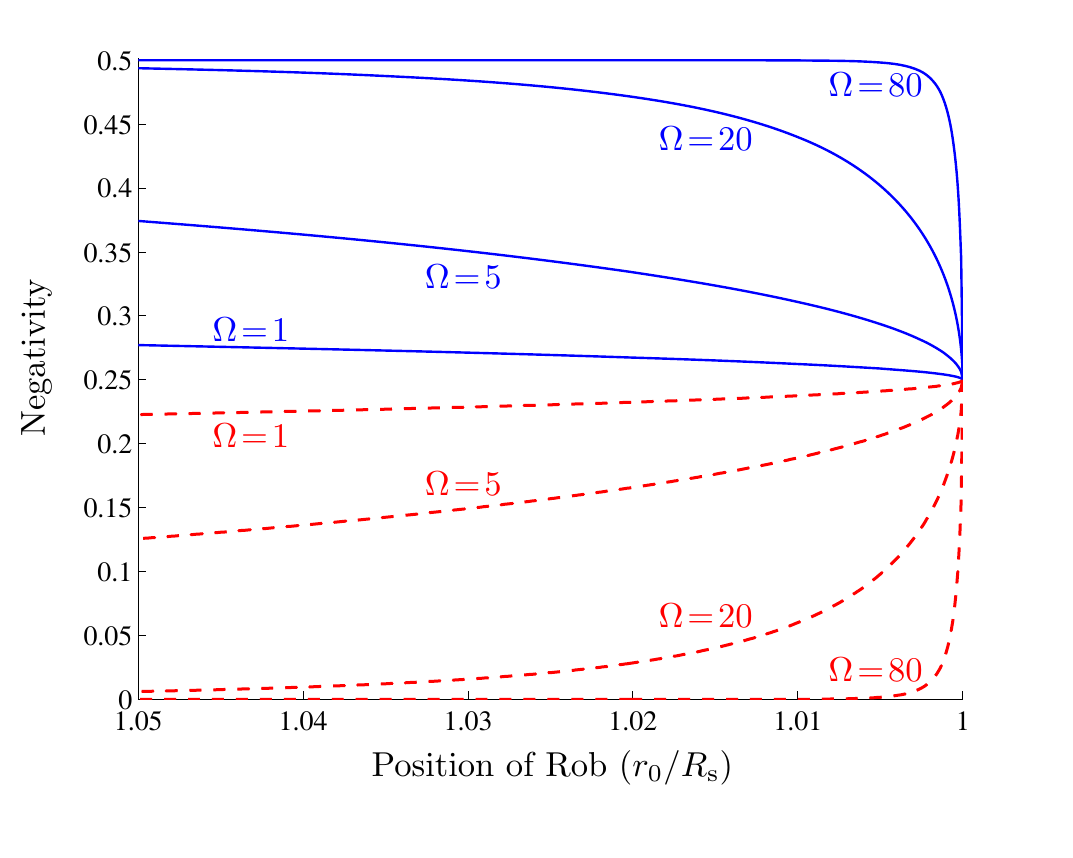}
\caption{Dirac field: Entanglement Alice-Rob (blue solid line) and
Alice-AntiRob (red dashed line). Universal conservation law for fermions
is shown for different values of $\Omega$. The entanglement
degradation in AR is quicker when $\Omega$ is smaller. The maximum
degradation is not total and its value is independent of $\Omega$.
}
\label{dncca}
\end{center}
\end{figure}

\begin{figure}[h]
\begin{center}
\includegraphics[width=.85\textwidth]{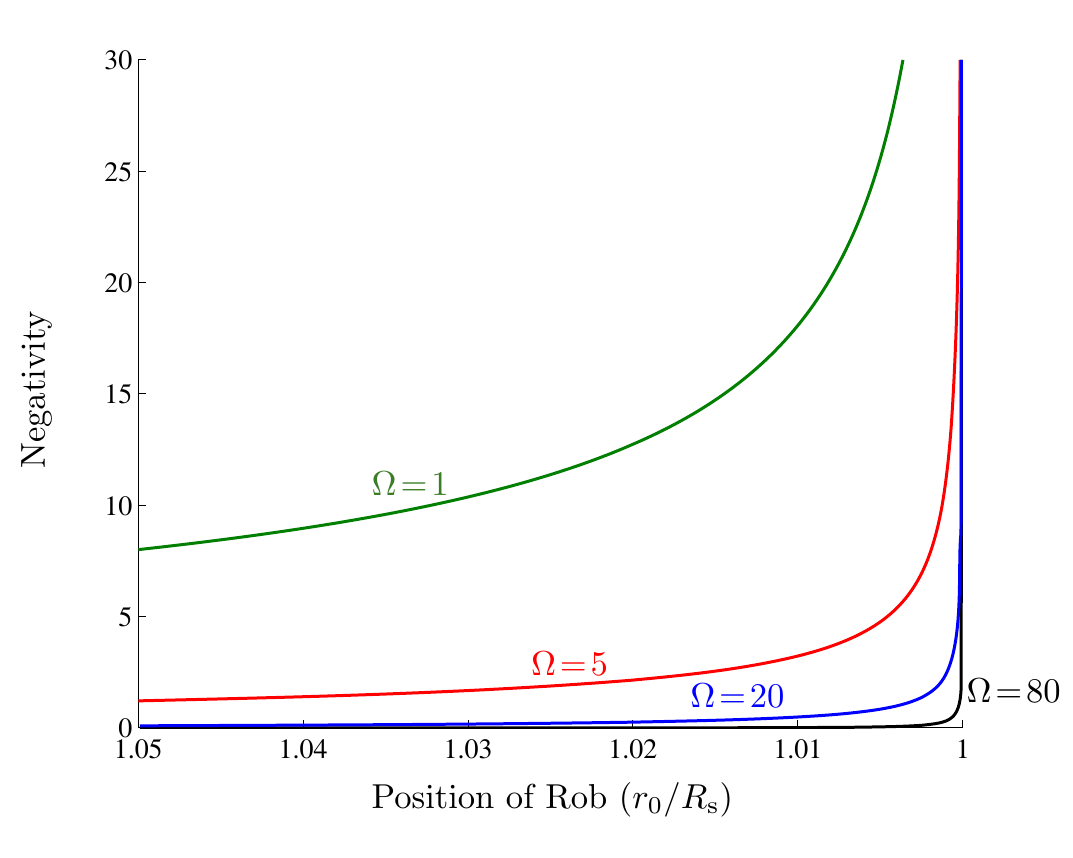}
\caption{Scalar field: Entanglement of the system Rob-AntiRob
(entanglement between regions I and II) as a function of the position of
Rob for different values of $\Omega$. Entanglement diverges as Rob
approaches   the Schwarzschild radius.}
\label{snrar}
\end{center}
\end{figure}

\begin{figure}[h]
\begin{center}
\includegraphics[width=.85\textwidth]{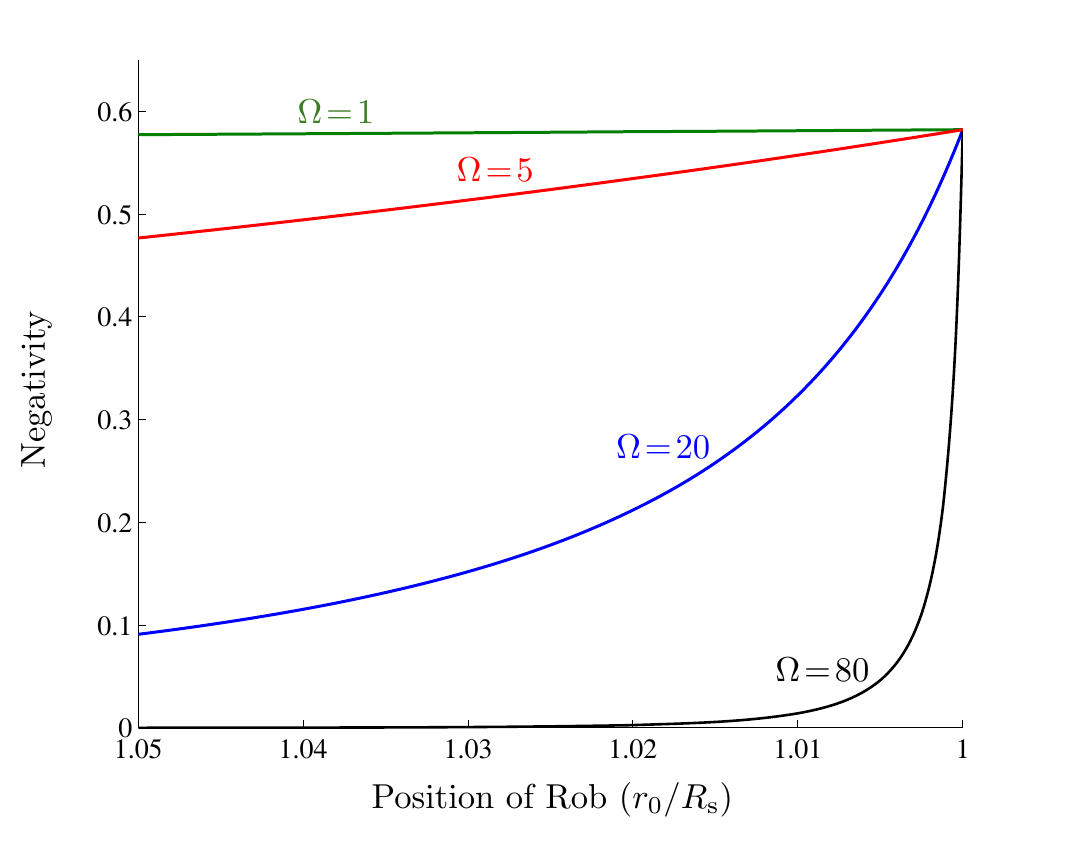}
\caption{Dirac field: Entanglement of the system Rob-AntiRob
(entanglement between regions I and II) as a function of the position of Rob for
different values of $\Omega$. Entanglement tends to a finite value as Rob
approaches   the Schwarzschild radius.}
\label{dnrar}
\end{center}
\end{figure}

We will use the negativity ($\mathcal{N}$) to account for the quantum
correlations between the different bipartitions of the system. Hence, to compute it, we will need the partial transpose of the bipartite
density matrices \eqref{traza2}. The details associated to the
diagonalisation of the partial transposed density matrices for each
subsystem are technically very similar to the Rindler case, and are not of
much interest for the purposes of this article. All the technical aspects of
such calculations can be found in chapter \ref{etanthrough} for Dirac and scalar fields.
The results of those calculations are shown in Figs. \ref{sncca} to
\ref{dnrar}. In Figs. \ref{sncca} and \ref{dncca} we can see the behaviour
of the negativity on the CCA bipartitions for different values of Rob's
frequency $\Omega$.

For the scalar field we can see that as Rob is closer to the event horizon
the entanglement shared between Alice and Rob decreases. In the limit in
which Rob is very close to the horizon, entanglement is completely lost.
With the study performed here we can see the functional
dependence of the entanglement with the distance to the horizon. As seen in the figures,
 the degradation phenomenon occurs in a narrow region very close
to the event horizon. If Rob is far enough from the black hole he will not
appreciate any entanglement degradation effects unless either the mass of the black hole or the frequency of the mode considered are extremely small. There must be, indeed, a minimum residual effect associated to the Hawking thermal bath experienced in the asymptotically flat region of the spacetime, far from the region in which this approximation is valid, but it is unnoticeably small. Certainly, as it will be seen in fig. \ref{realb} and the discussion below, even very close to the horizon no effective entanglement degradation 
 occurs for physically meaningful values of mass and frequency.  
 
If we keep the frequency measured by Rob $(\omegar)$ constant,
$\Omega$ will grow proportional to  the black hole mass. With this in
mind, fig. \ref{sncca} shows that the degradation is stronger for less
massive black holes. This result is consistent with the fact that the
Hawking temperature increases as the mass of the black hole goes to
zero. In the next section (specifically in fig. \ref{realb}) we will show
that this is not an effect of choosing natural units, when an observer is at
a fixed distance of a black hole, the degradation will be higher for less
massive black holes.

In any case, for the scalar field, the entanglement in the system AR is
completely degraded when one of the observers is resisting very close to
the event horizon of the black hole. Hence, in this scenario, no quantum
information resources can be used (for instance to perform quantum
teleportation or quantum computing) between a free-falling observer and
an observer arbitrarily close to an event horizon. Moreover, no
entanglement of any kind is created among the CCA bipartitions of the
system (the ones where classical communication is allowed). Therefore, all useful
quantum correlations between a free-falling observer and an observer at
the event horizon are lost due to the Hawking effect degrading all the
entanglement in the system.

For the Dirac field  (fig. \ref{dncca}) something very different happens.
We see that correlations in the bipartition AR decrease to a certain
finite limit, which means that there is entanglement survival even when
Rob is asymptotically close to the event horizon. This survival is a well
known phenomenon in the Rindler case \cite{AlsingSchul}. At the
same time that entanglement is destroyed in the AR bipartition,
entanglement is created in the complementary $\text{A}\bar{\text{R}}$
bipartition so that negativity in the CCA bipartitions fulfils a
conservation law regardless of the distance to the event horizon and the
mass of the black hole
\begin{equation}\label{conservan}
\mathcal{N}_\text{AR}+\mathcal{N}_{\text{A}\bar{\text{R}}}=\frac12.
\end{equation}
The nature of this entanglement and the survival of correlations, even in
the limit of positions arbitrarily close to the horizon, is discussed in chapters \ref{etanthrough} and \ref{boundedpop} for the Rindler case. When we deal with fermionic
fields there are correlations that come from the statistical fermionic
nature of the field which we cannot get rid of. The hypothesis is that this
entanglement, which is purely statistical, is the second quantised version
of the statistical entanglement disclosed in \cite{sta1}. Here we see that
the same conclusions drawn in that case can be perfectly applied to the
Schwarzschild black hole case.

About the dependence of the entanglement degradation on the frequency
of the Boulware mode, Fig. \ref{sncca}  shows that, for a scalar field, the
loss of entanglement between a free falling observer and an observer
outside but very close to the event horizon (AR) is greater for modes of
lower frequency. This makes sense because, energetically speaking, it is
cheaper to excite those modes and, therefore, they are more sensitive to
the Hawking thermal noise. For a Dirac field (Fig. \ref{dncca}) we see a
similar behaviour. However, the surviving
entanglement in the limit in which Rob is infinitely close to the event
horizon is not sensitive to the frequency of the mode considered;
remarkably,  the entanglement decays down to the same finite value for all
modes. This is in line with the idea that the entanglement that survives the
event horizon is merely due to statistical correlations,  and the only
information that survives when Rob is exactly at the horizon is the fact
that the field is fermionic as suggested in chapter \ref{etanthrough}.

From Figs. \ref{sncca} and \ref{dncca} we can also conclude that all the
relevant entanglement degradation phenomena is produced in the
proximities of the event horizon so that the Rindler approximation that
we are carrying out is valid  (Eq. \ref{conditioaproximatio2}). We can also
see that the degradation is small even in regions in which the
approximation  still holds. Therefore for longer distances from the
horizon the presence of event horizons is not expected to perturb
entangled systems.

In Figs. \ref{snrar} and \ref{dnrar} we can see the behaviour of the
negativity on the $\text{R}\bar{\text{R}}$ bipartition for scalar and
Dirac fields respectively. Here we see that quantum correlations between I and II are created as Rob is standing closer to the event horizon. In
other words,  as Rob is getting closer to the event horizon the partial
system $\text{R}\bar{\text{R}}$ gains quantum correlations. This result
shows that, when Rob is near the horizon, the field states in both sides of
the event horizon are not completely independent. Instead, they get more
and more correlated. However, this $\text{R}\bar{\text{R}}$
entanglement is useless for quantum information tasks because classical
communication between both sides of an event horizon is forbidden. It is
well known for the Rindler case that quantum correlations are created
between Rob and AntiRob when the acceleration increases.
Here we see the direct translation to the Kruskal scenario. The growth of
those correlations encodes information about the dimension of the Fock
space for each field mode as seen in chapter \ref{boundedpop}.

\subsubsection{Mutual information}

\begin{figure}[h]
\begin{center}
\includegraphics[width=.85\textwidth]{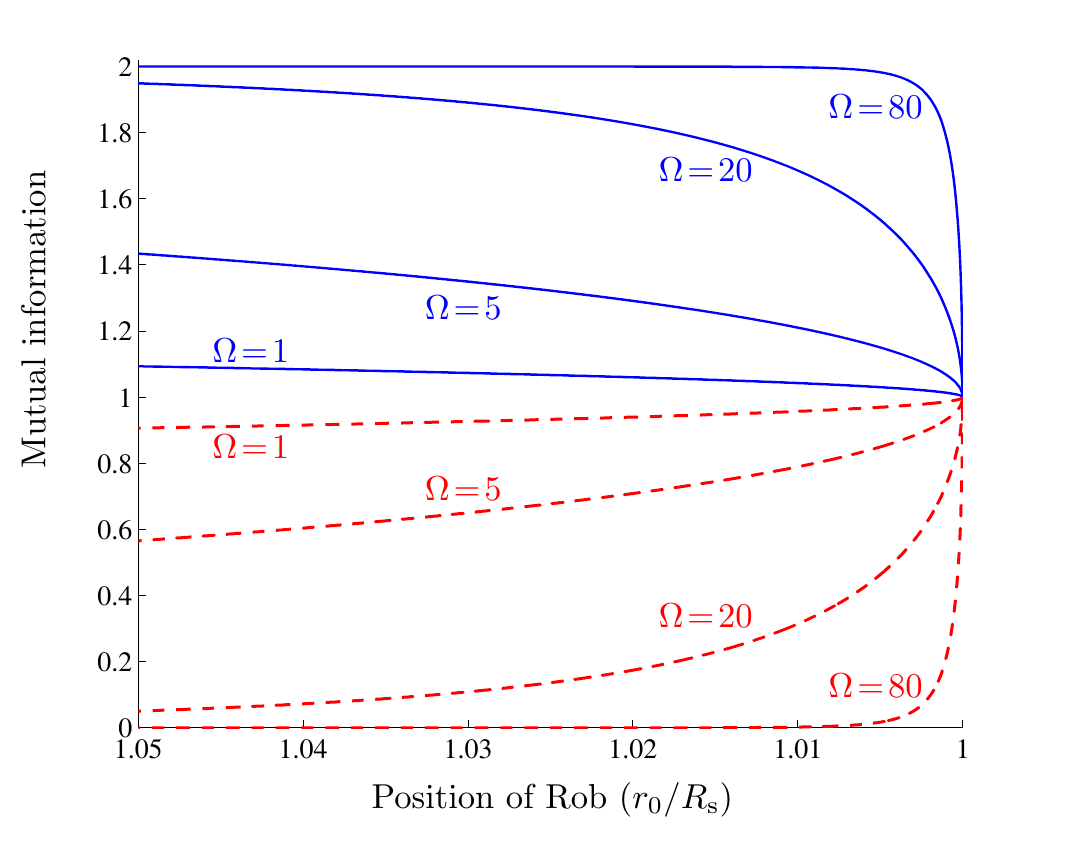}
\caption{Scalar field: Mutual information Alice-Rob (blue solid line) and
Alice-AntiRob (red dashed line).  Mutual information  AR
decreases as Rob is closer to the horizon and mutual information
$\text{A}\bar{\text{R}}$ grows.}
\label{smi}
\end{center}
\end{figure}

\begin{figure}[h]
\begin{center}
\includegraphics[width=.85\textwidth]{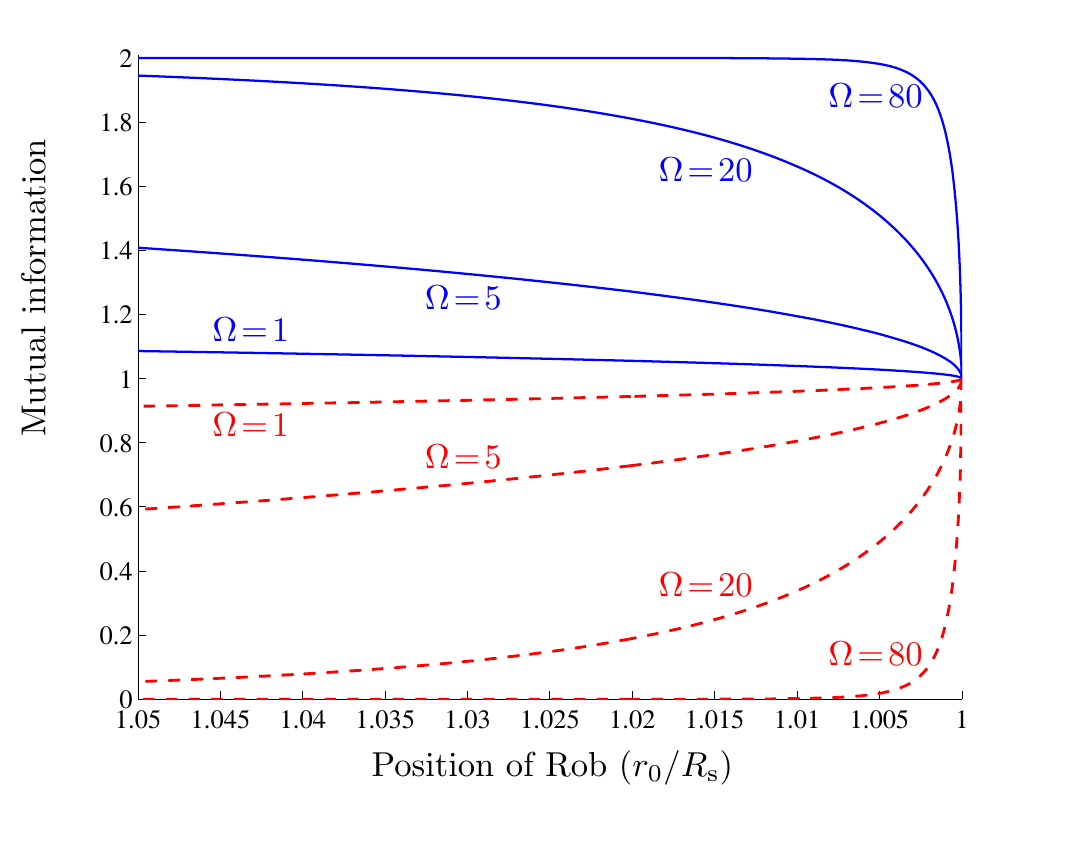}
\caption{Dirac field: Mutual information Alice-Rob (blue solid line) and
Alice-AntiRob (red dashed line).  Mutual information  AR
 decreases as Rob is closer to the horizon and mutual information $\text{A}\bar{\text{R}}$
  grows.}
\label{dmi}
\end{center}
\end{figure}

To compute the
mutual information  for each bipartition we will need the eigenvalues of
the corresponding density matrices. Again the technicalities of this
analysis can be found in previous chapters. The results for the
CCA bipartitions are shown in Figs. \ref{smi} and \ref{dmi}.

We see here that we obtain the black hole version of the mutual
information universal conservation law found in previous chapters for the
Rindler case. Namely, for any distance to the horizon or
black hole mass it is fulfilled that
\begin{equation}\label{conservami}
I_\text{AR}+I_{\text{A}\bar{\text{R}}}=2.
\end{equation}
Although, as we can see by comparing Figs. \ref{smi} and  \ref{dmi}, the
behaviour of the mutual information is very similar for both fermions and
bosons, the origin of this conservation law near the event horizon is
completely different.

For scalar fields this conservation near the horizon responds to a
conservation of  classical correlations only. This can be deduced from Fig.
\ref{sncca} which shows that quantum correlations drop very quickly as
the distance of Rob to the horizon decreases and, consequently, the only
correlations left  must be classical. However, the conservation of
classical correlations in the CCA bipartitions has to do with the
infiniteness of the dimension of the Hilbert space, as it is shown in
chapter \ref{boundedpop}. If the dimension of a bosonic field is limited to a finite value,
classical correlations also drop  as Rob is closer to the horizon (as
quantum correlations do).

On the other hand, a Dirac field has a built-in dimensional limit for the
Hilbert space of each mode imposed by Pauli exclusion principle. Although
previous chapters demonstrated that this limit in the dimension has nothing
to do with the behaviour of quantum correlations, it
does limit the creation of classical correlations. Analogously to what is
discussed in chapter \ref{boundedpop}, the origin for the conservation law
(\ref{conservami}) in the fermionic case is a direct consequence of the
quantum correlations conservation law \eqref{conservan} while for scalar fields it responds to a conservation of
classical correlations.

Mutual information for the $\text{R}\bar{\text{R}}$ bipartition does
not add any new result as it inherits the quantum correlations behaviour
showed in figs. \ref{sncca} and \ref{dncca}.

\subsection{Entanglement degradation dependence on the black hole mass}

In this section we will analyse the entanglement degradation for an
observer with the same characteristics in the presence of  different
black holes. To do so we are going to use the full dimensional quantities
$\omegar$ and $\Delta_0$.

We will consider that Rob's mode frequency is $\omegar=1.5$ Mhz, and
he is standing at a distance $\Delta_0=1$ cm and $\Delta_0=10$ cm
from the event horizon of  black holes with different masses, while he
shares an entangled state
\eqref{Haw1} or \eqref{Hawf} with a free-falling observer Alice.

\begin{figure}[h]
\begin{center}
\includegraphics[width=.85\textwidth]{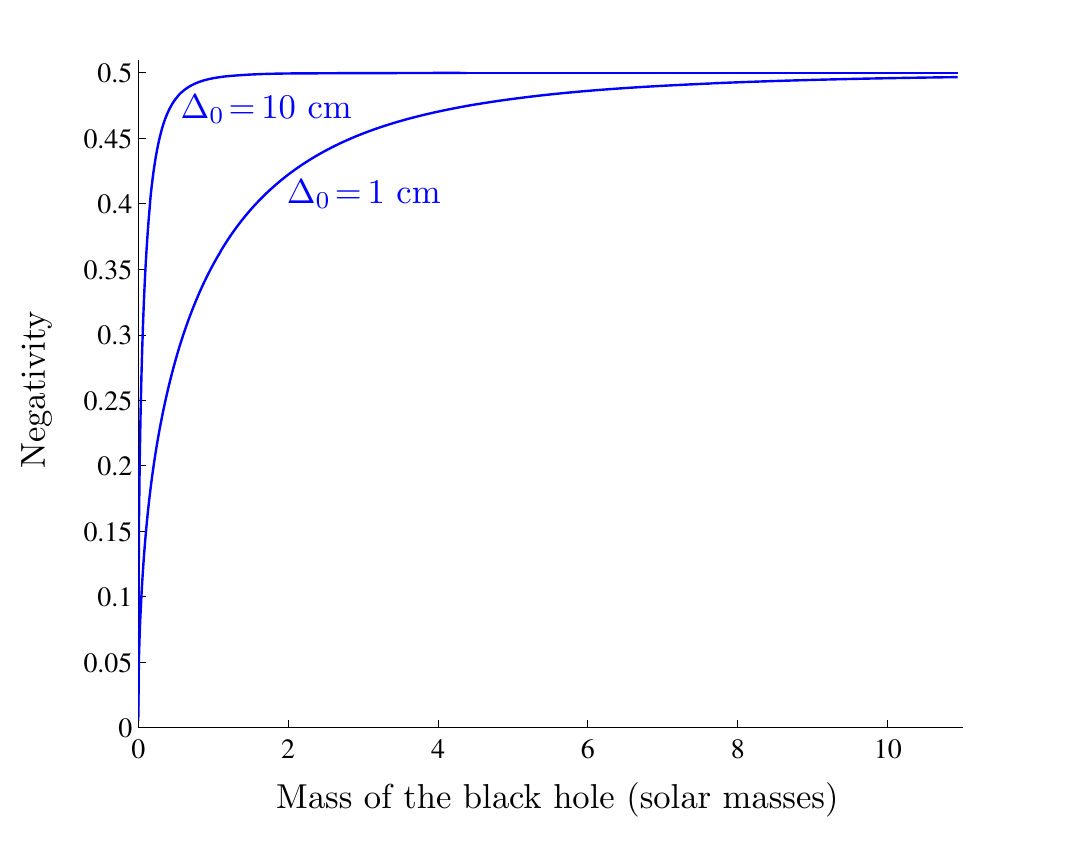}
\caption{Scalar field: Entanglement Alice-Rob when Rob stands at a distance
of 1 cm and 10 cm from the event horizon for a fixed frequency
$\omegar=1.5$ Mhz as a function of the black hole mass. Notice that,
for these values of $\Delta_0$, the approximation holds perfectly for any mass $m>10^{-5}$ solar masses.}
\label{realb}
\end{center}
\end{figure}

\begin{figure}[h]
\begin{center}
\includegraphics[width=.85\textwidth]{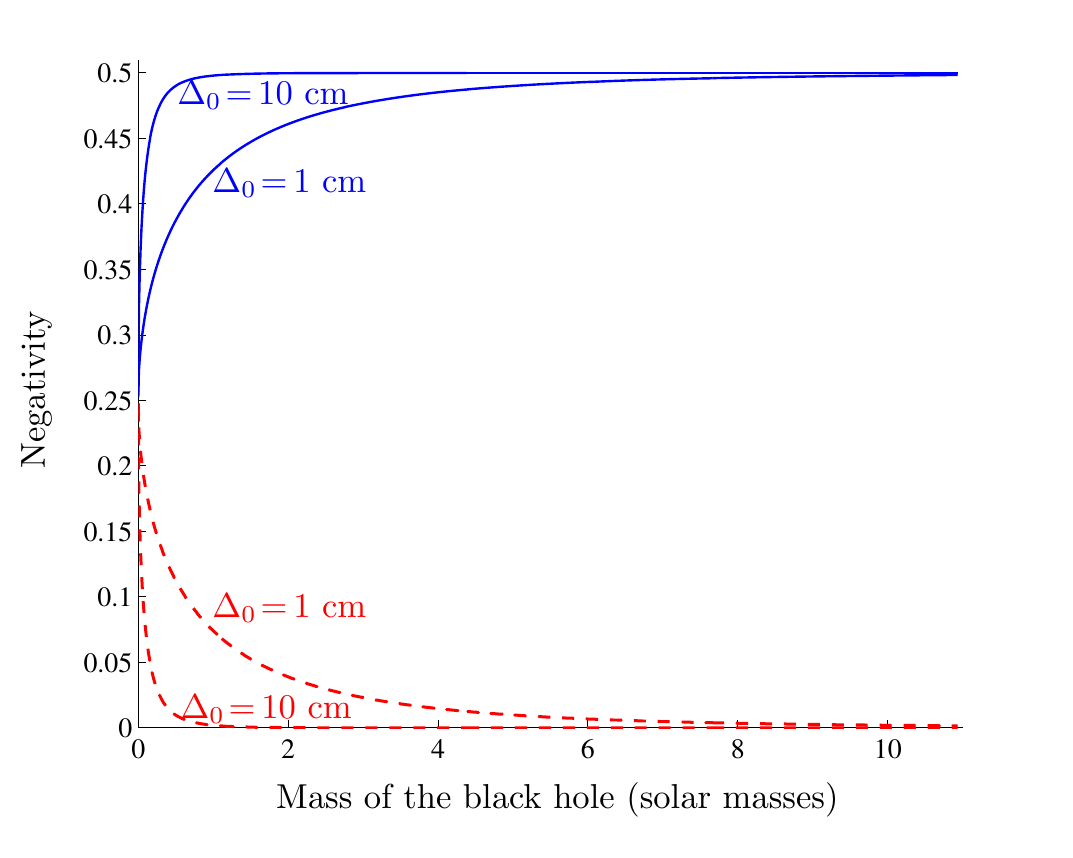}
\caption{Dirac field: Entanglement AR (blue continuous line) and
 $\text{A}\bar{\text{R}}$ (red dashed line) when Rob stands at a
 distance of 1 cm and 10 cm from the event horizon for a fixed frequency
  $\omegar=1.5$ Mhz as a function of the black hole mass.
  Notice that, for these values of $\Delta_0$, the approximation holds perfectly for any mass $m>10^{-5}$ solar masses.}
\label{realf}
\end{center}
\end{figure}

The quantum correlations that Rob and Alice share are shown in Figs.
\ref{realb} and \ref{realf} for scalar and Dirac fields, respectively.
From these figures we see that for a really close distance from the event
horizon, only small black holes would produce significant entanglement
degradation. Actually, the degradation decreases very quickly as the
black hole mass is increased.

Furthermore, we can see that the effects on the entanglement decrease
very quickly as the distance to the event horizon is increased. This shows
that quantum information tasks can be safely performed in universes that
present event horizons since only in the closest vicinity of the less massive
black holes the Hawking effect impedes the application of quantum
information protocols.

\section{Localisation of the states}\label{newsec5}

Along this work we have used a plane-wave-like basis to express the quantum state of the field for the inertial an accelerated observers. These plane wave modes are completely delocalised, and therefore, they are not the most natural election of modes if we want to think of the observers Alice and Rob as spatially localised to some degree. We will present the method to build such states, although this topic will be more deeply treated in chapter \ref{sma}.

A very similar analysis to the one carried out in sections \ref{sec2m6} and \ref{sec2m6} can be performed using a complete set of wave packet modes for both the Minkowski and Rindler solutions of the wave equation. These modes can be spatially localised and provide a clearer physical interpretation for Alice and Rob, which will eventually have to carry out measurements on the field. The way to build these wave packet modes can be found in chapter \ref{sma} and, among many others, in \cite{Birrell,Takagi,NavarroSalas}.

The elements of this basis are defined as a function of the plane wave modes \eqref{modmin} as
\begin{equation}
u^{\text{M}}_{\hat\omega,l}=\frac{1}{\sqrt{\epsilon}}\int_{\hat \omega}^{\hat \omega+\epsilon}d\nu\, e^{-i \nu l}u^{\text{M}}_{\nu},
\end{equation}
where $\hat\omega$ and $l$ label each wave packet.
 
 We can define creation an annihilation operators associated to these wavepackets $a_{\hat \omega,l,\text{M}}$, $a_{\hat\omega,l,\text{M}}^\dagger$  such that $a_{\hat\omega,l,\text{M}}$ annihilates the Minkowski vacuum and $a_{\hat\omega,l,\text{M}}^\dagger \ket{0}_\text{M}=\ket{1_{\hat\omega,l}}_\text{M}$ represents a wavepacket peaked for a frequency $\hat \omega$ and whose spatial localisation can be associated to the maximum of $u^{\text{M}}_{\omega,l}$ as a function of $\hat x$ and $\hat t$.
 
A similar analysis can be done for the Rindler basis
\begin{align}
\nonumber u^{\text{I}}_{\omega,l'}&=\frac{1}{\sqrt{\epsilon}}\int_{\omega}^{ \omega+\epsilon}d\nu\, e^{-i \nu l'}u^{\text{I}}_{\nu},\\
u^{\text{II}}_{\omega,l'}&=\frac{1}{\sqrt{\epsilon}}\int_{\omega}^{ \omega+\epsilon}d\nu\, e^{-i \nu l'}u^{\text{II}}_{\nu},
\end{align}
 $\omega$ and $l'$ label each wave packet.  We can define creation an annihilation operators associated to these wavepackets $a_{\omega,l',R}$, $a_{\omega,l',R}^\dagger$ (where $R=\text{I},\text{II}$) such that $a_{\omega,l',R}$ annihilates the region Rindler region $R$ vacuum and $a_{\omega,l',R}^\dagger \ket{0}_R=\ket{1_{\omega,l'}}_R$ represents a wavepacket peaked for a Rindler frequency $\omega$ and whose spatial localisation can be associated to the maximum of $u^R_{\omega,l'}$ as a function of $x$ and $t$.
 
 We can compute then the Bogoliubov transformation between the Minkowski wavepackets and the Rindler wavepackets \cite{NavarroSalas}
 \begin{equation}
a_{\hat\omega,l,\text{M}}=\alpha^{\text{I}*}_{\omega,l',\hat \omega,l}\,
a^{\phantom{\dagger}}_{\omega,l',\text{I}}-\beta^{\text{II}*}_{\omega,l',\hat \omega,l}\,
a^\dagger_{\omega,l',\text{II}}.
\end{equation}
Where the Bogoliubov coefficients are computed in the same fashion as for the plane wave case
\begin{equation}
\alpha^{\text{I}}_{\omega,l',\hat \omega,l}=\left(u^{\text{I}}_{\omega,l'},u^{\text{M}}_{\hat\omega,l}\right),
\quad\beta^{\text{II}}_{\omega,l',\hat \omega,l}=-\left(u^{\text{II}}_{\omega,l'},u^{\text{M}*}_{\hat\omega,l}\right).
\end{equation}

It is shown in \cite{NavarroSalas} that, apart from an irrelevant phase factor, the Bogoliubov coefficients are related with \eqref{bogo1} as follows
\begin{align}\label{bogonew}
\nonumber\alpha^{\text{I}}_{\omega,l',\hat \omega,l}&=\hat\alpha^{\text{I}}_{ij}\,\mathcal{G}_\alpha(\hat\omega,l,\omega,l'),\\
\beta^{\text{I}}_{\omega,l',\hat \omega,l}&=\hat\beta^{\text{II}}_{ij}\,\mathcal{G}_\beta(\hat\omega,l,\omega,l').
\end{align}
It is shown in \cite{NavarroSalas} that $\mathcal{G}_\alpha(\hat\omega,l,\omega,l')\approx \delta_{\omega\omega_\alpha}\delta_{ll_\alpha}$ and  $\mathcal{G}_\beta(\hat\omega,l,\omega,l')\approx \delta_{\omega\omega_\beta}\delta_{ll_\beta}$, where $l_\alpha=l_\alpha(l')$ and $\omega_\alpha =\omega_\alpha( \omega,l')$.

The key feature of this transformations is that they have again a diagonal form. As it can be read from \eqref{bogonew}, a Minkowski wavepacket $\ket{1_{\omega,l'}}_\text{M}$ is connected with a pair of Rindler wavepackets in regions I and II. Moreover, the functional form of the dependence of this coefficients with the acceleration is effectively the same. This analysis made for the Rindler and Minkowskian modes can be straightforwardly translated to the Boulware and Hartle-Hawking modes.  A completely analogous analysis can be done for the fermionic case.

Consequently all the conclusions extracted in this article for delocalised modes are also valid for the localised modes defined above.

\section{Discussion}\label{conclusions}

We have analysed the entanglement degradation produced in the vicinity
of a Schwarzschild black hole.

With this aim, we have carried out a detailed study of the Schwarzschild
metric in the proximity of the horizon, showing how we can adapt the
tools developed in the study of the entanglement degradation for
uniformly accelerated observers
to the black hole case. In particular, we have shown that, regarding
entanglement degradation effects, the Rindler limit of infinite acceleration reproduces
a black hole scenario in which Rob is arbitrarily close to the event horizon.
More importantly, we have shown the fine structure of this  limit, making
explicit the dependence of the entanglement degradation phenomena on the distance
to the horizon, the mass of the black hole, and the Boulware frequency
$\omegar$ of the entangled mode under consideration, while keeping
control of the approximation to make sure that the toolbox developed for
the Rindler case can be still rigorously used here.

By means of this analysis we have seen that all the interesting
 entanglement degradation phenomena due to the Hawking effect are produced very
close to the event horizon of the Schwarzschild black hole. The
entanglement degradation introduced by the Hawking effect becomes
quickly negligible as Rob is further away from the event horizon. In  other
words, quantum information tasks done far away from event horizons are
not perturbed by the existence of such horizons.

We have also shown that for a fixed Rob's mode frequency and at a fixed
distance from the event horizon the entanglement degradation is greater
for less massive black holes. This is consistent with the fact that the
Hawking temperature is higher for less massive black holes. Furthermore,
  the Hawking entanglement degradation is a universal
phenomenon in the sense that the degradation depends only on Rob's mode 
frequency and his distance to   the horizon in units natural to the black
hole (namely, the surface gravity for frequencies and the Schwarzschild
radius for distances). In these units, there is no extra dependence on the
black hole mass,  as expected.

We have been able to adapt all the conclusions drawn for the Rindler case
to the Schwarzschild scenario. In particular, we have seen that bosonic
and fermionic entanglement behave in a very different way in the
proximity of a black hole. As it was known for the Rindler case
in chapter \ref{etanthrough}, entanglement on the CCA bipartitions is completely lost for
the scalar field while there is a quantum correlation  conservation law for
the Dirac field.

In chapters \ref{onehalf} and \ref{multimode} it was shown that for two different kinds of fermionic
fields (Dirac fields or Grassmann scalars) and also for different
maximally entangled states (occupation number or spin Bell states) the
entanglement in the CCA bipartitions behaves exactly the same way. This
fact was used to argue that  it is statistics and not dimensionality that
determines  the behaviour  of correlations in the CCA bipartitions in the
case of uniformly accelerated observers. This study
proves that this argument is also valid for Schwarzschild black holes, not
only in the limit in which Rob is on the event horizon, but in the whole
region in which the interesting entanglement degradation phenomena are produced.
Therefore, the universal fermionic entanglement behaviour is also manifest
in the presence of a black hole.

For the Schwarzschild case, there also appears the  universal mutual
information conservation law found for both scalar and Dirac fields in
the Rindler case in chapter \ref{etanthrough}. In the fermionic case, it is due to a
conservation of quantum correlations while, for bosons, it only reflects
the conservation of classical correlations  that happens in the case of
infinite dimensional Hilbert spaces for each mode.

Moreover, as Rob is getting closer to the event horizon, quantum
correlations between modes on both sides of bifurcated the event horizon are
created, namely the correlations between field modes in region I and II
of the Kruskal spacetime grow up to a value determined by the dimension
of the Hilbert space of each mode, which is finite for the fermionic case
and infinite for the scalar field.

The problem of the localisation of the Rindler and Minkowski modes has also been analysed, showing that the results obtained here can be extrapolated to the case in which we consider a complete set of localised wave packets as a basis of the Fock space for the inertial and accelerated observers.

\part[Beyond the single mode approximation]{Beyond the single mode approximation}
\label{part2}

\chapter[Relativistic QI beyond the single-mode approximation]{Relativistic quantum information beyond the single-mode approximation\footnote{D. \!
Bruschi,\! J. Louko,\! E. Mart\'in-Mart\'inez,\! A. Dragan,\! I. Fuentes,\! Phys.\! Rev.\! A 82,\! 042332 (2010)}}\label{sma}

In previous chapters we have seen that entanglement between modes of bosonic or fermionic fields can be degraded from the perspective of observers moving  in uniform acceleration. In this chapter, we analyse the validity and correct interpretation of the single-mode approximation, commonly used in previous literature, and show that the approximation is justified only for a special family of states.  

The single-mode approximation, which was introduced in \cite{Alsingtelep,AlsingMcmhMil}, has been extensively used in the literature not only in discussions concerning entanglement but also in other relativistic quantum information scenarios \cite{Alicefalls,AlsingSchul,Bradler,highdim,chapucilla,chapucilla2,Edu2,Shapoor,matsako,Edu3,Ditta,Edu4,Geneferm,Edu5,DiracDiscord}. Although it has been briefly dealt with in section \ref{probexcitations}, we will analyse it here in full detail. 

As in previous chapters, the field, from the inertial perspective, is considered to be in a state where all modes are in the vacuum state except for two of them which are in a two-mode entangled state. For example, the Bell state analysed many times previously
\begin{equation}\label{maxent1}
\ket{\Psi}_\text{M}=\frac{1}{\sqrt2}\left(\ket{0_{\omega}}_\text{M}\ket{0_{\omega^{\prime}}}_\text{M}+\ket{1_{\omega}}_\text{M}\ket{1_{\omega^{\prime}}}_\text{M}\right),
\end{equation}
where M labels Minkowski states and $\omega$, $\omega^{\prime}$ are two Minkowski frequencies.  Two inertial observers, Alice and Bob, each carrying a monocromatic detector sensitive to frequencies $\omega$ and $\omega^{\prime}$ respectively,  would find maximal correlations in their measurements since the Bell state is maximally entangled. It is then interesting to investigate to what degree the state is entangled when described by observers in uniform acceleration. In the simplest scenario,  Alice is again considered to be inertial and an uniformly accelerated observer Rob is introduced, who carries a monocromatic detector sensitive to mode $\omega^{\prime}$.  To study this situation, the states corresponding to Rob must be transformed into the appropriate basis, in this case, the Rindler basis.  It is then when the  single-mode approximation is invoked to relate Minkowski single particle states $\ket{1_{\omega^{\prime}}}_\text{M}$  to states in Rindler space. 

We argue that the single-mode approximation is not valid for general states. However, the approximation holds for a family of peaked Minkowski wave packets provided constraints imposed by an appropriate Fourier transform are satisfied.  We show that the state analysed canonically in the literature corresponds to an entangled state between a Minkowski and a special type of Unruh mode. We therefore revise previous results for both bosonic and fermionic field entanglement.  The results are qualitatively similar to those obtained under the single-mode approximation.  We confirm that entanglement  is degraded with acceleration, vanishing in the infinite acceleration limit in the bosonic case and reaching a non-vanishing minimum for fermionic fields.  However, we find that in the fermionic case, the degree to which entanglement is degraded depends on the election of Unruh modes.

\section{Minkowski, Unruh and Rindler modes}\label{sec3m7}

For the sake of clarity and to introduce the new notation that we will need for this chapter, let us start with a brief review of section \ref{Rindbogo}, revisiting the transformation between the Fock bases natural to an inertial and an accelerated observer.

We consider a 
real massless scalar field $\phi$ in a 
two-dimensional Minkowski spacetime. 
The field equation is the massless 
Klein-Gordon equation, $\Box \phi=0$. 
The (indefinite) 
Klein-Gordon inner product reads 
\begin{align}
(\phi_1,\phi_2) = i \int_{\Sigma} \phi_1^* \overleftrightarrow{\partial_a}\phi_2 \, n^a  \, \text{d}\Sigma,
\end{align}
where $n^a$ is a future-pointing normal vector to the spacelike hypersurface $\Sigma$ and $\text{d}\Sigma$ 
is the volume element on~$\Sigma$. 

The Klein-Gordon equation can be solved in Minkowski coordinates
$(t,x)$ which are an appropriate choice for inertial observers. The
positive energy mode solutions with respect to the timelike Killing
vector field $\partial_t$ are given by
\begin{align}
\label{eq:masslessMmodes}
u_{\omega,\text{M}} (t,x) = \frac{1}{\sqrt{4\pi \omega}}
\exp[-i\omega(t-\epsilon x)],
\end{align}
where $\omega>0$ is the Minkowski frequency and the discrete index $\epsilon$ 
takes the value $1$ for modes with positive momentum (the right-movers) 
and the value $-1$ for modes with negative momentum (the left-movers). 
As the right-movers and the left-movers decouple, we have suppressed the index $\epsilon$ on the left-hand side of \eqref{eq:masslessMmodes} and we continue to do so in all the formulas. The mode solutions and their complex conjugates are normalised in the usual sense of Dirac delta-functions in $\omega$ as 
\begin{eqnarray}
\left(u_{\omega,\text{M}},  u_{\omega',\text{M}}\right)& =& 
\delta_{\omega\omega'},\nonumber\\
\left(u^{\ast}_{\omega,\text{M}},  u^{\ast}_{\omega',\text{M}}\right)& =& 
- \delta_{\omega\omega'} ,\nonumber\\ 
\left(u^{\ast}_{\omega,\text{M}},  u_{\omega',\text{M}}\right)& =& 0.
\end{eqnarray}

The Klein-Gordon equation can also be separated in coordinates
that are adapted to the Rindler family of 
uniformly accelerated observers. 
Let region I (respectively region II) denote the wedge $|t|<x$ 
($x<-|t|$). In each of the wedges, we introduce the Rindler
coordinates $(\eta,\chi)$ by \cite{Takagi} 
\begin{equation}
\label{Rindlertransformation}
\eta = \text{atanh} \! \left(\frac{t}{x}\right), 
~~~\chi = \sqrt{x^2-t^2},
\end{equation}
where $0<\chi<\infty$ and $-\infty < \eta < \infty$ individually in each
wedge. The curve $\chi = 1/a$, where $a$ is a positive constant
of dimension inverse length, is then the world line of a
uniformly-accelerated observer whose proper acceleration equals~$a$,
and the proper time of this observer is given by $\eta/a$ in I and by
$-\eta/a$ in~II\null.  Note that $\partial_\eta$ is a timelike Killing
vector in both I and~II, and it is future-pointing in I but
past-pointing in~II\null.

Separating the Klein-Gordon equation in regions I and II in Rindler coordinates yields the solutions
\begin{eqnarray}
u_{\Omega,\text{I}} (t,x) &=& \frac{1}{\sqrt{4\pi \Omega}}
{\left(\frac{x-\epsilon t}{l_{\Omega}}\right)}^{i\epsilon\Omega},\nonumber\\
u_{\Omega,\text{II}} (t,x) &=& \frac{1}{\sqrt{4\pi \Omega}}
{\left(\frac{\epsilon t - x}{l_{\Omega}}\right)}^{-i\epsilon\Omega}, 
\end{eqnarray}
where $\epsilon = 1$ again corresponds to right-movers and $\epsilon =
-1$ to left-movers, 
$\Omega$ is a positive dimensionless constant 
and $l_{\Omega}$ is a positive constant of dimension length. 
Since $\partial_\eta
u_{\Omega,\text{I}} = -i\Omega u_{\Omega,\text{I}}$ 
and 
$\partial_\eta 
u_{\Omega,\text{II}} = i\Omega u_{\Omega,\text{II}}$, 
$u_{\Omega,\text{I}}$ and $u_{\Omega,\text{II}}$ are the
positive frequency mode functions 
with respect to the future-pointing Rindler Killing
vectors $\pm \partial_\eta$ in their respective wedges. Notice that, for convenience, we have introduced here the dimensionless Rindler frequency $\Omega$. 
The dimensional frequency with respect to the
proper time of a Rindler observer located at $\chi = 1/a$ is given
in terms of the dimensionless $\Omega$ by $\Omega_a = a\Omega$. The modes are 
delta-normalised in $\Omega$ in their respective
wedges as usual.

Note that the choice of the constant $l_{\Omega}$ is equivalent to
specifying the phase of the Rindler modes. This choice is hence purely
a matter of convention, and it can be made independently for each
$\Omega$ and~$\epsilon$. We shall shortly specify the choice so
that the transformation between the Minkowski and Rindler modes
becomes simple.


A third basis of interesting solutions to the field equation 
is provided by the Unruh modes, 
defined by 
\begin{eqnarray}
\label{eq:unruhmodes}
u_{\Omega,\text{\text{R}}} &=& 
\cosh(r_\Omega)  u_{\Omega,\text{I}} + \sinh(r_\Omega)  u^{\ast}_{\Omega,\text{II}}, \nonumber\\
u_{\Omega,\text{\text{L}}} &=& 
\cosh(r_\Omega)  u_{\Omega,\text{II}} + \sinh(r_\Omega) u^{\ast}_{\Omega,\text{I}},
\end{eqnarray} 
where $\tanh r_\Omega= e^{-\pi\Omega}$. 
While the Unruh modes have a sharp Rindler frequency, an analytic continuation argument shows that they are
purely positive frequency linear combinations 
of the Minkowski modes~\cite{Unruh,Birrell}. 
It is hence convenient to examine the transformation 
between the Minkowski and Rindler modes in two stages:  
\begin{itemize}
\item The well-known transformation 
\eqref{eq:unruhmodes} between the Unruh and Rindler modes isolates the consequences of the differing Minkowski and Rindler definitions of positive frequency.
\item The less well-known transformation between the Minkowski and Unruh modes \cite{Takagi} shows that a monochromatic wave in the Rindler basis  corresponds to a non-monochromatic superposition in the Minkowski basis.
\end{itemize}
 It is these latter effects from which the new observations in this paper will stem. 


To find the Bogoliubov transformations that relate the bases, 
we expand the field in each of the bases as 
\begin{eqnarray}
\!\!\phi &=&\!\! \int_0^\infty\left( a_{\omega,\text{M}} u_{\omega,\text{M}} + a_{\omega,\text{M}}^\dag u^{\ast}_{\omega,\text{M}} 
\right) \text{d}\omega\nonumber\\*
=& &\!\!\!\!\!\! \!\!\!\!\!\! \int_0^\infty \!\!\!\left( A_{\Omega,\text{\text{R}}} u_{\Omega,\text{\text{R}}}\!+\!A^{\dagger}_{\Omega,\text{\text{R}}} u^{\ast}_{\Omega,\text{\text{R}}}\!+\! A_{\Omega,\text{\text{L}}} u_{\Omega,\text{\text{L}}} \!+\! A^{\dagger}_{\Omega,\text{\text{L}}} u^{\ast}_{\Omega,\text{\text{L}}} \right)\! \text{d}\Omega\nonumber\\ \nonumber \\
=& &\!\!\!\!\!\! \!\!\!\!\!\!\int_0^\infty\!\!\!\left( a_{\Omega,\text{I}} u_{\Omega,\text{I}}\!+\!a^{\dagger}_{\Omega,\text{I}} u^{\ast}_{\Omega,\text{I}}\! +\! a_{\Omega,\text{II}} u_{\Omega,\text{II}}\!+\!a^{\dagger}_{\Omega,\text{II}} u^{\ast}_{\Omega,\text{II}}\right)\! \text{d}\Omega,
\label{eq:fieldexpansions}
\end{eqnarray}
where $a_{\omega,\text{M}}$, $A_{\Omega,\text{\text{R}}}, A_{\Omega,\text{\text{L}}}$, and $a_{\Omega,\text{I}}, a_{\Omega,\text{II}}$ are the Minkowski, Unruh and Rindler annihilation operators, respectively. The usual bosonic commutation relations $[a_{\omega,\text{M}},a^{\dagger}_{\omega^{\prime},\text{M}}] =\delta_{\omega\omega^{\prime}}$,  $[A_{\Omega,\text{\text{R}}},A^{\dagger}_{\Omega^{\prime},\text{\text{R}}}]=[A_{\Omega,\text{\text{L}}},A^{\dagger}_{\Omega^{\prime},\text{\text{L}}}]=\delta_{\Omega\Omega^{\prime}}$ and $[a_{\Omega,\text{I}},a^{\dagger}_{\Omega^{\prime},\text{I}}] = [a_{\Omega,\text{II}},a^{\dagger}_{\Omega^{\prime},\text{II}}]=\delta_{\Omega\Omega^{\prime}}$ hold, and commutators for mixed $\text{\text{R}}$, $\text{\text{L}}$ and mixed $\text{I}$, $\text{II}$ vanish. The transformation between the 
Unruh and Rindler bases is given by~\eqref{eq:unruhmodes}. The transformation between the Minkowski and Unruh bases can be evaluated by taking appropriate inner products of formula \eqref{eq:fieldexpansions} with the mode functions~\cite{Takagi}, 
with the result 
\begin{eqnarray}
u_{\omega,\text{M}} &=& \int_{0}^{\infty} \left(\alpha^\text{\text{R}}_{\omega\Omega} u_{\Omega,\text{\text{R}}}+  \alpha^\text{\text{L}}_{\omega\Omega} u_{\Omega,\text{\text{L}}}\right) \text{d}\Omega, \nonumber \\
u_{\Omega,\text{\text{R}}} &=& \int_{0}^{\infty} (\alpha^\text{\text{R}}_{\omega\Omega})^{\ast} u_{\omega,\text{M}} \, \text{d}\omega, \nonumber \\
u_{\Omega,\text{\text{L}}} &=& \int_{0}^{\infty}(\alpha^\text{\text{L}}_{\omega\Omega})^{\ast} u_{\omega,\text{M}} \, \text{d}\omega,
\label{eq:Mink-v-Unruh}
\end{eqnarray}
where 
\begin{eqnarray}
\label{eq:alphas-raw}
\alpha^\text{\text{R}}_{\omega\Omega} &=&\frac{1}{\sqrt{2\pi\omega}}\sqrt{\frac{\Omega\sinh\pi\Omega}{\pi}}\Gamma(-i\epsilon\Omega){(\omega l_{\Omega})}^{i\epsilon\Omega} , \nonumber \\
\alpha^\text{\text{L}}_{\omega\Omega} &=& \frac{1}{\sqrt{2\pi\omega}}\sqrt{\frac{\Omega\sinh\pi\Omega}{\pi}}\Gamma(i\epsilon\Omega){(\omega l_{\Omega})}^{-i\epsilon\Omega}.
\end{eqnarray}
By the properties of the Gamma-function 
(\cite{NISTlibrary}, formula 5.4.3), we can take 
advantage of the arbitrariness of the constants 
$l_{\Omega}$ and choose them so that 
\eqref{eq:alphas-raw} simplifies to 
\begin{eqnarray}
\alpha^\text{\text{R}}_{\omega\Omega} &=&\frac{1}{\sqrt{2\pi\omega}}{(\omega l)}^{i\epsilon\Omega} , \nonumber \\
\alpha^\text{\text{L}}_{\omega\Omega} &=& \frac{1}{\sqrt{2\pi\omega}}{(\omega l)}^{-i\epsilon\Omega} , 
\label{eq:alphas}
\end{eqnarray}
where $l$ is an overall constant of dimension length, 
independent of $\epsilon$ and~$\Omega$. 

The transformations between the modes give rise to transformations between the corresponding field operators. From~\eqref{eq:Mink-v-Unruh}, 
the Minkowski and Unruh operators are related by 
\begin{eqnarray} 
\label{eq:Aaboth-transform}
a_{\omega,\text{M}} &=& \int_0^{\infty}\left[(\alpha^\text{\text{R}}_{\omega\Omega})^{\ast} A_{\Omega,\text{\text{R}}}+(\alpha^\text{\text{L}}_{\omega\Omega})^{\ast} A_{\Omega,\text{\text{L}}}\right] \text{d}\Omega , \nonumber \\
A_{\Omega,\text{\text{R}}} &=& \int_0^{\infty} \alpha^\text{\text{R}}_{\omega\Omega} a_{\omega,\text{M}} \, \text{d}\omega , \nonumber \\
A_{\Omega,\text{\text{L}}} &=& \int_0^{\infty} \alpha^\text{\text{L}}_{\omega\Omega} \, a_{\omega,\text{M}} \, \text{d}\omega,
\end{eqnarray}
and from~\eqref{eq:unruhmodes}, the Unruh and Rindler operators 
are related by 
\begin{eqnarray}
a_{\Omega,\text{I}} &=&  \cosh(r_\Omega)\,A_{\Omega, \text{R}} + \sinh(r_\Omega)\, A^\dag_{\Omega,\text{\text{L}}} , \notag \\
a_{\Omega,\text{II}} &=& \cosh(r_\Omega)\, A_{\Omega,\text{\text{L}}} + \sinh(r_\Omega)\, A^\dag_{\Omega,\text{\text{R}}}.
\label{eq:Mink-v-Unruh-operators}
\end{eqnarray}

We can now investigate how the vacua and excited states 
defined with respect to the different bases are related. Since the transformation between the Minkowski and Unruh bases does not mix the creation and annihilation operators, these two bases share the common Minkowski vacuum state $|0\rangle_{\text{M}}=|0\rangle_\text{U}=\prod_{\Omega}|0_{\Omega}\rangle_\text{U}$, where $A_{\Omega,\text{\text{R}}}|0_{\Omega}\rangle_\text{U}=A_{\Omega,\text{\text{L}}}|0_{\Omega}\rangle_\text{U}=0$. However, $|0\rangle_\text{U}$ does not 
coincide with the Rindler vacuum: 
if one makes the ansatz 
\begin{align}
\label{eq:UOmegavac-expansion}
|0_{\Omega}\rangle_\text{U}= \sum_n f_{\Omega}(n) \, |n_{\Omega}\rangle_\text{I} |n_{\Omega}\rangle_\text{II}, 
\end{align}
where $|n_{\Omega}\rangle_\text{I}$
is the state with $n$ Rindler $\text{I}$-excitations 
over the Rindler $\text{I}$-vacuum $|0_{\Omega}\rangle_\text{I}$, and 
similarly 
$|n_{\Omega}\rangle_\text{II}$
is the state with $n$ Rindler $\text{II}$-excitations 
over the Rindler $\text{II}$-vacuum $|0_{\Omega}\rangle_\text{II}$,
use of \eqref{eq:Mink-v-Unruh-operators} shows that the 
coefficient functions are given by 
$f_{\Omega}(n) =\tanh^n \! r_{\Omega}/\cosh r_\Omega$. 
$|0\rangle_\text{U}$ is thus a two-mode squeezed state of Rindler excitations over the Rindler vacuum for each~$\Omega$.

Although states with a completely sharp value of 
$\Omega$ are not normalisable, 
we may approximate 
normalisable wave packets that are sufficiently narrowly peaked in 
$\Omega$ by taking a fixed $\Omega$ 
and renormalising the Unruh and 
Rindler commutators to read 
$[A_{\Omega,\text{\text{R}}},A^{\dagger}_{\Omega,\text{\text{R}}}]=[A_{\Omega,\text{\text{L}}},A^{\dagger}_{\Omega,\text{\text{L}}}]=1$ and 
$[a_{\Omega,\text{I}},a^{\dagger}_{\Omega,\text{I}}] = [a_{\Omega,\text{II}},a^{\dagger}_{\Omega,\text{II}}]=1$, with the commutators for mixed $\text{\text{R}}$, $\text{\text{L}}$ 
and mixed $\text{I}$, $\text{II}$ vanishing. 
In this idealisation of sharp peaking in~$\Omega$, 
the most general creation operator
that is of purely positive Minkowski frequency and sharply peaked in~$\Omega$ can be written as 
a linear combination of the two Unruh creation operators, in the form 
\begin{align}
\label{eq:q-defs}
a_{\Omega,\text{U}}^\dag = q_{\text{L}} A^\dag_{\Omega,\text{\text{L}}} + q_{\text{R}}  A^\dag_{\Omega,\text{\text{R}}},
\end{align}
where $q_{\text{R}}$ and $q_{\text{L}}$ are complex numbers with 
${|q_{\text{R}}|}^2 + {|q_{\text{L}}|}^2 =1$. Note that 
$[a_{\Omega,\text{U}}, a_{\Omega,\text{U}}^\dag]=1$. 
From 
\eqref{eq:UOmegavac-expansion} and \eqref{eq:q-defs}
we then see that adding into Minkowski vacuum one 
idealised particle
of this kind, of purely positive Minkowski frequency, yields the state 
\begin{align} 
& a^\dag_{\Omega,\text{U}} |0_\Omega\rangle_\text{U} =  
\sum_{n=0}^\infty 
f_{\Omega}(n)\frac{\sqrt{n+1}}{\cosh r_{\Omega}}|\Phi^{n}_{\Omega}\rangle
,  \notag 
\\*
& |\Phi^n _{\Omega}\rangle = 
q_{\text{L}} \, |n_{\Omega}\rangle_\text{I}|(n+1)_{\Omega}\rangle_\text{II} +
q_{\text{R}} |(n+1)_{\Omega}\rangle_\text{I} |n_{\Omega}\rangle_\text{II} . 
\label{eq:singleOmegaex}
\end{align}

In previous studies on relativistic quantum information, it has been always considered a state of the form \eqref{eq:singleOmegaex} with $q_{\text{R}}=1$ and $q_{\text{L}}=0$ invoking the so called `single mode approximation'\footnote{In previous chapters we worked implicitly with this choice of Unruh modes without invoking any misleading approximation. Indeed, the unprimed basis defined in \eqref{modopsi} can be obtained by acting with \eqref{eq:q-defs} and $q_R=1$ on the Minkowski vacuum. The primed basis corresponds to $q_\text{L}=1$} whose problems were discussed in section \ref{probexcitations}. The above discussion shows that this choice for $q_{\text{R}}$ and $q_{\text{L}}$ is rather special; in particular, it breaks the symmetry 
between the right and left Rindler wedges. We shall address next how entanglement is modified for these sharp $\Omega$ states when both $q_{\text{R}}$ and $q_{\text{L}}$ are present, 
and we then turn to examine the assumption of sharp~$\Omega$.

\section{Entanglement revised beyond the single mode approximation\label{sec:entganglementrev}}

In the relativistic quantum information literature, the single mode approximation $a_{\omega,\text{M}}\approx a_{\omega,\text{U}}$ is considered to relate Minkowski and Unruh modes. The main argument for taking this approximation is that the distribution
\begin{equation}
\label{eq:Aa-transform}
a_{\omega,\text{M}} = \int_0^{\infty}\left[
(\alpha^\text{\text{R}}_{\omega\Omega})^{\ast} A_{\Omega,\text{\text{R}}}+(\alpha^\text{\text{L}}_{\omega\Omega})^{\ast} A_{\Omega,\text{\text{L}}}\right]\text{d}\Omega
\end{equation}
is peaked. However,  we can see from equations \eqref{eq:alphas}  that this distribution in fact oscillates and it is not peaked at all.  Entanglement in non-inertial frames can be studied provided we consider the state
\begin{equation}\label{maxent}
\ket{\Psi}=\frac{1}{\sqrt2}\left(\ket{0_\omega}_{\text M}\ket{0_\Omega}_\text{U}+\ket{1_\omega}_{\text{M}}\ket{1_\Omega}_\text{U}\right),
\end{equation}
where the states corresponding to $\Omega$ are Unruh states. In this case, a single Unruh frequency $\Omega$ corresponds to the same Rindler frequency. In the special case  $q_\text{\text{R}}=1$ and  $q_{\text{L}}=0$ we recover the results canonically presented in the literature \cite{Alicefalls,AlsingSchul,Edu4}. In this section, we will revise the analysis of entanglement in non-inertial frames for the general Unruh modes. However, since a Minkowski monochromatic basis seems to be a natural choice for inertial observers, we will show in the section \ref{sec:peaking} that the standard results also hold for Minkowski states as long as special Minkowski wavepackets are considered. 

Having the expressions for the vacuum and single particle states in the Minkowski, Unruh and Rindler basis enables us to return to the standard scenario for analysing the degradation of entanglement from the perspective of observers in uniform acceleration. Let us consider the maximally entangled state Eq.~\eqref{maxent} from the perspective of inertial observers. By choosing different  $q_\text{\text{R}}$ we can vary the states under consideration. An arbitrary Unruh single particle state has different right and left components where $q_\text{\text{R}}, q_\text{\text{L}}$ represent the respective weighs. When working with Unruh modes, there is no particular reason why to choose a specific $q_\text{\text{R}}$. In fact, and as as we will see later, feasible elections of Minkowski states are in general, linear superpositions of different Unruh modes  with different values of $q_\text{\text{R}}$.

The Minkowski-Unruh state under consideration can be viewed as an entangled state of a  tri-partite system.  The partitions correspond to the three sets of modes:  Minkowski modes with  frequency $\omega$ and  two sets of Unruh modes (left and right) with frequency $\Omega$.  Therefore, it is convenient to define the following bi-partitions: the Alice-Bob bi-partition corresponds to Minkowski and right Unruh modes while the Alice-AntiBob bipartition refers to Minkowski and left Unruh modes. We will see that the distribution of entanglement in these bi-partitions becomes relevant when analysing the entanglement content in the state from the non-inertial perspective. 

 We now want to study the entanglement in the state considering that the $\Omega$ modes are  described by observers in uniform acceleration.  Therefore, Unruh states must be transformed into the Rindler basis. The state in the Minkowski-Rinder basis is also a state of a tri-partite system.  Therefore, we define the Alice-Rob bi-partition as the Minkowski and region I Rindler modes while the Alice-AntiRob bi-partitions corresponds to Minkowski and region II Rindler modes.  In the limit of very small accelerations Alice-Rob and Alice-AntiRob approximate to Alice-Bob and Alice-AntiBob bi-partitions respectively. This is because, as shown in \eqref{eq:unruhmodes} and \eqref{eq:Mink-v-Unruh-operators},  region I (II) Rindler modes tend to R (L) Unruh modes in such limit. Likewise,  region II Rindler modes tend to L Unruh modes in the small acceleration limit.

The entanglement can be quantified using the negativity \eqref{negativitydef} as in previous chapters. In what follows we will study the entanglement between the Alice-Rob and Alice-AntiRob modes. For this we will go through the standard procedure learnt in previous chapters: after expressing Rob's modes in the Rindler basis, the Alice-Rob density matrix is obtained by tracing over the region II, with the result
\begin{equation}
\rho_{AR} = \frac{1}{2}\sum_{n=0}^\infty {\left[\frac{\tanh^n r_\Omega}{\cosh r_\Omega}\right]}^2 \rho_{AR}^n,
\end{equation}
where 
\begin{align}
\nonumber\rho_{AR}^n&=\proj{0n}{0n}+\frac{n+1}{\cosh^2r_{\Omega}}\Big(|q_\text{\text{R}}|^2\proj{1\,n+1}{1\,n+1}+|q_\text{\text{L}}|^2\proj{1n}{1n}\Big)\\*\nonumber&+\frac{\sqrt{n+1}}{\cosh r_{\Omega}}\Big(q_\text{\text{R}}\proj{1\,n+1}{0n}+ q_\text{\text{L}}\tanh r_{\Omega}\proj{1\,n}{0\, n+1}\Big)\\*&+\frac{\sqrt{(n+1)(n+2)}}{\cosh^2r_{\Omega}} q_\text{\text{R}}\,q_\text{\text{L}}^*\tanh r_{\Omega}\proj{1\, n+2}{1n}+(\text{H.c.})_{_{\substack{\text{non-}\\\text{diag.}}}}.
\end{align}
Here $(\text{H.c.})_{\text{non-diag.}}$ means Hermitian conjugate of only the non-diagonal terms. The Alice-AntiRob density matrix is obtained by tracing over region I. However, due to the symmetry in the Unruh modes between region I and II, we can obtain the Alice-AntiRob  matrix by interchanging $q_\text{\text{R}}$ and $q_\text{\text{L}}$.  The partial transpose $\sigma_\text{\text{R}}$ of $\rho_\text{\text{R}}$ with respect 
to Alice is given by
\begin{equation}
\sigma_{AR} = 
\frac{1}{2}
\sum_{n=0}^\infty {\big[{\left[\frac{\tanh^n r_\Omega}{\cosh r_\Omega}\right]}^2\big]}^2 \sigma^n_{AR},
\label{eq:sigmaR}
\end{equation}
where 
\begin{align}
\nonumber\sigma_{AR}^n&=\proj{0n}{0n}+\frac{n+1}{\cosh^2r_{\Omega}}\Big(|q_\text{\text{R}}|^2\proj{1\,n+1}{1\,n+1}+|q_\text{\text{L}}|^2\proj{1n}{1n}\Big)\\*
\nonumber&+\frac{\sqrt{n+1}}{\cosh r_{\Omega}}\Big(q_\text{\text{R}}\proj{0\,n+1}{1n}+ q_\text{\text{L}}\tanh r_{\Omega}\proj{0\,n}{1\, n+1}\Big)\\*
&+\frac{\sqrt{(n+1)(n+2)}}{\cosh^2r_{\Omega}} q_\text{\text{R}}\,q_\text{\text{L}}^*\tanh r_{\Omega}\proj{1\, n+2}{1n}+(\text{H.c.})_{_{\substack{\text{non-}\\\text{diag.}}}}.
\end{align}
 The eigenvalues of $\sigma_{AR}$ only depend on $|q_\text{R}|$ and $|q_\text{L}|$ and not on the relative phase between them. This means that the entanglement is insensitive to the election of this phase.

\begin{figure}[h]
\begin{center}
\includegraphics[width=.80\textwidth]{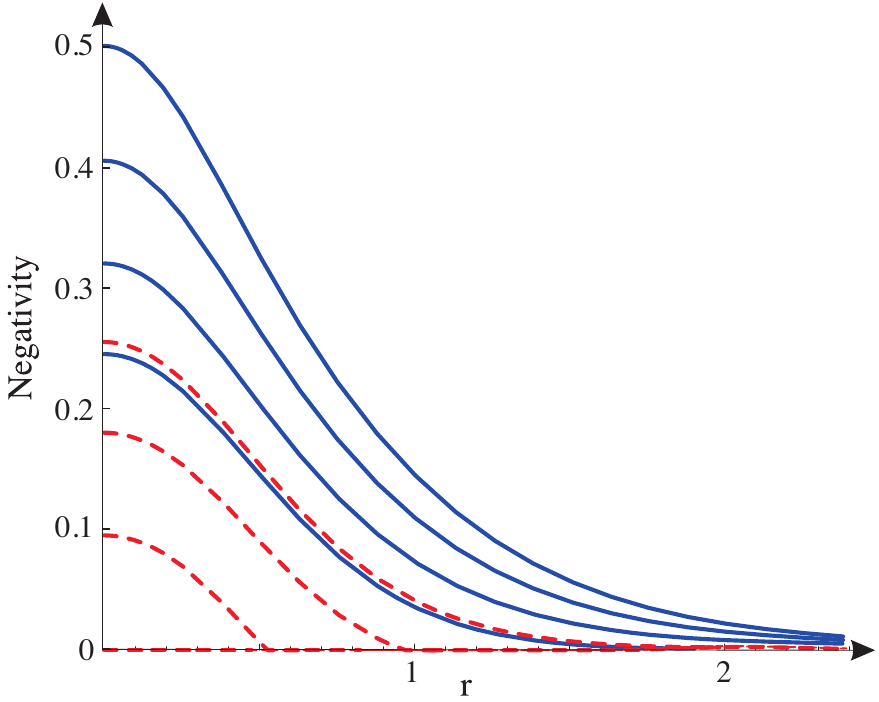}
\end{center}
\caption{Scalar field: negativity for the bipartition Alice-Rob (Blue continuous) and Alice-AntiRob (red dashed) as a function of $r_{\Omega}=\operatorname{artanh}{e^{-\pi\Omega_a/a}}$ for various choices of $|q_\text{\text{R}}|$. The blue continuous (red dashed) curves from top to bottom (from bottom to top) correspond, to $|q_\text{\text{R}}|=1,0.9,0.8,0.7$  respectively. All entanglement in both bipartitions vanishes in the infinite acceleration limit.}
\label{figbosons}
\end{figure}

The two extreme cases when $q_R=1$ and $q_L=1$ are analytically solvable since the partial transpose density matrix has a block diagonal structure as it was shown in previous works \cite{Alicefalls}. However, for all other cases,  the matrix is no longer block diagonal and the eigenvalues of the partial transpose density matrix are computed numerically. The resulting negativity between Alice-Rob and Alice-AntiRob modes is plotted in Fig. \ref{figbosons} for different values of $|q_\text{\text{R}}|=1,0.9,0.8,0.7$.  $|q_\text{\text{R}}|=1$ corresponds to the canonical case studied in the literature \cite{Alicefalls}. 

In the bosonic case, the entanglement between the Alice-Rob and Alice-AntiRob modes always vanishes in the infinite acceleration limit. Interestingly,  there is no fundamental difference in the degradation of entanglement  for different choices of $|q_\text{\text{R}}|$. The entanglement always degrades with acceleration at the same rate. There is no special Unruh state which degrades less with acceleration.

\section{Wave packets: recovering the single-mode approximation\label{sec:peaking}}

The entanglement analysis of section \ref{sec:entganglementrev}
assumes Alice's state
to be a Minkowski mode with a sharp Minkowski momentum and Rob's
state to be an Unruh mode with sharp Unruh frequency, 
such that Rob's linear combination of the two Unruh modes 
is specified by the two
complex-valued parameters $q_\text{\text{R}}$ and $q_\text{\text{L}}$ satisfying 
${|q_\text{\text{R}}|}^2 + {|q_\text{\text{L}}|}^2=1$. 
The Alice and Rob states are further
assumed to be orthogonal to each other, 
so that the system can be treated as
bipartite. We now discuss the sense in which these assumptions are a
good approximation to Alice and Rob states that can be built as
Minkowski wave packets.

Recall that a state with a sharp frequency, be it Minkowski or Unruh,
is not normalisable and should be understood as the idealisation of a
wave packet that contains a continuum of frequencies with an
appropriate peaking. Suppose that the Alice and Rob states are
initially set up as Minkowski wave packets, peaked about distinct
Minkowski momenta and having negligible overlap, so that the bipartite
assumption is a good approximation. The transformation between the
Minkowski and Unruh bases is an integral transform, given by 
\eqref{eq:Mink-v-Unruh}
and~\eqref{eq:alphas}: 
can the Rob state
be arranged to be peaked about a single Unruh frequency? If so, how
are the frequency uncertainties on the Minkowski and Unruh sides
related?

\subsection{Massless scalar field}\label{sec:peakingmassless}

We focus first on the massless scalar field of 
section~\ref{sec:entganglementrev}. 
The massive scalar field will be discussed 
in section~\ref{subsec:massivescalar}. 
We expect the analysis for fermions to be qualitatively
similar.

Consider a packet of Minkowski creation operators $a_{\omega,\text{M}}^\dag$
smeared with a weight function~$f(\omega)$. 
We wish to express this packet in terms of
Unruh creation operators $A_{\Omega,\text{\text{R}}}^\dag$ and $A_{\Omega,\text{\text{L}}}^\dag$ 
smeared with the weight
functions $g_{\text{R}}(\Omega)$ and $g_{\text{L}}(\Omega)$, so that 
\begin{align}
\int_{0}^{\infty}
&f(\omega) \, a_{\omega,\text{M}}^\dag \, d\omega
= 
\int_0^{\infty}
\left(
g_{\text{R}}(\Omega)A_{\Omega,\text{\text{R}}}^\dag+g_{\text{L}}(\Omega) A_{\Omega,\text{\text{L}}}^\dag
\right)
\text{d}\Omega. 
\label{eq:smeared-creator}
\end{align}
From \eqref{eq:Aaboth-transform} it follows that the smearing functions are related by
\begin{align}
g_\text{\text{R}}(\Omega) &= \int_{0}^{\infty}\alpha^\text{\text{R}}_{\omega\Omega}f(\omega) \, \text{d}\omega , \nonumber \\
g_\text{\text{L}}(\Omega) &= \int_{0}^{\infty}\alpha^\text{\text{L}}_{\omega\Omega}f(\omega) \, \text{d}\omega , \notag  \\
f(\omega) &= \int_{0}^{\infty}\left[(\alpha^\text{\text{R}}_{\omega\Omega})^{\ast}g_\text{\text{R}}(\Omega) 
+(\alpha^\text{\text{L}}_{\omega\Omega})^{\ast}g_\text{\text{L}}(\Omega)\right]\text{d}\Omega . 
\label{eq:f-gs-transform}
\end{align}
By~\eqref{eq:alphas}, equations \eqref{eq:f-gs-transform}
are recognised as a Fourier transform pair between the variable $\ln
(\omega\dimctwo) \in\BbbR$ on the Minkowski side and the variable
$\pm\Omega\in\BbbR$ on the Unruh side: the full real line on the Unruh
side has been broken into the Unruh frequency $\Omega \in\BbbR^+$ and
the discrete index R, L. 
All standard properties of
Fourier transforms thus apply. 
Parseval's theorem takes the form 
\begin{align}
\int_0^\infty {|f(\omega)|}^2 \, \text{d}\omega 
= 
\int_{0}^{\infty} 
\left(|g_\text{\text{R}}(\Omega)|^2+|g_\text{\text{L}}(\Omega)|^2 \right) 
\text{d}\Omega,
\end{align}
and the uncertainty 
relation reads 
\begin{align}
\label{eq:gen-uncertainty}
(\Delta \Omega)  
\bigl(\Delta \ln(\omega\dimctwo)\bigr) \ge \tfrac12, 
\end{align}
where $\Delta \Omega$ is understood by combining contributions from
$g_\text{\text{R}}(\Omega)$ and $g_\text{\text{L}}(\Omega)$ in the sense
of~\eqref{eq:f-gs-transform}.  Note that since equality in 
\eqref{eq:gen-uncertainty} holds only for Gaussians, any state in
which one of $g_\text{\text{R}}(\Omega)$ and ${}g_\text{\text{L}}(\Omega)$ vanishes will
satisfy \eqref{eq:gen-uncertainty} with a genuine inequality.

As a concrete example, with a view to optimising the peaking both in
Minkowski frequency and in
Unruh frequency, 
consider a Minkowski smearing function that is
a Gaussian in  $\ln(\omega\dimctwo)$, 
\begin{align}
\label{eq:gaussian-in-log}
f(\omega) = 
\left(\frac{\lambda}{\pi\omega^2}\right)^{\!\!1/4}
\exp\left\{- \tfrac12 \lambda
{\bigl[\ln(\omega/\omega_0)\bigr]}^2
\right\}
{(\omega/\omega_0)}^{-i\mu},
\end{align}
where $\omega_0$ and $\lambda$ are positive parameters and $\mu$ is a real-valued parameter. 
$\lambda$~and $\mu$ are dimensionless and $\omega_0$ has the dimension of inverse length. 
Note that $f$~is normalised, $\int_{0}^{\infty} {|f(\omega)|}^2 \, d\omega =1$. 
The expectation value and uncertainty of $\ln(\omega\dimctwo)$ are those of 
a standard Gaussian, $\langle \ln(\omega\dimctwo) \rangle = \ln(\omega_0\dimctwo)$ 
and $\Delta \ln(\omega\dimctwo) = {(2\lambda)}^{-1/2}$, 
while the expectation value and uncertainty of $\omega$ are given by 
\begin{eqnarray}
\langle \omega \rangle &=& \exp\left(\tfrac14\lambda^{-1}\right),\nonumber \\
\Delta\omega &=& \langle \omega \rangle {\left[\exp\bigl(\tfrac12\lambda^{-1}\bigr) -1 \right]}^{1/2}. 
\end{eqnarray}

The Unruh smearing functions are cropped Gaussians, 
\begin{align}
\label{eq:gs-gauss}
g_\text{\text{R}}(\Omega) &= \frac{1}{{(\pi\lambda)}^{1/4}}\exp\!\left[- \tfrac12 \lambda^{-1}{(\Omega - \epsilon\mu)}^2 \right] 
{(\omega_0\dimctwo)}^{i\epsilon\Omega},\nonumber \\
g_\text{\text{L}}(\Omega) &= \frac{1}{{(\pi\lambda)}^{1/4}}\exp\left[- \tfrac12 \lambda^{-1}{(\Omega + \epsilon\mu)}^2\right] 
{(\omega_0\dimctwo)}^{-i\epsilon\Omega}.
\end{align}
For $\epsilon\mu \gg \lambda^{1/2}$, $g_\text{\text{L}}(\Omega)$ is small and $g_\text{\text{R}}(\Omega)$ 
is peaked around $\Omega = \epsilon\mu$ with uncertainty ${(\lambda/2)}^{1/2}$; 
conversely, for $\epsilon\mu \ll -\lambda^{1/2}$, $g_\text{\text{R}}(\Omega)$ is small 
and $g_\text{\text{L}}(\Omega)$ is peaked around $\Omega = - \epsilon\mu$ with 
uncertainty ${(\lambda/2)}^{1/2}$. Note that in these limits, 
the relative magnitudes of $g_\text{\text{L}}(\Omega)$ and $g_\text{\text{R}}(\Omega)$ are 
consistent with the magnitude of the smeared Minkowski mode 
function $\int_{0}^{\infty}f(\omega) \, u_{\omega,\text{M}} \, d\omega$ 
in the corresponding regions of Minkowski space: 
a contour deformation argument shows that for $\epsilon\mu \gg \lambda^{1/2}$ 
the smeared mode function 
is large in the region $t+x>0$ and small in the region $t+x<0$, 
while for $\epsilon\mu \ll -\lambda^{1/2}$ it is large in the 
region $t-x>0$ and small in the region $t-x<0$. 

Now, let the Rob state have the smearing
function~\eqref{eq:gaussian-in-log}, and choose for Alice any state
that has negligible overlap with the Rob state, for example by taking
for Alice and Rob distinct values of~$\epsilon$. For $|\mu| \gg
\lambda^{1/2}$ and $\lambda$ not larger than of order unity, the
combined state is then well approximated by the single Unruh frequency
state of section \ref{sec:entganglementrev}
with $\Omega=|\mu|$ and with one of $q_\text{\text{R}}$ and
$q_\text{\text{L}}$ vanishing. In this case we hence recover the 
results in~\cite{Alicefalls}. 
To build a Rob state that is 
peaked about a single Unruh frequency with 
comparable $q_\text{\text{R}}$ and~$q_\text{\text{L}}$, 
so that the results of section 
\ref{sec:entganglementrev} are recovered, we may take 
a Minkowski smearing function that
is a linear combination of \eqref{eq:gaussian-in-log} and its complex
conjugate.

While the phase factor ${(\omega/\omega_0)}^{-i\mu}$ in the Minkowski
smearing function \eqref{eq:gaussian-in-log} is essential for
adjusting the locus of the peak in the Unruh smearing functions,
the choice of a logarithmic Gaussian for the magnitude appears not
essential. We have verified that similar results ensue with the
choices 
\begin{align}
f(\omega) = 
\frac{2^{\lambda}
{(\omega/\omega_0)}^{\lambda-i\mu} \exp(-\omega/\omega_0)}
{\sqrt{\omega\Gamma(2\lambda)}}
\end{align}
and 
\begin{align}
f(\omega) = 
\frac{{(\omega/\omega_0)}^{-i\mu}}{\sqrt{2\omega K_0(2\lambda)}}
\exp
\left[- \frac{\lambda}{2} 
\left(\frac{\omega}{\omega_0}+ \frac{\omega_0}{\omega}\right)
\right], 
\end{align}
for which the respective Unruh smearing functions can be expressed respectively in terms of
the gamma-function and a modified Bessel function. 

\subsection{Massive scalar field\label{subsec:massivescalar}}

For a scalar field of mass $m>0$, the 
Minkowski modes of the Klein-Gordon equation are
\begin{align}
\label{eq:massivescalar-Mmode}
u_{k,\text{M}} (t,x) = \frac{1}{\sqrt{4\pi \omega}}\exp(-i\omega t + i kx),
\end{align}
where $k\in\BbbR$ is the Minkowski momentum and $\omega \equiv \omega_k =\sqrt{m^2+k^2}$ is the Minkowski frequency. 
These modes are delta-normalised in $k$ as usual. 
The Rindler modes are \cite{Takagi}
\begin{eqnarray}
u_{\Omega,\text{I}} (t,x) &=& N_{\Omega}\exp\left[ -\frac{i\Omega}{2}\ln\left(\frac{x+t}{x-t}\right)\right],\nonumber \\
u_{\Omega,\text{II}} (t,x) &=& N_{\Omega}\exp\left[ -\frac{i\Omega}{2}\ln\left(\frac{-x+t}{-x-t}\right)\right],
\end{eqnarray}
where $N_{\Omega}= \frac{\sqrt{\sinh \pi\Omega}}{\pi}K_{i\Omega}\bigl(m \sqrt{x^2-t^2}\,\bigr)$ and 
$\Omega>0$ is the (dimensionless) Rindler frequency. These modes are delta-normalised in~$\Omega$. 
The Unruh modes $u_{\Omega,\text{\text{R}}}$ and $u_{\Omega,\text{\text{L}}}$ 
are as in~\eqref{eq:unruhmodes}. 
Note that in the Minkowski modes \eqref{eq:massivescalar-Mmode}
the distinction between 
the left-movers and the right-movers is in the sign of the label 
$k\in\BbbR$, but in the Rindler and Unruh modes the right-movers and the left-movers 
do not decouple, owing to the 
asymptotic behaviour of the solutions at the Rindler spatial infinity. 
The Rindler and Unruh modes do therefore not carry an index $\epsilon$ 
that would distinguish the right-movers and the left-movers. 

The transformation between the Minkowski and Unruh modes 
can be found by the methods of~\cite{Takagi}. 
In our notation, the transformation reads 
\begin{eqnarray}
\label{eq:alphas-massive}
u_{\Omega,\text{\text{R}}} &=& \int_{-\infty}^{\infty}(\alpha^\text{\text{R}}_{k\Omega})^{\ast}u_{k,\text{M}} \, \text{d}k , \nonumber \\
u_{\Omega,\text{\text{L}}} &=& \int_{-\infty}^{\infty}(\alpha^\text{\text{L}}_{k\Omega})^{\ast}u_{k,\text{M}} \, \text{d}k , \nonumber \\
u_{k ,\text{M}} &=& \int_{0}^{\infty}
\left(\alpha^\text{\text{R}}_{k\Omega} u_{\Omega,\text{\text{R}}} +\alpha^\text{\text{L}}_{k\Omega} u_{\Omega,\text{\text{L}}} \right)\text{d}\Omega, 
\end{eqnarray}
where
\begin{eqnarray}\label{eq:alphas-massive2}
\alpha^\text{\text{R}}_{k\Omega} &=& 
\frac{1}{\sqrt{2\pi\omega}}
{\left(\frac{\omega+k}{m}\right)}^{i\Omega} ,\nonumber  \\
\alpha^\text{\text{L}}_{k\Omega} &=&
\frac{1}{\sqrt{2\pi\omega}}
{\left(\frac{\omega+k}{m}\right)}^{-i\Omega}.
\end{eqnarray}
Transformations for the various operators read hence as 
in section \ref{sec:entganglementrev}
but with the replacements 
\begin{equation}
\label{eq:massless-to-massive}
\omega\to k, \quad
\int_0^\infty \text{d}\omega \longrightarrow \int_{-\infty}^\infty \text{d}k 
\end{equation}
and no $\epsilon$. In particular, 
\begin{eqnarray}
\label{eq:Aaboth-massive-transform}
A_{\Omega,\text{\text{R}}}&=&\int_{-\infty}^{\infty}  \alpha^\text{\text{R}}_{k\Omega} \, a_{k,\text{M}} \, \text{d}k \nonumber \\
A_{\Omega,\text{\text{L}}} &=& \int_{-\infty}^{\infty}  \alpha^\text{\text{L}}_{k\Omega} \, a_{k,\text{M}} \, \text{d}k  \\
a_{k,\text{M}} &=&  \int_{0}^{\infty}\left((\alpha^\text{\text{R}}_{k\Omega})^{\ast} A_{\Omega,\text{\text{R}}}+(\alpha^\text{\text{L}}_{k\Omega})^{\ast} A_{\Omega,\text{\text{L}}}\right)\text{d}\Omega.\nonumber
\end{eqnarray}

To consider peaking of Minkowski wave packets in the Unruh frequency, 
we note that the transform \eqref{eq:Aaboth-massive-transform} with
\eqref{eq:alphas-massive2} is now a Fourier transform between the
Minkowski rapidity $\tanh^{-1} (k/\omega) = \ln[(\omega+k)/m]\in\BbbR$
and $\pm\Omega\in\BbbR$. The bulk of the massless peaking discussion
of section \ref{sec:peakingmassless}
goes hence through with the replacements
\eqref{eq:massless-to-massive} and $\omega\dimctwo \to
(\omega+k)/m$. The main qualitative difference is that in the massive
case one cannot appeal to the decoupling of the right-movers and
left-movers when choosing for Alice and Rob states that have
negligible overlap.

\section{Unruh  entanglement degradation for Dirac fields}\label{secferme}

 Statistics plays a very important role in the behaviour of entanglement described by observers in uniform acceleration. While entanglement vanishes in the limit of infinite acceleration in the bosonic case \cite{Alicefalls}, it remains finite for Dirac fields \cite{AlsingSchul}. Therefore, it is interesting to revise the analysis of entanglement between Dirac fields for different elections of Unruh modes.
\subsection{Dirac fields}
In a parallel analysis to the bosonic case, we consider a Dirac field $\phi$ satisfying the
equation  $\{i\gamma^{\mu}(\partial_{\mu}-\Gamma_{\mu})+m\}\phi=0$
where $\gamma^{\mu}$ are the Dirac-Pauli matrices and $\Gamma_{\mu}$ are spinorial
affine connections\footnote{See Appendix \ref{appB}}. The field expansion in terms of the Minkowski solutions of the Dirac equation is
\begin{equation}\label{fermfieldminm7}
\phi=N_\text{M}\sum_k\left(c_{k,\text{M}}\, u^+_{k,\text{M}}+ d_{k,\text{M}}^\dagger \ u^-_{k,\text{M}}\right),
\end{equation}
Where $N_\text{M}$ is a normalisation constant and the label $\pm$  denotes respectively positive and negative energy solutions (particles/antiparticles) with respect to the Minkowskian Killing vector field $\partial_{t}$. The label $k$ is a multilabel including energy and spin $k=\{E_\omega,s\}$ where $s$ is the component of the spin on the quantisation direction. $c_k$ and $d_k$ are the particle/antiparticle operators that satisfy the usual anticommutation rule
\begin{equation}
\{c_{k,\text{M}},c_{k',\text{M}}^\dagger\}=\{ d_{k,\text{M}},d_{k',\text{M}}^\dagger\}=\delta_{kk'},
\end{equation}
and all other anticommutators vanishing. The Dirac field operator in terms of Rindler modes is given by
\begin{equation}\label{fermfieldm7}
\phi=N_\text{\text{R}}\sum_j\left(c_{j,\text{I}} u^+_{j,\text{I}}+ d_{j,\text{I}}^\dagger u^-_{j,\text{I}}+c_{j,\text{II}} u^+_{j,\text{II}}+ d_{j,\text{II}}^\dagger  u^-_{j,\text{II}}\!\right),
\end{equation}
where $N_\text{R}$ is, again, a normalisation constant.  $c_{j,\Sigma},d_{j,\Sigma}$ with $\Sigma=\text{I},\text{II}$ represent Rindler particle/antiparticle operators. The usual anticommutation rules again apply. Note that operators in different regions $\Sigma=\text{I},\text{II}$ do not commute but anticommute. $j=\{E_\Omega,s'\}$ is again a multi-label including all the degrees of freedom. Here  $u^\pm_{k,\text{I}}$ and $u^\pm_{k,\text{II}}$  are the positive/negative frequency solutions of the Dirac equation in Rindler coordinates with respect to the Rindler timelike Killing vector field in region $\text{I}$ and $\text{II}$, respectively. The modes  $u^\pm_{k,\text{I}}$, $u^\pm_{k,\text{II}}$  do not have support outside the right, left Rindler wedge. 
The annihilation operators $c_{k,\text{M}},d_{k,\text{M}} $ define the Minkowski vacuum $\ket{0}_\text{M}$ which must satisfy
\begin{equation}
c_{k,\text{M}}\ket{0}_\text{M}= d_{k,\text{M}}\ket{0}_\text{M}=0, \qquad \forall k.
\end{equation}
In the same fashion $c_{j,\Sigma},d_{j,\Sigma}$, define the Rindler vacua in regions $\Sigma=\text{I},\text{II}$
\begin{equation}
c_{j,\text{\text{R}}}\ket{0}_\Sigma=d_{j,\text{\text{R}}}\ket{0}_\Sigma=0, \qquad \forall j, \, \Sigma=\text{I},\text{II}.
\end{equation}	
The transformation  between the Minkowski and Rindler modes is given by
\begin{equation}
 u^+_{j,\text{M}}=\sum_k\left[\alpha^\text{I}_{jk}u^+_{k,\text{I}} + \beta^{\text{I}}_{jk} u^-_{k,\text{I}}+\alpha^\text{II}_{jk}u^+_{k,\text{II}}+\beta^{\text{II}}_{jk} u^-_{k,\text{II}}\right].
\end{equation}
The coefficients which relate both set of modes are given by the inner product\footnote{Note that the Dirac inner product properties are slightly different from the Klein-Gordon inner product \eqref{KGproperties}. Namely, $(u_1,u_2)^*=(u_1^*,u_2^*)=(u_2,u_1)$} coming from the continuity equation derived from the Dirac equation
\begin{equation}
(u_k,u_j)=\int d^3x\, u_k^\dagger u_j,
\end{equation}
so that he  Bogoliubov coefficients
are, after some elementary but lengthy algebra \cite{Jauregui,Langlois},
\begin{equation}
\alpha^\text{I}_{jk}=e^{i\theta E_\Omega}\frac{1+i}{2\sqrt{\pi E_\omega}}\,\frac{e^{\pi E_\Omega/2}}{\sqrt{e^{\pi E_\Omega}+e^{-\pi E_\Omega}}}\delta_{ss'},
\end{equation}
\begin{equation}
\beta^\text{I}_{jk}=-e^{-i\theta E_\Omega}\frac{1+i}{2\sqrt{\pi E_\omega}}\,\frac{e^{-\pi E_\Omega/2}}{\sqrt{e^{\pi E_\Omega}+e^{-\pi E_\Omega}}}\delta_{s s'},
\end{equation}
where $E_\Omega$ is the energy of the Rindler mode $k$, $E_\omega$ is the energy of the Minkowski mode $j$ and $\theta$ is a parameter defined such that it satisfies the condition $E_\Omega=m\cosh\theta$ and $|\bm k_\Omega|=m\sinh\theta$   (see \cite{Jauregui}).  One can verify that $\alpha^\text{II}=(\alpha^\text{I})^*$ and $\beta^\text{II}=(\beta^\text{I})^*$.  Defining $\tan r_{\Omega}=e^{-\pi E_\Omega}$
the coefficients become
\begin{eqnarray}
\alpha^\text{I}_{jk}=e^{i\theta E_\Omega}\frac{1+i}{2\sqrt{\pi E_\omega}}\,\cos r_{\Omega}\,\delta_{s s'},\nonumber\\
\beta^\text{I}_{jk}=-e^{-i\theta E_\Omega}\frac{1+i}{2\sqrt{\pi E_\omega}}\,\sin r_{\Omega}\,\delta_{s s'}.
\end{eqnarray}
Finally, taking into account that $c_{j,\text{M}}=(\phi, u^+_{j,\text{M}})$  we find the Minkowski particle annihilation operator to be
\begin{equation}
 c_{j,\text{M}}=\sum_{k}\left[\alpha^{\text{I}*}_{jk}c_{k,\text{I}} + \beta^{\text{I}*}_{jk} d_{k,\text{I}}^\dagger+\alpha^{\text{II}*}_{jk}c_{k,\text{II}}+\beta^{\text{II}*}_{jk}d_{k,\text{II}}^{\dagger}\right].
\end{equation}

We now consider the transformations between states in different basis. For this we define an arbitrary element of the Dirac field Fock basis for each mode as
\begin{equation}
\ket{F_k}=\ket{F_k}_\text{\text{R}}\otimes \ket{F_k}_\text{\text{L}},
\end{equation}
where 
\begin{equation}
\nonumber\ket{F_k}_\text{R}=\ket{n}^+_\text{I}\ket{m}^-_\text{II},\qquad \ket{F_k}_\text{L}=\ket{p}^-_\text{I}\ket{q}^+_\text{II}.
\end{equation}
Here the $\pm$  indicates particle/antiparticle. Operating with the Bogoliubov coefficients making this tensor product structure explicit we obtain
\begin{equation}\label{annihil}
c_{j,\text{M}}=N_j\sum_k\Bigg[\chi^*(C_{k,\text{\text{R}}}\otimes\openone_\text{\text{L}})+\chi(\openone_\text{\text{R}}\otimes C_{k,\text{\text{L}}})\Bigg],
\end{equation}
where
\begin{equation}\label{ene}
N_j=\frac{1}{2\sqrt{\pi E_\omega}}\qquad  \chi=(1+i)e^{i\theta E_{\Omega}},
\end{equation}
and the operators
\begin{eqnarray}\label{Unruhopm7}
\nonumber C_{k,\text{\text{R}}}&\equiv&\left(\cos r_k\, c_{k,\text{I}}-\sin r_k\, d^\dagger_{k,\text{II}}\right),\\*
C_{k,\text{\text{L}}}&\equiv&\left(\cos r_k\, c_{k,\text{II}}-\sin r_k\, d^\dagger_{k,\text{I}}\right)
\end{eqnarray}
are the so-called Unruh operators. 

It can be shown \cite{ch1} that for a massless Dirac field the Unruh operators have the same form as Eq. \eqref{Unruhopm7} however in this case $\tan r_{k}=e^{-\pi\Omega_a/a}$.

In the massless case, to find the Minkowski vacuum in the Rindler basis we consider the following ansatz
\begin{equation}
\ket{0}_\text{M}=\bigotimes_{\Omega}\ket{0_\Omega}_\text{M},
\end{equation}
where $\ket{0_\Omega}_\text{M}=\ket{0_\Omega}_\text{\text{R}}\otimes\ket{0_\Omega}_\text{\text{L}}$.
We find that
\begin{eqnarray}
\label{vauno}
\ket{0_{\Omega}}_\text{\text{R}}&=&\sum_{n,s}\left(F_{n,\Omega,s}\ket{n_{\Omega,s}}^+_\text{I}\ket{n_{\Omega,-s}}^-_\text{II}\right)\nonumber \\
\ket{0_{\Omega}}_\text{\text{L}}&=&\sum_{n,s}\left(G_{n,\Omega,s}\ket{n_{\Omega,s}}^-_\text{I}\ket{n_{\Omega,-s}}^+_\text{II}\right),
\end{eqnarray}
where the label $\pm$ denotes particle/antiparticle modes  and $s$ labels the spin. The minus signs on the spin label in region $\text{II}$ show explicitly that spin, as all the magnitudes which change under time reversal, is opposite in region I with respect to region II.

We obtain the form of the coefficients $F_{n,\Omega,s},G_{n,\Omega,s}$ for the vacuum by imposing that the Minkowski vacuum is annihilated by the particle annihilator for all frequencies and values for the spin third component.

\subsection{Grassmann scalars}

Since the simplest case that preserves the fundamental Dirac characteristics corresponds to Grassmann scalars, we study them in what follows. Moreover,  the entanglement in non-inertial frames between scalar fermionic fields has been extensively studied under the single mode approximation in the literature \cite{AlsingSchul}. In this case, the Pauli exclusion principle limits the sums \eqref{vauno} and only the two following terms contribute
\begin{eqnarray}
\label{vaunos}
\ket{0_\Omega}_\text{\text{R}}&=&F_0\ket{0_\Omega}^+_\text{I}\ket{0_\Omega}^-_\text{II}+F_1\ket{1_\Omega}^+_\text{I}\ket{1_\Omega}^-_\text{II},\nonumber \\
\ket{0_\Omega}_\text{\text{L}}&=&G_0\ket{0_\Omega}^-_\text{I}\ket{0_\Omega}^+_\text{II}+G_1\ket{1_\Omega}^-_\text{I}\ket{1_\Omega}^+_\text{II}.
\end{eqnarray}
Due to the anticommutation relations we must introduce the following sign conventions
\begin{align}
\ket{1_\Omega}^+_\text{I}\!\ket{1_\Omega}^-_\text{II}&=c^\dagger_{\Omega,\text{I}}d^\dagger_{\Omega,\text{II}}\!\ket{0_\Omega}^+_\text{I}\!\ket{0_\Omega}^-_\text{II}=-d^\dagger_{\Omega,\text{II}}c^\dagger_{\Omega,\text{I}}\!\ket{0_\Omega}^+_\text{I}\!\ket{0_\Omega}^-_\text{II},\nonumber\\*
\ket{1_\Omega}^-_\text{I}\!\ket{1_\Omega}^+_\text{II}&=d^\dagger_{\Omega,\text{I}}c^\dagger_{\Omega,\text{II}}\!\ket{0_\Omega}^-_\text{I}\!\ket{0_\Omega}^+_\text{II}=-c^\dagger_{\Omega,\text{II}}d^\dagger_{\Omega,\text{I}}\!\ket{0_\Omega}^-_\text{I}\!\ket{0_\Omega}^+_\text{II}.
\end{align}
We obtain the form of the coefficients by imposing that $c_{\omega,\text{M}}\ket{0_\Omega}_\text{M}=0$ which translates into $C_{\Omega,\text{\text{R}}}\ket{0_\Omega}_\text{\text{R}}=C_{\Omega,\text{\text{L}}}\ket{0_\Omega}_\text{\text{L}}=0$. Therefore
\begin{eqnarray}
\label{co1} C_{\Omega,\text{\text{R}}}\left(F_0\ket{0_\Omega}^+_\text{I}\ket{0_\Omega}^-_\text{II}+F_1\ket{1_\Omega}^+_\text{I}\ket{1_\Omega}^-_\text{II}\right)&=&0,\nonumber \\
\label{co2} C_{\Omega,\text{\text{L}}}\left(G_0\ket{0_\Omega}^-_\text{I}\ket{0_\Omega}^+_\text{II}+G_1\ket{1_\Omega}^-_\text{I}\ket{1_\Omega}^+_\text{II}\right)&=&0.
\end{eqnarray}
These conditions imply that
\begin{eqnarray}
F_1\cos r_{\Omega}&-&F_0\sin r_{\Omega}=0\Rightarrow F_1=F_0\tan r_{\Omega},\nonumber \\
G_1\cos r_{\Omega}&+&G_0\sin r_{\Omega}=0\Rightarrow G_1=-G_0\tan r_{\Omega},
\end{eqnarray}
which together with the normalisation conditions $\bra{0_\Omega}_\text{\text{R}}\ket{0_\Omega}_\text{\text{R}}=1$ and $\bra{0_\Omega}_\text{\text{L}}\ket{0_\Omega}_\text{\text{L}}=1$ yield
\begin{eqnarray}
F_0&=&\cos r_{\Omega},\qquad F_1=\sin r_{\Omega},\\
G_0&=&\cos r_{\Omega}\qquad G_1=-\sin r_{\Omega}.\nonumber
\end{eqnarray}
Therefore  the vacuum state is given by,
 \begin{align}\label{vacgrassman} 
\nonumber\ket{0_{\Omega}}&=\left(\cos r_{\Omega}\ket{0_{\Omega}}^+_\text{I}\ket{0_{\Omega}}^-_\text{II}+\sin r_{\Omega}\ket{1_{\Omega}}^+_\text{I}\ket{1_{\Omega}}^-_\text{II}\right)\\
&\otimes\left(\cos r_{\Omega}\ket{0_{\Omega}}^-_\text{I}\ket{0_{\Omega}}^+_\text{II}-\sin r_{\Omega}\ket{1_{\Omega}}^-_\text{I}\ket{1_{\Omega}}^+_\text{II}\right)\!,
\end{align}
which is compatible with the result obtained with the Unruh modes. For convenience, we introduce the following notation, \begin{equation}\label{shortnot}
\ket{n n' n'' n'''}_{\Omega}\equiv\ket{n_{\Omega}}^+_\text{I}\ket{n'_{\Omega}}^-_\text{II}\ket{n''_{\Omega}}^-_\text{I}\ket{n'''_{\Omega}}^+_\text{II},
\end{equation}
in which the vacuum state is written as,
\begin{equation}\label{shortvac}
\ket{0_{\Omega}}=\cos^2 r_{\Omega}\ket{0000}_{\Omega}-\sin r_{\Omega}\cos r_{\Omega}\ket{0011}_{\Omega}+\sin r_{\Omega}\cos r_{\Omega} \ket{1100}_{\Omega}-\sin^2 r_{\Omega} \ket{1111}_{\Omega}.
\end{equation}

The Minkowskian one particle state is obtained by applying the creation operator to the vacuum state $\ket{1_{j}}_{\text{U}}=c_{\Omega,\text{U}}^\dagger\ket{0}_\text{M}$,
where the Unruh particle creator is a combination of the two Unruh operators $C_{\Omega,\text{\text{R}}}^\dagger$ and $C_{\Omega,\text{\text{L}}}^\dagger$,
\begin{equation}\label{creat}
c_{k,\text{U}}^\dagger\!=\!q_\text{\text{R}}(C^\dagger_{\Omega,\text{\text{R}}}\otimes\openone_\text{\text{L}})+q_\text{\text{L}}(\openone_\text{\text{R}}\otimes C^\dagger_{\Omega,\text{\text{L}}}).
\end{equation}
Since
\begin{eqnarray}
C^\dagger_{\Omega,\text{\text{R}}}&\equiv&\left(\cos r_{\Omega}\, c^\dagger_{\Omega,\text{I}}-\sin r_{\Omega}\, d_{\Omega,\text{II}}\right),\\*
C^\dagger_{\Omega,\text{\text{L}}}&\equiv&\left(\cos r_{\Omega}\, c^\dagger_{\Omega,\text{II}}-\sin r_{\Omega}\, d_{\Omega,\text{I}}\right),\nonumber
\end{eqnarray}
with $q_\text{\text{R}},q_\text{\text{L}}$ complex numbers satisfying $|q_\text{\text{R}}|^2+|q_\text{\text{L}}|^2=1$,
we obtain,
\begin{eqnarray}
\ket{1_{\Omega}}^+_\text{\text{R}}&=&C^\dagger_{\Omega,\text{\text{R}}}\ket{0_{\Omega}}_\text{\text{R}}=\ket{1_{\Omega}}^+_\text{I}\ket{0_{\Omega}}^-_\text{II}\\*
\ket{1_{\Omega}}^+_\text{\text{L}}&=&C^\dagger_{\Omega,\text{\text{L}}}\ket{0_{\Omega}}_\text{\text{L}}=\ket{0_{\Omega}}^-_\text{I}\ket{1_{\Omega}}^+_\text{II}\nonumber
\end{eqnarray}
and therefore,
\begin{equation}
\ket{1_{k}}^+_\text{U}=q_\text{\text{R}}\ket{1_{\Omega}}_\text{\text{R}}\otimes\ket{0_{\Omega}}_\text{\text{L}}+q_\text{\text{L}}\ket{0_{\Omega}}_\text{\text{R}}\otimes\ket{1_{\Omega}}_\text{\text{L}}.
\end{equation}
In the short notation we have introduced the state reads,  
\begin{equation}\label{onegrassman}
\ket{1_{k}}^+_\text{U}= q_\text{\text{R}}\left[\cos r_{k}\ket{1000}_{\Omega}-\sin r_{\Omega}\ket{1011}_{\Omega}\right]
+q_\text{\text{L}}\left[\sin r_{\Omega}\ket{1101}_{\Omega}+\cos r_{\Omega}\ket{0001}_{\Omega}\right].
\end{equation}

\subsection{Fermionic entanglement beyond the single mode approximation}

 Let us now consider the following fermionic maximally entangled state
\begin{equation}\label{eq:states}
\ket{\Psi}=\frac{1}{\sqrt2}\left(\ket{0_\omega}_{\text M}\ket{0_\Omega}_\text{U}+\ket{1_\omega}^+_{\text{M}}\ket{1_\Omega}^+_\text{U}\right),
\end{equation}
which is the fermionic analog to \eqref{maxent} and where the modes labeled with $\text{U}$ are Grassmann Unruh modes. To  compute Alice-Rob partial density matrix  we trace over region II in in $\proj{\Psi}{\Psi}$ and obtain,  
\begin{align}\label{ek1}
\nonumber\rho_{AR}&=\frac{1}{2}\Big[C^4\proj{000}{000}+S^2C^2(\proj{010}{010}+\proj{001}{001})S^4\proj{011}{011}\\*
\nonumber&+|q_\text{\text{R}}|^2(C^2\proj{110}{110}+S^2\proj{111}{111})+\phantom{\Big[}\!\!|q_\text{\text{L}}|^2(S^2\!\proj{110}{110}\!+\!C^2\!\proj{100}{100})\\*
\nonumber&+q_\text{\text{R}}^*(C^3\!\proj{000}{110}+S^2C\!\proj{001}{111})-q_\text{\text{L}}^*(C^2S\!\proj{001}{100}+S^3\!\proj{011}{110})\\*
&-q_\text{R} q_\text{L}^*SC\!\proj{111}{100}\Big]+(\text{H.c.})_{_{\substack{\text{non-}\text{diag.}}}}
\end{align}
in the basis were $C=\cos r_{\Omega}$ and $S=\sin r_{\Omega}$. To compute the negativity, we first obtain the partial transpose density matrix (transpose only in the subspace of Alice or Rob) and find its negative eigenvalues. The partial transpose matrix is block diagonal and only the following two blocks contribute to negativity,
\begin{itemize}
\item $\{\ket{100},\ket{010},\ket{111}\}$
\end{itemize}
\begin{equation}
\frac12\left(\begin{array}{ccc}
 C^2|q_\text{\text{L}}|^2& C^3q_\text{\text{R}}^*&-q_\text{R}^* q_\text{L}SC\\
C^3q_\text{\text{R}}& S^2C^2& -q_\text{\text{L}} S^3\\
-q_\text{R} q_\text{L}^*SC & -q_\text{\text{L}}^* S^3&|q_\text{\text{R}}|^2S^2\\
\end{array}\right),
\end{equation}
\begin{itemize}
\item $\{\ket{000},\ket{101},\ket{011}\}$
\end{itemize}
\begin{equation}
\frac12\left(\begin{array}{ccc}
C^4&-q_\text{\text{L}}C^2S&0\\
-q_\text{\text{L}}^*C^2S& 0 & q_\text{\text{R}}^*S^2C \\
0& q_\text{\text{R}}S^2C & S^4 \\
\end{array}\right).
\end{equation}
were the basis used is $\ket{ijk}=\ket{i}_\text{M}\overbrace{\ket{j}_\text{I}^+\ket{k}^-_\text{I}}^{\text{Rob}}$.  Notice that although the system is bipartite, the dimension of the partial Hilbert space for Alice is lower than the dimension of the Hilbert space for Rob, which includes particle and antiparticle modes.
The eigenvalues only depend on $|q_\text{\text{R}}|$ and not on the relative phase between $q_\text{\text{R}}$ and $q_\text{\text{L}}$.

The density matrix for the Alice-AntiRob modes is obtained by tracing over region $\text{I}$ in $\proj{\Psi}{\Psi}$, \begin{align}\label{ek2}
\nonumber\rho_{A\bar R}&=\frac{1}{2}\Big[C^4\proj{000}{000}+S^2C^2(\proj{010}{010}+\proj{001}{001})+S^4\proj{011}{011}\\*
\nonumber&+|q_\text{\text{R}}|^2(C^2\proj{100}{100}+S^2\proj{101}{101})+|q_\text{\text{L}}|^2(S^2\proj{111}{111}\!+\!C^2\!\proj{101}{101})\\*
\nonumber&\phantom{\Big[}+q_\text{\text{L}}^*(C^3\!\proj{000}{101}+S^2C\!\proj{010}{111})+q_\text{\text{R}}^*(C^2S\!\proj{010}{100}+S^3\!\proj{011}{101})\\*
&+q_\text{R} q_\text{L}^*SC\!\proj{100}{111}\Big]+(\text{H.c.})_{_{\substack{\text{non-}\\\text{diag.}}}}.
\end{align}
In this case the blocks of the partial transpose density matrix which contribute to the negativity are, 
\begin{itemize}
\item $\{\ket{111},\ket{001},\ket{100}\}$
\end{itemize}
\begin{equation}
\frac12\left(\begin{array}{ccc}
 S^2|q_\text{\text{L}}|^2& S^3q_\text{\text{R}}^*&q_\text{R}^* q_\text{L}SC\\
S^3q_\text{\text{R}}& C^2S^2& q_\text{\text{L}} C^3\\
q_\text{R} q_\text{L}^*SC & q_\text{\text{L}}^* C^3&|q_\text{\text{R}}|^2C^2\\
\end{array}\right),
\end{equation}
\begin{itemize}
\item $\{\ket{011},\ket{110},\ket{000}\}$
\end{itemize}
\begin{equation}
\frac12\left(\begin{array}{ccc}
S^4&q_\text{\text{L}}S^2C&0\\
q_\text{\text{L}}^*S^2C& 0 & q_\text{\text{R}}^*C^2S \\
0& q_\text{\text{R}}C^2S & C^4 \\
\end{array}\right),
\end{equation}
where we have considered the basis $\ket{ijk}=\ket{i}_\text{M}\overbrace{\ket{j}_\text{II}^-\ket{k}^+_\text{II}}^{\text{Anti-Rob}}$. Once more, the eigenvalues only depend on $|q_\text{\text{R}}|$ and not on the relative phase between $q_\text{\text{R}}$ and $q_\text{\text{L}}$.

\begin{figure}[h]
\begin{center}
\includegraphics[width=.85\textwidth]{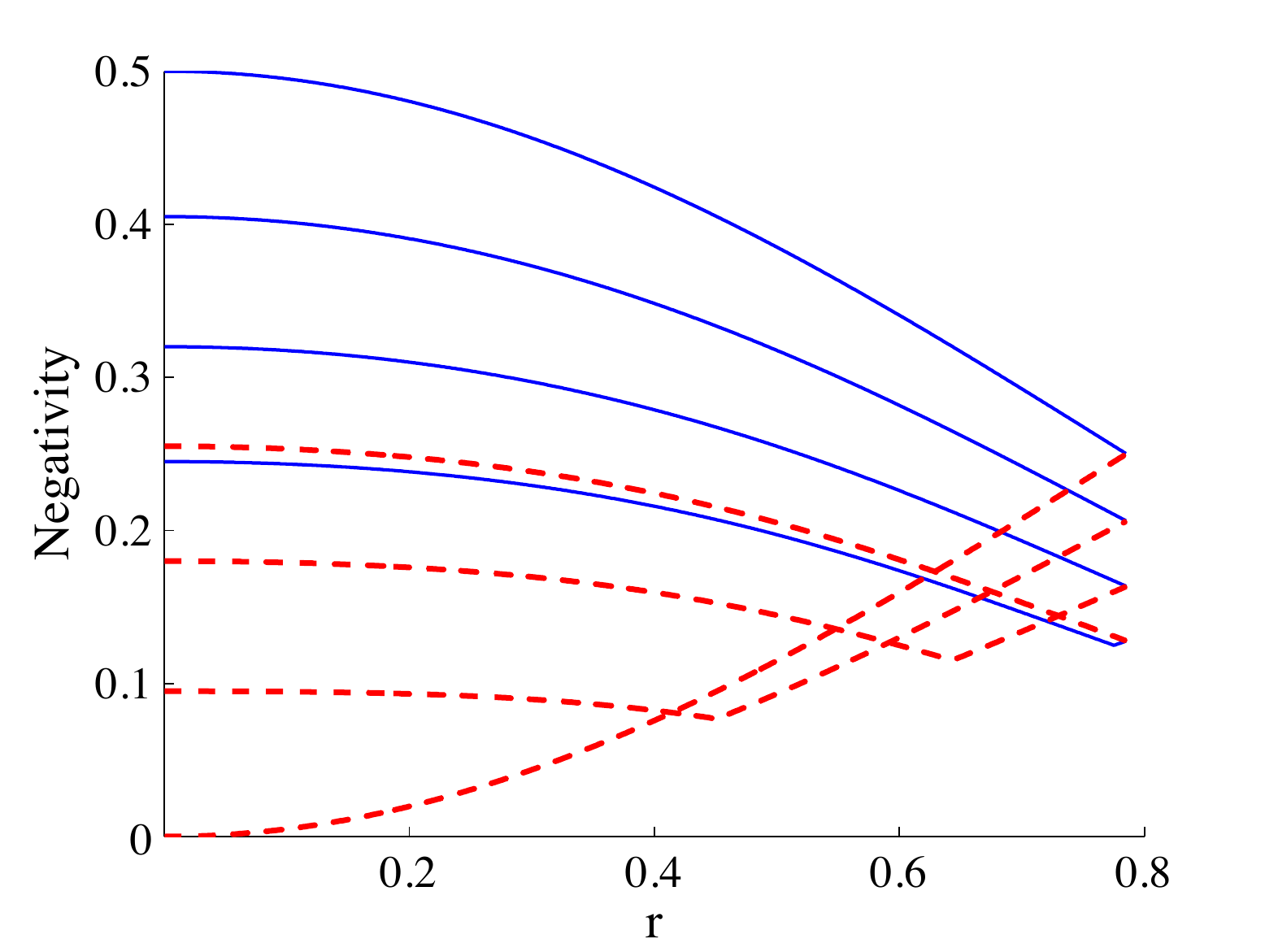}
\end{center}
\caption{Grassmann field: negativity for the bipartition Alice-Rob (Blue continuous) and Alice-AntiRob (red dashed) as a function of $r_{\Omega}=\arctan{e^{-\pi\Omega_a/a}}$ for various choices of $|q_\text{\text{R}}|$. The blue continuous (red dashed) curves from top to bottom (from bottom to top) correspond, to $|q_\text{\text{R}}|=1,0.9,0.8,0.7$  respectively. All the curves for Alice-AntiRob entanglement have a minimum, except from the extreme case $|q_R|=1$.}
\label{bundlefm7}
\end{figure}

In Fig.~\ref{bundlefm7} we plot the entanglement between  Alice-Rob (solid line) and Alice-AntiRob (dashed line) modes quantified by the negativity as a function of acceleration for different choices of $|q_\text{\text{R}}|$ (in the range \mbox{$1\ge|q_\text{\text{R}}|>1/\sqrt{2}$}).   

We confirm that the case $|q_\text{\text{R}}|=1$ reproduced the results reported in the literature \cite{AlsingSchul}. The entanglement between Alice-Rob modes is degraded as the acceleration parameter increases reaching a non-vanishing minimum value in the infinite acceleration limit $a\rightarrow\infty$. However,  while the entanglement Alice-Rob decreases, entanglement between the  Alice-AntiRob partition (dashed line) grows.  Interestingly, the quantum correlations between the bipartitions Alice-Rob and Alice-AntiRob fulfill a conservation law $\mathcal{N}(\text{Alice-Rob})+\mathcal{N}(\text{Alice-AntiRob})=1/2$. 
Note that the choice $|q_\text{\text{R}}|=0$ corresponds to an exchange of the Alice-Rob and Alice-AntiRob bipartitions.  In such case, the entanglement between Alice and Anti-Robs's modes degrades with acceleration while the entanglement between Alice and Rob's modes grows.  In fact, regarding entanglement, the role of the Alice-Rob and Alice-AntiRob partitions are exchanged when $|q_\text{\text{R}}|<|q_\text{\text{L}}|$. This is because there is an explicit symmetry between field excitations in the Rindler wedges.  Therefore, we will limit our analysis to  $|q_\text{\text{R}}|>|q_\text{\text{L}}|$.

 In the fermionic case different choices of $|q_\text{\text{R}}|$ result in different degrees of entanglement between modes. In particular, the amount of entanglement in the limit of infinite acceleration depends on this choice. Therefore, we can find a special Unruh state which is more resilient  to entanglement degradation. The total entanglement is maximal in the infinite acceleration limit in the case $|q_\text{\text{R}}|=1$ (or $|q_\text{\text{L}}|=1$) in which  $\mathcal{N}_{\infty}(\text{Alice-Rob})=\mathcal{N}_{\infty}(\text{Alice-AntiRob})=0.25$.  In this case, the entanglement lost between Alice-Rob modes is completely compensated by the creation of entanglement between Alice-AntiRob modes.

In the case $|q_\text{\text{R}}|=|q_\text{\text{L}}|=1/\sqrt2$ we see in Fig. \ref{bundlefm7} that the behaviour of both bipartitions is identical. The entanglement from the inertial perspective is equally distributed between between the Alice-Bob and Alice-AntiBob partitions and adds up to
$\mathcal{N}(\text{Alice-Bob})+\mathcal{N}(\text{Alice-AntiBob}) = 0.5$ which corresponds to the total entanglement
between Alice-Bob when $|q_\text{\text{R}}|=1$. In the infinite acceleration limit, the case $|q_\text{\text{R}}|=|q_\text{\text{L}}|$ reaches the minimum total entanglement.   To understand this we note that the entanglement in the Alice-Rob bipartition for  $|q_\text{\text{R}}|>|q_\text{\text{L}}|$ is always monotonic. However, this is not the case for the entanglement between the Alice-AntiRob modes.  Consider the plot in Fig. \ref{bundlefm7} for the cases $|q_\text{R}|<1$,  for small accelerations, entanglement is degraded in both bipartitons.  However, as the acceleration increases, entanglement between Alice-AntiRob modes is created compensating the entanglement lost between Alice-Rob. The equilibrium point between degradation and creation  is the minimum  that Alice-AntiRob entanglement curves present. Therefore, if $|q_R|<1$ the entanglement lost is not entirely compensated by the creation of entanglement between Alice-AntiRob resulting in a less entangled state in the infinite acceleration limit. 

 In the case $|q_\text{\text{R}}|=|q_\text{\text{L}}|=1/\sqrt2$ entanglement is always degraded between Alice-AntiRob modes resulting in the state, among all the possible elections of Unruh modes,  with less entanglement in the infinite acceleration limit.

\section{Discussion}\label{conclusions}

We have shown here that the single-mode approximation used in the relativistic quantum information literature, especially to analyse entanglement between field modes from the perspective of observers in uniform acceleration, does not hold. The single-mode approximation attempts to relate a single Minkowski frequency mode (observed by inertial observers) with a single Rindler frequency mode (observed by uniformly accelerated observers).

 We show that the state canonically analysed in earlier literature corresponds to a maximally entangled state of a Minkowski mode and a very specific kind of Unruh mode ($q_\text{R}=1$). We analyse the entanglement between two bosonic or fermionic modes in the case when, from the inertial perspective, the state corresponds to a maximally entangled state between a Minkowski frequency mode and an arbitrary Unruh frequency mode.
 
We find that the entanglement between modes described by an Unruh observer and a Rindler observer constrained to move in Rindler region I (Alice-Rob) is always degraded with acceleration.  In the bosonic case, the entanglement between the inertial modes  and region II Rindler  modes (Alice-AntiRob) are also degraded with acceleration.  We find that, in this case,  the rate of entanglement degradation is independent of the choice of Unruh modes.

 For the fermionic case the entanglement between the inertial and region I Rindler modes (Alice-Rob) is degraded as the acceleration increases reaching a minimum value when it tends to infinity. There is, therefore, entanglement survival in the limit of infinite acceleration for any choice of Unruh modes. However, we find an important difference with the bosonic case: the amount of surviving entanglement depends on the specific election of such modes.

We also find that the entanglement between inertial and region II Rindler modes (Alice-AntiRob) can be  created and degraded depending on the election of Unruh modes. This gives rise to different values of entanglement in the infinite acceleration limit.  Interestingly,  in the fermionic case one can find a state which is most resilient to entanglement degradation. This corresponds to the Unruh mode with  $|q_\text{\text{R}}|=1$ which is the Unruh mode considered in the canonical studies of entanglement \cite{Alsingtelep,Alicefalls,AlsingSchul,Edu2,Edu3,Edu4}. It  could be argued that this is the most natural choice of Unruh modes since for this choice ($|q_R|=1$) the entanglement for very small accelerations ($a\rightarrow0$) is mainly  contained in the subsystem Alice-Rob. In this case, there is nearly no entanglement between the Alice-AntiRob modes.  However, other choices of Unruh modes become relevant if one wishes to consider an entangled state from the inertial perspective which involves only Minkowski frequencies. We have shown that a Minkowski wavepacket involving a superposition of general Unruh modes can  be constructed in such way that the corresponding Rindler state involves a single frequency.  This result is especially interesting since it presents an instance where the single-mode approximation can be considered recovering the standard results in the literature. 

\chapter{Particle and anti-particle entanglement in non-inertial frames\footnote{E. Mart\'in-Mart\'inez and I. Fuentes. Phys. Rev. A 83, 052306 (2011)}}\label{parantpar}

\markboth{Chapter 9. Particle and anti-particle non-inertial entanglement}{\rightmark}

We have discussed in previous chapters that entanglement between modes of bosonic and fermionic fields is degraded from the perspective of observers moving in uniform acceleration. Interestingly,  entanglement is completely degraded in the infinite acceleration limit in the bosonic case while for fermionic fields a finite amount of entanglement remains in the limit. However, the reasons for these differences were not completely clear.  In this chapter we show that a redistribution of  entanglement between particle and anti-particle modes plays a key role for the preservation of fermionic field entanglement  in the infinite acceleration limit.

In our analysis we consider entangled states which involve particle and antiparticle field modes from the perspective of inertial observers.  Previous studies considered entangled  states involving exclusively particle modes from the inertial perspective.  To study particle and antiparticle entangled states we develop a generalisation of the formalism introduced in the previous chapter which relates general Unruh and Rindler modes. This formalism refines the single-mode approximation \cite{Alsingtelep,AlsingMcmhMil} which has been extensively used in the literature.  
 In particular, we will consider in our analysis a fermionic maximally entangled state which has no neutral bosonic analog.  This state which is  entangled in the particle/antiparticle degree of freedom can be produced, for example, in particle-antiparticle pair creation or in the production of Cooper pairs. The analysis of such states is only possible under the mode transformations we introduce here since the single mode approximation  does not hold in this case.

 Considering a more general set of states from the inertial perspective  allows us to understand that, in non-inertial frames, entanglement redistributes between particle and anti-particle modes.  This is a somewhat similar effect to that observed in the inertial case:  entanglement redistributes between spin and position degrees of freedom from the perspective of different inertial observers  \cite{peresterno2,Lamata}.  Interestingly, one can conclude that fermionic entanglement remains finite in the infinite acceleration limit due to this redistribution of entanglement, which does not occur in the bosonic case. Our results are in agreement with the results in previous chapters which show that main differences in the behaivour of entanglement in the bosonic and fermionic case are due to Fermi-Dirac and Bose-Einstein statistics, contrary to the idea that the dimension of the Hilbert space played an important role.

\section{Dirac field states for uniformly accelerated observers}\label{sec18m}

We consider a Dirac field in $1+1$ dimensions. The field can be expressed from the perspective of inertial and uniformly accelerated observers. In this section we introduce the transformations which relate the mode operators and states from both perspectives.  Such transformations have been introduced in chapter \ref{sma}  for particle states. Here we extend this results including transformations for anti-particle modes which will be needed in our analysis. 

Minkowski coordinates $(t,x)$ are an appropriate choice of coordinates to express the field from the perspective of inertial observers. However, in the uniformly accelerated case Rindler coordinates  $(\eta,\chi)$ must be employed.  The coordinate transformation is given 
 by \eqref{Rindlertransformation}.
 
The Dirac field $\phi$ satisfies the
equation  $\{i\gamma^{\mu}(\partial_{\mu}-\Gamma_{\mu})+m\}\phi=0$
where $\gamma^{\mu}$ are the Dirac-Pauli matrices and $\Gamma_{\mu}$ are spinorial
affine connections\footnote{See appendix \ref{appB}}. The field expansion in terms of the Minkowski and Rindler solutions of the Dirac equation was given in the previous chapter in equations \eqref{fermfieldminm7} and \eqref{fermfieldm7}
\begin{align}
\phi&=N_\text{M}\sum_k\left(c_{k,\text{M}}\, u^+_{k,\text{M}}+ d_{k,\text{M}}^\dagger \ u^{-}_{k,\text{M}}\right),\\*
\phi&=N_\text{\text{R}}\sum_j\left(c_{j,\text{I}} u^+_{j,\text{I}}+ d_{j,\text{I}}^\dagger u^{-}_{j,\text{I}}+c_{j,\text{II}} u^+_{j,\text{II}}+ d_{j,\text{II}}^\dagger  u^{-}_{j,\text{II}}\!\right).
\end{align}
The transformation between the Minkowski and Rindler particle and antiparticle modes is given by
\begin{align}
\nonumber u^+_{j,\text{M}}=&\sum_k\left[\alpha^\text{I}_{jk}u^+_{k,\text{I}} + \beta^{\text{I}}_{jk} u^-_{k,\text{I}}+\alpha^\text{II}_{jk}u^+_{k,\text{II}}+\beta^{\text{II}}_{jk} u^-_{k,\text{II}}\right],
\end{align}
\begin{align}
\nonumber u^{-}_{j,\text{M}}=&\sum_k\left[\gamma^\text{I}_{jk} u^+_{k,\text{I}} + \eta^{\text{I}}_{jk} u^{-}_{k,\text{I}}+\gamma^\text{II}_{jk}u^+_{k,\text{II}}+\eta^{\text{II}}_{jk} u^{-}_{k,\text{II}}\right].
\end{align}
As discussed in  section  \ref{secferme}, the coefficients which relate both set of modes are given by the Dirac inner product, so that he  Bogoliubov coefficients yield \cite{Jauregui,Langlois},
\begin{equation}\label{bogosolved}
\begin{array}{c}
\alpha^\text{I}_{jk}=e^{i\theta E_\Omega}\dfrac{1+i}{2\sqrt{\pi E_\omega}}\,\cos r_{\Omega}\,\delta_{s s'},\\[4mm]
\beta^\text{I}_{jk}=-e^{-i\theta E_\Omega}\dfrac{1+i}{2\sqrt{\pi E_\omega}}\,\sin r_{\Omega}\,\delta_{s s'},\\[4mm]
\gamma^\text{I}_{jk}=-\beta^\text{I*}_{jk}, \qquad \eta^\text{I}_{jk}=\alpha^\text{I*}_{jk},\\[4mm]
\alpha^\text{II}=(\alpha^\text{I})^*,\!\!\qquad \beta^\text{II}=(\beta^\text{I})^*,\!\!\qquad \gamma^\text{II}=(\gamma^\text{I})^*,\!\!\qquad \eta^\text{II}=(\eta^\text{I})^*,
\end{array}
\end{equation}
where $\tan r_{\Omega}=e^{-\pi E_\Omega}$, $E_\Omega$ is the energy of the Rindler mode $k$, $E_\omega$ is the energy of the Minkowski mode $j$ and $\theta$ is a parameter defined such that it satisfies the conditions $E_\Omega=m\cosh\theta$ and $|\bm k_\Omega|=m\sinh\theta$   (see \cite{Jauregui}). 
Finally, taking into account that $c_{j,\text{M}}=(\phi,u^+_{j,\text{M}})$ and $d^\dagger_{j,\text{M}}=(\phi,u^{-}_{j,\text{M}})$,  we find 
\begin{align}
\nonumber c_{j,\text{M}}=\sum_{k}\left[\alpha^{\text{I}*}_{jk}c_{k,\text{I}} + \beta^{\text{I}*}_{jk} d_{k,\text{I}}^\dagger+\alpha^{\text{II}*}_{jk}c_{k,\text{II}}+\beta^{\text{II}*}_{jk}d_{k,\text{II}}^{\dagger}\right],\\
d^\dagger_{j,\text{M}}=\sum_{k}\left[\gamma^{\text{I}*}_{jk}c_{k,\text{I}} + \eta^{\text{I}*}_{jk} d_{k,\text{I}}^\dagger+\gamma^{\text{II}*}_{jk}c_{k,\text{II}}+\eta^{\text{II}*}_{jk}d_{k,\text{II}}^{\dagger}\right].
\end{align}

We now consider the transformations between states in different basis. For this we define an arbitrary element of the Dirac field Fock basis for each mode as
\begin{equation} \label{tensprod}
\ket{F_k}=\ket{F_k}_\text{\text{R}}\otimes \ket{F_k}_\text{\text{L}},
\end{equation}
where 
\begin{equation}
\ket{F_k}_\text{R}=\ket{n}^+_\text{I}\ket{m}^-_\text{II},\qquad\ket{F_k}_\text{L}=\ket{p}^-_\text{I}\ket{q}^+_\text{II}.
\end{equation}
In the same fashion as in the previous chapter, here the $\pm$ signs  denote particle or antiparticle. It is now convenient to introduce a new basis for inertial observers which corresponds to a superposition of Minkowski monocromatic modes. As seen in the previous chapter, the reason for this is that the new modes (Unruh modes), and Rindler modes have a simple Bogoliubov transformation: each Unruh mode transforms to a single frequency Rindler mode.  This transformation is given by
\begin{align}\label{Unruhopm8}
\nonumber C_{k,\text{\text{R}}}=&\left(\cos r_k\, c_{k,\text{I}}-\sin r_k\, d^\dagger_{k,\text{II}}\right),\\*
\nonumber C_{k,\text{\text{L}}}=&\left(\cos r_k\, c_{k,\text{II}}-\sin r_k\, d^\dagger_{k,\text{I}}\right),\\*
\nonumber D^\dagger_{k,\text{\text{R}}}=&\left(\sin r_k\, c_{k,\text{I}}+\cos r_k\, d^\dagger_{k,\text{II}}\right),\\*
D^\dagger_{k,\text{\text{L}}}=&\left(\sin r_k\, c_{k,\text{II}}+\cos r_k\, d^\dagger_{k,\text{I}}\right),
\end{align}
were $ C_{k,\text{\text{R,L}}}$ and $ D_{k,\text{\text{R,L}}}$ are the Unruh mode operators. 

The corresponding transformation between Minkowski and Unruh modes is given by
\begin{align}
c_{j,\text{M}}=&N_j\sum_k\Bigg[\chi^*(C_{k,\text{\text{R}}}\otimes\openone_\text{\text{L}})+\chi(\openone_\text{\text{R}}\otimes C_{k,\text{\text{L}}})\Bigg],\\
\label{annihil} d^\dagger_{j,\text{M}}=&N_j\sum_k\Bigg[\chi(D^\dagger_{k,\text{\text{R}}}\otimes\openone_\text{\text{L}})+\chi^*(\openone_\text{\text{R}}\otimes D^\dagger_{k,\text{\text{L}}})\Bigg],
\end{align}
where
\begin{equation}\label{ene}
N_j=\frac{1}{2\sqrt{\pi E_\omega}},\qquad  \chi=(1+i)e^{i\theta E_{\Omega}}.
\end{equation}
Here we have expressed the tensor product structure \eqref{tensprod} explicitly.

 For massless fields it can be shown \cite{ch1} that  the Unruh operators have the same form as Eq. \eqref{Unruhopm8} however in this case $\tan r_{k}=e^{-\pi\Omega/a}$.

In the massless case, to find the Minkowski vacuum in the Rindler basis we consider the same ansatz as in the previous chapter
\begin{equation}
\ket{0}_\text{M}=\bigotimes_{\Omega}\ket{0_\Omega}_\text{M},
\end{equation}
where $\ket{0_\Omega}_\text{M}=\ket{0_\Omega}_\text{\text{R}}\otimes\ket{0_\Omega}_\text{\text{L}}$.
We find that
\begin{eqnarray}
\label{vauno1}
\ket{0_{\Omega}}_\text{\text{R}}&=&\sum_{n,s}\left(F_{n,\Omega,s}\ket{n_{\Omega,s}}^+_\text{I}\ket{n_{\Omega,-s}}^-_\text{II}\right)\nonumber \\
\ket{0_{\Omega}}_\text{\text{L}}&=&\sum_{n,s}\left(G_{n,\Omega,s}\ket{n_{\Omega,s}}^-_\text{I}\ket{n_{\Omega,-s}}^+_\text{II}\right),
\end{eqnarray}
where the notation and sign conventions are the same as in chapter \ref{sma}.

For the case of Grassmann scalars, we recall expression \eqref{vacgrassman}, obtained after imposing that $c_{\omega,\text{M}}\ket{0_\Omega}_\text{M}=0$ for all $\omega$ which is equivalent to $C_{\Omega,\text{\text{R}}}\ket{0_\Omega}_\text{\text{R}}=C_{\Omega,\text{\text{L}}}\ket{0_\Omega}_\text{\text{L}}=0$ for all $\Omega$.  The vacuum state then yields
 \begin{align}\label{vacgrassmanm81} 
\nonumber\ket{0_{\Omega}}&=\left(\cos r_{\Omega}\ket{0_{\Omega}}^+_\text{I}\ket{0_{\Omega}}^-_\text{II}+\sin r_{\Omega}\ket{1_{\Omega}}^+_\text{I}\ket{1_{\Omega}}^-_\text{II}\right)\\*
&\otimes\left(\cos r_{\Omega}\ket{0_{\Omega}}^-_\text{I}\ket{0_{\Omega}}^+_\text{II}-\sin r_{\Omega}\ket{1_{\Omega}}^-_\text{I}\ket{1_{\Omega}}^+_\text{II}\right)\!,
\end{align}
Using equation \eqref{annihil} we see that this vacuum state also satisfies $d_{\omega,\text{M}}\ket{0_\Omega}_\text{M}=0\; \forall\omega$, which is equivalent to $D_{\Omega,\text{\text{R}}}\ket{0_\Omega}_\text{\text{R}}=D_{\Omega,\text{\text{L}}}\ket{0_\Omega}_\text{\text{L}}=0\;\forall\Omega$.
For convenience, we introduce the following compact notation, \begin{equation}\label{shortnot1}
\ket{i j k l}_{\Omega}\equiv\ket{i_{\Omega}}^+_\text{I}\ket{j_{\Omega}}^-_\text{I}\ket{k_{\Omega}}^+_\text{II}\ket{l_{\Omega}}^-_\text{II}.
\end{equation}

Notice that this notation for our basis is slightly different from the one employed in the previous chapter (equation \eqref{shortnot}). We do this because it is more convenient for our purposes in this chapter. In this notation the vacuum state is written as,
\begin{equation}\label{shortvac1}
\ket{0_{\Omega}}=\cos^2 r_{\Omega}\ket{0000}_{\Omega}-\sin r_{j}\cos r_{\Omega}\ket{0110}_{\Omega}+\sin r_{\Omega}\cos r_{\Omega} \ket{1001}_{\Omega}-\sin^2 r_{\Omega} \ket{1111}_{\Omega}.
\end{equation}
instead of \eqref{shortvac}.

The Minkowskian one particle/antiparticls state is obtained by applying the creation operator of particle or antiparticle to the vacuum state $\ket{1_{j}}^+_{\text{U}}=c_{\Omega,\text{U}}^\dagger\ket{0}_\text{M}$, $\ket{1_{j}}^-_{\text{U}}=d_{\Omega,\text{U}}^\dagger\ket{0}_\text{M}$
where the Unruh particle/antiparticle creator is a combination of the two Unruh operators 
\begin{align}\label{creat1}
c_{\Omega,\text{U}}^\dagger=&q_\text{\text{R}}(C^\dagger_{\Omega,\text{\text{R}}}\otimes\openone_\text{\text{L}})+q_\text{\text{L}}(\openone_\text{\text{R}}\otimes C^\dagger_{\Omega,\text{\text{L}}}),\nonumber\\
d_{\Omega,\text{U}}^\dagger=&p_\text{\text{R}}(D^\dagger_{\Omega,\text{\text{R}}}\otimes\openone_\text{\text{L}})+p_\text{\text{L}}(\openone_\text{\text{R}}\otimes D^\dagger_{\Omega,\text{\text{L}}}).
\end{align}
$q_\text{\text{R}},q_\text{\text{L}},p_\text{\text{R}},p_\text{\text{L}}$ are complex numbers satisfying $|q_\text{\text{R}}|^2+|q_\text{\text{L}}|^2=1$,  $|p_\text{\text{R}}|^2+|p_\text{\text{L}}|^2=1$.

The parameters $p_\text{R,L}$  are not independent of $q_\text{R,L}$. We demand that the Unruh particle and antiparticle operators are referred to particle and antiparticle modes in the same Rindler wedges. Therefore to be coherent with a particular election of $q_\text{R}$ and $q_\text{L}$,  we have to choose $p_\text{L}=q_\text{R}$ and $p_\text{R}=q_\text{L}$, 
\begin{align}\label{creat2}
c_{k,\text{U}}^\dagger=&q_\text{\text{R}}(C^\dagger_{\Omega,\text{\text{R}}}\otimes\openone_\text{\text{L}})+q_\text{\text{L}}(\openone_\text{\text{R}}\otimes C^\dagger_{\Omega,\text{\text{L}}}),\nonumber\\
d_{k,\text{U}}^\dagger=&q_\text{\text{L}}(D^\dagger_{\Omega,\text{\text{R}}}\otimes\openone_\text{\text{L}})+q_\text{\text{R}}(\openone_\text{\text{R}}\otimes D^\dagger_{\Omega,\text{\text{L}}}).
\end{align}

The Unruh L and R field excitations are given by
\begin{align}
\nonumber\ket{1_{\Omega}}^+_\text{\text{R}}&=C^\dagger_{\Omega,\text{\text{R}}}\ket{0_{\Omega}}_\text{\text{R}}=\ket{1_{\Omega}}^+_\text{I}\ket{0_{\Omega}}^-_\text{II},\\*
\nonumber \ket{1_{\Omega}}^+_\text{\text{L}}&=C^\dagger_{\Omega,\text{\text{L}}}\ket{0_{\Omega}}_\text{\text{L}}=\ket{0_{\Omega}}^-_\text{I}\ket{1_{\Omega}}^+_\text{II},\\
\nonumber\ket{1_{\Omega}}^-_\text{\text{R}}&=D^\dagger_{\Omega,\text{\text{R}}}\ket{0_{\Omega}}_\text{\text{R}}=\ket{0_{\Omega}}^+_\text{I}\ket{1_{\Omega}}^-_\text{II},\\*
\ket{1_{\Omega}}^-_\text{\text{L}}&=D^\dagger_{\Omega,\text{\text{L}}}\ket{0_{\Omega}}_\text{\text{L}}=\ket{1_{\Omega}}^-_\text{I}\ket{0_{\Omega}}^+_\text{II},
\end{align}
and therefore,
\begin{align}
\nonumber \ket{1_{k}}^+_\text{U}&=q_\text{\text{R}}\ket{1_{\Omega}}^+_\text{\text{R}}\otimes\ket{0_{\Omega}}_\text{\text{L}}+q_\text{\text{L}}\ket{0_{\Omega}}_\text{\text{R}}\otimes\ket{1_{\Omega}}^+_\text{\text{L}},\\
\ket{1_{k}}^-_\text{U}&=q_\text{\text{L}}\ket{1_{\Omega}}^-_\text{\text{R}}\otimes\ket{0_{\Omega}}_\text{\text{L}}+q_\text{\text{R}}\ket{0_{\Omega}}_\text{\text{R}}\otimes\ket{1_{\Omega}}^-_\text{\text{L}}.
\end{align}
In the short notation we have introduced the state reads,  
\begin{align}\label{onegrassman}
\nonumber\ket{1_{k}}^+_\text{U}&=q_\text{\text{R}}\left[\cos r_{k}\ket{1000}_{\Omega}-\sin r_{\Omega}\ket{1110}_{\Omega}\right]+q_\text{\text{L}}\left[\cos r_{\Omega}\ket{0010}_{\Omega}+\sin r_{\Omega}\ket{1011}_{\Omega}\right],\nonumber\\[4mm]
\ket{1_{k}}^-_\text{U}&= q_\text{\text{L}}\left[\cos r_{k}\ket{0001}_{\Omega}-\sin r_{\Omega}\ket{0111}_{\Omega}\right]+q_\text{\text{R}}\left[\cos r_{\Omega}\ket{0100}_{\Omega}+\sin r_{\Omega}\ket{1101}_{\Omega}\right].
\end{align}

\section{Particle and Anti-particle entanglement in non-inertial frames}\label{sec2}

Having the expressions for the vacuum and single particle states in the  Unruh and Rindler bases enables us to analyse the degradation of entanglement from the perspective of observers in uniform acceleration.  Let us consider the following maximally entangled states from the inertial perspective
\begin{align}
\label{1e}\ket{\Psi_+}&=\frac{1}{\sqrt2}\left(\ket{0_\omega}_{\text M}\ket{0_\Omega}_\text{U}+\ket{1_\omega}^\sigma_{\text{M}}\ket{1_\Omega}^+_\text{U}\right),\\*
\label{2e}\ket{\Psi_-}&=\frac{1}{\sqrt2}\left(\ket{0_\omega}_{\text M}\ket{0_\Omega}_\text{U}+\ket{1_\omega}^\sigma_{\text{M}}\ket{1_\Omega}^-_\text{U}\right),\\*
\label{3e}\ket{\Psi_1}&=\frac{1}{\sqrt2}\left(\ket{1_\omega}^+_{\text M}\ket{1_\Omega}^-_\text{U}+\ket{1_\omega}^-_{\text{M}}\ket{1_\Omega}^+_\text{U}\right),
\end{align}
where the modes labeled with $\text{U}$ are Grassmann Unruh modes  and the label $\sigma=\pm$ denotes particle and antiparticle modes. The first two states correspond to entangled states with particle and antiparticle Unruh excitations, respectively. These two states are analogous to the bosonic state $\frac{1}{\sqrt2}(\ket{0}_M\ket{0}_U+\ket{1}_M\ket{1}_U)$ which is entangled in the  occupation number degree of freedom.  The third state has no analog in the neutral bosonic scenario since in this case the state is entangled in the particle/antiparticle degree of freedom. In spite of the fact that fermionic entanglement in non-inertial frames  has been extensively studied in the literature \cite{AlsingSchul},  states \eqref{2e} and \eqref{3e} have not been considered before.

We consider Alice to be an inertial observer with a detector sensitive to $\omega$ modes while her partner Rob who is in uniform acceleration carries a detector sensitive to $\Omega$ modes.  To  study the entanglement in the states  from their perspective we must rewrite the $\Omega$ modes in terms of Rindler modes.  Therefore, Unruh states must be transformed into the Rindler basis. The state in the Minkowski-Rinder basis becomes effectively  a tri-partite system.  As it is commonplace in the literature,  we define the Alice-Rob bipartition as the Minkowski and region I Rindler modes, while the Alice-AntiRob bipartition corresponds to Minkowski and region II Rindler modes.  To study distillable entanglement we will employ the negativity $\mathcal{N}$ as in previous chapters. 

Two cases of interest will be considered. In the first case we assume that Alice and Rob have detectors which do not distinguish between particles and antiparticles. In this case,  particles and antiparticles together are considered to be a subsystem. In the second case we consider that Rob and AntiRob have detectors which are only sensitive to particles (antiparticles) therefore, antiparticle (particle) states must be traced out. Our results will show that when Rob is accelerated, the entanglement redistributes between particles and antiparticles as a function of his acceleration. This effect is a unique feature of fermionic fields and plays an important role in the behaviour of fermionic  entanglement in the infinite acceleration limit.

\subsection{Entanglement in states $\ket{\Psi_+}$ and $\ket{\Psi_-}$ }

\subsubsection{Full sensitivity}

Let us revisit the results presented in the previous chapter where Rob was sensitive to both, particle and antiparticle modes. To compute Alice-Rob partial density matrix  in \eqref{1e}  we trace over region II in $\proj{\Psi_+}{\Psi_+}$ and obtain,  
\begin{align}\label{ARd}
\nonumber\rho^+_{\text{AR}}&=\frac{1}{2}\Big[C^4\proj{000}{000}+S^2C^2(\proj{010}{010}+\proj{001}{001})+S^4\proj{011}{011}\\*
\nonumber&+\phantom{\Big[}|q_\text{\text{R}}|^2(C^2\proj{110}{110}+S^2\proj{111}{111})+|q_\text{\text{L}}|^2(S^2\!\proj{110}{110}+C^2\!\proj{100}{100})\\*
\nonumber&+q_\text{\text{R}}^*(C^3\!\proj{000}{110}+S^2C\!\proj{001}{111})-q_\text{\text{L}}^*(C^2S\!\proj{001}{100}+S^3\!\proj{011}{110})\\*
&-q_\text{R} q_\text{L}^*SC\!\proj{111}{100}\Big]+(\text{H.c.})_{_{\substack{\text{non-}\text{diag.}}}},
\end{align}
where $C=\cos r_{\Omega}$ and $S=\sin r_{\Omega}$. 

The density matrix for the Alice-AntiRob modes is obtained by tracing over region~$\text{I}$,
\begin{align}\label{AaRd}
\nonumber\rho^+_{\text{A}{\bar{\text{R}}}}&=\frac{1}{2}\Big[C^4\proj{000}{000}+S^2C^2(\proj{001}{001}+\proj{010}{010})+S^4\proj{011}{011}\\*
\nonumber&+|q_\text{\text{R}}|^2(C^2\proj{100}{100}+S^2\proj{110}{110})+|q_\text{\text{L}}|^2(S^2\proj{111}{111}+C^2\!\proj{110}{110})\\*
\nonumber&+\phantom{\Big[}q_\text{\text{L}}^*(C^3\!\proj{000}{110}+S^2C\!\proj{001}{111})+q_\text{\text{R}}^*(C^2S\!\proj{001}{100}+S^3\!\proj{011}{110})\\*
&+q_\text{R} q_\text{L}^*SC\!\proj{100}{111}\Big]+(\text{H.c.})_{_{\substack{\text{non-}\\\text{diag.}}}}.
\end{align}

\begin{figure}[h]
\begin{center}
\includegraphics[width=.80\textwidth]{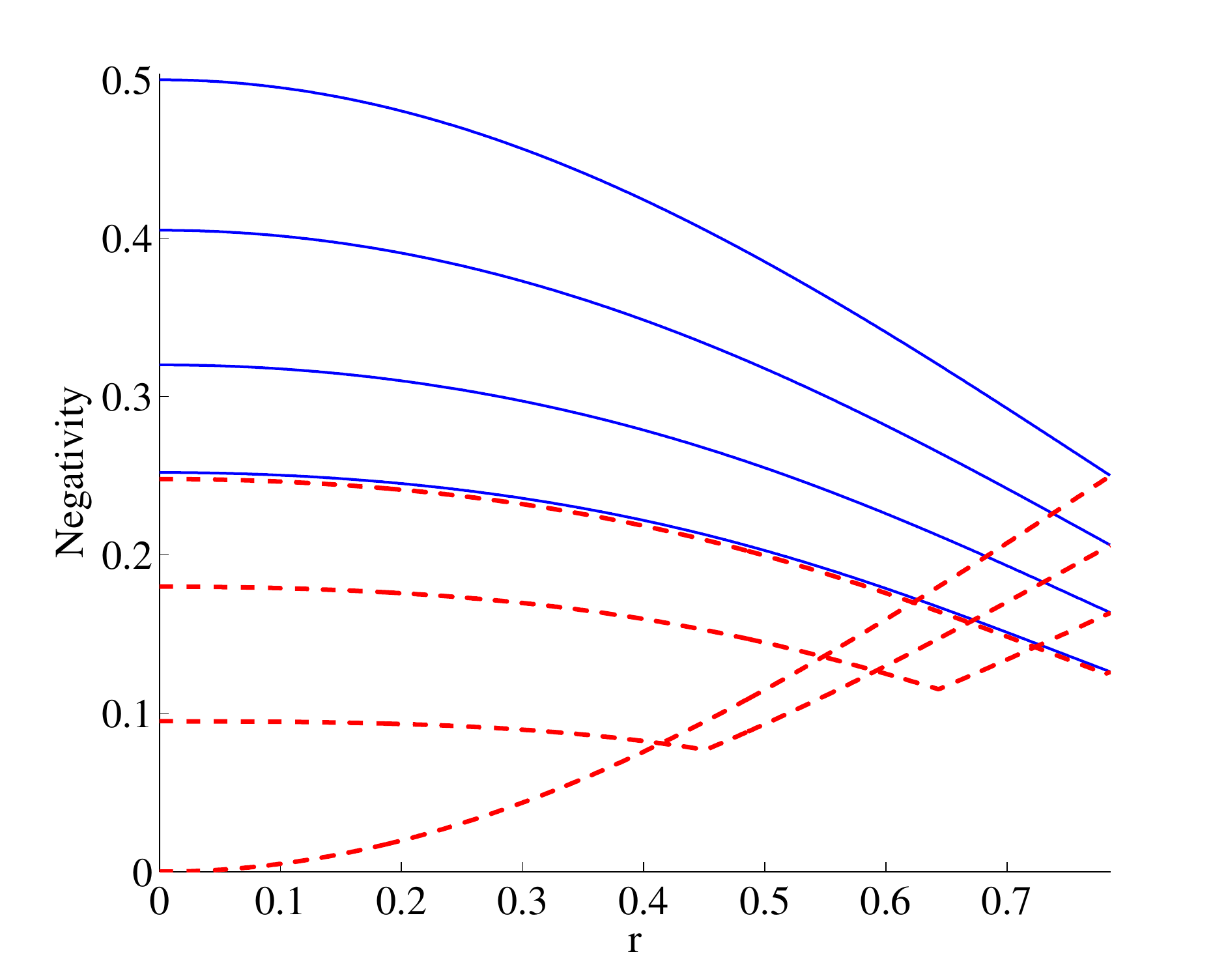}
\end{center}
\caption{Full sensitivity Negativity for the Alice-Rob (Blue continuous) and Alice-AntiRob (red dashed) bipartitions as a function of $r_{\Omega}=\arctan{e^{-\pi\Omega/a}}$ for various choices of $|q_\text{\text{R}}|$. The blue continuous (red dashed) curves from top to bottom (from bottom to top) correspond, to $|q_\text{\text{R}}|=1,0.9,0.8,0.7$  respectively.}
\label{N0}
\end{figure}

These are the same matrices \eqref{ek1} and \eqref{ek2} but with the different notation employed in this chapter. Negativity for the bipartitions AR and $\text{A}{\bar{\text{R}}}$ were calculated in chapter \ref{sma}, and are shown again in Figure \ref{N0}.

As discussed in the previous chapter, when the acceleration increases the entanglement between Alice-AntiRob modes is created compensating the entanglement lost between Alice-Rob. An analysis of the quantum entanglement between Alice's modes and particle/antiparticle modes of Rob and AntiRob will be useful to disclose why correlations present this behaviour. 

\subsubsection{Sensitivity to particles}

We now analyse the entanglement when Rob's and AntiRob's detectors are not able to detect antiparticles.  In this case the entanglement  is between their particle modes and Alice's subsystem. Since Rob cannot detect antiparticles we must trace over all antiparticle states in \eqref{ARd}: $\rho^+_{\text{AR}^+}=\sum_{n=0,1}\bra{n}_{\text{I}}^-\rho^+_{\text{AR}}\ket{n}_{\text{I}}^-$. This yields
\begin{align}
\nonumber\rho^+_{\text{AR}^+}&=\frac{1}{2}\Big[C^2\proj{00}{00}+S^2\proj{01}{01}+q_\text{\text{R}}^*C\proj{00}{11})+(|q_\text{\text{R}}|^2+|q_\text{\text{L}}|^2S^2)\proj{11}{11}\\*
&+|q_\text{\text{L}}|^2C^2\proj{10}{10})\Big]+(\text{H.c.})_{_{\substack{\text{non-}\text{diag.}}}},
\end{align}
which is the partial state of Alice and the particles sector of Rob.

The partial transpose $(\rho^+_{\text{AR}^+})^{pT}$ has only one block whose negative eigenvalue contributes to negativity
\begin{itemize}
\item Basis: $\{\ket{10},\ket{01}\}$
\end{itemize}
\begin{equation}
\frac12\left(\begin{array}{cc}
|q_\text{\text{L}}|^2C^2&q_\text{\text{R}}^*C\\
q_\text{\text{R}}C& S^2  
\end{array}\right).
\end{equation}

\begin{figure}[h]
\begin{center}
\includegraphics[width=.85\textwidth]{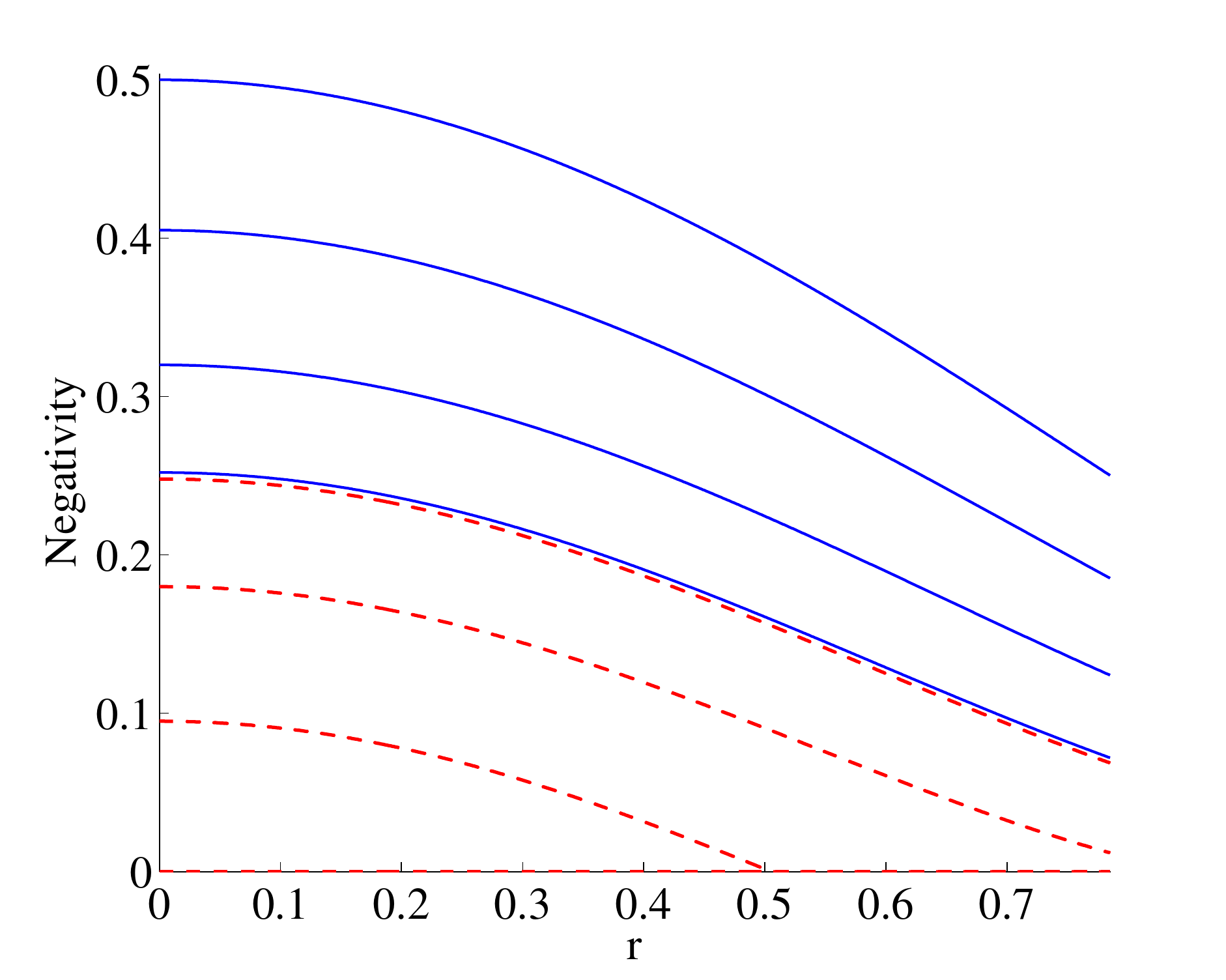}
\end{center}
\caption{Negativity in the particle sector (Rob and AntiRob can only detect particle modes) for the bipartition Alice-Rob (Blue continuous) and Alice-AntiRob (red dashed) for the state \eqref{1e} as a function of $r_{\Omega}=\arctan{e^{-\pi\Omega/a}}$ for various choices of $|q_\text{\text{R}}|$. The blue continuous (red dashed) curves from top to bottom (from bottom to top) correspond, to $|q_\text{\text{R}}|=1,0.9,0.8,0.71$  respectively. For $|q_R|=1$ Alice-AntiRob curve is zero $\forall a$.}
\label{Npp}
\end{figure}

The same procedure can be carried out for the system $A\bar R$ tracing over the antiparticle sector in \eqref{AaRd} obtaining 
\begin{align}
\nonumber\rho^+_{\text{A}{\bar{\text{R}}}^+}&=\frac{1}{2}\Big[C^2\proj{00}{00}+S^2\proj{01}{01}+q_\text{\text{L}}^*C\proj{00}{11}+(|q_\text{L}|^2+|q_\text{R}|^2S^2)\proj{11}{11}\\*
&+|q_\text{\text{R}}|^2C^2\proj{10}{10}\Big]+(\text{H.c.})_{_{\substack{\text{non-}\\\text{diag.}}}}.
\end{align}
The partial transpose $(\rho^+_{\text{A}{\bar{\text{R}}}^+})^{pT}$ has only one block whose negative eigenvalue contributes to negativity
\begin{itemize}
\item Basis: $\{\ket{10},\ket{01}\}$
\end{itemize}
\begin{equation}
\frac12\left(\begin{array}{cc}
|q_\text{\text{R}}|^2C^2&q_\text{\text{L}}^*C\\
q_\text{\text{L}}C& S^2  
\end{array}\right).
\end{equation}
Entanglement for $\rho^+_{\text{AR}^+}$ and $\rho^+_{\text{A}{\bar{\text{R}}}^+}$ is plotted in Fig. \ref{Npp}.

\subsubsection{Sensitivity to antiparticles}

A similar calculation can be carried out considering that Rob and AntiRob detectors are only sensitive to antiparticles, i.e. tracing over particle states. In this case we obtain that $\rho^+_{\text{AR}^-}=\sum_{n=0,1}\bra{n}_{\text{I}}^+\rho^+_{\text{AR}}\ket{n}_{\text{I}}^+$, and therefore, from \eqref{ARd},
\begin{align}
\nonumber\rho^+_{\text{AR}^-}&=\frac{1}{2}\Big[C^2\proj{00}{00}+S^2\proj{01}{01}-q_\text{\text{L}}^*S\proj{01}{10}+(|q_\text{\text{L}}|^2+|q_\text{\text{R}}|^2 C^2)\proj{10}{10}\\*
&+|q_\text{\text{R}}|^2 S^2\proj{11}{11})\Big]+(\text{H.c.})_{_{\substack{\text{non-}\text{diag.}}}}
\end{align}
is the partial state of Alice and the particles sector of Rob.

The only block giving negative eigenvalues is
\begin{itemize}
\item Basis: $\{\ket{11},\ket{00}\}$
\end{itemize}
\begin{equation}
\frac12\left(\begin{array}{cc}
|q_\text{\text{R}}|^2S^2&-q_\text{\text{L}}^*S\\
-q_\text{\text{L}}S& C^2  
\end{array}\right).
\end{equation}

The density matrix for the Alice-AntiRob antiparticle modes is obtained by tracing over the particle sector in \eqref{AaRd}
\begin{align}
\nonumber\rho^+_{\text{A}{\bar{\text{R}}}^-}&=\frac{1}{2}\Big[C^2\proj{00}{00}+S^2\proj{01}{01}+(|q_\text{\text{R}}|^2+|q_\text{\text{L}}|^2C^2)\proj{10}{10}\\*
&+|q_\text{\text{L}}|^2S^2\!\proj{11}{11}+q_\text{\text{R}}^*S\proj{01}{10})\Big]+(\text{H.c.})_{_{\substack{\text{non-}\\\text{diag.}}}}.
\end{align}

\begin{figure}[h]
\begin{center}
\includegraphics[width=.85\textwidth]{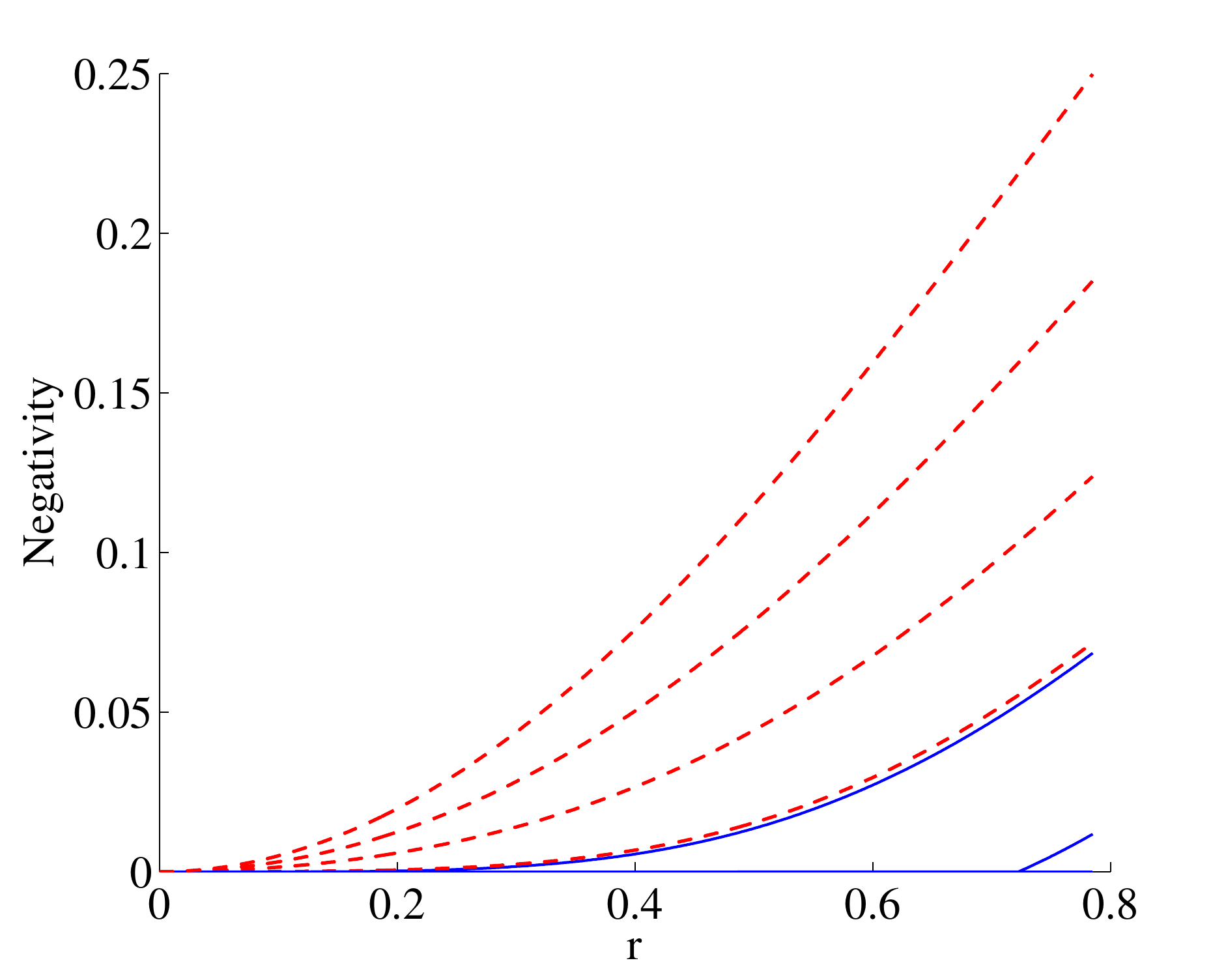}
\end{center}
\caption{Negativity in the antiparticle sector (Rob and AntiRob can only detect antiparticle modes) for the bipartition Alice-Rob (Blue continuous) and Alice-AntiRob (red dashed) for the state \eqref{1e} as a function of $r_{\Omega}=\arctan{e^{-\pi\Omega/a}}$ for various choices of $|q_\text{\text{R}}|$. The red dashed curves from top to bottom correspond, to $|q_\text{\text{R}}|=1,0.9,0.8,0.71$  respectively. The two blue curves seen correspond (from bottom to top) to $|q_\text{\text{R}}|=0.8,0.71$ respectively.}
\label{Npm}
\end{figure}

Again only one block of the density matrix contributes to negativity
\begin{itemize}
\item Basis: $\{\ket{11},\ket{00}\}$
\end{itemize}
\begin{equation}
\frac12\left(\begin{array}{cc}
|q_\text{\text{L}}|^2S^2&q_\text{\text{R}}^*S\\
q_\text{\text{R}}S& C^2  
\end{array}\right).
\end{equation}
Entanglement for $\rho^+_{\text{AR}^-}$ and $\rho^+_{\text{A}{\bar{\text{R}}}^-}$ is plotted in Fig. \ref{Npm}.

The analysis of entanglement in the state \eqref{2e} is done in a completely analogous way. We find that the entanglement behaves exactly the same way as in state \eqref{1e} only that the role of particles is replaced by anti-particles. Therefore negativities  are related in the following way:
\begin{equation*}\mathcal{N}_{\text{AR}^+}^+=\mathcal{N}_{A R^-}^-,\quad\mathcal{N}_{\text{A}{\bar{\text{R}}}^+}^+=\mathcal{N}_{\text{A}{\bar{\text{R}}}^-}^-,\end{equation*}
\qquad\\[-13mm]
\begin{equation*}\mathcal{N}_{\text{AR}^-}^+=\mathcal{N}_{A R^+}^-,\quad\mathcal{N}_{\text{A}{\bar{\text{R}}}^-}^+=\mathcal{N}_{\text{A}{\bar{\text{R}}}^+}^-,\end{equation*}
\begin{equation}\mathcal{N}_{\text{AR}}^+=\mathcal{N}_{A R}^-,\quad\mathcal{N}_{\text{A}{\bar{\text{R}}}}^+=\mathcal{N}_{A \bar R}^-.\end{equation}

We see in Figure \ref{N0} that the total entanglement between Alice and AntiRob starts decreasing and presents a minimum before starting to grow again for higher accelerations. If $|q_R|<1$ the entanglement in the limit $a\rightarrow0$ is distributed between the bipartitions $AR$ and $A\bar R$ as shown in chapter \ref{sma}. The entanglement lost in the bipartition  $A\bar R$ is not entirely compensated by the creation of entanglement in  $A\bar R$  and therefore, this results in a state containing less entanglement in the infinite acceleration limit. 

Interestingly, the correlations between Alice and the particle sector of Rob and AntiRob always decrease while the correlations between Alice and the antiparticle sector of Rob and AntiRob always grow. This behaviour explains why entanglement always survive the infinite acceleration limit for any election of $q_R$ and $q_L$. As Rob accelerates there is a process of entanglement transfer between the particle and antiparticle sector of his Hilbert space. The same happens with AntiRob, such that for neither  $AR$ nor $A\bar R$ the entanglement vanishes for any value of the acceleration.

For the simplest case $|q_R|=1$ we see that all the entanglement is initially ($a\rightarrow0$) in the particle sector of the bipartition $AR$. As the acceleration increases the entanglement is transferred to the antiparticle sector of the bipartition $A\bar R$ such that, in the limit of infinite acceleration entanglement has been equally distributed between these two bipartitions.  

The tensor product structure of the particle and antiparticle sectors plays an important role in the behaviour of entanglement in the infinite acceleration limit.  In the case of neutral scalar fields there are no antiparticles and entanglement is completely degraded. Note that in the case of charged bosonic fields there are indeed charged conjugate antiparticles. However, in this case the Hilbert space has a similar structure to the uncharged field  \cite{Alicefalls}. The existence of  bosonic antiparticles simply adds another copy of the same Hilbert space and no entanglement transfer is possible between particle and antiparticles. 

Now, if we move to less trivial cases where $q_R\neq 1$ the situation gets more complicated. In these scenarios we initially start with some entanglement in the particle sector of $A\bar R$, and there can also be an entanglement transfer between this sector and $AR$ antiparticle sector. However  no entanglement at all is transferred to the antiparticle sector of the subsystem $AR$ unless the acceleration reaches a threshold given by
\begin{equation}\label{thres}
  \cos^2r =\frac{|q_L|^2}{|q_R|^2}=\frac{1}{|q_R|^2}-1.
  \end{equation}
The maximum value of $\cos \left[r(a)\right]$ is $\cos \left[r(a\rightarrow\infty)\right]\rightarrow1/\sqrt2$  therefore for $|q_R|^2>2/3$  entanglement  is  not  transferred to the antiparticle sector of Alice-Rob for any value of the acceleration while the opposite (transfer from AR to $\text{A}{\bar{\text{R}}}$) always occurs. This explains why when $q_R\neq1$ the entanglement loss in the bipartition AR in the limit $a\rightarrow\infty$ (understood as the difference between the initial entanglement in $AR$ and the final entanglement)  is smaller than in the extreme case $q_R=1$.
 
It is evident that the choice of Unruh modes influence the transfer of entanglement between particle and antiparticle sectors. When the acceleration is larger than \eqref{thres},  when $|q_\text{R}|$ grows closer  to $1/\sqrt{2}$ more entanglement is transferred from the particle sector of $A\bar R$ to the antiparticles of $AR$. In the limit $q_\text{R}=1/\sqrt2$ the same amount of entanglement  is transferred to the antiparticle sector of both $AR$ and $A \bar R$.  

The particle entanglement (always monotonically decreasing) resembles the behaviour of bosonic entanglement studied in previous chapters. Bosonic entanglement is monotonically decreasing for both $AR$ and $A\bar R$ subsystems. In the bosonic case there are not antiparticles and, hence, there is no possibility of entanglement transfer to antiparticle sectors of $A\bar R$. This is the origin of the differences in entanglement behaviour for bosons and fermions.

\subsection{Entanglement in state $\ket{\Psi_1}$}

This state has no neutral bosonic analog since it is entangled in the particle/antiparticle degree of freedom. Therefore, the analysis of entanglement in this state revels interesting features which are of genuinely fermionic nature.

To study this type of state we employ the generalisation of the formalism developed in chapter \ref{sma} which relates general Unruh modes with Rindler modes. This formalism refines the single-mode approximation introduced in \cite{Alsingtelep,AlsingMcmhMil}  which has been extensively used in the literature.  For this type of state the single mode approximation used in  \cite{AlsingSchul}  does not hold and attempting to use it may lead to misleading results: applied just as presented in \cite{AlsingSchul} one would find the erroneous result that maximally entangled states from the inertial perspective appear disentangled  from the accelerated perspective, irrespectively of the value of acceleration. Using the mode transformation introduced in section \ref{sec18m} leads to sensible results: acceleration behaves regularly for accelerated observers and approaches a maximally entangled state in the inertial limit. 

For convenience we will introduce a new notation for this case. For Alice, we denote the states by $\ket{+}$ if they correspond to particles and $\ket{-}$ for antiparticles. Therefore the state is written as
\begin{equation}
\ket{\Psi_1}=\frac{1}{\sqrt2}\left(\ket{+}_{\text M}\ket{1_\Omega}^-_\text{U}+\ket{-}_{\text{M}}\ket{1_\Omega}^+_\text{U}\right)
\end{equation}
The density matrix for the subsystem Alice-Rob is obtained from $\proj{\Psi_1}{\Psi_1}$  tracing over region II
\begin{align}
\nonumber\rho^1_{\text{AR}}&=|q_\text{R}|^2C^2\proj{+10}{+10}+|q_\text{R}|^2S^2\proj{+11}{+11}+|q_\text{L}|^2C^2\proj{+00}{+00} \\
\nonumber &+|q_\text{L}|^2S^2\proj{+10}{+10}-SCq_\text{R}q_\text{L}^*\proj{+11}{+00}+|q_\text{L}|^2C^2\proj{-00}{-00}\\
\nonumber&+|q_\text{L}|^2S^2\proj{-01}{-01}+|q_\text{R}|^2S^2\proj{-11}{-11}+SCq_\text{R}^*q_\text{L}\proj{-00}{-11}\\
&+|q_\text{R}|^2C^2\proj{-01}{-01}+(|q_\text{R}|^2C^2-|q_\text{L}|^2S^2)\proj{+10}{-01}+(\text{H.c.})_{_{\substack{\text{non-}\\\text{diag.}}}},
\end{align}
and for the Alice-AntiRob bipartition we obtain
\begin{align}
\nonumber\rho^1_{\text{A}{\bar{\text{R}}}}&=|q_\text{R}|^2C^2\proj{+00}{+00}+|q_\text{R}|^2S^2\proj{+10}{+10}+|q_\text{L}|^2S^2\proj{+11}{+11}\\
\nonumber&+|q_\text{L}|^2C^2\proj{+10}{+10} +SCq_\text{R}q_\text{L}^*\proj{+00}{+11}+|q_\text{L}|^2C^2\proj{-01}{-01}\\
\nonumber&+|q_\text{L}|^2S^2\proj{-11}{-11}+|q_\text{R}|^2C^2\proj{-00}{-00}+|q_\text{R}|^2S^2\proj{-01}{-01}\\
&-SCq_\text{R}^*q_\text{L}\proj{-11}{-00}+(|q_\text{L}|^2C^2-|q_\text{R}|^2S^2)\proj{+10}{-01}+(\text{H.c.})_{_{\substack{\text{non-}\\\text{diag.}}}}.
\end{align}

Assuming that the observers cannot distinguish between particles and antiparticles yields matrices where only one block of the partial transpose density matrix gives negative eigenvalues
\begin{itemize}
\item Basis: $\{\ket{-10},\ket{+01}\}$
\end{itemize}
\begin{equation}
\frac12\left(\begin{array}{cc}
0& (|q_\text{R}|^2C^2-|q_\text{L}|^2S^2)\\
(|q_\text{R}|^2C^2-|q_\text{L}|^2S^2)&0
\end{array}\right),
\end{equation}

In this case the negativity is given by \[\mathcal{N}^1_{\text{AR}}=\frac12\left||q_\text{R}|^2C^2-|q_\text{L}|^2S^2|\right|.\]

\begin{figure}[h]
\begin{center}
\includegraphics[width=.85\textwidth]{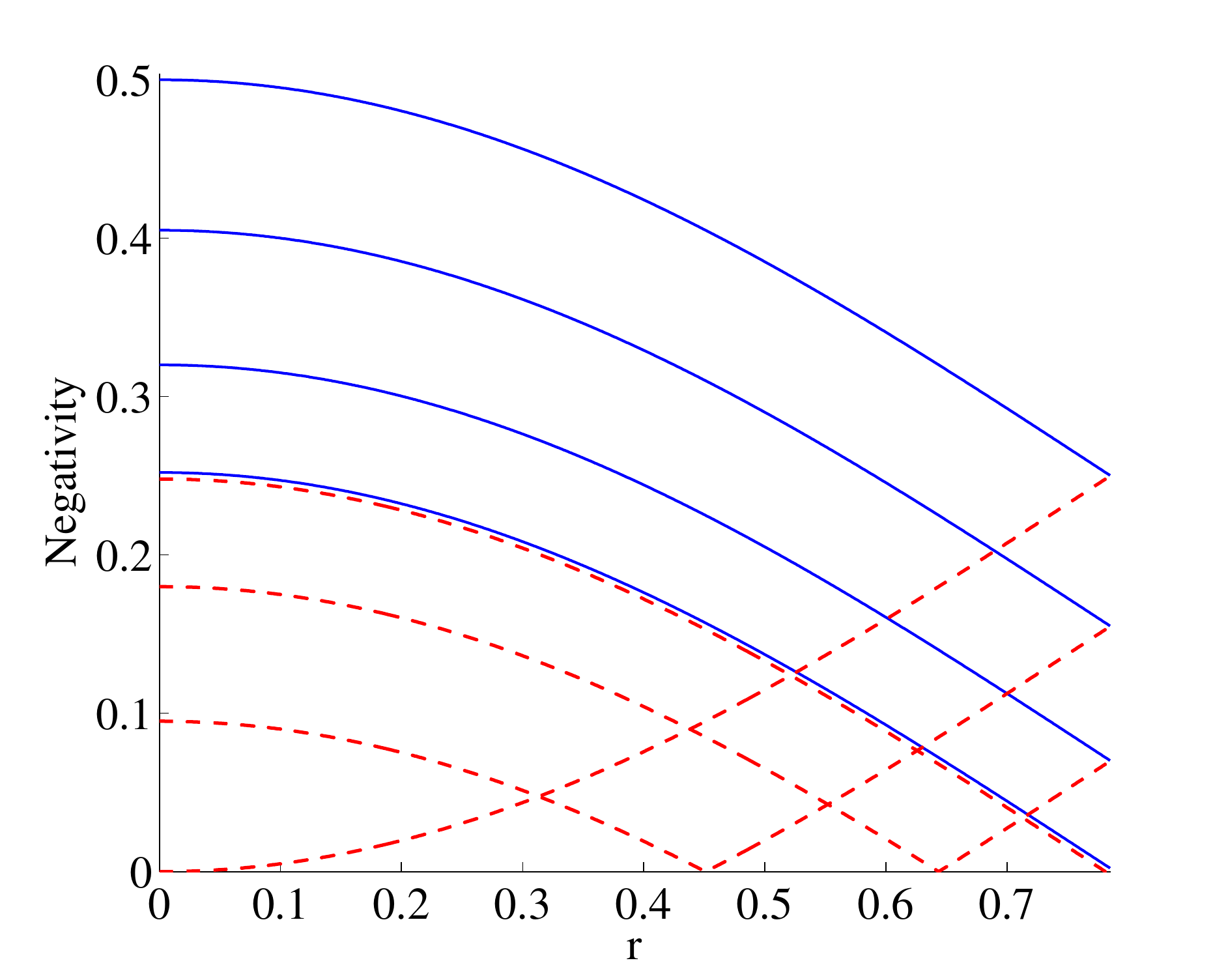}
\end{center}
\caption{Particle-antiparticle maximally entangled state \eqref{3e}: Negativity for the bipartition Alice-Rob (Blue continuous) and Alice-AntiRob (red dashed) as a function of $r_{\Omega}=\arctan{e^{-\pi\Omega/a}}$ for various choices of $|q_\text{\text{R}}|$. The blue continuous (red dashed) curves from top to bottom (from bottom to top) correspond, to $|q_\text{\text{R}}|=1,0.9,0.8,0.7$  respectively. The curves for Alice-AntiRob entanglement have a minimum where the negativity vanishes.}
\label{bundlefm8}
\end{figure}

A similar result is obtained for the system Alice-AntiRob $\rho^1_{\text{A}{\bar{\text{R}}}}$. In this case the only block of the partial transpose that contributes to negativity is
\begin{itemize}
\item Basis: $\{\ket{-10},\ket{+01}\}$
\end{itemize}
\begin{equation}
\frac12\left(\begin{array}{cc}
0& (|q_\text{L}|^2C^2-|q_\text{R}|^2S^2)\\
(|q_\text{L}|^2C^2-|q_\text{R}|^2S^2)&0
\end{array}\right),
\end{equation}
resulting in
\[\mathcal{N}^1_{\text{A}{\bar{\text{R}}}}=\frac12\left||q_\text{L}|^2C^2-|q_\text{R}|^2S^2|\right|.\]
Both negativities are shown in Fig. \ref{bundlefm8}.

Interestingly, when Rob and AntiRob are not able to detect either particle or antiparticle modes the entanglement in the state vanishes. The partial density matrices for $AR$ and $A\bar R$ in this case yield
\begin{align}
\nonumber\rho^1_{\text{AR}^+}&=(|q_\text{R}|^2+|q_\text{L}|^2S^2)\proj{+1}{+1}+|q_\text{L}|^2C^2\proj{+0}{+0} +(|q_\text{L}|^2+|q_\text{R}|^2C^2)\proj{-0}{-0}\\
&+|q_\text{R}|^2S^2\proj{-1}{-1}+(\text{H.c.})_{_{\substack{\text{non-}\\\text{diag.}}}},
\end{align}
\begin{align}
\nonumber\rho^1_{\text{A}{\bar{\text{R}}}^+}&=(|q_\text{L}|^2+|q_\text{R}|^2S^2)\proj{+1}{+1}+|q_\text{R}|^2C^2\proj{+0}{+0} +(|q_\text{R}|^2+|q_\text{L}|^2C^2)\proj{-0}{-0}\\*
\nonumber&+|q_\text{L}|^2S^2\proj{-1}{-1}+(\text{H.c.})_{_{\substack{\text{non-}\\\text{diag.}}}},
\end{align}
\begin{align}
\nonumber\rho^1_{\text{AR}^-}&=(|q_\text{L}|^2+|q_\text{R}|^2C^2)\proj{+0}{+0}+|q_\text{L}|^2C^2\proj{-0}{-0} +(|q_\text{R}|^2+|q_\text{L}|^2S^2)\proj{-1}{-1}\\*
&+|q_\text{R}|^2S^2\proj{+1}{+1}+(\text{H.c.})_{_{\substack{\text{non-}\\\text{diag.}}}},
\end{align}
\begin{align}
\nonumber\rho^1_{\text{A}{\bar{\text{R}}}^-}&=(|q_\text{R}|^2+|q_\text{L}|^2C^2)\proj{+0}{+0}+|q_\text{R}|^2C^2\proj{-0}{-0} +(|q_\text{L}|^2+|q_\text{R}|^2S^2)\proj{-1}{-1}\\*
&+|q_\text{L}|^2S^2\proj{+1}{+1}+(\text{H.c.})_{_{\substack{\text{non-}\\\text{diag.}}}},
\end{align}
for which negativity is strictly zero. The entanglement in this state is of a different nature as the entanglement in states $\ket{\Psi_+}$ and $\ket{\Psi_-}$,  therefore a direct comparison of the behaviour of entanglement cannot be done. The total entanglement here is associated to correlations between particles and antiparticles and, therefore, if we trace out either the particles or antiparticles  we effectively remove all the correlations codified in this degree of freedom. 

In the case when the detectors do not distinguish between particles and antiparticles we found that the entanglement in the Alice-AntiRob bipartition is degraded with acceleration vanishing at a critical point. For higher accelerations entanglement  begins to grow again. Namely, the entanglement on the bipartition Alice-AntiRob vanishes for a specific value of the acceleration if $|q_\text{R}|<1$. This value of the acceleration is given by \begin{equation}\tan^2 r=\frac{1}{|q_\text{R}|^2}-1.\end{equation}
What is more, the surviving entanglement in the limit $a\rightarrow\infty$ is 
\begin{equation}\mathcal{N}^1_{\text{AR}}(a\rightarrow\infty)=\mathcal{N}_{\text{A}{\bar{\text{R}}}}(a\rightarrow\infty)=\frac14(|q_\text{R}|^2-|q_\text{L}|^2),\end{equation}
therefore,  when $|q_\text{R}|=|q_\text{L}|=1/\sqrt{2}$ no entanglement survives in the limit of infinite acceleration. 

This shows that entanglement has a non-vanishing minimum value in the infinite acceleration limit (regardless of the election of Unruh modes) only  when there is transfer of entanglement between particles and antiparticles. Otherwise, it is possible to find an Unruh mode whose entanglement vanishes in the infinite acceleration limit as in the bosonic case. We therefore conclude that the entanglement transfer between particle and antiparticle sectors plays a key role in explaining the behaviour of entanglement in the infinite acceleration limit.

\section{Discussion}\label{conclusions}

Including antiparticles in the study of fermionic entanglement allowed us to understand key features which explain the difference in behaviour of entanglement  in the fermionic and bosonic case.  Namely, we have shown that there is an entanglement redistribution between the particle and antiparticle sectors when Rob is in uniform acceleration. This entanglement transfer is not possible in the bosonic case and, therefore, the differences between the bosonic and fermionic cases arise. In particular, we have shown that this entanglement tradeoff gives rise to a non-vanishing minimum value of fermionic entanglement in the infinite acceleration limit for any choice of Unruh modes.

We also exhibit a special fermionic state for which entanglement transfer between particle and antiparticle states is not possible. Interestingly, in this case we can find a specific choice of Unruh modes such that entanglement vanishes in the infinite acceleration limit.  Incidentally, this choice ($|q_\text{R}|=|q_{\text{L}}|=1/\sqrt2$) minimises the surviving entanglement of states \eqref{1e} and \eqref{2e}. We showed that it is the tradeoff between the particles and antiparticles sector what protected them from a complete entanglement loss.

Our analysis is based on an extension to antiparticles of the formalism introduced in chapter \ref{sma} which relates Unruh and Rindler modes.  This allowed us to analyse a more general family of fermionic maximally entangled states for which the single-mode approximation does not hold.  

This study sheds light in the understanding of  relativistic entanglement: the differences in bosonic and fermionic statistics give rise to differences in entanglement behaviour. This provides a deep insight on the mechanisms which makes fermionic entanglement more resilient to Unruh-Hawking radiation.

\chapter{Entanglement amplification via the Unruh-Hawking effect\footnote{M. Montero and E. Mart\'in-Mart\'inez. arXiv:1011.6540}}\label{generatio}

We have studied in previous chapters the influence of the so-called Unruh and Hawking effects on quantum entanglement. In all these previous studies it was shown how starting with entangled states from an inertial perspective we end up with a less entangled state when one of the observers is non-inertial. Due to those results, it has been always considered that the Unruh effect would only be able to degrade entanglement.

This belief was reinforced by the intuitive argument that since the Unruh effect acts in a similar way to a thermal bath with $T\propto a$,  entanglement should be degraded in a similar way as by thermal decoherence. This argument is flawed because the Unruh thermal state is derived for the Minkowski vaccuum state, not for states containing excitations. Nevertheless, it was supported  by the fact that all previous works under the single mode approximation (SMA) found that entanglement in the AR bipartition was a monotonically decreasing function of the acceleration. Also, for maximally entangled states beyond the SMA, the same monotonic behaviour of AR entanglement was found (see figures \ref{figbosons} and \ref{bundlefm7}). So it still seemed true that acceleration tends to degrade entanglement. When studying non-maximally entangled states these trends would be expected to continue holding in principle.

In this chapter we show an unexpected outcome of the Unruh and Hawking effects: the appearance of entanglement when one of the observers of a bipartite system undergoes a constant acceleration. We prove here that there are some states, shared by two observers, whose degree of entanglement increases as one of the observers accelerates. The phenomenon is thoroughly studied here for Grassmann scalar and bosonic scalar fields, it is thus not a peculiarity of fermionic statistics (as it was the entanglement survival in the infinite acceleration limit) but a universal phenomenon. 

This entanglement amplification is promising in order to detect quantum effects due to acceleration (and therefore gravity). Entanglement, unlike other phenomena (such as thermal noise) does not admit a classical description. Thus, its observation would account for a pure quantum origin of the aforementioned effects. 

On the other hand entanglement is very sensitive to any interaction with the environment, which tends to degrade it. This made it very difficult for any experiment relying on entanglement degradation \cite{Alicefalls} to find evidence for these effects. By the same token, experiments studying entanglement creation are safer from these flaws: if a small amount of entanglement is created, no matter how damped by decoherence it may be, the only possible origin is an acceleration-induced quantum effect. The entanglement amplification phenomenon provides a novel way to distinguish genuine quantum effects of gravity from classically induced ones, something worth considering when trying to detect the Unruh and Hawking effects in analog gravity set-ups \cite{garay}.  

The main reason why this phenomenon has gone unnoticed so far is the reliance in the single mode approximation (SMA)  that many previous works assumed. And, on the other hand when going beyond the single mode approximation, only maximally entangled states were studied. We need to study non-maximally entangled states beyond SMA to find the effect.

\section{The setting}

Let us consider a system composed of an inertial observer, Alice, who watches an inertial mode of a quantum field (either a Grassmann or bosonic scalar  field) and a uniformly accelerated Rindler observer either in region I or II of Rindler spacetime. As in previous chapters we will call this observer Rob if he is in region I and AntiRob if he is in region II. Rob (or AntiRob) watches an Unruh mode of the quantum field (see chapter \ref{sma}), which is entangled with Alice's. As we saw in chapter \ref{sma}, consideration of two different kinds of Unruh modes is necessary for a complete description of an arbitrary solution to the field equation.

 We will only consider Unruh modes of a given Rindler frequency $\omega$ as seen by Rob or AntiRob (who moves with proper acceleration $a$) but which are arbitrary superpositions of left and right mover modes, so that the creation and annihilation operators that we are considering here are  of the general form \eqref{eq:q-defs} for a scalar field and \eqref{creat} for the Grassmann scalar field. This is to say
 \begin{align}C_\omega=q_\text{L}C_{\omega,\text{L}}+\qr C_{\omega,\text{R}}\label{umodes},\end{align}
where $\vert q_\text{L}\vert^2+\vert\qr\vert^2=1$, $\qr \ge q_\text{L}$ and $C_{\omega,X}$ for $X=\text{L},\text{R}$ are
\begin{eqnarray}\label{bogoboson}
 C_{\omega,\text{R}}&=&\cosh r_{\text{b},\omega}\, a_{\omega,\text{I}} - \sinh r_{\text{b},\omega}\, a^\dagger_{\omega,\text{II}},\\*
 C_{\omega,\text{L}}&=&\cosh r_{\text{b},\omega}\, a_{\omega,\text{II}} - \sinh r_{\text{b},\omega}\, a^\dagger_{\omega,\text{I}}, \end{eqnarray}
where $a,a^\dagger$ are particle operators for the scalar field, and
\begin{eqnarray}\label{Unruhop}
\nonumber C_{\omega,\text{\text{R}}}&=&\left(\cos r_{\text{f},\omega}\, c_{\omega,\text{I}}-\sin r_{\text{f},\omega}\, d^\dagger_{\omega,\text{II}}\right),\\*
C_{\omega,\text{\text{L}}}&=&\left(\cos r_{\text{f},\omega}\, c_{\omega,\text{II}}-\sin r_{\text{f},\omega}\, d^\dagger_{\omega,\text{I}}\right),
\end{eqnarray}
where $c,c^\dagger$ and $d,d^\dagger$ are respectively particle and antiparticle operators, for the Grassmann case.

The family of bipartite states in which we will observe entanglament amplification is of the form 
\begin{align}\label{geGras} \ket{\Psi}&= P \ket{0}_\text{A} \left[\alpha\ket{1} +\sqrt{1-\alpha^2}\ket{0}\right] + \sqrt{1-P^2}\ket{1}_\text{A} \left[\beta\ket{1}+ \sqrt{1-\beta^2}\ket{0}\right]
.\end{align}
Here, the subscript `A' refers to Alice's inertial mode, and $\ket{1}=C^\dagger_\omega\ket{0}$ is the Unruh particle excitation. All these states have an implicit dependence on Rob's acceleration $a$ when expressed in the Rindler basis through the parameter defined by $\tan r_{\text{f},\omega}=e^{-\pi c\,\omega/a}$ in the fermionic case, and $\tanh r_{\text{b},\omega}=e^{-\pi c\,\omega/a}$ in the bosonic case.

As usual \cite{Alicefalls,AlsingSchul}, we transform the state \eqref{geGras} into the Rindler basis following the same conventions as in chapter \ref{sma}. Although the system is obviously bipartite, shifting to the Rindler basis for the second qubit the mathematical description of the system admits a straightforward tripartition: Minkowskian modes, Rindler region I modes, and Rindler region II  modes. 

The density matrix for the state, which includes modes on both wedges of the spacetime along with Minkowskian modes, is built from \eqref{geGras}. Namely, $\rho^{\text{AR}\bar{\text{R}}}=\proj{\Psi}{\Psi}$.

As discussed previously, an accelerated observer in region I is causally disconnected from region II (and vice-versa). For this reason when we consider the bipartite system Alice-Rob we need to trace over the modes that only have support in region II and to which Rob is causally disconnected. Equivalently, we would have to trace over modes in region I if we consider that the accelerated observer is in region II. The density matrices for the bipartite systems Alice-Accelerated observer are 
\begin{align}
\label{AR2}\rho^{\text{AR}}&=\tr_{\text{II}}\rho^{\text{AR}\bar{\text{R}}}=\sum_{n} \bra{n}_{\text{II}}\rho^{\text{AR}\bar{\text{R}}}\ket{n}_{\text{II}},\\*
\label{AAR2}\rho^{\text{A}\bar{\text{R}}}&=\tr_{\text{I}}\rho^{\text{AR}\bar{\text{R}}}=\sum_{n} \bra{n}_{\text{I}}\rho^{\text{AR}\bar{\text{R}}}\ket{n}_{\text{I}}.
\end{align}

The density matrix for the subsystem Alice-Rob is, therefore, given by
\begin{align}\label{rhog}\rho^{\text{A}{\text{R}}}&=\tr_\text{II}(\ket{\Psi}\bra{\Psi})=P^2\ket{0}_\text{A}\bra{0}_\text{A}\left[\alpha^2\tr_\text{II}(\ket{1}_\text{U}\bra{1}_\text{U})\right.+(1-\alpha^2)\tr_\text{II}(\ket{0}\bra{0})\nonumber\\&\!+\!\left.\alpha\sqrt{1\!-\!\alpha^2}\left(\tr_\text{II}(\ket{1}_\text{U}\bra{0})\!+\!\tr_\text{II}(\ket{0}\bra{1}_\text{U})\right)\right]\!+\!(1\!-\!P^2)\ket{1}_\text{A}\bra{1}_\text{A}\left[\beta^2\tr_\text{II}(\ket{1}_\text{U}\bra{1}_\text{U})\right.\nonumber\\&+(1-\beta^2)\tr_\text{II}(\ket{0}\bra{0})+\left.\beta\sqrt{1-\beta^2}\left(\tr_\text{II}(\ket{1}_\text{U}\bra{0})+\tr_\text{II}(\ket{0}\bra{1}_\text{U})\right)\right]\nonumber\\&+P\sqrt{1-P^2}\ket{1}_\text{A}\bra{0}_\text{A}\left[\alpha\beta\tr_\text{II}(\ket{1}_\text{U}\bra{1}_\text{U})\right.+\sqrt{1-\alpha^2}\sqrt{1-\beta^2}\tr_\text{II}(\ket{0}\bra{0})\nonumber\\&+\left.\beta\sqrt{1-\alpha^2}\tr_\text{II}(\ket{1}_\text{U}\bra{0})\right.\left.\alpha\sqrt{1-\beta^2}\tr_\text{II}(\ket{0}\bra{1}_\text{U})\right]+(\text{H.c.})_\text{Alice-Nondiag}\end{align}
for both cases, scalar field and Grassmann field. The kets and bras labeled with an A correspond to Alice and those inside the traces correspond to Rob.  by $(\text{H.c.})_\text{Alice-Nondiag}$ we mean the Hermitian conjugate of the terms with different Alice indices. 

The relevant matrices in \eq{rhog} are different for the scalar and the Grassmann case. Specifically, for the Grassmann scalar case
\begin{align}\tr_\text{II}(\ket{0}\bra{0})&=C^4\proj{00}{00}+S^2C^2(\proj{10}{10}+\proj{01}{01})+S^4\proj{11}{11},\nonumber\\\tr_\text{II}(\ket{1}_\text{U}\bra{1}_\text{U})&=\qr^2\left[C^2\proj{10}{10}+S^2\proj{11}{11}\right]+\ql^2\left[S^2\proj{10}{10}+C^2\proj{00}{00}\right]\nonumber\\
&-\qr\ql SC(\proj{11}{00}+\proj{00}{11}),\end{align}
and
\begin{align}\tr_\text{II}(\ket{1}_\text{U}\bra{0})&=\qr\left[C^3\proj{10}{00}+S^2C\proj{11}{01}\right]-\ql\left[S^3\proj{10}{11}+SC^2\proj{00}{01}\right].\end{align}
Where we have defined for brevity $S\equiv\sin r_{\text{f},\omega}$, $C\equiv\cos{\text{f},\omega}$ and $\ket{ij}$ is a state with $i$ particles and $j$ antiparticles in region I, obtained after tracing out the modes in region II.

The partial transpose with respect to Alice can be obtained by simply swapping the Alice states in \eq{rhog}. The resultant matrix can be diagonalised numerically to compute the negativity.

We shall now compute the density matrix after tracing out region II for the scalar field. For a general state \eqref{geGras} , the traced density matrix is given again by \eq{rhog}, but now the relevant matrices have a different form, namely
\begin{align}\nonumber\tr_\text{II}(\ket{0}\bra{0})&=\sum_{n=0}^\infty f(n)^2\proj{n}{n},\\\tr_\text{II}(\ket{1}_\text{U}\bra{1}_\text{U})&=\frac{1}{\cosh^2\ro}\sum_{n=0}^\infty f(n)^2(n+1)\left(\ql^2\proj{n}{n}+\qr^2\proj{n+1}{n+1}\right)\nonumber\\&+
\ql\qr f(n)f(n+1)\sqrt{(n+1)(n+2)}\left(\proj{n}{n+2}+\proj{n+2}{n}\right)\end{align}
and
\begin{align}\tr_\text{II}(\ket{1}_\text{U}\bra{0})&=\frac{1}{\cosh\ro}\sum_{n=0}^\infty \ql f(n)f(n+1)\sqrt{(n+1)}\proj{n}{n+1}\nonumber\\&+\qr f(n)^2\sqrt{n+1}\proj{n+1}{n}\end{align},
where $\ket{n}$ are field modes in region I and
\begin{equation}
f(n)=\frac{\tanh^n r_\Omega}{\cosh r_\Omega}.
\end{equation}
The resultant matrix is diagonalised numerically taking proper care of convergence issues to compute the negativity that we use to quantify the entanglement between Alice and Rob. Analogous operations would lead to Alice-AntiRob density matrix tracing over I instead of II as discussed in previous chapters (See for instance chapter \ref{sma}).

\section{Entanglement amplification}

\begin{figure}[h] 
\begin{center}
\includegraphics[width=.85\textwidth]{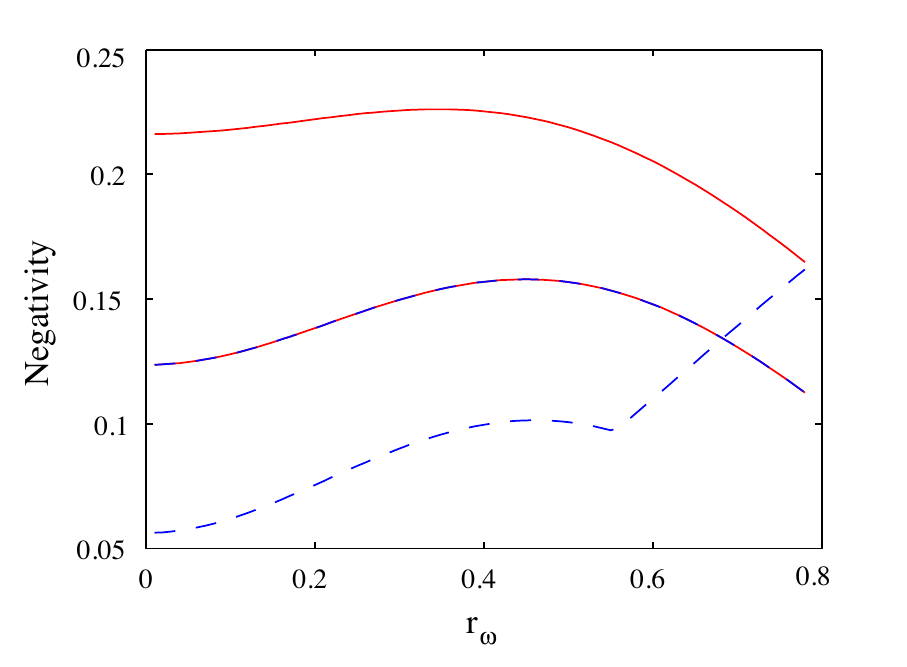}
\end{center}
\caption{Entanglement amplification in Alice-Rob and Alice-AntiRob bipartitions for the Grassmann scalar field. Negativity curves for $AR$ (red continuous lines) and $A\bar R$ (blue dashed lines) as a function of $r_\omega$ for $\qr=0.85$ and $\qr=1/\sqrt{2}$ (where both contributions are equal). The state considered is of the general form \eqref{geGras}, with the choice of parameters $P=0.4$, $\alpha=0$, $\beta=1$. }
\label{ent}
\end{figure}

For $\qr\neq1$ and a rather simple choice of parameters in \eqref{geGras} (for instance, $P=0.4$, $\alpha=0$, $\beta=1$) the surprise appears: there can be entanglement amplification due to acceleration, as seen in fig. \ref{ent}. Furthermore, this amplification becomes more evident as $\qr$ approaches the extremal value $\qr=1/\sqrt{2}$.  AR and $\text{A}\bar{\text{R}}$ behave the same way in this case since the symmetry between regions I and II is not explicitly broken (see eq. \eqref{umodes} and chapter \ref{sma}). As $\qr$ tends to 1 (limit analogous to the SMA), the effect vanishes. 

More importantly, it is also possible to obtain high entanglement amplification considering almost separable states. In some of these cases the negativity is a increasing monotone function of $r_\omega$ (as for instance in \eqref{geGras} when $P=0.1$, $\alpha=0.8695$, $\beta=0.909$). Therefore, there are states for which the Unruh effect does exactly the opposite of what was expected. That is to say, entanglement is monotonically created rather than being monotonically destroyed.

\begin{figure}[h] 
  \begin{center}
\includegraphics[width=.85\textwidth]{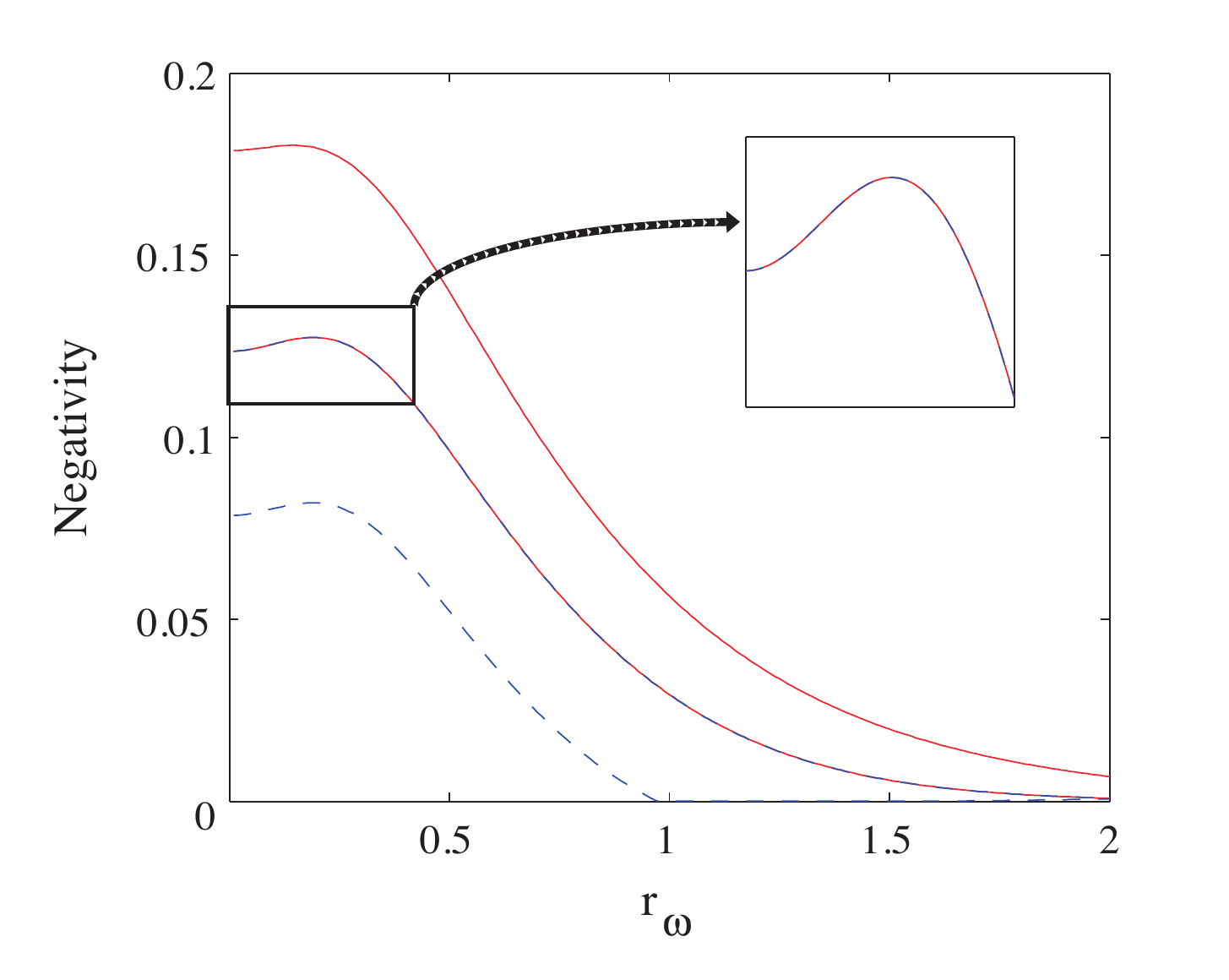}
\end{center}
\caption{Entanglement creation for the bosonic scalar field. Negativity for $AR$ (red continuous) and $A\bar R$ (blue dashed) as a function of $r_\omega$ for $\qr=0.85$ and $\qr=1/\sqrt{2}$ (where both contributions are equal). The state considered is of the general form \eqref{geGras}, with $P=0.4$, $\alpha=0$, $\beta=1$.}
\label{ent2}
\end{figure}

The phenomenon of entanglement creation also shows up for the bosonic scalar field, much in the same way as it did in the Grassmann case. The main difference is that  in the bosonic case entanglement is bound to vanish in the infinite acceleration limit, in concordance with previous results. Therefore, entanglement can be created only for a finite range of accelerations, as shown in Fig.\ref{ent2}. For $\qr=1/\sqrt{2}$, Alice-Rob negativity attains a maximum of $0.127$ at $r_\omega=0.191$. This is  3.1 \% above inertial level. Considering frequencies of order $\approx 1$  GHz, which correspond to reasonable experimental possibilities \cite{Adessada} the acceleration corresponding to this value of $r_\omega$ is $a\approx 10^{17}g$, much closer to experimental feasibility than previous proposals \cite{ChenTaj} which suggested accelerations of $\approx 10^{25}g$.

In order to study the experimental implications of this phenomenon, let us introduce a specific scenario. Consider the family of states \eq{geGras} for fixed $P$ and $\beta \gtrsim 0.2$. This family shows an unbounded relative increase of entanglement  as one approaches the limit $\alpha\rightarrow \beta$ (separable limit). This means that  there are states for which an arbitrary small acceleration produces an arbitrary large relative increase in negativity. The same happens as  we approach a separable state taking $P\rightarrow0$ for certain values of $\alpha$ and $\beta$. This behaviour is quite general and appears for both fermionic and bosonic fields. However the more relative entanglement increase (signal-to-background ratio) we want to achieve, the more separable the inertial states should be. Preparing and controlling these quasi-separable states would be the experimental challenge to detect the Unruh effect by means of these techniques.

\begin{figure}[h]
\begin{center} 
\includegraphics[width=.85\textwidth]{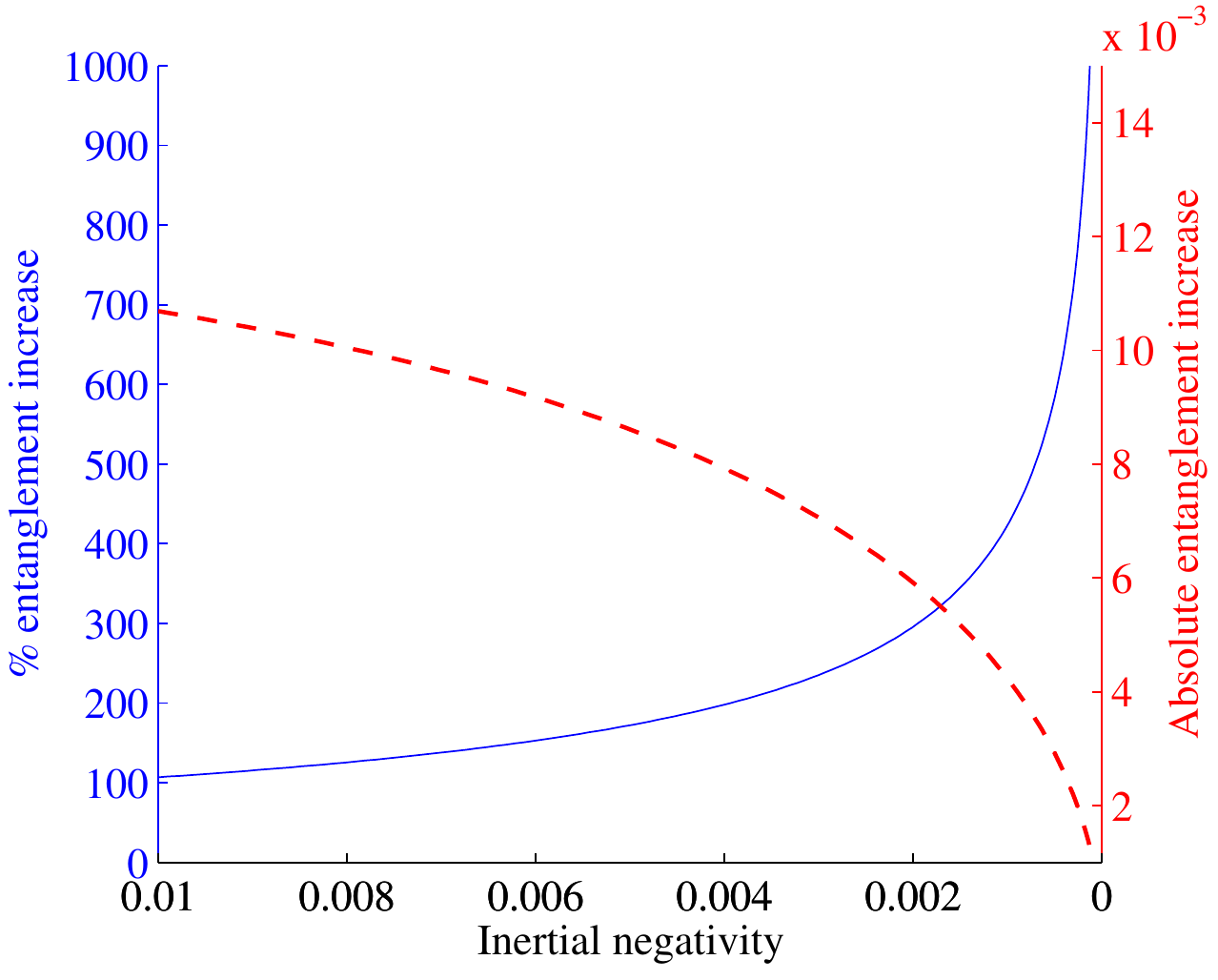}
\end{center}
\caption{Relative (blue continuous) and absolute (red dashed) entanglement creation  as a function of the inertial negativity for a Grassmann field  state of the family \eqref{geGras} with $P=0.4$, $\beta=0.8$ and different values of $\alpha$ for $\qr=1/\sqrt{2}$ and  $r_\omega=0.15$ (accelerations from $a\approx 5\cdot10^{13}g $ to $5\cdot10^{16} g$ for frequencies from $1 \text{ MHz}$ to $1 \text{ GHz}$). Notice an unbounded growth of the `signal-to-background' ratio.}
\label{unbo}
\end{figure}

Any such experiment would be naturally interested in negativity behaviour in the vicinity of $r_\omega=0$, easier to obtain in laboratory conditions. This means that in order to maximise experimental feasibility we are interested in states whose negativity shows a quick growth for small $r_\omega$. We study the relative increase of $AR$ negativity with respect to its inertial value for the family of states \eqref{geGras} at fixed $r_\omega$. As an example we choose  $r_\omega=0.15$ which corresponds to accelerations from $a\approx 5\cdot10^{13}g $ to $5\cdot10^{16} g$ for frequencies from $1 \text{ MHz}$ to $1 \text{ GHz}$. This unbounded entanglement creation can be seen in Fig. \ref{unbo}.


We obtain huge `signal-to-background' ratios and the better the negativity can be experimentally determined the bigger this ratio can become. If we relied on entanglement degradation to detect the Unruh effect \cite{Alicefalls}, the percental change in negativity would be bounded by 100 \%. With the plethora of new states presented in this work, this relative change can be made arbitrarily high.

The same analysis carried out for Unruh modes can be repeated if we consider that the excitations $\ket{1}$ are Gaussian wavepackets from the inertial perspective. As detailedly shown in section \ref{sec:peaking}, Gaussian wavepackets of Minkowski modes transform into Gaussian wavepackets of Rindler modes. We can consider at the same time peaked wavepackets in the Minkowskian basis and in the Rindler basis such that the analysis would be completely analogous to the monochromatic case. In this case different choices of $q_\text{R}$ and $q_\text{L}$ represent different spatial/momentum localisation of the Gaussian wavepackets. This means that in principle we can define two local bases (for Alice and Rob) in which this entanglement creation phenomenon is present.

\section{Discussion}

We have shown that the Unruh effect can not only degrade quantum entanglement but also amplify it, banishing previous fundamental misconceptions such as the belief that the Unruh and Hawking effects are sources of entanglement degradation. We have demonstrated that there are families of states whose entanglement can be increased by an arbitrarily high relative factor.

 One could ask why entanglement seems to be created in this case when the natural way of thinking may suggest that under partial tracing correlations can only be lost. This has to do with a mechanism related to the change of basis Unruh-Rindler.  Observing an entangled state of the field from the perspective of an accelerated observer implies two processes: 1) a generation of entanglement due to the Bogoliubov relationships implied in the change of basis that was shadowed under the SMA and 2) an erasure of correlations due to the tracing over one of the Rindler regions. We saw that going beyond the single mode approximation these competing trends explain why for a certain acceleration the amplification of entanglement is maximal. In previous works under the SMA it was simply not possible to see these two mechanisms in action.

Furthermore, with these results we move the experimental difficulties from generating and sustaining high accelerations to the preparation and measurement of quasi-separable entangled states. Hence, we are presenting a new way to detect the Unruh effect. As a matter of fact, our results are independent of the specific implementation to detect the entanglement magnification,  hence they can be exported to a huge variety of experimental set-ups as for instance analog gravity settings.

Besides, these results can be readily exported to a setting consisting in an observer hovering at certain distance close to the event horizon of an Schwarzschild black hole, following chapter \ref{blackhole1}, and therefore the same conclusions drawn for the Unruh effect are also valid for the Hawking effect.

\part[Non-stationary spacetimes and field dynamics]{Entanglement in non-stationary scenarios and field dynamics}
\label{part3}

\chapter{Quantum entanglement produced in the formation of a black hole\footnote{E. Mart\'in-Mart\'inez, L. J. Garay, J. Le\'on. Phys. Rev. D, 82, 064028 (2010)}}\label{stellarcollapse}

\markboth{Chapter 11. Quantum entanglement in the formation of a black hole}{\rightmark}

In this chapter we analyse the issue of entanglement production in a
dynamical gravitational collapse. With this aim, we consider both a bosonic
(scalar) and a fermionic (Grassmann scalar) field which initially are  in the
vacuum state and compute their asymptotic time evolution under   the
gravitational interaction in a stellar collapse. The vacuum state  evolves to
an  entangled state of  modes in the future null infinity (which give rise to
Hawking radiation \cite{Hawking}) and modes that do not reach it
because they fall into the forming event horizon.

We will argue that the initial vacuum state  in the asymptotic past does
not have any quantum entanglement with information about the future black hole, and that it evolves to a state
that is physically entangled  as a consequence of the creation of the
event horizon. This entanglement depends on the mass of the black hole
and the frequency of the field modes. In particular, for very small
frequencies or very small black holes, a maximally entangled state can
be produced.

The entanglement generated in a gravitational collapse thus
appears as a quantum resource for non-demolition methods aiming to
extract information about the field modes which fall into the horizon
from the outgoing Hawking radiation. These methods would be most
relevant for cases such as the formation of micro-black holes and the
final stages of an evaporating black hole when the mass is getting smaller
and, therefore, quantum correlations generated between the Hawking
radiation and the infalling modes grow to become even maximal, as we will show.

We have seen that fermions and bosons  have qualitatively
different behaviours in phenomena such as the Unruh entanglement
degradation.
Here, we will show that for fermions the generation of entanglement due
to gravitational collapse is more robust than for bosons. We will see that this robustness
becomes more and more evident from the peak of the thermal spectrum of Hawking
radiation towards the ultraviolet.

Previous literature (see for example
\cite{Balbinot,NavarroSalas,BalbinotII,serena} among many others)
showed that Hawking radiation is correlated with the field state falling
into the collapsing star. However neither the analysis of the associated
entanglement entropy as a function of the black hole parameters nor the
comparison between fermionic and bosonic behaviour have been carried
out so far. The study of these issues,  the nature of the entanglement
produced in a gravitational collapse and, more important, its dependence
on the nature of the quantum field statistics (bosonic/fermionic)  is decisive in
order to gain a deeper understanding about quantum entanglement in
general relativistic scenarios as it was proven for other setups such as
acceleration horizons, eternal black holes and (as we will see in the next chapter) expanding universes.

Since entanglement is a pure quantum effect,  understanding its behaviour
in these scenarios can well be relevant to discern the genuine quantum
Hawking radiation from classical induced emission in black hole analogs
\cite{Unruhan} (see, for example, Ref. \cite{serena}), where both
classical and quantum perturbations obey the same evolution laws.  It will
also follow from our study  that fermionic modes could be more suitable
for this task since they are more reliable in encoding entanglement
information.

Finally,  we will argue that the entanglement between the infalling and the
Hawking radiation modes neither existed as a quantum information
resource nor could have  been acknowledged by any observer before the
collapse occurs, namely in the asymptotic past. This is important in order
to understand the dynamics of the creation of correlations in the
gravitational collapse scenario since these correlations are exclusively
due to quantum entanglement, as discussed in the literature
\cite{Balbinot,NavarroSalas,BalbinotII,serena}.

\section{Gravitational Collapse}

Gravitational collapse is the inward fall of a body due to the influence of its own gravity. In stars, gravitational force is counterbalanced by the internal pressure increased by the nuclear reactions taking place in their cores. If the inwards pointing gravitational force, however, is stronger than all outward pointing forces, this equilibrium is disturbed and a collapse occurs until the internal pressure might rise sufficiently to counterbalance gravity again.

When stars run out of nuclear fuel, thermodynamic pressure cannot stop the star from collapsing. If the star has a mass approximately $M\le 1.4$ solar masses it will reach a point in which the electron degeneracy pressure counterbalances the gravitational collapse. These are the astrophysical objects called `white dwarfs'. If the star has a mass between $1.4$ and $3$ solar masses, electron degeneracy is not enough to stop the gravitational collapse, then electrons and protons interact forming neutrons. Neutronic degeneracy pressure may be enough to counterbalance the gravitational pull. These are the so-called `neutron stars'. If the star mass is above 3 solar masses no known force in nature can stop the collapse and then a black hole is formed.

\begin{figure}[h]
\begin{center}
\includegraphics[width=.9\columnwidth]{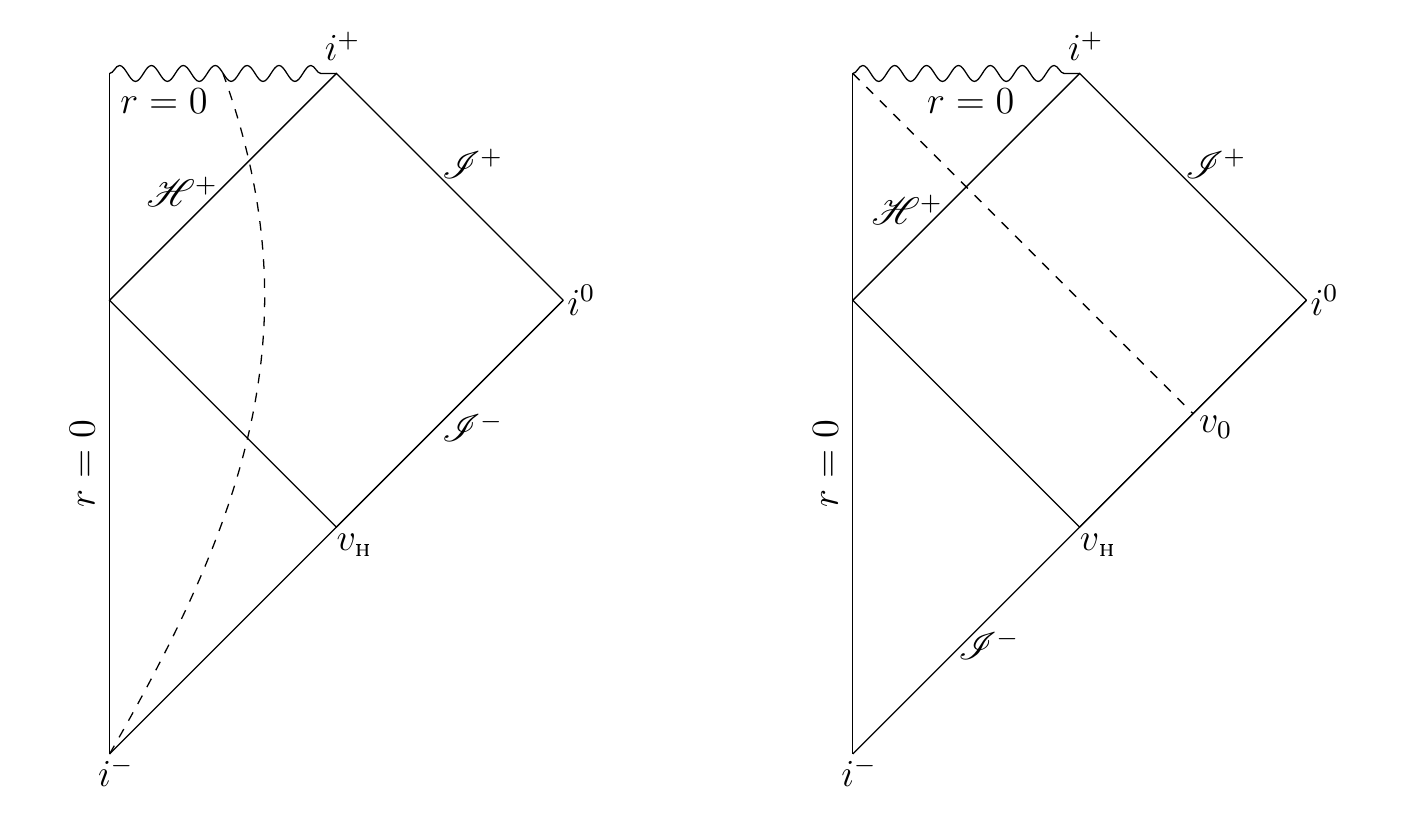}
\end{center}
\caption{Carter-Penrose diagrams for gravitational collapse: Stellar collapse (left)
and Vaidya spacetime (right).}
\label{fig:vaidya}
\end{figure}

In order to analyse the entanglement production induced by gravitational
collapse we will consider the Vaidya dynamical solution to
Einstein equations (see e.g. Ref. \cite{NavarroSalas})  that, despite its
simplicity, contains all the ingredients relevant to our study. Refinements
of the model to make it more realistic only introduce subleading
corrections. The Vaidya spacetime (Fig.~\ref{fig:vaidya}), is conveniently
described in terms of ingoing Eddington-Finkelstein coordinates by the
metric
\begin{equation}
\diff s^2=-\left(1-\frac{2M(v)}{r}\right)\diff v^2+2\diff v\diff r+r^2 \,
\diff \Omega^2,
\end{equation}
where $r$ is the radial coordinate, $v$ is the ingoing null coordinate, and
$M(v)=m\theta(v-v_0)$. For $v_0<v$ this is nothing but the ingoing
Eddington-Finkelstein representation for the Schwarzschild metric
whereas for $v<v_0$ it is just Minkowski spacetime. This metric
represents a radial  ingoing collapsing shockwave of radiation. As it can
be seen in Fig. \ref{fig:vaidya}, $v_\textsc{h}=v_0-4m$ represents the
last null ray that can escape to the future null infinity $\mathscr{I}^+$
and hence that will eventually form the event horizon.

Let us now introduce two different bases of solutions to the
Klein-Gordon equation in this collapsing spacetime. On the one hand, we
shall define the `in' Fock basis in terms of ingoing positive frequency
modes, associated with the natural time parameter $v$ at the null past
infinity $\mathscr{I}^-$, which is a Cauchy surface:
\begin{equation}
\uin{\omega}\sim\frac{1}{4\pi r\sqrt{\omega}}e^{-i\omega v}.
\end{equation}
On the other hand, we can also consider an alternative Cauchy surface in
the future to define another basis. In this case, the asymptotic future
$\mathscr{I}^+$ is not enough and we also need the future event horizon
$\mathscr{H}^+$. The `out' modes defined as being outgoing
positive-frequency in terms  of the natural time parameter
$\eta_{\text{out}}$ at  $\mathscr{I}^+$ are
\begin{equation}
\uout{\omega}\sim\frac{1}{4\pi r\sqrt{\omega}}e^{-i\omega \eta_{\text{out}}},
\end{equation}
where $\eo=v-2r^*_{\text{out}}$ and $r^*_\text{out}
$ is the radial tortoise coordinate in the Schwarzschild region.  It can be
shown (see e.g. Ref. \cite{NavarroSalas}) that, for late times
$\eo\to\infty$ at $\mathscr{I}^+$, these modes $\uout{\omega}$ are
concentrated near $v_\textsc{h}$ at $\mathscr{I}^-$ and have the
following behaviour:
\begin{equation}\label{uout2}
\uout{\omega}\approx\frac{1}{4\pi r \sqrt{\omega}}
e^{-i\omega\left(v_\textsc{h}-4m\ln\frac{|v_\textsc{h}-v|}{4m}\right)}
\theta(v_\textsc{h}-v).
\end{equation}
These modes have only support in the region $v < v_\textsc{h}$. This is
evident as only the light rays that depart from $v < v_\textsc{h}$ will
reach the asymptotic region $\mathscr{I}^+$ since the rest will fall into
the forming horizon defined by $v=v_\textsc{h}$. This is the only
relevant regime, as far as entanglement production is concerned.

For the modes defined at  $\mathscr{H}^+$ (denoted `hor'), there is no such
natural time parameter. A simple way to choose these modes is defining
them as the modes that in the asymptotic past $\mathscr{I}^-$ behave in
the same way as $\uout{\omega}$ but defined for $v > v_\textsc{h}$,
that is to say, as modes that leave the asymptotic past but do not reach
the asymptotic future  since they will fall into the horizon. This criterion
is the simplest that clearly shows the generation of quantum
entanglement between the field in the horizon and the asymptotic region.
In any case, since we will trace over all modes at the horizon, the choice
of such modes does not affect the result. Therefore, we define the
incoming modes crossing the horizon  by reversing the signs of
$v_\textsc{h}-v$ and $\omega$ in
\eqref{uout2} so that near $\mathscr{I}^-$ these modes are
\begin{equation}
\uh{\omega}\sim\frac{1}{4\pi r \sqrt{\omega}}
e^{i\omega\left(v_\textsc{h}-4m\ln\frac{|v_\textsc{h}-v|}{4m}\right)}
\theta(v-v_\textsc{h}).
\end{equation}

We are now ready to write the annihilation operators of bosonic field
modes in the  asymptotic past   in terms of the corresponding creation
and annihilation operators defined in terms of modes in the future:
\begin{equation}\label{ain}
\ain{\omega'}=\int d\omega \Big[\alpha^*_{\omega\omega'}
\big(\aout{\omega}-\tanh r_{\omega}\,\ah{\omega}^\dagger\big)+\alpha_{\omega\omega'}e^{i\varphi}\big(\ah{\omega}-\tanh r_{\omega}\,
\aout{\omega}^\dagger\big)\Big],
\end{equation}
where  $\tanh r_{\omega} = e^{-4\pi m\omega}$. The vacuum $\ket{0}_\text{in}$ is annihilated by \eqref{ain} for all
frequencies $\omega'$, this is precisely the scenario described in section \eqref{nonstaint}. One sees that the precise
values of $\varphi$ and  $\alpha_{\omega\omega'}$ are not relevant for this analysis and the `in' vacuum, of the form \eqref{dynoresult}, is expanded as follows in terms of the `out-hor' basis:
\begin{equation}
\ket{0}_\text{in}=N \exp\left(\sum_{\omega}\tanh r_{\omega}\,
\ah{\omega}^\dagger \aout{\omega}^\dagger\right)
\ket{0}_{\text{hor}}\ket{0}_{\text{out}},
\end{equation}
where $N=\big(\prod_{\omega}\cosh r_{\omega}\big)^{-1}$ is a
normalisation constant. We can rewrite this state in terms of modes
$\ket{n_{\omega}}$ with frequency $\omega$ and occupation number
$n$ as
\begin{equation}\label{sque}
\ket{0}_\text{in}=\prod_{\omega}\frac{1}{\cosh r_{\omega}}
\sum_{n=0}^\infty (\tanh r_{\omega})^{n}\ket{n_{\omega}}_\text{hor}
\ket{n_{\omega}}_{\text{out}}.
\end{equation}

\section{Analysing entanglement}

The `in' vacuum \eqref{sque} in the `out-hor' basis is a two-mode squeezed state.  Therefore, it is a pure  entangled
state of the modes   in the asymptotic future and the modes falling
across  the event horizon. Given the tensor product structure no
entanglement is created between different frequency modes. Hence, we
will concentrate the analysis in one single arbitrary frequency $\omega$.

\begin{figure}[h]
\begin{center}
\includegraphics[width=.90\columnwidth,height=.5\columnwidth]{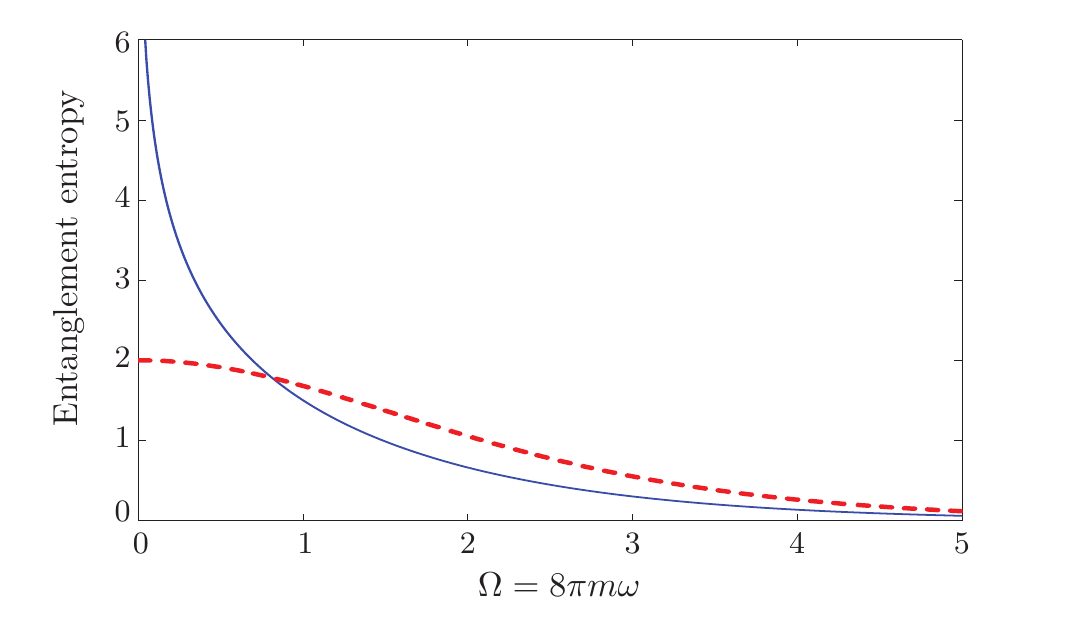}
\end{center}
\caption{Entanglement between bosonic (continuous blue) and fermionic (red dashed)
field modes in $\mathscr{H}^+$ and in $\mathscr{I}^+$.
The lesser the mass of the star or the mode frequency, the higher the
entanglement reached.}
\label{fig2}
\end{figure}

We can compute the entropy of entanglement for this state, which is the
 natural entanglement measure for a bipartite pure state,
defined as the Von Neumann entropy of  the reduced state obtained upon
tracing over one of the subsystems of the bipartite state. To compute it
we need the partial state
$\rho_{\text{out}}=\tr_{\text{hor}}(\ket{0}_{\text{in}}\!\!\bra{0})$,
which turns out to be
$\rho_{\text{out}}=\prod_\omega\rho_{\text{out},\omega}$, where
\begin{equation}
\rho_{\text{out},\omega}=\frac{1}{(\cosh r_{\omega})^2}
\sum_{n=0}^\infty (\tanh r_{\omega})^{2n}
\ket{n_{\omega}}_{\text{out}}\!\!\bra{n_{\omega}}.
\end{equation}
This is, indeed, a thermal radiation state  whose temperature is nothing
but the Hawking temperature $(8\pi m)^{-1}$, as  it can be easily
seen. However, this is only the partial state of the field, not the complete
quantum state, which is globally  entangled. If we compute the entropy of
entanglement
$S_{\textsc{e},\omega}=\tr(\rho_{\text{out},\omega}\log_2
\rho_{\text{out},\omega})$ for each frequency, after some calculations, we obtain
\begin{equation}
 S_{\textsc{e},\omega}=
\left(\cosh r_\omega\right)^2\log_2\left(\cosh r_\omega\right)^2-(\sinh r_\omega)^2\log_2\left(\sinh r_\omega\right)^2,
\end{equation}
which is displayed in Fig.~\ref{fig2}. As \eqref{sque} is a pure state, all
the correlations between modes at the horizon and modes in the
asymptotic region are due to quantum entanglement.

Analogously we can compute the entanglement for fermionic
fields. If we consider a spinless Dirac field (either one dimensional or a
Grassmann scalar), the analysis is entirely analogous considering
now both particle and antiparticle modes. We assume again that the
initial state of the field is the vacuum that, after some 
calculations completely analogous to those in chapter \ref{sma}, can be expressed in the Fock basis at the asymptotic
future and the `hor' modes:
\begin{align}\label{squef}
\ket{0}_\text{in}&=\prod_{\omega}\Big[(\cos\tilde r_{\omega})^2
\ket{00}_\text{hor}\ket{00}_{\text{out}}-\frac{\sin 2\tilde r_{\omega}}{2}
\big(\ket{01_\omega}_\text{hor}\ket{1_\omega0}_{\text{out}}
 - \ket{1_\omega0}_\text{hor}
 \ket{01_\omega}_{\text{out}}\big)
 \nonumber\\
 &-(\sin\tilde r_{\omega})^2
 \ket{1_\omega1_\omega}_\text{hor}
 \ket{1_\omega1_\omega}_{\text{out}}\Big],
\end{align}
where  $\tan \tilde r_\omega= e^{-4\pi m\omega} $. Here, we are using
the double Fock basis, the first figure inside each ket representing
particles and the second  antiparticles.

We can compute the entropy of entanglement of this pure state.
The partial density matrix in the asymptotic future
$\rho_{\text{out}}=\tr_{\text{hor}}(\ket{0}_{\text{in}}\!\!\bra{0})=
\prod_\omega \rho_{\text{out},\omega}$, is given by
\begin{align}
\rho_{\text{out},\omega}&= (\cos\tilde r_\omega)^4
\ket{00}_{\text{out}}\!\!\bra{00}+\frac{(\sin 2\tilde r_\omega)^2}{4}
\Big(\ket{1_{\omega}0}_{\text{out}}\!\!\bra{1_{\omega}0}
+\ket{01_{\omega}}_{\text{out}}\!\!\bra{01_{\omega}}\Big)
\nonumber\\
&+(\sin \tilde r_\omega)^4 \ket{1_{\omega}1_{\omega}}_{\text{out}}\!\!
\bra{1_{\omega}1_{\omega}},
\end{align}
which   is again a thermal state with Hawking temperature $(8\pi
m)^{-1}$, and
\begin{equation}
S_{\textsc{e},\omega}\!=\!-2\big[(\cos \tilde r_\omega)^2\log_2
(\cos \tilde r_\omega)^2+
(\sin \tilde r_\omega)^2\log_2(\sin \tilde r_\omega)^2\big],
\end{equation}
which is also displayed in Fig.~\ref{fig2}.

Figure~\ref{fig2} shows that entanglement decreases as the mass of
the black hole or the frequency of the mode increase. When comparing
bosons with  fermions one must have in mind that the entropy of
entanglement is bounded by (the logarithm of) the dimension of the
partial Hilbert space (`out' Fock space in our case). Therefore, due to
Pauli exclusion principle, the maximum entropy of entanglement for
fermions is $S_{\textsc{e},\omega}=2$, which corresponds to a
maximally entangled state. On the other hand, for bosons, the
entanglement is distributed among the superposition of all the occupation
numbers and the entropy can grow unboundedly, reaching the
maximally entangled state in the limit of infinite entropy. In this sense,
the entanglement generated in the fermionic case is more useful and
robust due to the limited dimension of the Fock space for each fermionic
mode.

This result can be traced back to the inherent differences between
fermions and bosons. Specifically, it is Pauli exclusion principle which
makes fermionic entanglement more reliable.  Similar results about
reliability of entanglement for fermions will be also found in the
expanding universe scenarios in the next chapter.  This responds to the high
influence of statistics in entanglement behaviour in general relativistic
settings as it was investigated in part \ref{part1} of this thesis. On the other hand,
Vaidya spacetime has all the fundamental features of a stellar collapse
and shows how the entanglement is created by the appearance of an
event horizon. Hence, in other collapsing scenarios or including the
sub-leading grey-body factor corrections, these fundamental statistical
differences will not disappear. The qualitatively different behaviour of
entanglement for bosons and fermions is not an artifact of choosing a
particular collapse scenario but is due to fundamental statistical
principles.

In the above analysis we have considered plane wave modes, which are
completely delocalised. However, an entirely analogous analysis can be easily
carried out using localised Gaussian states, with the same
results about quantum entanglement behaviour (see Chapter \ref{sma}).

\section{Discussion}

We have shown that the formation of an event horizon generates
entanglement. If we start from the vacuum state in the asymptotic past,
after the gravitational collapse process is complete we end up with a
state in the asymptotic future which shares pure quantum correlations
with the field modes which fall into the horizon. One could think that this
entanglement was already present before the   collapse, arguing that (as
proved in \cite{Vacbell}) the vacuum state of a quantum field can be
understood as an entangled state of space-like separated regions. In
other words, if we artificially divided  the Cauchy surface in which the
vacuum state is determined into two parts, we would have a quantum
correlated state between the two partitions. In principle  we could have
done a bipartition of the vacuum state in $\mathscr{I}^-$ such that it
would reflect entanglement between the partial state of the vacuum for
$v<v_\textsc{h}$ and the corresponding partial state for
$v>v_\textsc{h}$. However, it is not until the collapse occurs that we
have the information about what $v_\textsc{h}$ is. So, achieving
beforehand the right bipartition (trying to argue that the entanglement
was already in the vacuum state) would require a complete knowledge of
the whole future and, consequently, there is no reason `a priori' to do
such bipartition. The entanglement, eventually generated by the collapse,
will remain unnoticed to early observers, who are deprived of any means
to acknowledge and use it for quantum information tasks. It is well known
that if we introduce artificial bipartitions of a quantum system, its
description can show entanglement as a consequence of the partition.
However, not being associated with a physical bipartition this
entanglement does not codify any physical information. (One example of
this kind of non-useful entanglement is statistical entanglement between
two undistinguishable fermions~\cite{sta1}).

Gravitational collapse selects a specific partition  of the initial vacuum
state by means of the creation of an event horizon. In the asymptotic
past there was no reason to consider a specific bipartition of the vacuum
state, whereas in the future there is a clear physically meaningful
bipartition: what in $\mathscr{I}^-$ was expressed as a separable state,
now becomes expressed in terms of modes that correspond to the future
null infinity and the ones which fall across the event horizon. This means
that gravitational collapse defines a particular physical way to break the
arbitrariness of bipartitioning the vacuum into different  subsystems.
This gravitational production of entanglement would be a physical
realisation  of the potentiality of the vacuum state to be an entangled
state and is therefore a genuine entanglement creation process.

We have computed the explicit functional form of this entanglement and
its dependence on the mass of the black hole (which determines the
surface gravity). For more complicated scenarios (with charge or angular
momentum), it will depend on these parameters as well.

For small black holes, the outgoing Hawking radiation tends to be
maximally entangled with the state of the field falling into the horizon
for both bosons and fermions. This means that if a hypothetical high
energy process generates a micro-black hole, a projective measurement
carried out on the emitted radiation (as, for instance, the detection of
Hawking radiation) will `collapse' the quantum state of the field that is
falling into the event horizon and give us certainty about the outcome of
possible measurements carried out in the vicinity of the horizon (even beyond it).
Furthermore, at least theoretically speaking, the available quantum
information resources would be maximum and, therefore, one could
perform quantum information tasks such as  quantum teleportation with
maximum fidelity from the infalling modes to the modes in the asymptotic
future $\mathscr{I}^+$ if the observer of the infalling modes managed
to dispatch an outgoing classical signal  before crossing the horizon. On
the other hand, low frequency modes become more entangled than the
higher ones. So, the infrared part of the Hawking spectrum would provide
more information about the state at the horizon than the ultraviolet.

Arguably, similar conclusions can be drawn for the final stages of an
evaporating black hole: as the mass of the black hole diminishes, the
temperature of the Hawking radiation spectrum increases, and
therefore, the quantum state of the field tends to a maximally entangled state
 in the limit $m\rightarrow0$.

We have seen that the entanglement generated in fermionic fields is more
robust than for bosons. Although the entropy of entanglement in the zero
mass limit is greater in the bosonic case due to the higher dimension of
the partial Hilbert space, we have argued that the information is more
reliably encoded in the limited Fock space of fermionic fields.
Furthermore, as we consider higher frequency modes, fermionic
entanglement proves to be much more   easily created by the collapse.
What is more, the turning point in which the entropy of entanglement for
fermions becomes numerically larger than for bosons is actually near the
peak of the thermal emission (Fig.~\ref{fig2}). This means, that, in
general, a measurement carried out on Hawking radiation of fermionic
particles will give us more information about the near-horizon field state.
This might also be useful in analog gravity realisations as we have already
discussed,  specifically in systems where the field excitations are
fermionic (see e.g. Ref. \cite{Volovik}), which would be, as shown, at an
advantage over the bosonic cases. To account for this quantum
entanglement in analog experiments one should  carry out measurements
of the quantum correlations between the emitted thermal spectrum and
the infalling modes and detect Bell inequalities violations. This is easier as
it gets closer to the maximally entangled case.

\chapter{Entanglement of Dirac fields in an expanding Universe\footnote{I. Fuentes, R.B. Mann, E. Mart\'in-Mart\'inez, S. Moradi. Phys. Rev. D, 82, 045030 (2010)}}\label{expandingU}

In this chapter we study the creation of entanglement
between Dirac modes due to the expansion of a Robertson-Walker spacetime.
A general study of entanglement  in curved spacetime is problematic because particle states cannot always be defined in a meaningful way. However,
 it has been possible
 to learn about certain aspects of entanglement in curved spacetimes
that have asymptotically flat regions as discussed in section \ref{nonstaint}, \cite{caball,schacross,ShiYu,TeraUeda2}. Such
studies show that entanglement can be created by the dynamics of the
 underlying spacetime \cite{caball,Steeg} as well as destroyed by the loss of
information in the presence of a spacetime horizon \cite{Alicefalls,schacross}.

Such investigations not only deepen our understanding of entanglement but also offer the prospect of employing entanglement as a tool to learn about curved spacetime.  For example, the entanglement generated between bosonic modes due to the expansion of a model 2-dimensional universe was shown to contain information about its history \cite{caball}, affording the
possibility of deducing cosmological parameters of the underlying
spacetime from the entanglement. This novel way of obtaining
information about cosmological parameters could provide new insight into
the early universe  both theoretically (incorporating into cosmology entanglement as a purely quantum effect produced by gravitational interactions in an expanding universe) and experimentally (either by development of methods to measure entanglement between modes of the background fields or by measuring entanglement creation in  condensed matter analogs of expanding  spacetime \cite{analog1,analog2}). Other interesting results show that entanglement plays a role
in the thermodynamic properties of Robertson-Walker type spacetimes
\cite{Lousto} and can in principle be used to distinguish between different spacetimes
 \cite{Steeg} and probe spacetime fluctuations \cite{dowling}.

Here we consider entanglement between modes of a Dirac field in
a  2-dimensional Robertson-Walker universe.  We find
that the entanglement generated by the expansion of the
universe for the same fixed conditions is lower than for the bosonic case \cite{caball}.
However we also find that fermionic entanglement  codifies more information about the underlying spacetime structure. These contrasts are commensurate with 
the flat spacetime case, in which entanglement in fermionic systems was found to
be more robust against acceleration than that in bosonic systems
as discussed in chapters above.

\section{Dirac field in a $d$-dimensional Robertson-Walker universe}\label{sec2}

As we mentioned before, entanglement between modes of a quantum field in curved spacetime can be investigated in special cases where the spacetime has at least two asymptotically flat regions. Such is the case of a family of Robertson-Walker universes where spacetime is flat in the distant past and in the far future. In this section, following the work done by Bernard and Duncan \cite{BernardDuncan,dun1}, we find the state of a Dirac field in the far future that corresponds to a vacuum state in the remote past.

Consider a Dirac field $\psi$ with mass $m$ on a $d$-dimensional spatially flat Robertson-Walker spacetime with line element, \b
ds^2=C(\eta)(-d\eta^2+dx_idx^i).\e $x_i$ are the spacial coordinates and the temporal coordinate $\eta$ is called the conformal
time to distinguish it from the cosmological time $t$.
The metric is conformally flat, as are
all Robertson-Walker metrics.  The dynamics of the field is given by the covariant
form of the Dirac equation on a curved background\footnote{See appendix \eqref{appB}},\b \label{eq:dirac}
\{i\gamma^{\mu}(\partial_{\mu}-\Gamma_{\mu})+m\}\psi=0, \e
where $\gamma^{\mu}$ are the curved Dirac-Pauli matrices and $\Gamma_{\mu}$ are spinorial
affine connections. The curved Dirac-Pauli matrices satisfy the condition, \b
\gamma^{\mu}\gamma^{\nu}+\gamma^{\nu}\gamma^{\mu}=2g^{\mu\nu}, \e  where $g^{\mu\nu}$ is the spacetime metric.
In the flat case where the metric is given by $\eta^{\alpha\beta}$, the constant special relativistic matrices are defined by, \b
\bar{\gamma}^{\alpha}\bar{\gamma}^{\beta}+\bar{\gamma}^{\beta}\bar{\gamma}^{\alpha}=2\eta^{\alpha\beta}
.\e The relation between curved and flat ${\gamma}$ matrices is given by $\gamma_{\mu}=e_{\mu}^{\;\;\alpha}\bar{\gamma}_{\alpha}$ where
$e_{\mu}^{\;\;\alpha}$ is the vierbein (tetrad) field satisfying
the relation $e_{\mu}^{\;\;\alpha}e_{\nu}^{\;\;\beta}\eta_{\alpha\beta}=g_{\mu\nu}$.

In order to find the solutions to the Dirac equation Eq.
(\ref{eq:dirac}) on this spacetime, we note that $C(\eta)$ is
independent of x. We exploit the resulting spatial translational
invariance and separate the solutions into 
\b 
\psi_k(\eta,x)
=e^{i\bm{k\cdot x}}C^{(1-d)/4}\left(\bar{\gamma} ^{0}\partial
_{\eta
}+i\bar{\bm{\gamma}}\cdot{{\bm k}}-mC^{1/2}\right)\phi_k(\eta)
,\e
where $k^2=|{{\bm k}}|^2= \sum_{i=1}^{d-1}k_i^2 $.
Inserting this into the Dirac equation, we obtain the following
coupled equations   \b \label{eq:couple} \left(\partial_{\eta
}^{2}+m^{2}C\pm
im\dot{C}C^{-1/2}+|{{\bm k}}|^{2}\right)\phi_k^{(\pm)}=0, \e
using the fact that the eigenvalues of $\bar{\gamma}^0$ are $\pm
1$. In order to quantise the field and express it in terms of
creation and annihilation operators, positive and negative frequency modes
must be identified. This cannot be done globally. However
positive and negative frequency modes can be identified in the far
past and future where the spacetime admits timelike Killing vector
fields $\pm\partial/\partial \eta$. Provided $C(\eta)$ is constant
in the far past $\eta\rightarrow -\infty$ and
far future $\eta\rightarrow +\infty$, the asymptotic solutions of 
(\ref{eq:couple}) will be $\phi^{(\pm)}_{\text{in}}\sim
e^{\pm i\omega_{\text{in}}\eta}$ and $\phi^{(\pm)}_{\text{out}}\sim e^{\pm
i\omega_{\text{out}}\eta}$  respectively, where 
 \begin{eqnarray}
\omega_{\text{in}}&=&(|{\bm k}|^2+\mu_{\text{in}}^2)^{1/2}, \\
\omega_{\text{out}}&=&(|{\bm k}|^2+\mu_{\text{out}}^2)^{1/2}, \nonumber\\
\mu_{\text{in}}&=&m\sqrt{C(-\infty)},\nonumber\\\mu_{\text{out}}&=&m\sqrt{C(+\infty)}. \nonumber\end{eqnarray}

The action of the Killing vector field on the asymptotic solutions allows us to identify $\phi^{(\mp)\ast}_{\text{in}}$ and $\phi^{(\mp)\ast}_{\text{out}}$ as negative frequency  solutions. The sign flip is due to the explicit factor $i$ in (\ref{eq:couple}).
A consequence of the linear transformation
properties of such functions is that
the Bogolubov transformations associated with the transformation between in and out solutions take the simple form \cite{dun1}
\b
\phi_{\text{in}}^{(\pm)}(k)= \alpha_{k}^{(\pm)}\phi^{(\pm)}_{\text{out}}(k)+\beta_{k}^{(\pm)}\phi^{(\mp)*}_{\text{out}}(k)
,\e
where $\alpha^{\pm}_k$ and $\beta^{\pm}_k$ are Bogoliubov coefficients.

The curved-space spinor solutions of the Dirac equation are defined by (with corresponding $U_{\text{out}}$, $V_{\text{out}}$ and $K_{\text{out}}$),
\begin{align}
\nonumber U_{\text{in}}(\bm{k},\lambda;x,\eta)&\equiv K_{\text{in}}(k)[C(\eta)]^{(1-d)/4}\Big[-i\partial_\eta+i{\bm{k}}\cdot\bar{\gamma}-m\sqrt{C(\eta)}\Big]\phi_{k}^{in(-)(\eta)}e^{i\bm{k}\cdot
{\bm{x}}}u(0,\lambda),\\*
\nonumber V_{\text{in}}(\bm{k},\lambda;x,\eta)&\equiv
K_{\text{in}}(k)[C(\eta)]^{(1-d)/4}\Big[i\partial_\eta-i{\bm{k}}\cdot\bar{\gamma}-m\sqrt{C(\eta)}\Big]\phi_{k}^{in(+)\ast(\eta)}e^{-i\bm{k}\cdot
{\bm{x}}}v(0,\lambda),
\end{align}
where $K_{\text{in}}\equiv-(1/|k|)((\omega_{\text{in}}-\mu_{\text{in}})/2\mu_{\text{in}})^{1/2}$ and $u(0,\lambda)$, $v(0,\lambda)$ are flat space spinors satisfying
\begin{eqnarray}
\gamma^0u(0,\lambda)&=&-iu(0,\lambda),\nonumber\\
\gamma^0v(0,\lambda)&=& iv(0,\lambda),\nonumber
\end{eqnarray}
for $1\leq\lambda\leq2^{d/2-1}$. The field in the `in' region can
then be expanded as,
\begin{equation}
 \psi(x)=\frac{1}{\sqrt{(2\pi)^{1-d}}}\int
d^{d-1}k\left[\frac{\mu_{\text{in}}}{\omega_{\text{in}}}\right]\sum_{\lambda=1}^{d/2-1}
\big[a_{\text{in}}({\bm k},\lambda)U_{\text{in}}({\bm k},\lambda;\bm{x},\eta)+
 b^{\dag}_{\text{in}}({\bm k},\lambda)V_{\text{in}}({\bm k},\lambda;\bm{x},\eta)\big],
\end{equation} 
with a similar expression for the `out' region.  The `in'
and `out' creation and annihilation operators for particles and
anti-particles obey the usual anticommutation relations. Using the
Bogoliubov transformation one can expand the out  operators in
terms of ``in'' operators 
\begin{align}
a_{\text{out}}(\bm k,\lambda)&=\nonumber\left(\frac{\mu_{\text{in}}\omega_{\text{out}}}{\omega_{\text{in}}\mu_{\text{out}}}\right)^{\frac{1}{2}}
\frac{K_{\text{in}}}{K_{\text{out}}}\Bigg(\alpha_{k}^{(-)}a_{\text{in}}(\bm k,\lambda)+\beta_{k}^{(-)\ast}\sum_{\lambda'}
X_{\lambda\lambda'}(-{\bm k})b^{\dag}_{\text{in}}(-{\bm k},\lambda')\Bigg),\\*
 b_{\text{out}}(\bm k,\lambda)&=\left(\frac{\mu_{\text{in}}\omega_{\text{out}}}{\omega_{\text{in}}\mu_{\text{out}}}\right)^{\frac{1}{2}}
\frac{K_{\text{in}}}{K_{\text{out}}}\Bigg(\alpha_{k}^{(-)}b_{\text{in}}(\bm k,\lambda)\beta_{k}^{(-)\ast}\sum_{\lambda'}
X_{\lambda\lambda'}(-{\bm k})a^{\dag}_{\text{in}}(-{\bm k},\lambda')\Bigg),
\end{align}
where \b
X_{\lambda\lambda'}(-{\bm k})=-2\mu_{\text{out}}^2K^2_{\text{out}}\bar{u}_{\text{out}}(-{\bm k},\lambda')v(0,\lambda)
\e and \b
K_{in/out}=\frac{1}{|k|}\left(\frac{\omega_{in/out}(k)-\mu_{in/out}}{\mu_{in/out}}\right)^{1/2}
.\e This yields the following relationship between Bogoliubov
coefficients, 
\b
\left|\alpha^{(-)}_{k}\right|^2-2\mu_{\text{out}}^2K_{\text{out}}^2\left(1-\frac{\omega_{\text{out}}}{\mu_{\text{out}}}\right)\left|\beta_{k}^{(-)}\right|^2
=\frac{\mu_{\text{out}}}{\mu_{\text{in}}}\frac{\omega_{\text{in}}}{\omega_{\text{out}}}\left(\frac{K_{\text{out}}}{K_{\text{in}}}\right)^2
.\e We consider the special solvable case presented in \cite{dun1}
$C(\eta)=(1+\epsilon(1+\tanh\rho\eta))^2$, where $\epsilon,\rho $
are positive real parameters controlling the total volume and
rapidity of the expansion, respectively. This model reproduces the activation of a regime of quick expansion of the Universe that asymptotically tends to a stationary one (can be regarded as basic model of inflation). In this case the
solutions of the Dirac equation  that  in the remote past reduce to
positive frequency modes are,
\begin{align}
\nonumber\phi^{(\pm)}_{\text{in}}&=\exp \left( -i\omega _{+}\eta-\frac{
i\omega _{-}}{\rho }\ln [2\cosh \rho \eta ]\right)\\*
& \times F_{1}\left(
1+\frac{
i(\omega _{-}\pm m\epsilon)}{\rho },\frac{i(\omega
_{-}\mp m\epsilon)}{\rho },1-\frac{
i\omega _{\text{in}}}{\rho },\frac{1+\tanh (\rho \eta )}{2}\right),
\end{align}
where $F_{1}$ is the ordinary hypergeometric function. Similarly, one may
find a complete set of modes of the field behaving as
positive and negative frequency modes in the far future, 
\begin{align}
\nonumber \phi^{(\pm)}_{\text{out}}&=\exp \left( -i\omega _{+}\eta-\frac{%
i\omega _{-}}{\rho }\ln [2\cosh \rho \eta ]\right)\\*
& \times F_{1}\left(
1+\frac{
i(\omega _{-}\pm m\epsilon)}{\rho },\frac{i(\omega
_{-}\mp m\epsilon)}{\rho },1+\frac{
i\omega _{\text{out}}}{\rho },\frac{1-\tanh (\rho \eta )}{2}\right),
\end{align}
 where $\omega _{\pm }=(\omega _{\text{out}}\pm \omega _{\text{in}})/2 $. The spacetime
obtained by considering this special form of $C(\eta)$ was
introduced by Duncan \cite{dun1}. It is easy to see that it
corresponds to a Minkowskian spacetime in the far future and past,
i.e., $ C\rightarrow (1+2\epsilon)^2$ in the `out' region and
$C\rightarrow 1 $ at the `in' region.

If we define $|\gamma^{-}|^2 \equiv \left|\beta^{(-)}_{k}/\alpha^{(-)}_{k}\right|^2$, for this spacetime we get that 
\begin{equation}
\left|\gamma^{-}\right|^2 =\frac{(\omega_- +m\epsilon)(\omega_+ +
m\epsilon)}{( \omega_-
- m\epsilon)(\omega_+ -m\epsilon)}\frac{\sinh\left[\frac{\pi}{\rho} (\omega_- -
m\epsilon)\right]\sinh\left[\frac{\pi}{\rho} (\omega_- + m\epsilon)\right]}{\sinh\left[\frac{\pi}{\rho}(\omega_+
+m\epsilon)\right]\sinh\left[\frac{\pi}{\rho}(\omega_+ -m\epsilon)\right]}.\label{gamma-F}
\end{equation}

An analogous procedure can be followed for scalar fields \cite{caball}. The time dependent
Klein-Gordon equation in this spacetime is given by \b
\left(\partial_{\eta }^{2}+k^2+C(\eta) m^2\right)\chi_k(\eta)=0.
\e After some algebra, the solutions of the Klein-Gordon equation
behaving as positive frequency modes as $\eta \rightarrow -\infty
(t\rightarrow -\infty )$, are found to be
\begin{align}
\chi_{\text{in}}(\eta)&=\exp\left(-i\omega_+\eta-
 \nonumber \frac{i\omega_-}{\rho}\ln[2\cosh\rho\eta]\right)\\*
&\times F\left(\frac{1}{2}-\frac{i\bar{\omega}}{2\rho}+\frac{i\omega_-}{\rho},
\frac{1}{2}+\frac{i\bar{\omega}}{2\rho}+\frac{i\omega_-}{\rho},
1-i\frac{\omega_{\text{in}}}{\rho},\frac{1+\tanh(\rho\eta)}{2}\right).
\end{align}
Similarly we have
\begin{align}
\chi_{\text{out}}(\eta)&=\exp\left(-i\omega_+\eta-
 \frac{i\omega_-}{\rho}\ln[2\cosh\rho\eta]\right)\nonumber\\*
&\times F\left(\frac{1}{2}-\frac{i\bar{\omega}}{2\rho}+\frac{i\omega_-}{\rho},
\frac{1}{2}+\frac{i\bar{\omega}}{2\rho}+\frac{i\omega_-}{\rho},
1-i\frac{\omega_{\text{out}}}{\rho},\frac{1-\tanh(\rho\eta)}{2}\right),
\end{align}
 where $\bar{\omega}=(m^2(2\epsilon+1)^2-\rho^2)^{1/2}$. Computing the quotient of the Bogoliubov coefficients for this bosonic case,  we find
 \b
|\gamma_B^{-}|^2=\frac{\cosh\frac{\pi}{\rho}\bar{\omega}
+\cosh\frac{2\pi}{\rho}\omega_-}
{\cosh\frac{\pi}{\rho}\bar{\omega}
+\cosh\frac{2\pi}{\rho}\omega_+}.  \label{gammaB}
\e

\section{Entanglement generated due to the expansion of the Universe}\label{sec3}

It is possible to find the state in the far future that
corresponds to the vacuum state in the far past. By doing that we will show that the vacuum state of the field in the asymptotic past evolves to an entangled state in the asymptotic future. The entanglement generated by the expansion codifies information about the parameters of the expansion, this information is more easily obtained from fermionic fields than bosonic, as we will show below.

Since we want to study fundamental behaviour we will consider the 2-dimensional case, which has all the fundamental features of the higher dimensional settings. 
 
 Using the relationship between particle operators in asymptotic times,
 \begin{equation}
b_{\text{in}}(k)=\left[{\alpha^{-}_{k}}^{\ast}b_{\text{out}}(k)+{\beta^{-}_{k}}^{\ast}%
\chi(k)a_{\text{out}}^{\dagger}(-k)\right],
\end{equation}
 we can obtain the asymptotically past vacuum state in terms of the asymptotically future Fock basis. Demanding that
$b_{\text{in}}(\bar{k},\lambda)|0\rangle_{\text{in}}=0$ we can find the `in'
vacuum in terms of the `out' modes. Due to the form of the Bogoliubov transformations the `in' vacuum must be of the form 
\begin{equation}
|0\rangle_{\text{in}}=\prod_{k}(A_{0}|0
\rangle_{\text{out}}+A_{1}|1_{k} 1_{-k}\rangle_{\text{out}} ),
\end{equation}
where  to compress notation $\ket{1_{-k}}$ represents an antiparticle mode with momentum $-k$ and $\ket{1_k}$ a particle mode with momentum $k$. Here we wrote
the state for each frequency in the Schmidt decomposition. Since
different $k$ do not mix it is enough to consider only one
frequency.  Imposing $b_{\text{in}}(\bar{k})|0\rangle_{\text{in}}
=0$ we obtain the following condition on the vacuum coefficients
\begin{equation}
{\alpha^{-}_{k}}^*A_{1}|1_{-k}\rangle+
{\beta^{-}_{k}}^*\chi(\bar{k} )A_{0}|1_{-k}\rangle=0,
\end{equation}
giving
\begin{equation}
A_{1}=-\frac{ {\beta^{-}_{k}}^*}{{\alpha^{-}_{k}}^*}\chi(\bar{k}
)A_{0}=-\gamma^{-\ast}\chi(\bar{k})A_{0},
\end{equation} where
\b\gamma^{-\ast}({k})=\frac{
{\beta^{-}_{k}}^*}{{\alpha^{-}_{k}}^*}.\e 

Now, imposing the vacuum
normalisation we have that
\begin{eqnarray}
1&=&_{\text{in}}\langle 0|0\rangle_{\text{in}}=|A_{0}|^{2}(1+|\gamma^{-}({k})\chi(\bar{k}%
)|^{2}).
\end{eqnarray}
These results lead to the following expression for the vacuum state 
\begin{equation}
|0\rangle_{\text{in}}=\prod_{k}\frac{|0\rangle_{\text{out}}-\gamma^{-\ast}({k})\chi(\bar{k})|1_{k}
1_{-k}\rangle_{\text{out}} }{\sqrt{1+|\gamma^{-}({k})\chi(\bar{k})|^{2}}}
 \label{vacin}
\end{equation}
which, in the asymptotic future, is an entangled state of particle modes and antiparticle modes with opposite momenta.

Since the state is pure, the entanglement is quantified by the
von-Neumann entropy given by
$S(\rho_k)=\mathrm{Tr}(\rho_k\log_2\rho_k)$ where $\rho_k$ is the
reduced density matrix of the state for mode $k$.  Tracing over the antiparticle 
modes with momentum $-k$ (or alternatively, particle modes with momentum $k$)  we obtain 
\b
\rho_{k}=\frac{1}{(1+|\gamma_k^{-}\chi(\bar{k})|^{2})}(|0\rangle%
\langle0|+|\gamma_k^{-\ast}({k})\chi(\bar{k})|^{2}|1_{k}
\rangle\langle1_{k}|). \e The von Neumann entropy of this state is
simply
\b
S(\rho_{k})=
\log(1+|\gamma_k^{-}\chi (\bar{k})|^{2})
-\frac{|\gamma_k^{-\ast}\chi (\bar{k})|^{2}\log(|\gamma_k^{-\ast}\chi (\bar{k})|^{2})}{(1+|\gamma_k^{-}\chi (%
\bar{k})|^{2})}.
\label{Entropy} \e Using the following identity
\b
|\chi_{\lambda\lambda}|^{2}=
2\mu_{\text{out}}K^{2}_{\text{out}}(\mu_{\text{out}}-\omega_{\text{out}})=
 \left[\frac{%
\mu_{\text{out}}}{|k|}\left(1-\frac{\omega_{\text{out}}}{\mu_{\text{out}}}\right)\right]^{2} \e we
rewrite the entanglement entropy as
\begin{align}
\nonumber S(\rho_{k})
&=\log\left(1+\frac{\mu^{2}_{\text{out}}}{|k|^{2}}\left(1-\frac{\omega_{\text{out}}}{%
\mu_{\text{out}}}\right)^{2}|\gamma_k^{-}|^{2}\right)-\frac{\frac{\mu^{2}_{\text{out}}}{|k|^{2}}\left(1-\frac{\omega_{\text{out}}}{\mu_{\text{out}}}%
\right)^{2}|\gamma_k^{-\ast} |^{2}}{(1+\frac{\mu^{2}_{\text{out}}}{|k|^{2}}%
\left(1-\frac{\omega_{\text{out}}}{\mu_{\text{out}}}\right)^{2}|\gamma_k^{-}
|^{2})}\\
&\times\log\left(\frac{\mu^{2}_{\text{out}}}{|k|^{2}}\left(1-\frac{\omega_{\text{out}}}{\mu_{\text{out}}%
}\right)^{2}|\gamma_k^{-\ast}(\bar{k})|^{2}\right).  \label{Ent2}
\end{align}

Using (\ref{Ent2}) we find that the fermionic entanglement entropy is
\begin{equation}
S_F= \log\left(\frac{1+|\gamma_F^{-}|^{2}}{|\gamma_F^{-}|^{
\frac{2|\gamma_F^{-}|^{2}}{|\gamma_F^{-} |^{2}+1}}}\right)
\label{SF}
\end{equation}
where $|\gamma_F| = |\gamma_k^{-}\chi (\bar{k})| $. Note that for massless fields
($m=0$)  the entanglement vanishes since $\omega_-=0$ and $\gamma^-$=0.
Comparing  our result to the bosonic
case studied in \cite{caball} we find\footnote{where the expression for $\gamma_B$ in (\ref{gammaB}) differs from that in ref.  \cite{caball} due to the different scale factor of the FLRW metric used to compute it. }
\begin{equation}
S_B=\log\left(\frac{|\gamma_B^{-}|^{%
\frac{2|\gamma_B^{-}|^{2}}{|\gamma_B^{-}
|^{2}-1}}}{1-|\gamma_B|^{2}}\right). \label{Eq:entang}
\end{equation}

The difference between the bosonic and fermionic cases means that the response of
entanglement  to the dynamics of the expansion of the Universe depends on the nature of the quantum field. We see from \eqref{vacin} that  each fermionic field mode is always in a qubit state (the exclusion principle imposes a dimension-2 Hilbert space for the partial state). However in the bosonic case \cite{caball} the Hilbert space for each mode is of infinite dimension, as every occupation number state of the `out' Fock basis participates in the `in' vacuum.
In both cases the entanglement increases monotonically with the expansion rate $\rho$ and the total volume expansion parameter $\epsilon$. It is possible to find analytically the asymptotic values that both fermionic and bosonic entanglement reach at infinity. For example, when $k=m=\rho=1$ we find that as $\epsilon\rightarrow\infty$
\b\label{yields}
\gamma_B^{-} \to e^{-\pi\sqrt{2}},  \qquad  \gamma_F \to e^{-\pi\sqrt{2}} \frac{e^{\pi\sqrt{2}}-e^{\pi}}{e^{\pi\sqrt{2}+1}-1}
\e
respectively yielding
\b
 S_{B}(\epsilon\rightarrow\infty)\approx 0.0913, \qquad    S_{F}(\epsilon\rightarrow\infty)\approx 0.0048.
 \e

To interpret these results one must realise that the entanglement entropy is bounded by $S_E < \log_2 N$ where $N$ is the Hilbert space dimension of the partial state. The  fermionic upper limit $S_E=1$ corresponds to a maximally entangled state. For bosons the unbounded dimension of the Hilbert space implies the entropy of entanglement is not bounded by  unity  \cite{caball}. This means that \eqref{yields} does not guarantee that we can extract more information from bosons than from fermions. In fact, we shall now demonstrate that it is exactly the opposite.

\section{Fermionic entanglement and the expansion of the Universe}\label{sec4}

\begin{figure}[H]
\begin{center}
\includegraphics[width=.525\textwidth]{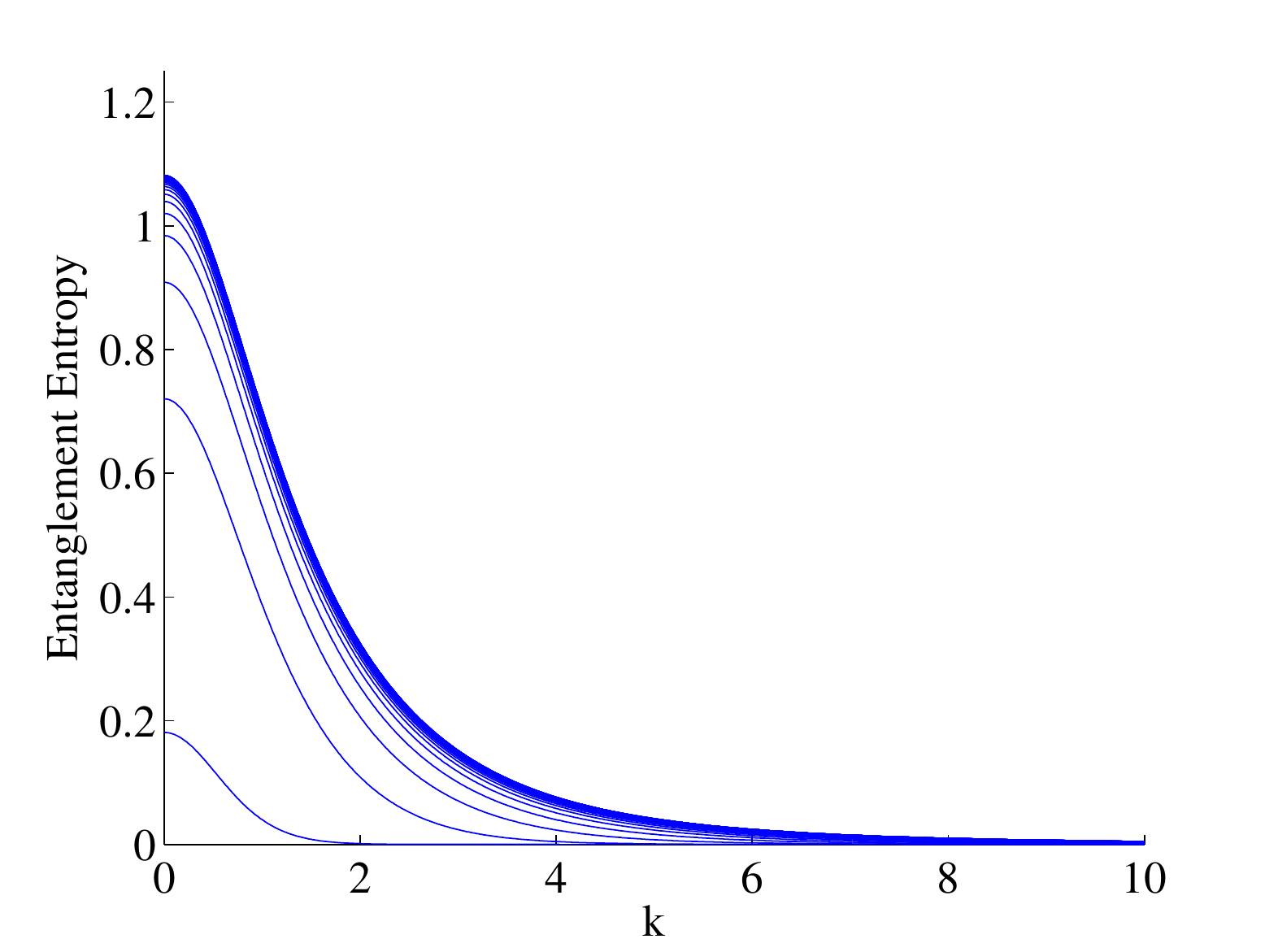}\!\!\!\!\!\!\!\!\!\!\!\!\!\!
\includegraphics[width=.525\textwidth]{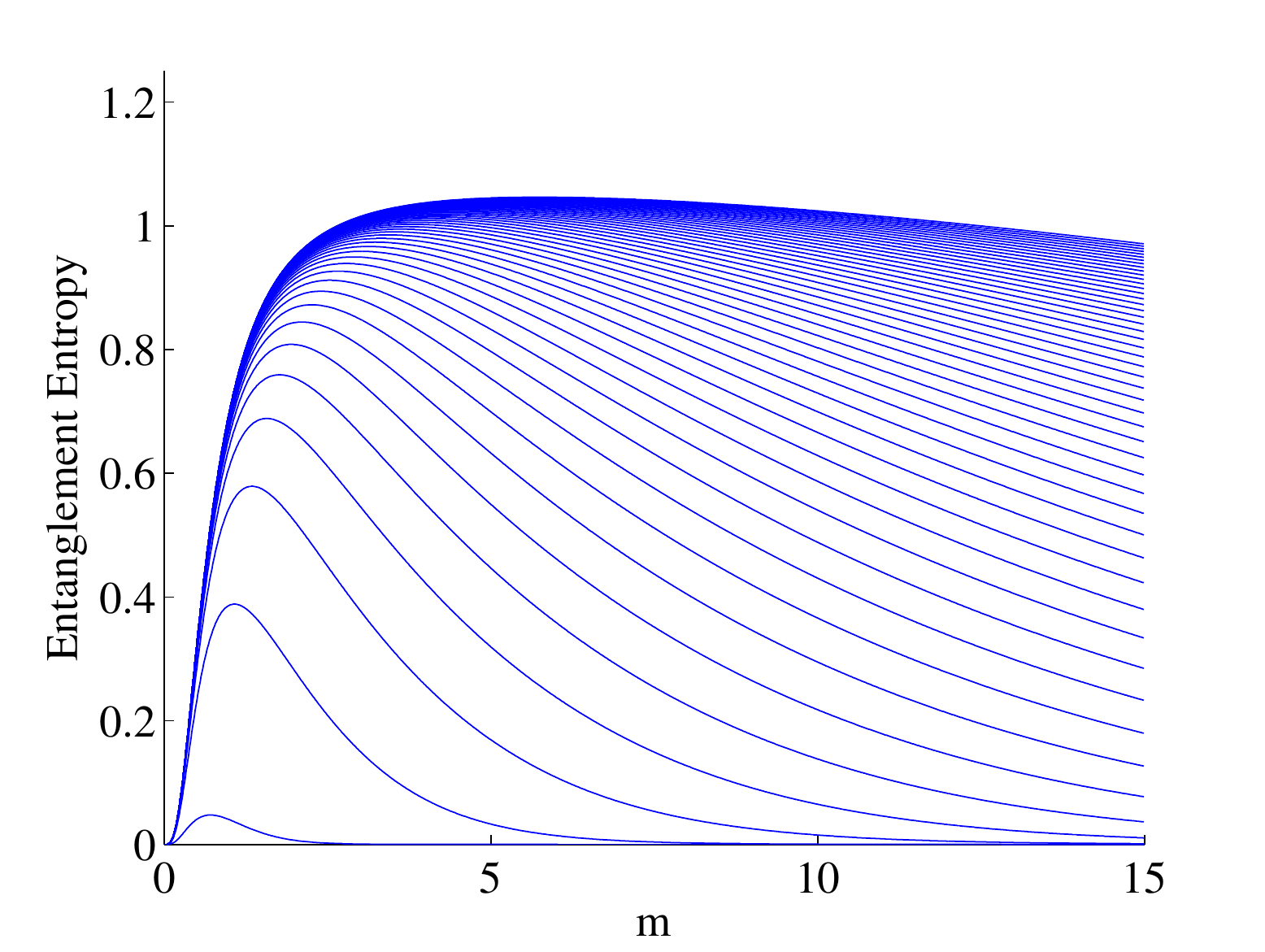}
\end{center}
\caption{Bosonic field: $S_E$ for a fixed mass $m=1$ as a function of $|\bm k|$ (left) and for a fixed $|\bm k|=1$ as a function of $m$ (right) for different rapidities $\rho=1,\dots,100$. An asymptotic regime is reached when $\rho\rightarrow\infty$. $\epsilon$ is fixed $\epsilon=1$}
\label{bosons}
\end{figure}

\begin{figure}[H]
\begin{center}
\includegraphics[width=.61\textwidth]{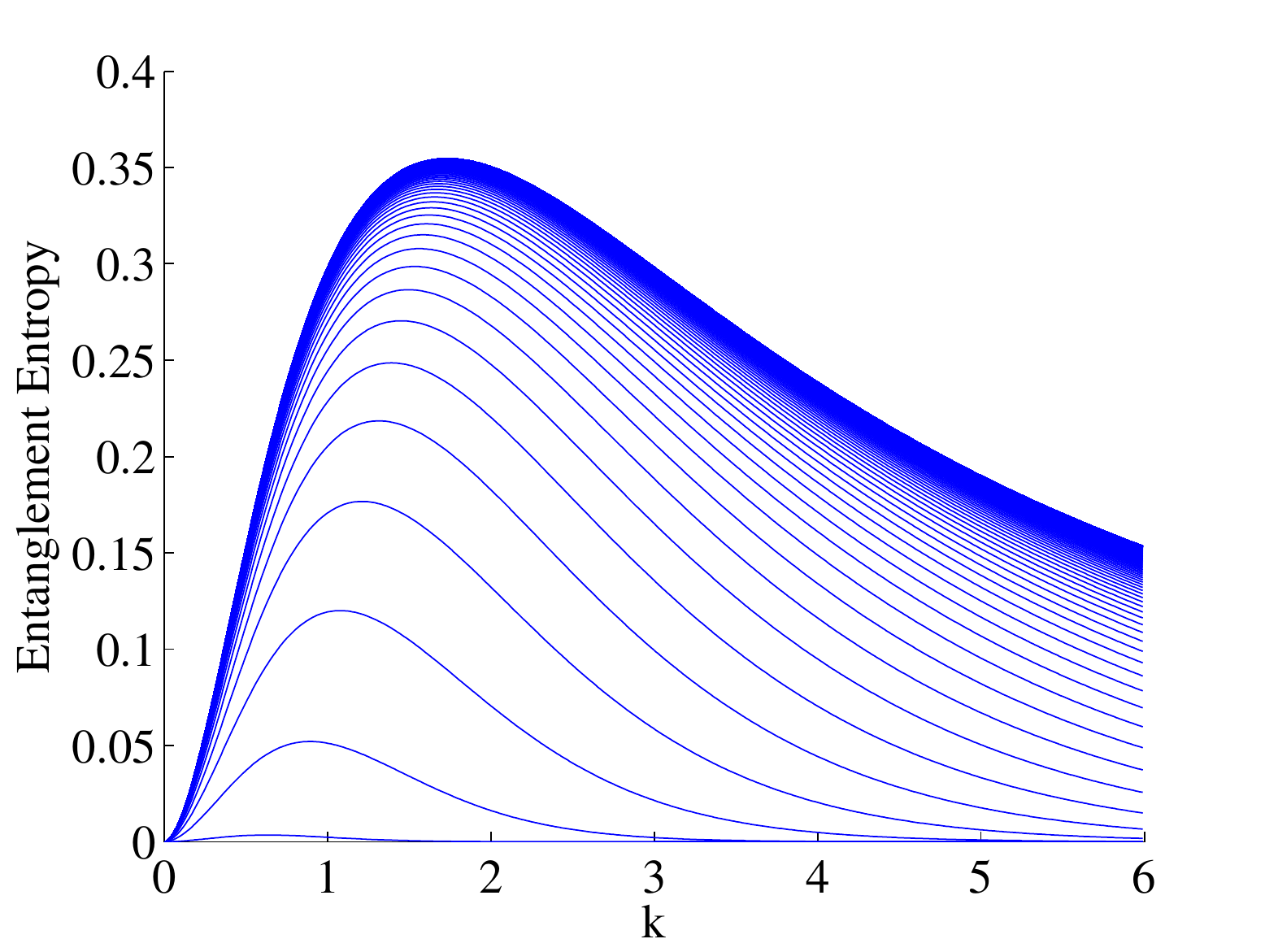}
\includegraphics[width=.61\textwidth]{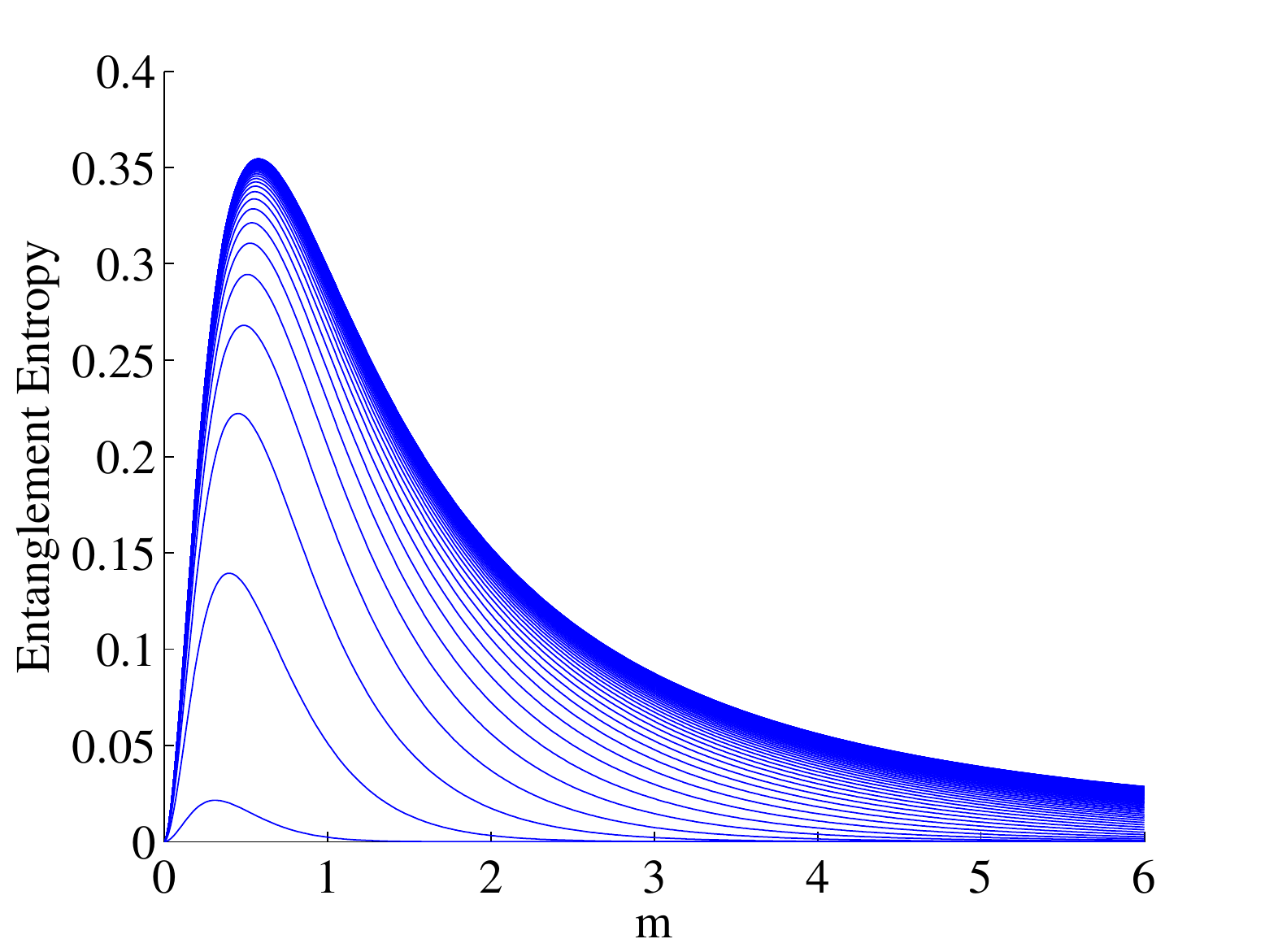}
\end{center}
\caption{ Fermionic field: $S_E$ for a fixed mass $m=1$ as a function of $|\bm k|$ (top) and for a fixed $|\bm k|=1$ as a function of $m$ (bottom) for different rapidities $\rho=1,\dots,100$. $\epsilon$ is fixed $\epsilon=1$. The behaviour as $|\bm k|$ varies is radically different from the bosonic case.}
\label{peaked}
\end{figure}

As seen in figures \ref{bosons} and \ref{peaked} the entanglement behaviour is completely different for bosons (Fig. \ref{bosons}) and fermions (Fig. \ref{peaked}). Although the behaviour as the mass of the field varies seems qualitatively similar, the variation with the frequency of the mode is completely different. 

The entanglement dependence on $|\bm k|$ for bosons is monotonically decreasing whereas for fermions, the global spacetime structure `selects' one value of $|\bm k|$ for which the expansion of the spacetime generates a larger amount of entanglement (peak in Figure \ref{peaked}). We shall see that this choice of a privileged mode is sensitive to the expansion parameters. This may be related to the fermionic nature of the field insofar as the exclusion principle impedes entanglement for too small $|\bm k|$.

Regardless of its origin, we can take advantage of this special behaviour for fermionic fields to use the expansion-generated entanglement to engineer a method to obtain information about the underlying spacetime more efficiently than for bosons. 

\subsection{Using fermionic fields to extract information from the ST structure}

\begin{figure}[H]
\begin{center}
\includegraphics[width=.75\textwidth]{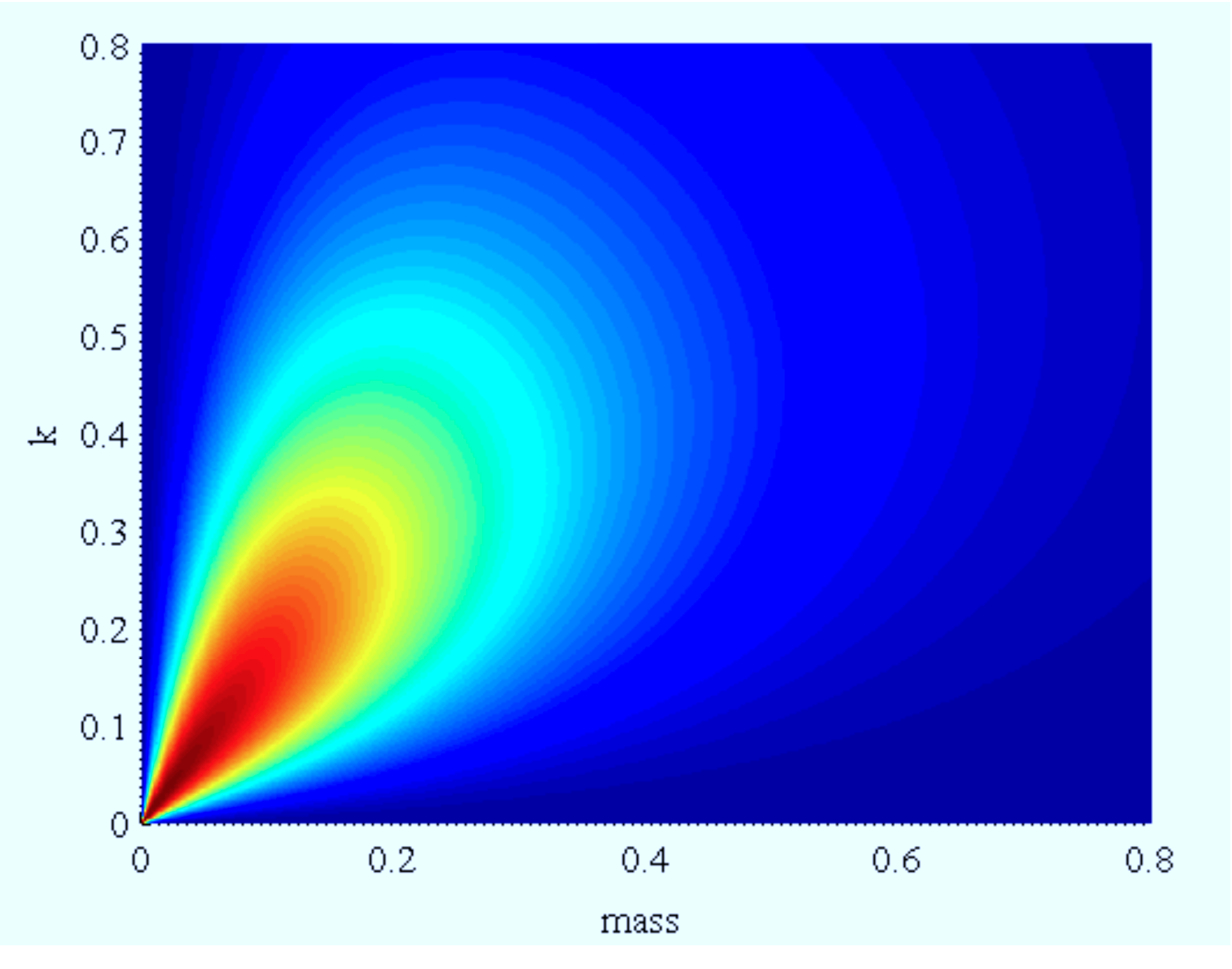}
\includegraphics[width=.75\textwidth]{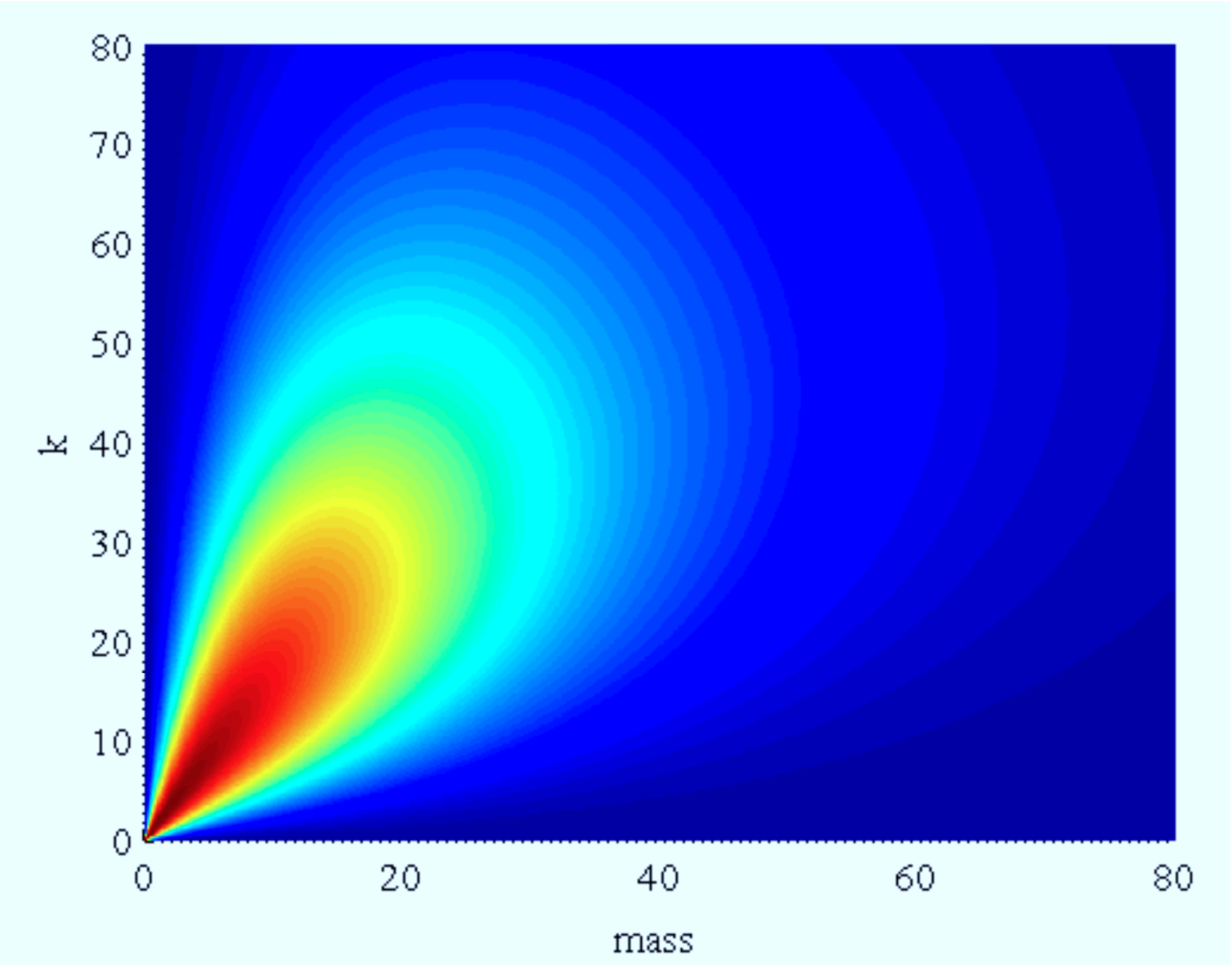}
\end{center}
\caption{ $S_E(m,|\bm k|)\!$ for $\!\epsilon=1,\! \rho=1\!$ (top) $\!\rho=100\!$ (bottom). Red is higher}
\label{fig3d2}
\end{figure}

Doing a conjoint analysis of the mass and momentum dependence of the entropy we can exploit the characteristic peak that $S_E(m,|\bm k|)$ presents for fermionic fields to obtain information from the underlying structure of the spacetime better than we can do with a bosonic field. Let us first show both dependences simultaneously. Figure \ref{fig3d2} shows the entropy of entanglement as a function of our field parameters ($|\bm k|$ and mass)  for different values of the rapidity.

\begin{figure}[h]
\begin{center}
\includegraphics[width=.51\textwidth]{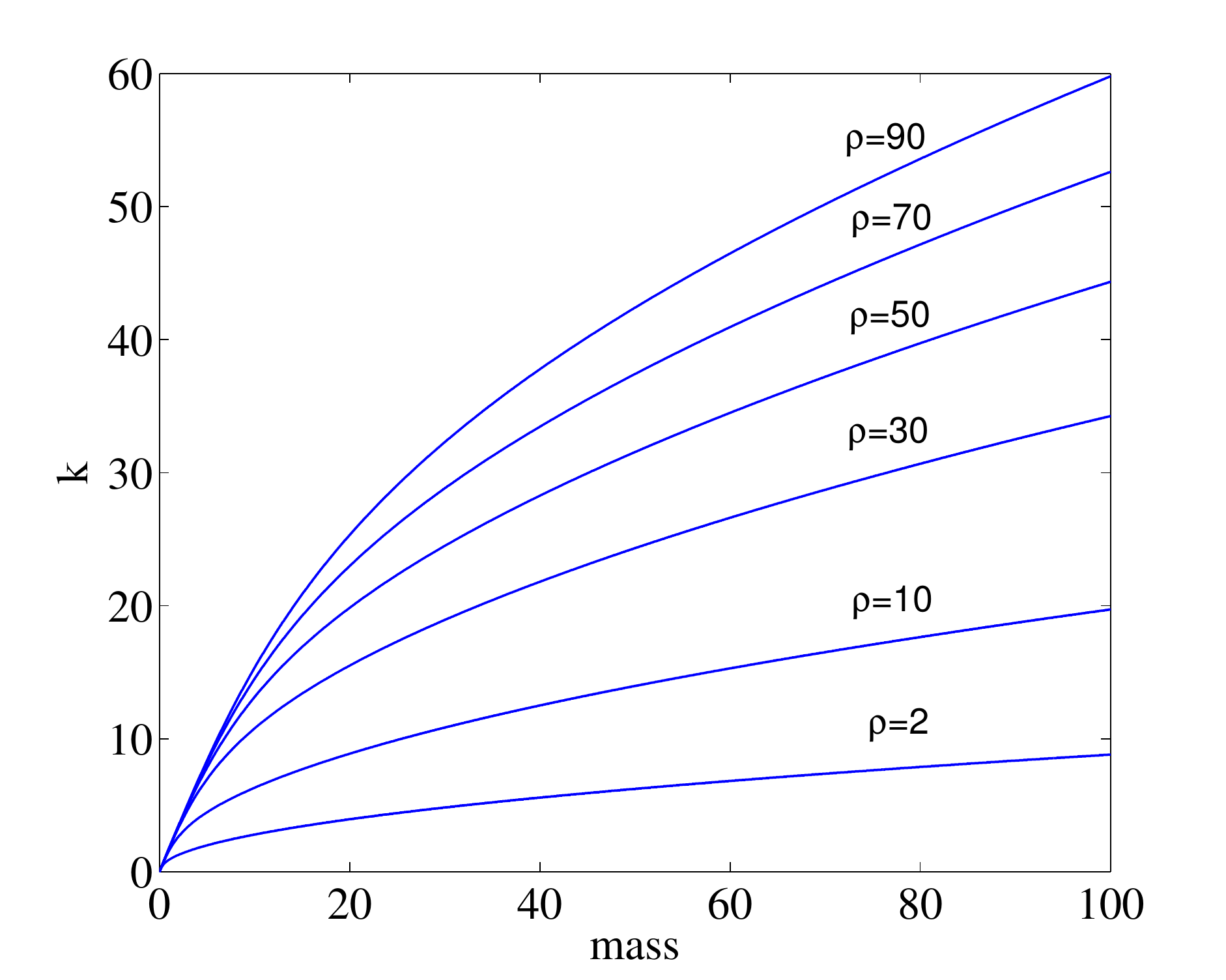}\!\!\!\!\!\!\!
\includegraphics[width=.51\textwidth]{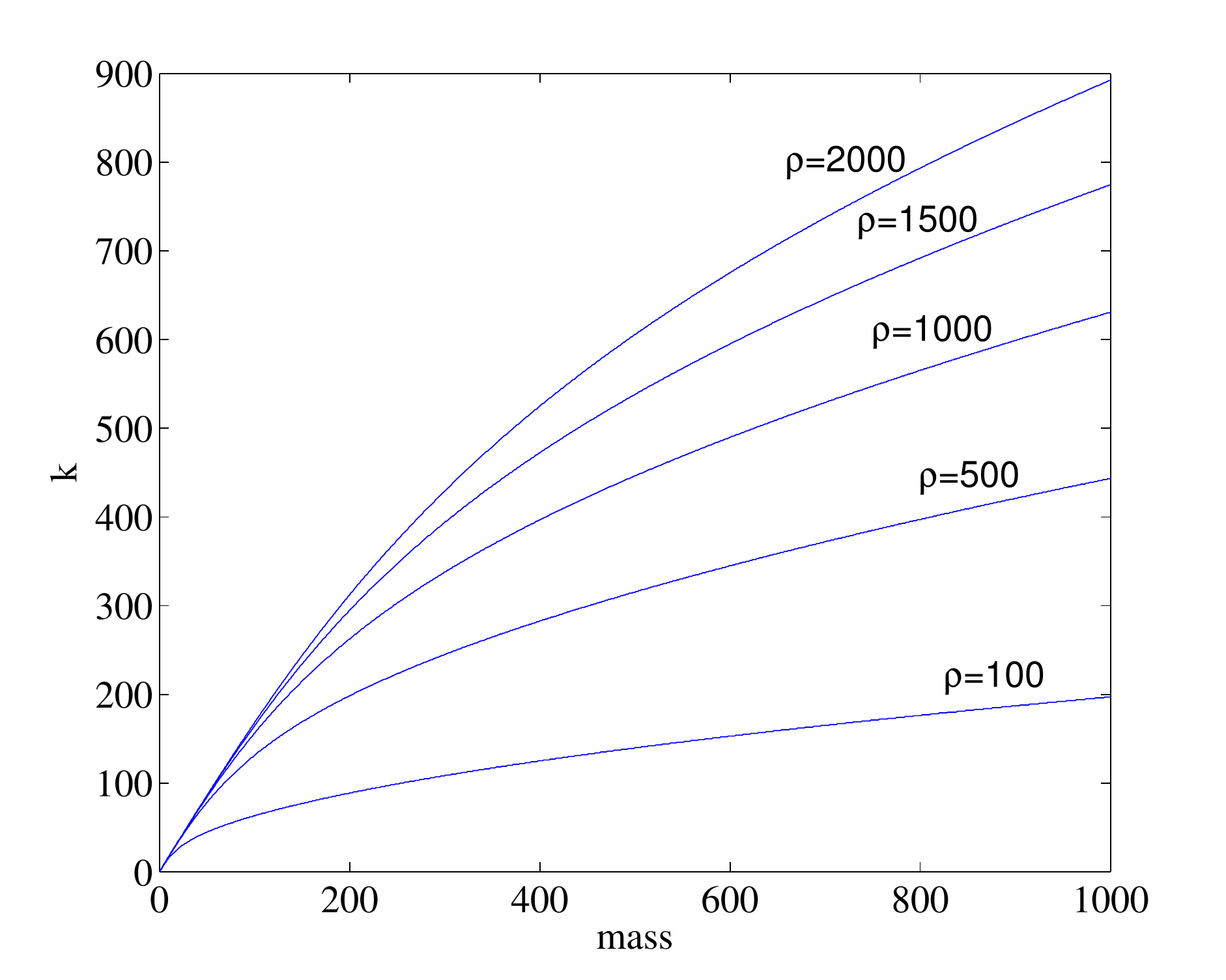}
\end{center}
\caption{$\epsilon=1$, optimal $|\bm k|$ curves (maximum entanglement mode) as a function of the field mass for $\rho=2,10,\dots,2000$}
\label{Fig8}
\end{figure}

We see from figure \ref{fig3d2} that there is no saturation as $\rho\rightarrow\infty$. Instead, as $\rho$ is increased the plot is just rescaled. This is crucial in order to be able to trace back the metric parameters from  entanglement creation.  
We also see from Figure \ref{fig3d2}  that, for a given  field mass, there is an optimal value of $|\bm k|$ that maximises the entropy. In Figure \ref{Fig8} we represent this optimal $|\bm k|$ as a function of the mass for different values of $\rho$, showing how the mode which gets the most entangled as a result of the spacetime expansion changes with the mass field for different rapidities.

From the figure we can readily notice two important features
\begin{itemize}
\item The optimal $|\bm k|$ curve is very sensitive to $\rho$ variations and  there is no saturation (no accumulation of these lines) as $\rho$ is increased.
\item There is always a field mass for which the optimal $|\bm k|$ clearly distinguishes arbitrarily large values of $\rho$.
\end{itemize}

\begin{figure}[H]
\begin{center}
\includegraphics[width=.75\textwidth]{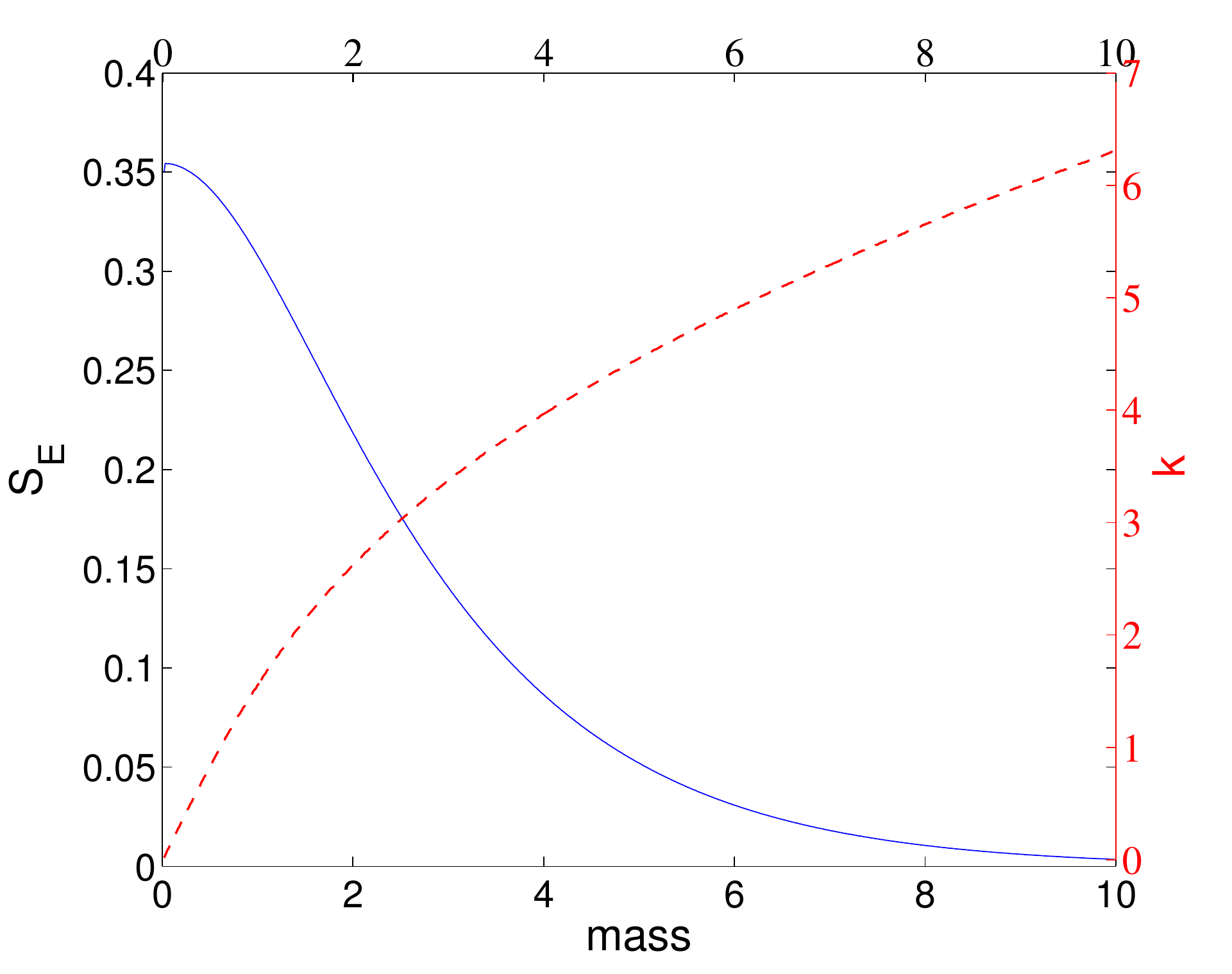}
\includegraphics[width=.75\textwidth]{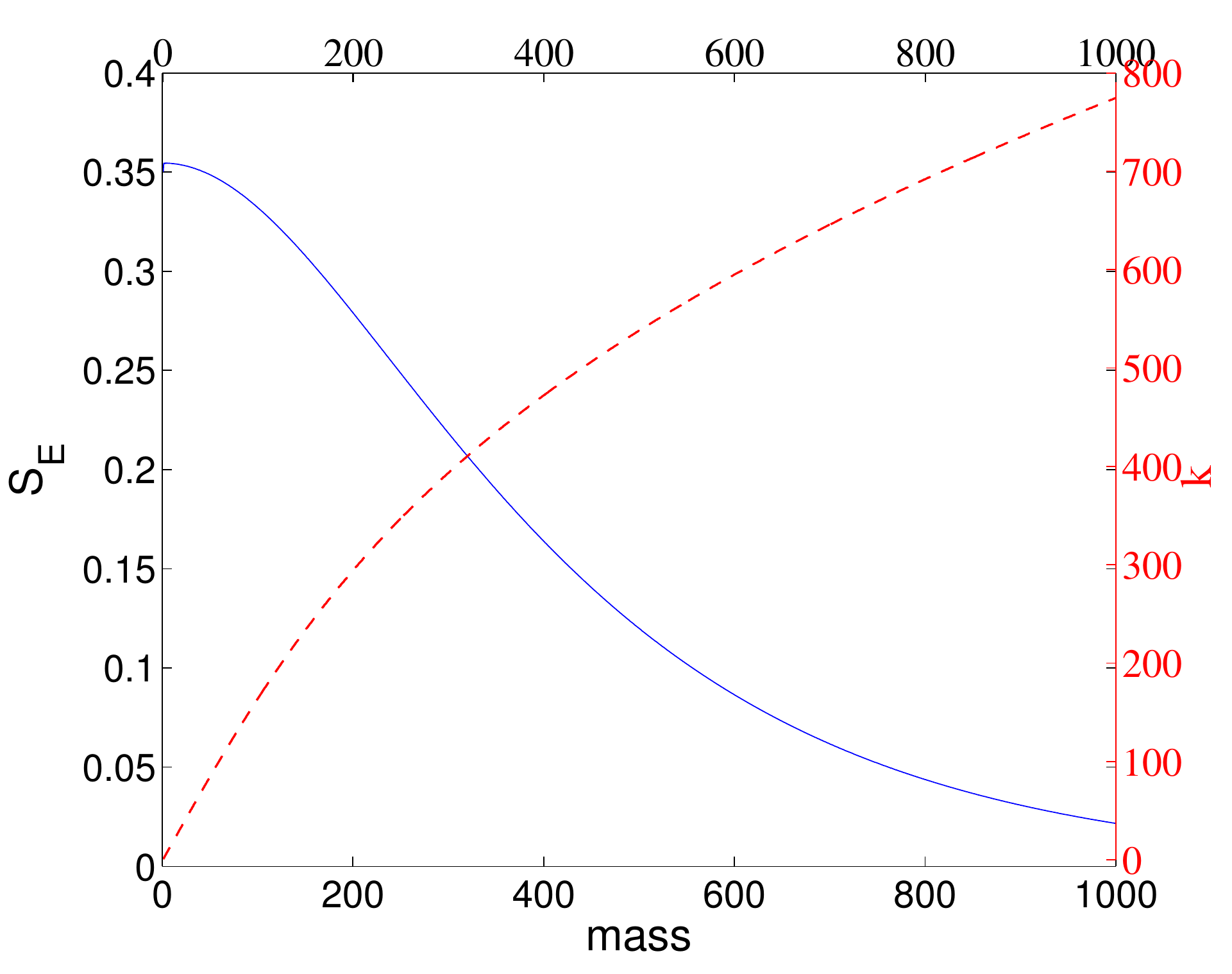}
\end{center}
\caption{ $S_E$ (blue continuous) and $k$ (red dashed) in the line of optimal $k$ as a function of mass for $\rho=10$ (top) and $\rho=1500$ (bottom). $\epsilon$ is fixed as $\epsilon=1$}
\label{Fig10}
\end{figure}

In Figure \ref{Fig10} we can see a consequence of the re-scaling (instead of saturation) of $S_E(m,|\bm k|)$ when $\rho$ varies. In this figure we show simultaneously the entropy in the optimal curve and the value of the optimal $|\bm k|$ as a function of the mass of the field for two different values of $\rho$, showing that if $\rho$ results to be very large,   entanglement decays more slowly for higher masses.

\begin{figure}[h]
\begin{center}
\includegraphics[width=.89\textwidth]{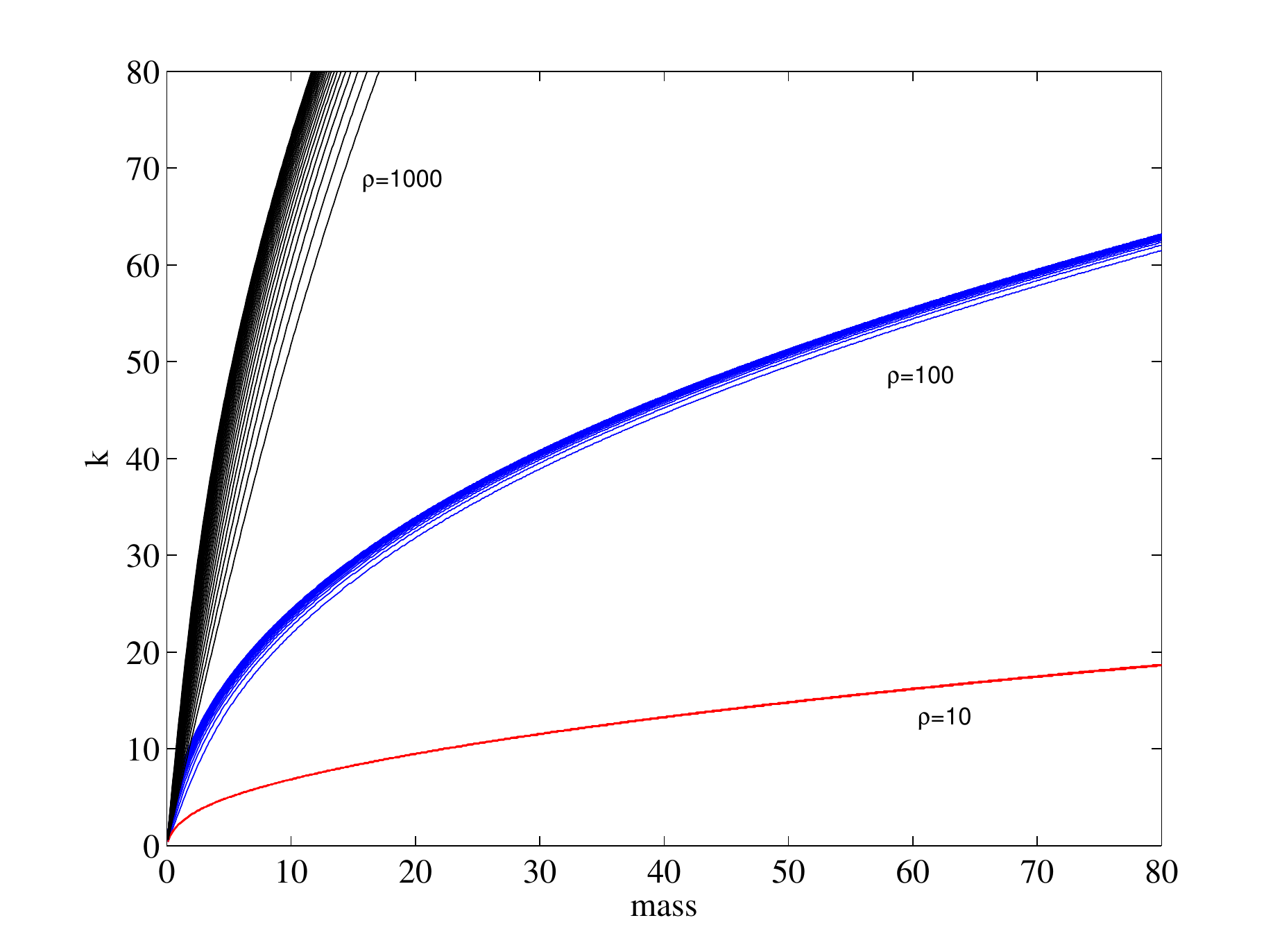}
\end{center}
\caption{ Optimal $|\bm k|$ as a function of the mass of the field for different vlues of $\epsilon=6,9,12,\dots,99$ and for $\rho=10,100,1000$ showing rapid saturation in $\epsilon$. For higher $\epsilon$ this curves are completely insensitive to $\epsilon$ variations, being very little sensitive for smaller values $\epsilon<10$}
\label{Figsat}
\end{figure}

The relationship between mass and the optimal frequency is very sensitive to variations in $\rho$, presenting no saturation. Conversely Figure \ref{Figsat} shows that the optimal $|\bm k|$ curve is almost completely insensitive to $\epsilon$. 

All the different $\epsilon$ curves are very close to each other. We can take advantage of this to estimate the rapidity independently of the value of $\epsilon$ using the entanglement induced by the expansion on fermionic fields.

\subsection{Optimal $|k|$ tuning method}

\subsubsection{Part I: Rapidity estimation protocol}

Given a field of fixed mass, we obtain the entanglement for different modes $k_1,\dots,k_n$ of the field. Then the mode $k_i$ that returns the maximum entropy will codify information about the rapidity $\rho$, as seen in Figure \ref{Fig8}.  One advantage of this method is that there is no need to assume a fixed $\epsilon$ to estimate $\rho$, since the tuning curves (Fig. \ref{Fig8}) have low sensitivity to $\epsilon$ (Fig. \ref{Figsat}). Furthermore this method does not saturate for  higher values of $\rho$ since we can use heavier fields to overcome the saturation observed in Figure \ref{peaked}. While one might expect  that heavier masses would mean smaller  maximum entropy, Figure \ref{Fig10} shows  that if $\rho$ is  high enough
to force us to look at heavier fields to improve its estimation, the  amount of entanglement will also be high enough due to the scaling properties of $S_E(|\bm k|,m)$. We can therefore safely use more massive fields to do estimate $\rho$ since they better codify its value.  

Hence  we have a method for extracting information about $\rho$ that is not affected by the value of $\epsilon$.  Information about $\rho$ is quite clearly encoded in the optimal $|\bm k|$ curve, which is a direct consequence of the peaked behaviour of $S_E(|\bm k|,m)$.

\subsubsection{Part II: Lower bound for $\epsilon$ via optimal $|k|$ tuning}

\begin{figure}[h]
\begin{center}
\includegraphics[width=.89\textwidth]{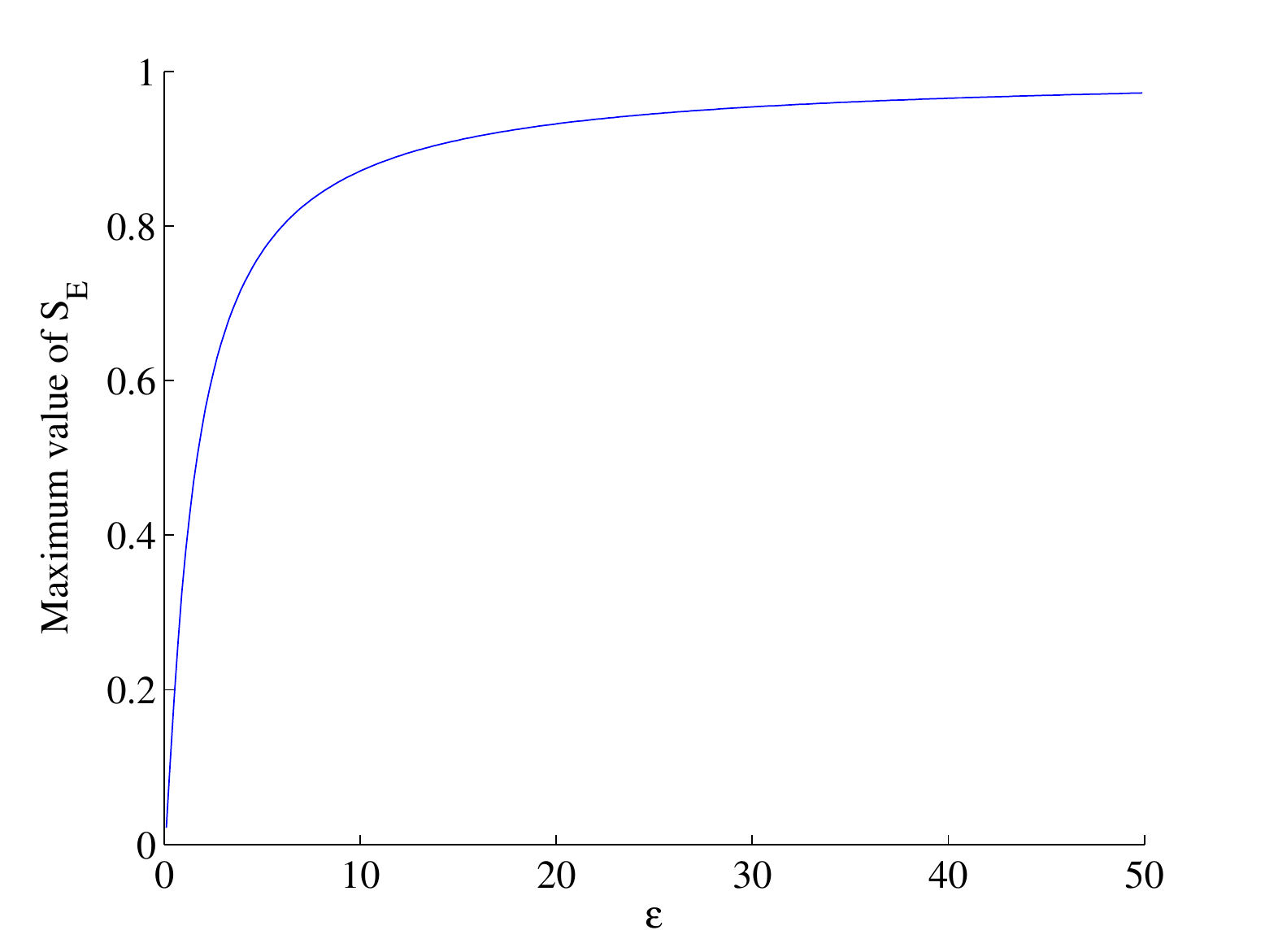}
\end{center}
\caption{$S_E^{\text{max}}(\epsilon)$: Maximum entanglement achievable (optimal $m$ and $|\bm k|$) as a function of $\epsilon$. It does not depend at all on $\rho$}
\label{epsmax}
\end{figure}

We can see from Figure \ref{Fig10} that for different values of  $\rho$  the maximum value for the entanglement at the optimal point (optimal $k$ and optimal $m$)  is always $S_E^{\text{max}}\approx0.35$. However, this is only for $\epsilon=1$. Consider now $\epsilon\neq 1$. In Figure \ref{epsmax} we can see how the maximum entanglement that can be achieved for optimal frequency and mass varies with the volume parameter $\epsilon$. Indeed, the maximum possible entanglement that the optimal mode can achieve is a function of only $\epsilon$ and is independent of $\rho$.
Hence  information about $\epsilon$ is encoded in the maximum achievable fermionic entanglement. Consequently we can find a method for obtaining a lower bound for the total volume of the expansion of the spacetime regardless of the value of the rapidity. 

In this fashion we obtain a lower bound for $\epsilon$ since the entanglement measured for the optimal mode is never larger than  the maximum achievable entanglement represented in Figure \ref{epsmax}, $S_E(|\bm k|,m) \le S_E^{\text{max}}$. For  instance if the entanglement in the optimal mode is $S_E>0.35$ this will tell us that $\epsilon>1$, whereas if $S_E>0.87$ then we can infer that $\epsilon>10$. Note that as $\epsilon$ increases the entanglement in the optimal $|k|$ mode for the optimal mass field  approaches that of a maximally entangled state when $\epsilon\rightarrow\infty$ .

Although this method presents saturation when $\epsilon\rightarrow\infty$ (being most  effective for $\epsilon \le 20$) its insensitivity to $\bm{\rho}$   means that the optimal $|\bm k|$ method gives us two independent methods for estimating $\rho$ and $\epsilon$. In other words, all the information about the parameters of the expansion (both volume and rapidity) is encoded in the entanglement for the optimal frequency $|\bm k|$.

\subsection{Interpretation for the dependence of $S_E$ on $|{\bm k}|$}

We have seen (Figure \ref{bosons}) that for bosons a monotonically decreasing entanglement is observed as $|\bm k|$ increases. By contrast, in the fermionic case we see that there are privileged $|\bm k|$ for which entanglement creation is maximum. These modes are far more prone  to entanglement than any others. 

To interpret  this  we can regard the optimal value of $|\bm k|$ as being associated with a characteristic wavelength (proportional to $|\bm k|^{-1}$) that is increasingly  correlated with the characteristic length of the Universe. 
As $\rho$  increases the peak of the entanglement entropy shifts towards higher $|\bm k|$, with smaller characteristic lengths. Intuitively,
 fermion modes with  higher characteristic lengths are less sensitive to the underlying spacetime because the exclusion principle impedes the excitation of `very long' modes (those whose $|\bm k|\rightarrow0$).

What about small $|\bm k|$ modes?
As shown in \cite{caball} and in Figure \ref{bosons},  the entanglement generation for bosons is higher when $|\bm k|\rightarrow0$. This makes sense because modes of smaller $|\bm k|$ are more easily excited as the spacetime expands (it is energetically much `cheaper'  to excite smaller $|\bm k|$ modes).
Entanglement generation for fermions, somewhat counterintuitively, decreases for $|\bm k|\rightarrow 0$.   However if 
we naively think of fermionic and bosonic excitations in a box we can appreciate the distinction. We can put an infinite number of bosons with the same quantum numbers into the box. Conversely, we cannot put an infinite number of fermions in the box due to the Pauli exclusion principle. This `degeneracy pressure'  impedes those `very long' modes (of small $|\bm k|$) from being entangled by the underlying structure of the spacetime.

\section{Discussion}\label{conclusionsexp}

We have shown that the expansion of the Universe (in a model 2-dimensional setting) generates an entanglement in quantum fields that is qualitatively different for fermions and bosons.  This result is commensurate with previous studies demonstrating significant differences between the entanglement of bosonic and fermionic fields  \cite{Alicefalls,AlsingSchul,Adeschul} and the ones found in previous chapters.

 We find that the entanglement generated by the expansion of the Universe as a function of the frequency of the mode peaks in the fermionic case, while  it decreases monotonically in the bosonic case. 
For bosons the most sensitive modes are those  whose $|\bm k|$ is close to zero. However for fermions modes of low $|\bm k|$ are insensitive to the underlying metric. There is an optimal value of $|\bm k|$ that is most prone to expansion-generated entanglement.  This feature may be a consequence of the Pauli exclusion principle, though we have no definitive proof of this.
 
 We have also demonstrated that information about the spacetime expansion parameters  is encoded
  in the entanglement between fermionic particle and antiparticle modes of opposite momenta.  This can be extracted from
the peaked behaviour of the entanglement shown in Figure \ref{peaked}, a feature  absent in the bosonic case.  
Information about the rapidity of the expansion ($\rho$) is codified in the frequency of the maximally entangled mode, whereas the information about the volume of the expansion ($\epsilon$) is codified in the amount of entanglement generated for this optimal mode. As  $\epsilon$ tends to infinity the maximum possible $S_E^{\text{max}}$ in the optimal mode approaches the maximally entangled state.

Hence the expansion parameters of spacetime are better estimated from cosmologically generated fermionic entanglement.   Furthermore, these results   show that fermionic entanglement is affected by the underlying spacetime structure in a very counterintuitive way and in a radically different manner than in the bosonic case.  The manner and extent to which these results carry over to $d$-dimensional spacetime remains a subject for future study.

\chapter{Berry's phase-based Unruh effect detection at lower accelerations\footnote{E. Mart\'in-Mart\'inez, I. Fuentes and R. B. Mann. arXiv:1012.2208}} \label{BerryPh}

\markboth{Chapter 13. Berry's phase-based Unruh effect detection}{\rightmark}

Finding indisputable corroboration of the Unruh effect is one of the main experimental goals of our time \cite{experiments,Crispino}. The effect is one of the best known predictions of quantum field theory incorporating general relativity. However, its  very existence has been object of a lengthy controversy  \cite{sceptic}.  Its observation would provide not only an end to such discussion but also experimental support for the Hawking radiation and black hole evaporation given the deep connection between these phenomena \cite{Hawking,Davies}.  Detection of the Unruh effect would therefore have an immediate impact in many fields such as astrophysics \cite{Astronature,Astrophysics}, cosmology \cite{Cosmo}, black hole physics \cite{Bholes}, particle physics \cite{Base}, quantum gravity \cite{Qg} and relativistic quantum information \cite{Alsingtelep,Alicefalls,QI}.

As seen along this thesis, in the Unruh effect \cite{Unruh0,Crispino} the vacuum state of a quantum field corresponds to a thermal state when described by uniformly accelerated observers. Its direct detection  is unfeasible with current technology since the Unruh temperature is smaller than 1 Kelvin even for accelerations
as high as $10^{21}$ $\text{m}/\text{s}^2$.  Sustained accelerations higher than $10^{26}$ $\text{m}/\text{s}^2$ are required in the best proposal so far  to diretcly detect the effect \cite{ChenTaj,Crispino}.

Efforts on finding evidence of the Unruh and Hawking effects also include proposals in analog systems such as fluids \cite{Unruhan}, Bose-Einstein condensates \cite{garay}, optical fibers \cite{optfib}, slow light \cite{slowlight},  superconducting circuits \cite{supercond} and trapped ions \cite{ions}.  Even in such systems,  analog effects produce temperatures of the order of nanokelvin that remain difficult to detect.

 In this chapter we show that the state of an accelerated detector coupled to the field acquires a Berry phase \cite{Berryoriginal,aharonov} due to its movement in spacetime. This geometric phase, which is a function of the detector's trajectory, encodes information about the Unruh temperature and it is observable for accelerations as low as $10^{17}$ $\text{m}/\text{s}^2$. Such acceleration must be sustained only for a few nanoseconds. Our results enormously simplify the challenge of measuring the Unruh effect with present technology since producing extremely high accelerations and measuring low temperatures were the main obstacles involved in its detection and we reduce the accelerations needed by a factor $10^9$.  The results presented here are independent of specific experimental implementations; however, we propose a possible scheme for the detection of this phase.

Interestingly, it has gone unnoticed that Berry's phase can be employed to detect the Unruh effect. Berry showed that an eigenstate of a quantum system acquires a phase, in addition to the usual dynamical phase, when the parameters of its Hamiltonian are varied in a cyclic and adiabatic fashion \cite{Berryoriginal}.  In the case of a point-like detector interacting with a quantum field, the movement of the detector in spacetime produces, under certain conditions, the cyclic and adiabatic evolution that gives rise to Berry's phase. In what follows we will show that the Berry phase for an inertial detector
differs from that of an accelerated one.  This difference arises due to the Unruh effect: one detector interacts with the vacuum state while the second interacts with a thermal state. The Berry phase of an accelerated detector depends on the Unruh temperature. Surprisingly, we find that this phase is observable for detectors moving with relatively low accelerations, making the detection of the Unruh effect  accessible with current technology.

\section{The setting}

In our analysis, we consider a massless scalar field in the vacuum state from the perspective of inertial observers moving in a flat ($1+1$)-dimensional spacetime. The same state of the field corresponds to a thermal state from the perspective of uniformly accelerated observers.  The temperature of this thermal state is the so-called Unruh temperature $T_U=\hbar a / (2\pi c \omega)$ where $a$ is the observer's acceleration.
\begin{figure}[h]
\begin{center}
\includegraphics[width=.70\textwidth]{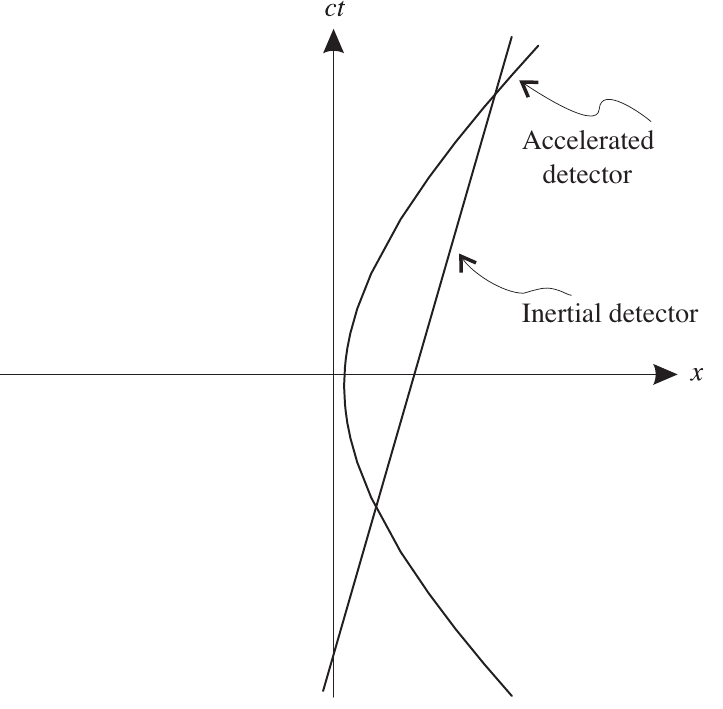}
\caption{Trajectories for an inertial and accelerated detector.}
\label{setup}
\end{center}
\end{figure}
In order to show evidence of this effect we consider a point-like detector endowed with an internal structure which couples linearly to a scalar field $\hat\phi(x(t))$ at a point $x(t)$ corresponding to the world line of the detector. The interaction Hamiltonian is of the form $H_I\propto  \hat m(t) \hat \phi(x(t))$ where $\hat m(t)$ is the monopole momentum of the detector. We have chosen the detector to be modeled by a  harmonic oscillator with frequency $\Omega_b$. In this case the operator $\hat m\propto (b^\dagger + b)$ corresponds to the detector's position quadrature where $b^{\dagger}$ and $b$ are ladder operators .

 Considering that the detector couples only to a single mode of the field with frequency  $|k|=\Omega_a$, the field operator takes the form
 \begin{equation}
 \hat\phi(x(t))\approx\hat\phi_k(x(t))\propto \left[a\, e^{i(kx-\Omega_a t)}+a^\dagger\, e^{-i(kx-\Omega_a t)}\right],
 \end{equation}
  where $a^{\dagger}$ and $a$ are creation and anihilation operators associated to the field mode $k$.  The Hamiltonian is therefore
\begin{equation}\label{goodham2}
H_{T}=\Omega_a\, a^\dagger a +\Omega_b\, b^\dagger b+ \lambda (b+b^\dagger)(a^\dagger e^{i\varphi}\!+ a\, e^{-i\varphi}),
\end{equation}
where $a$, $a^{\dagger}$ are creation and annihilation operators for the field and $b$, $b^{\dagger}$ are ladder operators of the harmonic oscillator's internal degrees of freedom. The phase $\varphi$  in \eqref{goodham2} is a function of time corresponding to the trajectory of the detector in spacetime.  Although calculations involving  Unruh-DeWitt detectors usually employ the interaction or the Heisenberg pictures (as transition probabilities are more conveniently calculated), in \eqref{goodham2} we employ a mixed picture where the detector's operators are time independent. This situation is mathematically more convenient for Berry phase calculations; the results are, of course, picture independent. 

  Note that,  in the particular case in which the detector models an atom, our interaction Hamiltonian corresponds to an Unruh-DeWitt-type detector \cite{Crispino},  with the proviso  that  the coupling $\lambda$ is such that the atom couples only to one mode of the field. In a realistic scenario the coupling $\lambda (\Omega_a,\Omega_b)$ can be considered as a peaked distribution. In the case that this function can be contrived to approach a delta function, we can assume  that only one mode of the field is coupled to the detector.  

This situation can be engineered, for instance, employing a cavity. Considering that the cavity field modes have very different frequencies and one of them is close to the detector's natural frequency,   the  detector effectively interacts only with this single mode. It is well known that introducing a cavity is problematic since the boundary conditions inhibit the Unruh effect. However, this problem is solved by allowing the cavity to be transparent to the field mode the detector couples to.  Therefore this single mode is a global mode.  In a realistic situation, the cavity would be transparent to a frequency window which is experimentally controllable. It is then an experimental task to reduce the window's width as required.

 The Hamiltonian \eqref{goodham2} can be diagonalised analytically.  The eigenstates are given by $U^{\dagger}|N_a N_b\rangle$ where $U=S_a(u,\theta_a)  S_b(v,\theta_b)  D_{ab}(s,\phi)\hat S_b(p) R_a(\varphi) $ and $|N_a N_b\rangle$ are eigenstates of the diagonal Hamiltonian\footnote{Note that $H_0$ is not the free part of the Hamiltonian $H_\text{T}$. The free part of $H_\text{T}$ is $\Omega_a\, a^\dagger a+\Omega_b\, b^\dagger b$.}  $H_0(\omega_a,\omega_b)=\omega_a\, a^\dagger a+\omega_b\, b^\dagger b$ which determines the energy spectrum of the system. The operators
\begin{align}
\nonumber D_{ab}=&D_{ab}(\chi)=\exp\big(\chi a^\dagger b - \chi^* a b^\dagger\big),\\*
\nonumber S_a=&S_{a}(\alpha)=\exp\big(\alpha^* {a^\dagger}^2 - \alpha a^2\big),\\*
\nonumber S_b=&S_{b}(\beta)=\exp\big(\beta^* {b^\dagger}^2 - \beta b^2\big),\\*
\nonumber  \hat S_b=&\hat S_{b}(p)=\exp\big[p\, ({b^\dagger}^2 - b^2)\big] ,\\*
R_a=&R_{a}(\varphi)=\exp\big(-i\varphi\, {a^\dagger a}\big)
\end{align}
are well-known in quantum optics  and called two-mode displacement (beam-splitter), single-mode squeezing and phase rotation operator, respectively. We define $\chi=s\,e^{i\phi}$, $\alpha=\frac12u\, e^{i\theta_a}$, $\beta=\frac12 v\,e^{i\theta_b}$.  To verify that these are the eigenstates of the Hamiltonian one must act on the diagonal Hamiltonian $H_0(\omega_a,\omega_b)=\omega_a a^\dagger a + \omega_b b^\dagger b$ with the unitary $U$ and identify terms with the Hamiltonian \eqref{goodham2}. By doing this we find constraints that fix the parameters $u,s,p,\phi,\theta_a,\theta_b$, obtaining the dependence of the coupling constant $\lambda$ and the frequencies  $\Omega_a$ and  $\Omega_b$ on  the parameters $v,\omega_a,\omega_b$. 

Specifically, in order to satisfy the equation
\[H_{T}(\omega_a,\omega_b,\alpha,\beta,\chi,\varphi)=U^{\dagger}H_0U\]
the following constraints must be satisfied
\begin{align*}\label{constr}
s=\operatorname{atan}\sqrt{\dfrac{\omega_a\,\sinh 2u}{\omega_b\,\sinh 2v}},\;\, \phi=\theta_a\!=  0,\;\,\theta_b=\!\pi,\;\, u=C-v,
\end{align*}
where $C=(1/2)\ln\left(\omega_a/\omega_b\right)$ with  $\omega_a/\omega_b>e^{2v}$.
The dependence of the Hamiltonian parameters on the parameters in the unitaries which diagonalise it are given by
\begin{align}&\Omega_a=\frac{\sinh 2v\left[\cosh \left[2(C-v)\right] +\frac{\sinh\left[2(C-v)\right]}{\tanh 2v}\right]}{{\omega_a^{-1}}\sinh 2v+\omega_b^{-1}\,\sinh \left[2(C-v)\right]},\\*[3mm]
&\Omega_b = \sqrt{\hat\Omega_b^2-4Z^2},\\*[3mm]
&\lambda=e^{p 
}\frac{\sqrt{\omega_a\omega_b\,\sinh [2(C-v)]\,\sinh 2v}}{\omega_b\,\sinh 2v+\omega_a\,\sinh [2(C-v)]}\left[\omega_a\,\cosh \left[2(C-v)\right] - \omega_b\cosh 2v\right],\\*[3mm]
&\varphi = kx-\Omega_a t,
\end{align}
where $2p=\operatorname{atanh}\big[-2Z/\hat \Omega_b\big]$ and
\begin{align}
\hat\Omega_b&= \frac{\sinh 2v\left[\omega_a^2\, \frac{\sinh \left[4\left(C-v\right)\right]}{2\sinh 2v}+\omega_b^2\,\cosh 2v\right]}{\omega_b\,\sinh 2v+\omega_a\,\sinh [2(C-v)]}\left[\omega_a\,\cosh \left[2(C-u)\right] - \omega_b\cosh 2u\right],\\*[3mm]
Z&=\frac12 \frac{\sinh 2v\left(\omega^2_a\frac{\sinh^2 \left[2(C-v)\right]}{\sinh 2v}- \omega_b^2 \sinh 2v\right)}{\omega_b\,\sinh 2v+\omega_a\,\sinh [2(C-v)]}.
\end{align}

The phase parameter $\varphi$ varies as a function of time, due to the time evolution along the detector's trajectory.  In the case of an inertial detector, $\varphi=|\Omega_a|x/c-\Omega_a t$ where Minkowski coordinates $(t,x)$ are a convenient choice in this case. Note that the movement of the detector in spacetime generates a change in the interaction Hamiltonian between the field and the atom. The change is cyclic; after a time $\Delta t\sim\Omega_a^{-1}$ the parameter $\varphi$ completes a $2\pi$ cycle completing a closed trajectory in parameter space.

Note that here we have considered the world line of an inertial observer in Minkowski coordinates $(t,x)$. For the accelerated case, the world line of the detector is given in Rindler coordinates and therefore, $\varphi = k\xi-\Omega_a \tau$.  The unitary $U$ has the same form in the Minkowski and Rindler basis, however, the operators involved must be considered in the appropriate basis. We will make this distinction explicit by naming $U_M$ and $U_R$ the unitaries involving Minkowski and Rindler operators, respectively.

Different choices of the parameters $v, \omega_a,\omega_b$  will produce specific values of $\Omega_a(v,\omega_a,\omega_b)$, $\Omega_b(v,\omega_a,\omega_b)$, $\lambda(v,\omega_a,\omega_b)$. In principle any desired experimental set-up can be attained for any values of $\Omega_a,\Omega_b$ and $\lambda$.

We consider that before the interaction between the field and the detector is switched on, the field is in the vacuum state and the detector in the ground state. Therefore, the system is in the state $\ket{0_f0_d}$. Employing the sudden approximation, we find that after the coupling is suddenly switched on\footnote{Suddenly switching on the coupling is known to be problematic since it can give rise to divergent results. However, in this case such problems are avoided by considering that the atom couples to a single mode of the field.}  the state of the system is
\begin{equation}\label{adiab}
\ket{0_f0_d}=\sum_{n,m} \ematriz{n_fm_d}{U}{00}U^\dagger \ket{n_fm_d}.
\end{equation}
In the coupling regimes we consider  $\ematriz{n_fm_d}{U}{00}= \bra{n_fm_d}{S_aS_bD_{ab}\hat S_b R_a}\ket{00}\approx \delta_{n_f0}\delta_{m_d0}$. Therefore, in this case, turning on the interaction does not excite the atom and the state of the system is  $U^\dagger\ket{0_f0_d}$. 

This fundamental eigenstate does not become degenerate and the energy gap between the ground and first excited state is time-independent.  For small but realistic coupling constant $\lambda$, energy conservation ensures a negligible probability for the system to evolve into an excited state.  In this case, the evolution due to the movement of the detector in spacetime  can be easily proven to be adiabatic since, if the field is initially in the vacuum state, for realistic couplings the ground state of the Hamiltonian $H(t_0)$, will evolve, after a time  $t-t_0$ to the ground state of the Hamiltonian $H(t)$\footnote{Conservation of energy guarantees that the probability of excitation of the detector in this case is virtually zero given small $\lambda$, so the adiabatic approximation that we need here (the vacuum evolves to the vacuum very approximately) holds very well.}. 

\section{Berry phase acquired by inertial detectors}

After the coupling is suddenly switched on and the state of the system is  $U^\dagger\ket{0_f0_d}$, the movement of the detector in spacetime, which can be considered cyclic and adiabatic, generates a Berry phase. 

The Berry phase acquired by the eigenstate $\ket{\psi(t)}$  of a system whose Hamiltonian depends on $k$ cyclicly and adiabatically varying parameters $R_1(t),\dots,R_k(t)$ is given by
\begin{equation}\label{Berry}
i\gamma=\oint_R\, \bm A \cdot  \text{d}\bm R,
\end{equation}
where
\begin{equation}\bm A=\left(\!\begin{array}{c}
\bra{\psi(t)}\partial_{R_1}\ket{\psi(t)}\\
\bra{\psi(t)}\partial_{R_2}\ket{\psi(t)}\\
 \vdots\\
 \bra{\psi(t)}\partial_{R_k}\ket{\psi(t)}
\end{array}\!\right)\end{equation}
and $R$ is a closed trajectory in the parameter space\footnote{See references \cite{Berryoriginal,aharonov} or, among many others, B. R. Holstein, American Journal of Physics. 57, 1079 (1989)}.\\

We calculate the Berry phase acquired by an eigenstate of the Hamiltonian \eqref{goodham2} under cyclic and adiabatic evolution of parameters  $(v,\varphi,\omega_a,\omega_b)$. $A_\varphi$ simplifies to $A_\varphi= \bra{N_f N_d}S_aS_bD_{ab}R_a\partial_{v} ( R_a^\dagger D^\dagger_{ab} S_b^\dagger S_a^\dagger)\ket{N_f N_d}$ and it is the only non-zero component of $\bm A$. Therefore,\begin{align}i\gamma_I= \oint_{\varphi\in[0,2\pi)}\!\!\!\!\!\!\!\!\!\!\!\!\!\!\! \bm A \cdot  \text{d}\bm R=\int_{0}^{2\pi}\text{d}\varphi\, A_{\varphi}.\end{align}
The Berry phase acquired by an eigenstate $U^\dagger\ket{N_f N_d}$ is
\begin{align}
 \gamma_{I_{N_fN_d}}\!=&\,2\pi\bigg[\frac{\omega_a\, N_d \cosh(2v)\sinh[2(C-v)]}{\omega_a\sinh[2(v)]+\omega_b\sinh(2v)}+\frac{\omega_b\, N_f \sinh(2v)\cosh[2(C-v)]}{\omega_a\sinh[2(C-v)]+\omega_b\sinh(2v)}+T_{00}\bigg],
\end{align}
where 
\begin{equation}
T_{00}=\frac{\omega_a \sin^2 v \sinh [2(C-v)]+\omega_b\,\sinh(2v)\sinh^2 (C-v)}{\omega_a\, \sinh[2(C-v)]+\omega_b\,\sinh (2v)}.
\end{equation}
In the specific case of the ground state\footnote{One should be careful with the adiabatic approximation if the state considered were not the ground + vacuum} ($N_f=N_d=0$).
\begin{align}
\gamma_{I_{00}}&=2\pi\,T_{00}.
\end{align}
Note that the ground state is non-degenerate and the gaps between energy levels are independent of time. 

The Berry phase acquired by the state $U^\dagger|00\rangle$ after a cyclic and adiabatic evolution of $\varphi$ for a given coupling frequency $\lambda$ is given by (See \cite{Berryoriginal})
\begin{align}
\frac{\gamma_{I_{00}}}{2\pi}=&\,\frac{\omega_a \sin^2 v \sinh [2(C-v)]+\omega_b\,\sinh(2v)\sinh^2 (C-v)}{\omega_a\, \sinh[2(C-v)]+\omega_b\,\sinh (2v)}.
\end{align}
Here the label $I$ denotes that the phase corresponds to the inertial detector. Note that the phase is identical for all inertial trajectories.  In what follows, we  show that, as a direct consequence of the Unruh effect, the phase is different for accelerated detectors.

\section{Berry phase acquired by accelerated detectors}

Computing the Berry phase in the accelerated case is slightly more involved.  A convenient choice of coordinates for the accelerated detector are Rindler coordinates $(\tau,\xi)$. In this case $\varphi=|\Omega_a|\xi-\Omega_a\tau$ and the evolution is cyclic after a time $\Delta \tau = \Omega_a^{-1}$.  Adiabaticity can also be ensured in this case since the probability of excitation is negligible for the accelerations we will later consider \cite{Scully,Crispino}.  

 We assume that it is exactly the same detector which couples to the field in the inertial and accelerated cases. Therefore, the detector couples to the same proper frequency (the frequency in the reference frame of the detector). It is important to point out that these frequencies are not the same from the perspective of any inertial observer. As mentioned above, the Hamiltonian \eqref{goodham2} has the same form in both scenarios, only that in the inertial case $a,a^\dagger$ are the Minkowski operators while for the accelerated detector, they correspond to Rindler operators.  For accelerated observers  the state of the field is not pure but mixed, a key distinction from  the inertial case.   Expressing the state of the field and  detector  in the basis of an accelerated observer, the state $\proj{0_f}{0_f}$ transforms  to  the thermal Unruh state $\rho_f$ \cite{Unruh0,Alicefalls}. 
 
Therefore, before turning on the interaction between the field and the detector, the system is in the mixed state  $\rho_f\otimes{\proj{0_d}{0_d}}$. When the interaction is suddenly switched on, a general state $\ket{N_f 0_d}$ evolves, in our coupling regime, very close to a superposition of eigenstates  $U_R^\dagger\ket{i_f j_d}$ where $N_f=i_f+j_d$. If immediately after switching on the interaction we verify that the detector is still in its ground state (by making a projective measurement) we can assure that the state of the joint system is  $\rho_T= U_R^\dagger \left(\rho_f\otimes{\proj{0_d}{0_d}}\right)U_R$.
We are now interested in the Berry phase acquired by the same detector if it follows a uniformly accelerated trajectory.  If the state of the field is in the vacuum state $\proj{0_f}{0_f}$ from the perspective of inertial observers, the corresponding state from the perspective of an accelerated observer is a thermal state (See section \ref{tue})
\b\rho_f=\frac{1}{\cosh^2r}\sum_n\tanh^{2n}{r}\proj{n_f}{n_f}\e
where $r=\operatorname{arctanh}\big(e^{-\pi\Omega_a c/a}\big)$ and $a$ is the proper acceleration of the detector.

We now use the formalism developed to compute the Berry phase acquired by mixed states \cite{Vlatko}. Let us consider a non-pure state of the form $\rho=\sum_i \omega_i \proj{i}{i}$ where $\ket{i}$ are eigenstates of the Hamiltonian. After a cyclic and adiabatic evolution the state acquires a geometric phase $\gamma = \operatorname{Re}\eta$  where
\begin{equation}\label{mixed}
e^{i\eta} = \sum_i\omega_ie^{i\gamma_i}.
\end{equation}
Here $\gamma_i$ is the Berry phase acquired by the eigenstate $\ket{i}$. 

Considering the state $\rho_T$ under cyclic and adiabatic evolution
\b e^{i\eta}=\frac{1}{\cosh^2 r}\sum_{n} \tanh^{2n}r\, e^{i\gamma_{I_{n0}}}=\frac{e^{i\gamma_{I_{00}}}}{\cosh^2 r-e^{2\pi\,i G}\sinh^2 r},\e
where
\b G= \frac{\omega_b\,  \sinh(2v)\cosh[2(C-v)]}{\omega_a\sinh[2(C-v)]+\omega_b\sinh(2v)}\e
depends on the characteristics of the detector and the coupling. Hence, the Berry phase acquired by the state of the accelerated detector is
\b \gamma_a=\gamma_{I_{00}}-\operatorname{Arg}\left(\cosh^2 r - e^{2\pi\,i G } \sinh^2 r\right),\e
where $\gamma_{I}$ is the inertial Berry phase and we recall $r=\arctan\big(e^{-\pi\Omega_a c/a}\big)$. Notice that $\gamma_a$ is the same no matter the sign of the acceleration.

\section{Measuring the phase. Detecting the Unruh effect}

We now compare the Berry phase acquired by the detector in the inertial and accelerated cases.
After a complete cycle in the parameter space (with a proper time $\Omega_a^{-1}$) the phase difference between an inertial and an accelerated detector is
$\delta =\gamma_{I}-  \gamma_a$.  

\begin{figure}[h]
\begin{center}
\includegraphics[width=.75\textwidth]{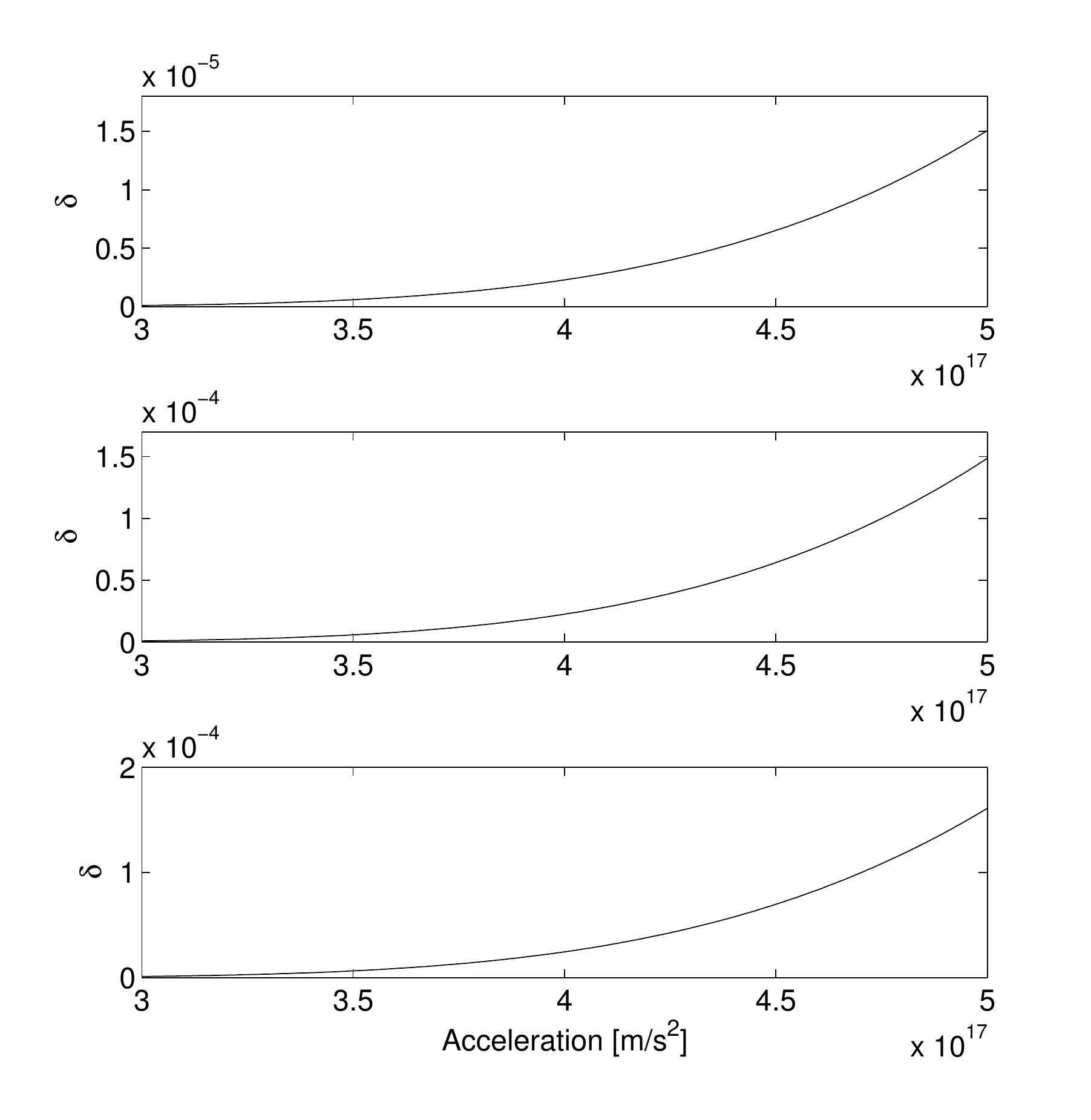}
\caption{$\delta$ for each cycle as a function of the acceleration for three different scenarios. First scenario (top): $\Omega_a\simeq 2.0$ GHz  $\Omega_b\simeq 2.0$ GHz $\lambda\simeq$ 34  Hz.
Second scenario (middle): $\Omega_a\simeq 2.0$ GHz  $\Omega_b\simeq 2.0$ GHz  $\lambda\simeq$  0.10  KHz.
Third scenario (bottom): $\Omega_a\simeq 2.0$ GHz  $\Omega_b\simeq 2.0$ GHz  $\lambda\simeq$ 0.25  KHz.}
\label{deph}
\end{center}
\end{figure}

In Figure \ref{deph} we plot the phase difference $\delta$ as a function of the acceleration corresponding to choosing physically relevant frequencies of atom transitions \cite{revat,Scully} coupled to the electromagnetic field (in resonance with the field mode they are coupled to) for three different coupling strengths: 
\begin{itemize}
\item $\Omega_a\simeq 2.0$ GHz  $\Omega_b\simeq 2.0$ GHz $\lambda\simeq$ $34$  Hz. 
\item $\Omega_a\simeq 2.0$ GHz  $\Omega_b\simeq 2.0$ GHz  $\lambda\simeq$  $0.10$   KHz. 
\item $\Omega_a\simeq 2.0$ GHz  $\Omega_b\simeq 2.0$ GHz  $\lambda\simeq$ $0.25$  KHz. 
\end{itemize}
The third case, where the coupling frequency $\lambda \simeq 10^{-7}\,\Omega_a$, corresponds to typical values for atoms in  free space with dipolar coupling  to the field \cite{revat}. For a single cycle (after 3.1 nanoseconds) the phase difference is large enough to be detected. The visibility of the interference pattern in this case is given by $V=\sqrt{\tr\left[\proj{0_f0_d}{0_f0_d}(\rho_f\otimes{\proj{0_d}{0_d}})\right]}=\cosh^{-1}r\simeq 1$. Note that the visibility is approximately  unity in all the situations we consider due to the relatively low accelerations considered. 

Since the Berry phase accumulates, we can enhance the phase difference by evolving the system through more cycles. By allowing the system to evolve for the right amount of time, it is possible to produce a maximal phase difference of $\delta=\pi$ (destructive interference). For example, considering an acceleration  of $a\approx 4.5\cdot10^{17}$ $\text{m}/\text{s}^2$  a maximal phase difference would be produced after $30000$ cycles.  Therefore, given the frequencies considered in our examples, one must allow the system to evolve for 95 microseconds. 

Note that for an acceleration of $a\approx 10^{17}$ $\text{m}/\text{s}^2$ the atom reaches speeds of $\approx 0.15\,c$ after a time $t\approx \Omega_a^{-1}$. 
If we do not vary the acceleration then the longer we allow the system to evolve in order to obtain a larger phase difference, the more relativistic the atom becomes. Therefore,  depending on the particular experimental implementation considered to measure the effect, a compromise between the desired phase difference and feasibility of handling relativistic atoms must be considered.  This is indeed an experimental challenge that can be overcome by means of different techniques. For example,  since the phase accumulates independently of the sign of the acceleration, one could consider alternating periods of positive and negative acceleration in order to reduce the final speed reached by the atom. 

The Berry phase acquired by an eigenstate of a system is always a global phase. In order to detect the phase, it is necessary to prepare an interferometric experiment. Indeed, considering a detector in a superposition of an inertial and accelerated trajectory would allow for the detection of the phase. In this context, any experimental set-up in which such a superposition can be implemented would serve our purposes. A possible scenario can be found in the context of atomic interferometry, technology that has already been successfully employed to measure with great precision general relativistic effects such as time dilation due to Earth's gravitational field \cite{atomGR}.

Consider the detector to be an atom which is introduced into an atomic interferometer after being prepared in its ground state. In one arm of the interferometer we let the atom move inertially while in the other arm we consider a mechanism  which produces a uniform acceleration of the atom. Such mechanism could consist of laser pulses that should be prolonged for fractions of nanoseconds.  Such laser technology to produce high accelerations is already available \cite{laser}.  Reaching the desired accelerations is technologically feasible -- however  the detector must survive such accelerations, at least long enough to conclude the interference experiment. For this, the laser pulses must be engineered to create the deep potential wells necessary to accelerate the atom transferring only kinetic energy without exciting it.  As long as the atom does not collide with other atoms this seems feasible \cite{ruso}.  An alternative to this is to consider ions or atomic nuclei as detectors which can be accelerated by applying a potential difference in one arm of the interferometer.  Although the suggested experimental set-up is obviously not exempt from technical difficulties, the problems derived from the experimental challenges involved are expected to be solvable with  present or near-future technology.

\section{Dynamical phase control}

To measure the Berry phase the relative dynamical phase between the inertial and accelerated paths must be cancelled. In principle the dynamical phase can be controlled and cancelled in any experimental setup. The way to do this depends on the specific experiment considered. 
Fortunately, there is considerable experimental freedom to engineer ways of canceling these phases.  In this section we discuss, as an example, the case of an atomic interferometry experiment. 
 
The control of the dynamical phase boils down to adjusting the relative path length of the interferometer.  Using commercial length metrology equipment it is possible to measure lengths with a precision of $\Delta L\approx 10^{-11}$ m  \footnote{Magnetscale Corporation $\text{Laserscale}^{\text{\textregistered}}$ http://www.gebotech.de/pdf/\\LaserscaleGeneralCatalog\_en\_2010\_04.pdf }.  A detector acquires different dynamical phases when its trajectory is  inertial or accelerated,  therefore the path lengths must be adjusted such that the relative dynamical phase difference cancels or is an integer multiple of $2\pi$.

Since dynamical phases oscillate and geometric ones accumulate, it becomes convenient to let the system evolve through many cycles.  In this way one can make sure that the Berry phase becomes much larger than any precision limit imposed on the length adjustment. If in a given situation the Berry and the dynamical phase differences are of the same order of magnitude, one can consider paths $n$ times longer such that the resulting Berry phase is multiplied by a factor of $n$. 

One can alternatively consider controlling the fly time in the accelerated path by changing the relative velocity of the atoms through the two paths  alternating periods of positive and negative acceleration. In this case, the Berry phase difference adds up (since it always has positive sign). 

In what follows we calculate explicitly with what precision the relative dynamical phase can be controlled.  Consider an atom moving with speed  $v$ (in the laboratory frame) along an inertial trajectory of length $L$. In the case $v\ll c$  the fly time is
$T=L/v$. Changing the path length with a precision of  $\Delta L$ translates to changing the fly time with a precision of
\[\Delta T = \left|\frac{\partial T}{\partial L}\Delta L\right|=\frac{\Delta L}{v}.\]
Since the dynamical phase $\phi$ is proportional to
$\Omega T$,
the resulting precision in adjusting the dynamical phase  is given by
\[\Delta \phi=\Omega \Delta T=\frac{\Omega \Delta L}{v}.\]
Considering for example, $v\approx 100$ m/s, $\Omega\approx 2$ GHz, and  $\Delta L\approx 1 \cdot 10^{-11}$ m we obtain $\Delta \phi=2\cdot10^{-4}$. This precision is enough to distinguish this phase from the Berry phase in a single cycle. The precision can further be increased by adjusting the path length of the accelerated atom.

The speed of the accelerated atom (in the laboratory frame) as a function of proper time is  given by
$v=c\,\tanh\left(\frac{a\tau}{c}\right)$. The path length for the accelerated atom in the laboratory frame $L_a$ is given by the integral of  the speed over proper fly time $T$,
\[L_a=\int_{0}^T c\,\tanh\left(\frac{a\tau}{c}\right) \text{d}\tau=\frac{c^2}{a}\ln\left[\cosh\left(\frac{aT}{c}\right)\right].\]
Therefore, the proper fly time as a function of the path length $L_a$ is
\[ T=c a^{-1}\operatorname{arccosh}\left[\exp\left(\frac{L_a a}{c^2}\right)\right].\]
The precision to control the proper fly time becomes  \[\Delta T=\frac{\Delta L_a}{c}\frac{\exp\left(\frac{L_aa}{c^2}\right)}{\sqrt{\exp\left(\frac{L_aa}{c^2}\right)-1}\sqrt{\exp\left(\frac{L_aa}{c^2}\right)+1}},\]
resulting in a precision to adjusting the dynamical phase of \[\Delta \phi=\Omega \Delta T=\Omega\frac{\Delta L_a}{c}\frac{\exp\left(\frac{L_aa}{c^2}\right)}{\sqrt{\exp\left(\frac{L_aa}{c^2}\right)-1}\sqrt{\exp\left(\frac{L_aa}{c^2}\right)+1}}.\]
\smallskip

For accelerations of $10^{17}$ $\text{m}/\text{s}^2$, $\Omega\approx 2$ GHz,  $\Delta L=10^{-11}$ m and proper lengths $L_a=1$ $\mu$m, $L_a=10$ cm we obtain, respectively\begin{eqnarray}
&\Delta \phi_{L_a=1\mu\text{m}}=2.25\cdot 10^{-8},&\\[3mm]
&\Delta \phi_{L_a=10\text{cm}}=1.49\cdot10^{-10}.&
\end{eqnarray}
We therefore conclude that it is possible to adjust the dynamical phase to several orders of magnitude greater in precision than
that of the Berry phase acquired in a single cycle of adiabatic evolution.

\section{Discussion}

With this proposal to detect the Unruh effect we overcome the difficulties associated with measuring temperatures as small as $10^{-4}$ K. We have shown that the Unruh effect leaves its footprint in the geometric phase acquired by the the joint state of the field and the detector for time scales of about $5\times10^{-10}$ s. The effect is observable for accelerations as low as $10^{17}$ $\text{m}/\text{s}^2$ and can be maximally enhanced by allowing the system to evolve a few microseconds. 

Notice that the Berry phase accumulates in each cycle of adiabatic evolution, and that this is true regardless the sign of the acceleration. This means that in any experiment we could alternate periods of positive and negative acceleration and still the Berry phase will grow, solving the problems of the high relative velocities between atoms going through the two different paths of the interferometer.

Our theoretical setting is general and independent of any particular implementation, paving the way for future experimental proposals. For instance, by considering detector frequencies in the MHz regime,  the method would allow detection of the Unruh effect for accelerations as low as $10^{-14}$ $\text{m}/\text{s}^2$ . For this, other multilevel harmonic systems could be employed as detectors, such as fine structure transitions where frequencies are closer to MHz regime.

\part*{Conclusions}

{\renewcommand{\thechapter}{}\renewcommand{\chaptername}{}
\addtocounter{chapter}{0}
\chapter*{Conclusions}\markboth{\sl CONCLUSIONS}{\sl CONCLUSIONS}}
\addcontentsline{toc}{part}{Conclusions}

This thesis is centred in the study of entanglement and quantum information problems in the background of general relativistic settings. In our exploration  of this brand new field called relativistic quantum information  we have obtained results in three different categories:

\begin{itemize}
\item[--]On the fundamental side, analysing the impact of statistics (fermionic or bosonic) on the behaviour of field entanglement in non-inertial frames; building the formalism to deal with entanglement of different degrees of freedom (spin, occupation number, ...) and studying how correlations behave in the proximities of an event horizon. 
\item[--]On the applied side, showing how entanglement can be useful to study the expansion of the Universe, the process of stellar collapse or to serve as a witness of the Unruh and Hawking effect which have not been detected yet.
\item[--]On the experimental proposals side, using the knowledge gained in quantum information to suggest experiments to detect quantum effects provoked by gravity: in analog gravity experiments, proposing ways to use entanglement to distinguish between quantum and classical Hawking effect; or taking advantage of the geometric phases acquired by moving detectors to directly measure the Unruh effect for accelerations much smaller than previous proposals.
\end{itemize}

\section*{Specific outcomes}

\begin{list}{\labelitemi}{\leftmargin=1em}
 \item We have shown the relationship between statistics and entanglement behaviour in non-inertial frames. We have studied fermionic cases beyond those in the literature (which only focused on spinless Grassmannian fields). We have formulated questions about the differences between fermionic and bosonic entanglement that helped us understand the origin of such differences. As a result of these studies we have disproved previous beliefs concerning the reason for the differences between bosonic and fermionic entanglement in non-inertial frames, showing that the Hilbert space dimension of the system has nothing to do with those differences.
\begin{itemize}
\item[--] We have extended the study of entanglement behaviour in non-inertial frames to spin 1/2 fields, analysing entanglement of spin Bell states from the perspective of non-inertial observers and comparing it with occupation number entanglement.
\item[--] A method to consistently erase the spin information from field states has been presented. With this method occupation number entanglement can be studied independently of the spin of the field considered.
\item[--] We have analysed entanglement behaviour in different kinds of states of fermionic fields, different spins and different dimension of the Hilbert space. We obtained universal laws for entanglement and mutual information that show that the Hilbert space dimension does not play any role in the  entanglement behaviour between inertial and accelerated observers.
\item[--] A comparative study between fermions and bosons in non-inertial frames has been presented, clearly exposing the differences between these two statistics not only for entanglement but also for the rest of correlations, classical and quantum, investigated by means of the mutual information. We payed special attention to all the possible bipartite systems that emerge from a spacetime with apparent horizons.
\item[--] Non-inertial bosonic field entanglement has been analysed in a scenario where we impose a bound on occupation number. We have seen that limiting the Hilbert space dimension has no qualitative effect on bosonic entanglement. We have compared a limited dimension bosonic field entangled state with its fermionic analog, showing that statistics (which imposes certain structure in the density matrix for the fermionic case via Pauli exclusion principle) is responsible for the differences between fermions and bosons.
\end{itemize}
\item We introduced a formalism to rigorously analyse entanglement behaviour between free-falling observers and observers resisting at a finite distance from the event horizon of a Schwarzschild black hole. We have shown at what distance from the event horizon of a typical solar mass black hole the Hawking effect starts to seriously disturb our ability to perform quantum information tasks.
\item The problems of the so-called `single mode approximation', used for years in all the literature on relativistic quantum information, have been exposed.  We have proved that it is not valid in general and that its meaning was misunderstood in those cases in which it is valid. We have shown and discussed the appropriate physical interpretation of such an approximation.  
 \item The first non-trivial results beyond the single mode approximation have been presented:
 \begin{itemize}
 \item[--] We have shown that for fermionic field entangled states there is an entanglement tradeoff between the particle and antiparticle sector of the different regions of the spacetime which is crucial to understand the phenomenon of fermionic entanglement survival in the limit of infinite Unruh temperature.
 \item[--] Contrary to the extended belief in the whole field of relativistic quantum information, we have shown that beyond the single mode approximation and for certain states of both fermionic and bosonic fields entanglement can be amplified instead of degraded.
 \end{itemize}
 \item We have analysed two dynamical scenarios in which the gravitational interaction generates quantum entanglement in the fields dwelling in the background of some interesting non-stationary spacetimes:
 \begin{itemize}
 \item[--] It has been shown how a stellar collapse generates entanglement. We have analysed the differences between fermionic and bosonic fields, showing that the former are better candidates to serve as a tool to detect genuine Hawking radiation in analog gravity experiments. We have also shown that for extremal black holes (microblackholes or the final stages of an evaporating black hole) the Hawking radiation emitted by the event horizon is in a maximally entangled state with the radiation falling into the black hole.
 \item[--] The fact that the expansion of the Universe generates entanglement in quantum fields has been analysed. Specifically we have analysed the differences between the bosonic and the fermionic case, showing that the entanglement created in fermionic fields contains more information about the history of the expansion and it is more reliable for a hypothetical experiment to obtain information about the parameters of the expansion. We have analysed a specific solvable model of inflationary-type expansion showing a protocol to extract complete information about the volume and rapidity of the expansion by means of fermionic entanglement. 
 \end{itemize}
 \item We have proved that a single mode detector moving through spacetime (even if it is at rest) acquires a geometric phase. We have shown that the phase is the same for any inertial detector but, due to the Unruh effect, it depends on the acceleration of the detector. As a consequence of this result we propose a generic way to use this geometric phase to detect the Unruh effect for accelerations as small as $10^{-9}$ times previous proposals. We conclude presenting a concrete experimental setup based on atomic interferometry using present-day technology. 
 \end{list}
\cleardoublepage

\appendix

\chapter{Appendix: Klein-Gordon and Dirac equations in curved spacetimes}
\label{appB}

In this brief appendix we introduce the Klein-Gordon and Dirac equations in the background of curved spacetimes. We also present the basis to understand what is a Grassmann field as a 1+1 Dirac field.

\section{Klein Gordon equation in curved spacetimes}

In flat spacetime the Klein-gordon equation has the well-known form
\begin{equation}
(\Box-m^2) \phi =0.
\end{equation}
where the D'Alembert operator is defined as $\Box=\partial_\mu\partial^{\mu}$. 

To extend this equation to general spacetimes the first step would be promoting the partial derivatives in the D'Alembert operator to covariant derivatives, we define such a general version of this operator as
\begin{equation}
\Box_g=\nabla^\mu \nabla_\mu=g^{\mu\nu}\nabla_\mu\nabla_\nu=\frac{1}{\sqrt{|g|}}\partial_\mu\left(\sqrt{|g|}g^{\mu\nu}\partial_\nu\right)
\end{equation}
where $g$ is the determinant of the metric tensor.

One can consider a free scalar field minimally coupled (one which does not transform under change of coordinates: $\phi'(x')=\phi(x)$ and therefore the field equation can be simply be written as
\begin{equation}
(\Box_g-m^2) \phi =0.
\end{equation}

However, this is not the most general kind of field one could have considered. This equation is a particular case of the Euler-Lagrange equations coming from the more general Klein-Gordon Langrangian density 
\begin{equation}
\mathcal{L}=\frac12\sqrt{\left|g\right|}\left(g^{\mu\nu}\partial_\mu\phi \partial_\nu \phi +m^2\phi^2-\xi R\phi^2\right),
\end{equation}
where the dimensionless constant $\xi$ couples the field to the scalar curvature. The more general Klein-Gordon equation is of the form 
\begin{equation}
(\Box_g-m^2+\xi R) \phi =0.
\end{equation}
The inclusion of this extra term, coupling the field with the curvature, is often included in the Lagrangian as a counter-term necessary to renormalise the theory when we include interaction terms such as $\sqrt{|g|}\lambda\phi^4$. In principle there is no physical reason to include such a term in the free case. Notice, as a curiosity, that for 4 dimensions and when $m=0$ if we chose $\xi=1/6$ (conformal coupling) the field equations are invariant under conformal transformations\footnote{The conformally related actions differ only by a surface term}. The case $\xi=0$ is known as `minimal coupling', and it is the case we adhere to. 

Notice also that there are physical reasons to induce that the constant $\xi$ cannot be large because if $\xi\neq0$ then the Lagrangian being proportional to $R\phi^2$  would cause the effective gravitational constant to vary with time and position as a result of the variations in $\phi$. In any case, the introduction of such term would act as an effective mass term (although dependent on spacetime). To describe all the casuistics we study here this is irrelevant.

\section{Dirac equation in curved spacetimes}

Let us define the flat spacetime Dirac matrices $\{\gamma^0,\gamma^1,\gamma^2,\gamma^3\}$
which have the following properties
\begin{equation}
\{\gamma^a,\gamma^b\}=2\eta^{ab},
\end{equation}
where $\eta^{ab}$ is the usual flat Minkowskian metric. Then the Dirac equation in Minkowski spacetime can be written as
\begin{equation}\label{Dirac3}
(i\gamma^{a}\partial_{a}+m)\psi=0,
\end{equation}

To write the Dirac equation in curved spacetimes we need to introduce the vierbein. Namely, an orthonormal set of four vector fields that serve as a local reference frame of the tangent Lorentzian manifold in each point of spacetime such that
\begin{equation}
g^{\mu\nu}=e^\mu_ae^\nu_b\eta^{ab}
\end{equation}
The vierbein enables us to convert local Lorentz indices to general indices.

With the help of the vierbein we can write the Dirac matrices $\gamma^\mu$ in a general spacetime as a function of the local gamma matrices
\begin{equation}
\gamma^\mu=e^\mu_a\gamma^a.
\end{equation}

 This curved spacetime gamma matrices fulfil
 \begin{equation}
\gamma^{\mu}\gamma^{\nu}+\gamma^{\nu}\gamma^{\mu}=2g^{\mu\nu}.
 \end{equation}

Now we have to be careful when defining the covariant derivative: we have a spinor bundle defined over the spacetime manifold. The spin connection can be expressed in terms of the Levi-Civita connection $\hat\Gamma^\nu_{\sigma\mu}$ as
\begin{equation}
\omega_\mu^{ab}=e^a_\nu\partial_\mu e^{\nu b}+ e^{a}_\nu e^{\sigma b}\hat\Gamma^{\nu}_{\sigma\mu}
\end{equation}
Now we want to define the covariant derivative that satisfies
\begin{equation}
D_{[\mu}e^a_{\nu]}=\partial_{[\mu}e^a_{\nu]}+\omega^a_{b[\mu}e^{b}_{\nu]}=0
\end{equation}
with that covariant derivative we write the Dirac equation as
\begin{equation}(i\gamma^{\mu}D_\mu+m)\psi=0\end{equation}
or explicitly
\begin{equation}
[i\gamma^{\mu}\left(\partial_\mu+\Gamma_\mu\right)+m]\psi=0
\end{equation}
where
\begin{equation}
\Gamma_\mu=\frac14\omega^{ab}_\mu[\gamma_{a},\gamma_b].
\end{equation}

\section{Grassmann fields}

The so-called Grassmann scalar field  has been very often considered in relativistic quantum information literature . This is a fermionic field
with no internal degrees of freedom that captures the essence of fermionic entanglement behaviour in general relativistic scenarios. This kind of field has been extremely useful to study the general features of entanglement in fermionic fields. This is because it is the simplest field that capture all the characteristics of the entanglement behaviour of fermionic field. However, as seen in this thesis, there is a universality principle that guarantees that all the results for Grassmann fields are exportable to Dirac fields which Grassmannian analogs.

However, such field is not free from problems when trying to trace it back to a Lorentz-covariant field theory. This is so because the only covariant equation of motion for scalar fields is the Klein-Gordon equation which, when quantised, results in bosonic statistics. This fact might result somehow confusing when one reflects about its use and physical meaning in relativistic quantum information settings.

There are, however, some physical scenarios in which Grassmann fields have complete physical meaning. Maybe the most obvious is considering quantisation of a Dirac field in one spatial dimension. The Lorentz group representation $SO(1,1)$ consists of only one boost and have no rotations. In this context  there are no internal degrees of freedom and the resulting theory is a Grassmann scalar field. This is particularly relevant for analog
gravity and quantum simulation implementations on experimental setups such as trapped ions \cite{SimulJuan}.

Indeed, a Dirac field in 1+1 dimensions  is naturally spinless, a possible gamma matrices representation in this case takes the form
\begin{equation}
\gamma^0=\sigma_x\qquad\gamma^1=-i\sigma_y
\end{equation}
where $\sigma_i$ are the usual Pauli matrices and the Dirac equation takes the usual form
\begin{equation}\label{Dirac3}
(i\gamma^{a}\partial_{a}+m)\psi=0,
\end{equation}
where now $\psi$ is a two component spinor, one for the positive and one for negative energy solutions branch. 

There are some other scenarios in which the use of Grassmann fields can be mathematically acceptable and a useful tool: we could regard the Grassmann scalar field as a Dirac field with a fixed spin-z component. Thomas precession prevents this setup from being Lorentz-covariant,
but if we choose the acceleration to be in the spin quantisation direction this phenomenon does not occur.

\cleardoublepage

\bibliographystyle{amsplain3}
\bibliography{references2}

\end{document}